\documentclass[%
 reprint,
superscriptaddress,
 amsmath,amssymb,
 aps,
floatfix,
longbibliography,
]{revtex4-2}

\usepackage{color}
\usepackage{inputenc}
\usepackage[dvipsnames]{xcolor}
\usepackage{cancel}
\usepackage{tikz}
\usepackage{graphicx} 
\usepackage{dcolumn}  
\usepackage{amsmath,amssymb}
\usepackage{epstopdf}
\usepackage{xparse}
\usepackage{physics}
\usepackage{appendix}
\usepackage{cases}
\usepackage{caption}
\usepackage{subcaption}
\usepackage{booktabs}
\usepackage{hyperref}
\hypersetup{
    colorlinks = true,
    citecolor = blue
}
\allowdisplaybreaks

\def\T{\tau}

\def\beal{\begin{align}}
\def\eal{\end{align}}
\def\kb{k_\textrm{B}}

\def\bea{\begin{eqnarray}}
\def\eea{\end{eqnarray}}
\def\ben{\begin{equation}}
\def\een{\end{equation}}
\def\benu{\begin{enumerate}}
\def\enu{\end{enumerate}}

\def\bei{\begin{itemize}}
\def\eei{\end{itemize}}
\def\beit{\begin{itemize}}
\def\eit{\end{itemize}}
\def\benu{\begin{enumerate}}
\def\enu{\end{enumerate}}

\def\n{n}

\def\sss{\scriptscriptstyle\rm}

\def\1var{(\bx_1...\bx\N)}

\def\br{{\bf r}}
\def\bR{{\bf R}}

\def\bx{{\bf x}}

\def\s{_{\sss S}}

\def\N{_{\sss N}}

\def\ext{_{\rm ext}}

\NewDocumentCommand{\clmb}{m m}{|#1-#2|}
\newcommand{\me}{m_\textrm{e}}
\newcommand{\Ne}{N_\textrm{e}}
\NewDocumentCommand{\NeI}{m}{N_\textrm{e}^{(#1)}}
\renewcommand{\vec}{\mathbf}
\newcommand{\RVS}{R_\textrm{VS}}
\newcommand{\free}{\Omega}
\newcommand{\Epot}{\expval{E_\textrm{pot}}}
\newcommand{\Ekin}{\expval{E_\textrm{kin}}}

\renewcommand{\S}{Section}
\renewcommand{\P}{Part}

\newcommand{\nb}{n^{\sigma}_\textrm{b}}
\newcommand{\nub}{n^{\sigma}_{\textrm{ub}}}
\newcommand{\Nb}{N^{\sigma}_{\textrm{b}}}
\newcommand{\Nub}{N^{\sigma}_{\textrm{ub}}}
\newcommand{\Nubb}{N_{\textrm{ub}}}
\newcommand{\nup}{n^\uparrow}
\newcommand{\ndown}{n^\downarrow}
\NewDocumentCommand{\isum}{ O{i} }{\sum_{#1=1}^{\Ne}}

\newcommand{\Hatom}{\hat{H}_\textrm{el}^\textrm{at}}

\newcommand{\Ylm}{Y_l^m(\theta,\phi)}
\NewDocumentCommand{\rnl}{O{} O{r} O{n} O{l}}{X_{#3 #4} ^{#1}(#2)}
\NewDocumentCommand{\pnl}{O{} O{r} O{n} O{l}}{P_{#3 #4} ^{#1}(#2)}
\NewDocumentCommand{\rnld}{O{r} O{n} O{l}}{X_{#2 #3} '(#1)}
\newcommand{\vs}{v_\textrm{s}^{\T,\sigma}}

\NewDocumentCommand{\wf}{O{i,\epsilon} O{\sigma}}{\psi_{#1}^{#2}}
\NewDocumentCommand{\epstilde}{O{nl} O{,\sigma}}{\bar{\epsilon}^{\T{#2}}_{#1}}
\NewDocumentCommand{\fnl}{O{n} O{l} O{\sigma}}{f_{#1 #2}^{\sigma}}
\newcommand{\Rws}{R_\textrm{VS}}
\newcommand{\Nn}{N_\textrm{n}}
\NewDocumentCommand{\yiJ}{O{i}}{\vec{y}_{#1J}}
\NewDocumentCommand{\yi}{O{i}}{\vec{y}_{#1}}
\newcommand{\RJ}{\hat{\vec{R}}_J}

\newcommand{\rr}{\mathbf{r}}

\newcommand{\EK}[1]{\textcolor{purple}{#1}} 
\newcommand{\change}[1]{\textcolor{black}{#1}}
\newcommand{\newchange}[1]{\textcolor{black}{#1}}

\begin{document}


\title{First-principles derivation and properties of density-functional average-atom models}

\author{T. J. Callow}
\email{t.callow@hzdr.de}
\affiliation{Center for Advanced Systems Understanding (CASUS), D--02826 G\"orlitz, Germany}
\affiliation{Helmholtz--Zentrum Dresden--Rossendorf, D--01328 Dresden, Germany}

\author{S. B. Hansen}
\affiliation{Sandia National Laboratories, Albuquerque, New Mexico 87185, USA}

\author{E. Kraisler}
\email[]{eli.kraisler@mail.huji.ac.il}
\affiliation{Fritz Haber Center for Molecular Dynamics and Institute of Chemistry, The Hebrew University of Jerusalem, 9091401 Jerusalem, Israel}

\author{A. Cangi}
\email{a.cangi@hzdr.de}
\affiliation{Center for Advanced Systems Understanding (CASUS), D--02826 G\"orlitz, Germany}
\affiliation{Helmholtz--Zentrum Dresden--Rossendorf, D--01328 Dresden, Germany}


\date{\today}

\begin{abstract}

Finite-temperature Kohn--Sham density-functional theory (KS-DFT) is a widely-used method in warm dense matter (WDM) simulations and diagnostics. Unfortunately, full KS-DFT-molecular dynamics models scale unfavourably with temperature and there remains uncertainty regarding the performance of existing approximate exchange-correlation (XC) functionals under WDM conditions. Of particular concern is the expected explicit dependence of the XC functional on temperature, which is absent from most approximations. Average-atom (AA) models, which significantly reduce the computational cost of KS-DFT calculations, have therefore become an integral part of WDM modelling. In this paper, we present a derivation of a first-principles AA model from the fully-interacting many-body Hamiltonian, carefully analysing the assumptions made and terms neglected in this reduction. We explore the impact of different choices within this model --- such as boundary conditions and XC functionals --- on common properties in WDM, for example equation-of-state data\change{, ionization degree and the behaviour of the frontier energy levels}.  
Furthermore, drawing upon insights from ground-state KS-DFT, we \change{discuss the} 
likely sources of error in KS-AA models and possible strategies for \change{mitigating} 
such errors.

\end{abstract}

\maketitle


\section{Introduction}\label{sec:intro}
Warm dense matter (WDM) is an energetic phase of matter exhibiting characteristics of solids, liquids, gases and plasmas \cite{GDRT14}. Thus, a better understanding of WDM can solve crucial problems at the intersection of several disciplines \cite{DOE09,MBRKA09}. The most important application of WDM research is the modelling and design of processes in inertial confinement fusion \cite{LABGGHKLS04,AM04,MS05,KDFMLWG11,HMGS11,BH16}; additionally, WDM simulations enhance our understanding of the earth's core \cite{AG98,NH04}; various astrophysical phenomena \cite{RDR06,F09} (including properties of exoplanets \cite{NFKR11,KKNF12}, giant gas planets \cite{KD09,LHR09,LHR11,KDBLCSMR15} and brown and white dwarfs \cite{HGLBSMF97,CBFS00}); and unexplored material properties such as novel chemistry \cite{TRB08,VN11}, non-equilibrium effects \cite{PHK06,EHH09}, phase transitions \cite{KSM07} and mechanical properties of solids \cite{GWGV10}. Furthermore, accurate theoretical modelling of WDM is important in processing and understanding data from large experimental facilities \cite{MBRKA09,SR09,SMCHMT10,SEJ14,GFGN16,TBG17}.


The theoretical description of WDM is particularly challenging: on the one hand, established plasma physics methods do not sufficiently account for quantum effects and strong coupling in WDM; on the other hand, the length, time and temperature scales of WDM often render popular approaches from condensed-matter physics computationally impractical. More formally, these difficulties can be understood in terms of several dimensionless parameters; in particular, the Coulomb coupling parameter $\Gamma_{i,e}$ and electron degeneracy parameter $\Theta_e$, which are defined as
\begin{gather}
    \Gamma_{i,e}  = \frac{\Epot}{\Ekin}\overset{\text{WDM}}{\sim}1; \\
    \label{eq:Theta_e}
    \Theta_{e} = \frac{\kb\T}{E_\textrm{F}}\overset{\text{WDM}}{\sim}1,
\end{gather}
where $\Ekin$ and $\Epot$ are respectively the average kinetic and potential energies; $\T$ the temperature; $E_\textrm{F}$ the Fermi energy; and the subscripts $i$ and $e$ respectively refer to nuclei and electrons. These parameters are of order unity in the WDM phase, $\Gamma,\Theta\sim 1$, which corresponds to a phase of matter lying somewhere between a classically ionized plasma and a strongly-correlated condensed-matter system. 

\begin{figure}
\centering
\begin{tikzpicture}

    \node[anchor=south west,inner sep=0] (image) at (0,0) {\includegraphics{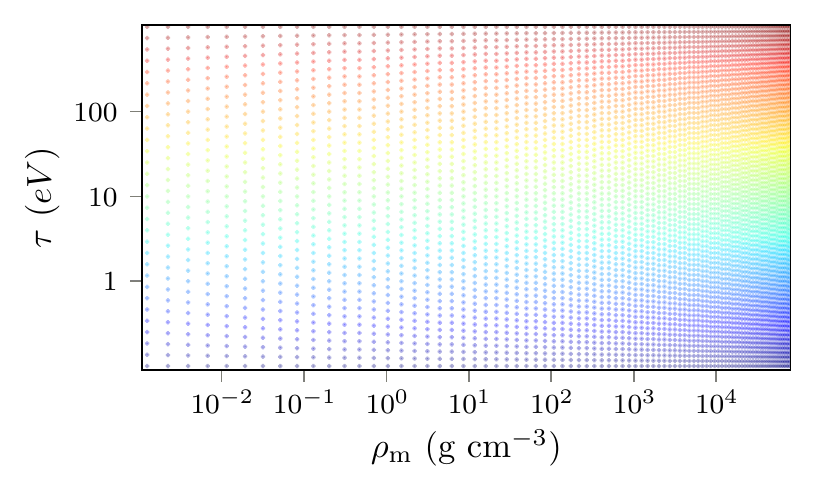}};
    \begin{scope}[x={(image.south east)},y={(image.north west)}]
    \draw[rotate around={20:(0.59,0.61)},line width=.5pt](0.59,0.61) ellipse (80pt and 30pt);
    \draw[line width=1pt, dashed, rounded corners, Sepia](0.23,0.25) rectangle (0.6,0.68) {};
    \draw[line width=1pt, dotted,fill=white](0.45,0.72) circle (21pt) {};
    \draw[line width=1pt,dash dot] (0.56,0.49) ellipse (17 pt and 22 pt);  \draw[line width=1pt,fill=brown] (0.6,0.35) ellipse (6 pt and 10 pt);
    \draw[line width=0.5pt,fill=white,rounded corners](0.43,0.26) rectangle (0.57,0.33) {};
    \node[text width=0.2, align=left] at (0.39,0.72) {White dwarfs$^*$};
    \node[] at (0.7,0.72) {\Large{WDM}};
    \node[] at (0.56,0.56) {Exo-};
    \node[] at (0.565,0.48) {planets};
    \node[] at (0.38,0.46) {\textcolor{Sepia}{Our model}};
    \node[] at (0.5,0.295) {CMP$^\dagger$};
    \node[] at (0.33,0.9) {Classical plasmas};
    \node[text width=0.2] at (0.64,0.35) {Planetary cores};
    \end{scope}
    \end{tikzpicture}
    \caption{A rough schematic of the temperature and density space spanned by the WDM regime, including some typical phenomena and the region we shall target \change{(dashed rectangle)}. 
    Data from Refs.~\onlinecite{DOE09,F09,DGB18,BDMZHKFRV20}. \\
    \footnotesize{$^*$Envelope (atmosphere) of white dwarfs\\ $^\dagger$ Condensed-matter physics}}
    \label{fig:wdm_params}
\end{figure}

At typical WDM temperatures and densities, the electron coupling parameter, $\Gamma_e$, is approximately equal to the density parameter $r_s$ \footnote{known also as the Wigner--Seitz radius or Brueckner parameter}, defined as \cite{BDMZHKFRV20}
\begin{equation}
    \label{eq:gamma_e}
    \Gamma_{e}\approx r_s = \left(\frac{3}{4\pi\n_e}\right)^{1/3},
\end{equation}
where $n_e$ is the number density of free electrons. We note that the Fermi energy for a non-interacting system of electrons can also be expressed using the above definition of $r_s$ \footnote{We note there are sometimes different conventions for the definitions of $E_\textrm{F}$ and $r_s$ (see for example Refs.~ \cite{BDMZHKFRV20} and \cite{DGB18}), but where we refer to these parameters we use the above definitions and adopt the convention of Ref.~\cite{NFP99} in which the free electron density is equated with the valence electron density.},
\begin{equation} \label{eq:E_fermi}
    E_\textrm{F}=\frac{\hbar^2}{2\me}(3\pi^2 n_e)^{2/3}= \frac{\hbar^2}{2\me} \left(\frac{9\pi}{4}\right)^{2/3} \frac{1}{r_s^2}.
\end{equation}
From the above relationship, and because $\Ekin\approx E_\textrm{F}$ at low temperatures, the approximation $\Gamma_\textrm{e}\approx r_s$ \eqref{eq:gamma_e} is valid in WDM (for $r_s\sim 1$). Finally, the classical Coulomb coupling parameter of nuclei is defined as
\begin{equation}\label{eq:Gamma_i}
\Gamma_i = \frac{(Ze)^2}{a_i\kb\T},
\end{equation}
with $Ze$ the nuclear charge, and $a_i$ the mean inter-ionic distance. In Fig.~\ref{fig:wdm_params}, we show a rough schematic of some typical phenomena in the WDM regime, and the approximate region of WDM phase space which we later explore with our model.

Notwithstanding the difficulties mentioned earlier, applying Kohn-Sham density functional theory (KS-DFT) \cite{HK64,KS65} has recently led to promising results in the simulation of WDM \cite{D03,HRD08}. KS-DFT is a well-established, successful method \cite{DG90,PY94,FNM03,HRPYCP11,B12,HST16} for predicting the electronic structure of materials, \change{from single atoms and small molecules to nanoparticles, periodic solids and large biomolecules. Within KS-DFT the interacting many-body problem is tackled by mapping it onto a fictitious, non-interacting system \cite{KS65} which yields the same electronic density as the interacting problem. Usually, the systems are treated} at zero temperature. 
The formal generalization of KS-DFT to finite temperature was established by Mermin \cite{M65}.  In WDM simulations, KS-DFT is used to calculate forces on nuclei, which are then time-evolved through molecular dynamics techniques \cite{MSDML04,HRD08,HFR11}.  The primary target is to calculate the equation of state (EOS) which relates the atomic density, energy, temperature and pressure of a material.  The EOS data can be used, for example, to compute the Hugoniot curve \cite{HRD08}, which describes the possible final state from a given initial state after a shock wave, relevant to many WDM experiments.  Furthermore, the electrical and thermal conductivities of WDM are calculated via Onsager coefficients \cite{HFR11} from KS orbitals and eigenvalues.


In principle, the KS method is an exact approach, but in practice, the exchange-correlation (XC) energy functional must be
approximated. There is no systematic approach for the development of XC functionals, and thus a plethora of zero-temperature approximations exist; see, for example, Refs.~\onlinecite{P99,PS01,CMY12,MBSPL17,T21} 
for discussions on this subject. In the extension to finite temperatures, functional construction is further complicated by the fact that, in principle, the XC functional should depend \emph{explicitly} on the temperature \cite{KV83}. However, the nature of this explicit temperature-dependence remains unclear and, in fact, is usually neglected in standard calculations \cite{KSDT14,SJB16}.
Hence, the temperature dependence is only crudely included through the implicit
temperature-dependence in the density, as the KS orbitals are occupied according to Fermi--Dirac statistics.  The theoretical development of temperature-dependent XC approximations has recently found new momentum \cite{PPFS11,DT11,PPGB14,DT16,JB16,SJB16,BSGJ16,SSB18,SB20}. Investigations of the electron liquid \cite{PD00} and uniform electron gas \cite{GR80,DT81,LM83,SD13} provide insights into constructing local \cite{BCDC13,KSDT14,DGSMFB16,GDSMFB17} and generalized gradient approximations \cite{SD14,KDT18} to the temperature-dependent XC contribution.  An exact inclusion of the exchange energy at finite temperature \cite{LMW06,GCG10} has also been achieved using the optimized-effective-potential method, however, with the drawback of higher computational cost than that of the standard KS method. These are constructive steps towards developing accurate, reliable and computationally affordable temperature-dependent XC functionals.

The aforementioned computational cost is a further challenge for finite-temperature KS-DFT, because a large number of KS states must be accounted for under WDM conditions \cite{KSCD14}. Despite this limitation, and the above-cited difficulties with developing suitable XC approximations, KS-DFT is currently the predominant first-principles method for simulations of large systems in this thermodynamic regime; alternative approaches such as orbital-free DFT (OF-DFT) \cite{LC05,kst14,WRCPHG13} or path-integral Monte--Carlo (PIMC) \cite{MC00,FBEF01,M09,DM12,DGB18} tend either to not be sufficiently accurate (OF-DFT \cite{BLC05,ZKWCZ16}) or too expensive (PIMC, especially at lower temperatures). 

Consequently, the development of methods which obtain close to KS-DFT accuracy at reduced computational cost is an active area of research. Recently, there have been some promising developments in this area, such as the development of surrogate models using machine learning \cite{ECMSTR20}, stochastic methods \cite{CRNB18,CRNPRB20,WC20}, and approaches to reduce the cost of core-electron calculations \cite{MZ08,ZZKZH16}.

In this paper, we consider an alternative approach known as an average-atom (AA) model: the premise of such a model is that the full system of interacting electrons and nuclei is partitioned into a set of Voronoi spheres, each containing a central nucleus, and the full electronic calculation is reduced (under certain approximations which we discuss later) to a calculation for a single atom. This concept, which has clear computational advantages, has a long history in plasma physics and electronic structure theory. The earliest AA models \cite{WS33,WS34,SK35,FMT49,La55} were based on the Thomas--Fermi (TF) approximation \cite{T27,F27} and modifications thereof; subsequent models built on this premise by adopting a mixed KS-DFT and TF approach for the bound and continuum electrons respectively \cite{R72}, then treating the full spectrum (discrete and continuous) via KS-DFT \cite{Li_1979}, and later incorporating effects from outside the central atom such as ionic correlations \cite{DP82,C85,PFD90,R91}. AA models continue to be extensively developed and used under a variety of different approaches and assumptions \change{ \cite{Y02,BF04,JGB06,SHWI07,WSSI06,BC07,FBCR10,SGSKR08,PB11,JNC12,SS12,SS13,MWHD13,STJZS14,SSADC_14,SS_14,SGSG19,DKR_20,MBVSM21}; we also mention here the recent works in Refs.~\onlinecite{BSZS20,SS20} which attempt to bridge the gap between AA models and full KS-DFT via novel approaches.}


AA models are a well established and successful tool in the plasma physics community as they produce results of useful accuracy at a fraction of the cost of full KS-DFT simulations. However, their evolution from the initial TF based models has been largely driven by organic developments, and there have been relatively few attempts to derive an AA model starting from a fully quantum-mechanical perspective (with some notable exceptions, for example \change{ Refs.~\onlinecite{C85,PFD90,BF04,BC07,SS12}}). 
As a result, AA models do not always follow the same conventions: for example, AA models which differ in their choice of boundary condition for solving the KS equations have been observed to yield quite different results \cite{JNC12}. By contrast, solid-state KS-DFT codes have differences in their numerical implementation, but generally follow a common set of theoretical assumptions and thus are expected to yield the same results for the same set of inputs (atomic configuration, temperature, etc) \cite{Lejaeghereaad3000}. Establishing a similar framework and hierarchy for AA models would improve understanding of the limits under which they might be expected to give reasonable results in the WDM regime, in particular for $\T\lesssim E_F/\kb$.

This motivates the main results of this paper, namely \change{a systematic, }
first-principles derivation of an AA model \change{starting from the many-body Hamiltonian of coupled electrons and nuclei}, and the comparison of some fundamental results when the model is solved using KS-DFT. The paper is structured as follows: in \S~\ref{sec:theory}, we start with the full many-body Hamiltonian of electrons and nuclei and reduce it to an effective atomic Hamiltonian for a classical nucleus and surrounding electron density. \change{We mention explicitly all the approximations that have to be done during the derivation and discuss their possible impact.} We then introduce finite-temperature KS theory and apply it to minimize the grand free potential for this AA model, which requires particular consideration of the boundary conditions and treatment of unbound (continuum) electrons. Following a discussion on numerical implementation in \S~\ref{sec:numerics}, we compute some common properties for a range of temperatures, densities, boundary conditions and XC functionals in \S~\ref{sec:results}. Finally, we discuss the implications of our results for AA models, and finite-temperature KS-DFT more generally, in \S~\ref{sec:discussion}.

\section{Theory} \label{sec:theory}
In this section, we first reduce the many-body Hamiltonian of interacting electrons and nuclei to an effective single-atom Hamiltonian (\P~\ref{subsec:theory_pt_1}), analysing the assumptions and approximations used in this reduction. 
Then, in \P~\ref{subsec:theory_pt_2}, we briefly review the grand canonical ensemble which can be used to describe a quantum system of electrons in thermal equilibrium with a reservoir (the nuclei); and in \P~\ref{subsec:theory_pt_3} we explain how the problem of minimizing the grand free energy by solving for all the interacting states is greatly simplified by finite-temperature KS-DFT. Next, in \P~\ref{subsec:theory_pt_4}, we apply KS-DFT to the reduced average-atom Hamiltonian and deduce suitable boundary conditions for the electron density within this model. Finally, in \P~\ref{subsec:theory_pt_5}, we discuss our treatment of the unbound electrons and how the boundary condition on the density can be realised through a number of different boundary conditions for the bound KS orbitals, from which we consider two simple choices.

\subsection{\change{Many-body Hamiltonian of coupled electrons and nuclei in the dilute gas limit}}\label{subsec:theory_pt_1}

We begin with the full many-body Hamiltonian of interacting electrons and nuclei,
\begin{gather}
\label{eq:H_tot}
\hat{H} = \hat{H}_{\sss nuc} + \hat{H}_{\sss el} + \hat{H}_{\sss el,nuc}\,.
\end{gather}
The individual components of the Hamiltonian are defined as
\begin{align}
\hat{H}_{\sss nuc}
=& \sum_{I=1}^{\Nn} \left( -\frac{\grad_I^2}{2M} \right) + \frac{1}{2} \sum_{I=1}^{\Nn} \sum_{\substack{J=1\\J\neq I}}^{\Nn} \frac{Z^2}{|\bR_I-\bR_J|}\,,\\
\hat{H}_{\sss el}
=& \sum_{i=1}^{\Nn \cdot N_e} \left( -\frac{\grad_i^2}{2} \right)
+ \frac{1}{2} \sum_{i=1}^{\Nn \cdot N_e} \sum_{\substack{j=1\\j\neq i}}^{\Nn \cdot N_e} \frac{1}{|\br_i-\br_j|}\,,\\
\label{eq:H_el_nuc}
\hat{H}_{\sss el,nuc}
=& -\sum_{I=1}^{\Nn}\sum_{i=1}^{\Nn \cdot N_e}\frac{Z}{|\br_i-\bR_I|}\,.
\end{align}


In the above, we have assumed that the system is composed of a single element with nuclear  \change{mass $M$, charge $Z$ and electron number $N_\textrm{e}$.  $\Nn$ denotes the number of nuclei in the system,thus} $\Nn\cdot N_\textrm{e}$ \change{is} the total number of electrons. 
$\br_i$ and $\bR_I$ \change{are} the positions of the $i^{\textrm{th}}$ electron and the $I^{\textrm{th}}$ nucleus respectively. We have also assumed that there is no external field applied to the system. Note that here and below we adopt Hartree atomic units, $\hbar=e=\me=a_0=1$.


We work within the Born--Oppenheimer (BO) approximation \cite{BO27}, which assumes the electrons react instantaneously to any changes in the positions of the nuclei due to their relatively small masses, $\me\ll M$. In fact, though the BO approximation is used extensively in WDM simulations 
\change{with} KS-DFT \cite{GDRT14}, it is likely to be prone to inaccuracies in the WDM regime, due to strong non-adiabatic effects from excited states, core electron chemistry, and so forth \cite{AMG10,LGRMWG19}. While we persist with the BO approximation in our derivation, we 
discuss some possibilities for incorporating non-adiabatic couplings between electrons and nuclei in \S~\ref{sec:discussion}.

Having fixed the nuclear co-ordinates in the Hamiltonian with the BO approximation, we transform the vectors $\br_i$ (which act on the electron wavefunction) as $\br_i = \bR_I + \vec{x}_{iI}$, where $\vec{x}_{iI}$ is determined by performing a Voronoi decomposition of space. In other words, each vector $\br_i$ is now defined relative to the closest nuclear co-ordinate $\vec{R}_I$. \newchange{We wish to make clear that this is simply a relabelling of the terms in the Hamiltonian and the notion of a ``closest nucleus'' does not imply any assumptions regarding the electron density distribution.} Following this transformation, and ignoring the nuclear kinetic energy to the BO approximation, the Hamiltonian can be re-written as
\begin{widetext}
\begin{align}
    \hat{H} =&\sum_{I=1}^{\Nn} \Bigg[ \frac{1}{2} \sum_{J\neq I}^{\Nn} \frac{Z^2}{\clmb{\bR_I}{\bR_J}} 
     \label{eq:Ham_1}
    + \sum_{i=1}^{\NeI{I}} \Bigg (-\frac{ \nabla^2_{iI}}{2} 
     + \sum_{J=1}^{\Nn} \Bigg\{   -\frac{Z}{|\bR_J-\bR_I-\bx_{iI}|} +  \frac{1}{2}\sum_{\substack{j=1\\(jJ) \neq (iI)}}^{\NeI{J}}\frac{1}{|\bR_J-\bR_I+\bx_{jJ}-\bx_{iI}|} \Bigg\}\Bigg)\Bigg]\,,
\end{align}
where $\NeI{I}$ denotes the number of \change{vectors $\vec{x}_{iI}$ closest to} the $I^\textrm{th}$ nucleus. Next, we decompose the electron-nuclear and electron-electron interactions in the above expression into those parts which contain interactions between charges in a single Voronoi cell, and those involving inter-cell interactions. The Hamiltonian \eqref{eq:Ham_1} thus becomes
\begin{align}
\nonumber
        \hat{H} =\sum_{I=1}^{\Nn} &\Bigg\{  \sum_{i=1}^{\NeI{I}}\Bigg[-\frac{ \nabla^2_{iI}}{2} - \frac{Z}{|\bx_{iI}|} + \frac{1}{2} \sum_{\substack{j=1\\j\neq i}}^{\NeI{I}} \frac{1}{|\bx_{iI}-\bx_{jI}|} \Bigg] \\
        &+ \sum_{J\neq I}^{\Nn} \Bigg[\frac{1}{2}\frac{Z^2}{|\bR_I-\bR_J|} +\sum_{i=1}^{\NeI{I}} \Bigg( -\frac{Z}{|\bR_J - \bR_I - \bx_{iI} |} + \frac{1}{2} \sum_{j=1}^{\NeI{J}} \frac{1}{|\bR_J-\bR_I +\bx_{jJ}-\bx_{iI}|} \Bigg) \Bigg ]\Bigg\}\,.
     \label{eq:Ham_2}
\end{align}


Next, we consider the decomposition of the Hamiltonian into two parts: an ``average'' term $\hat{H}_\textrm{av}$, and an ``inhomogeneous'' term $\hat{H}_\textrm{in}$, i.e. $\hat{H}=\hat{H}_\textrm{av}+\hat{H}_\textrm{in}$. The average component $\hat{H}_\textrm{av}$ is constructed by considering the expectation value $\mel{\Psi}{\hat{H}}{\Psi}$ (\newchange{where $\Psi$ is an anti-symmetric wave-function}), for the particular case in which the nuclei are distributed exactly evenly in space. In this case, the electron density is identical in each of the Voronoi partitions, and therefore the expectation value $\mel{\Psi}{\hat{H}}{\Psi}$ is equal to the expectation value of the average Hamiltonian $\mel{\Psi}{\hat{H}_\textrm{av}}{\Psi}$. \newchange{We then make two further assumptions regarding the average Hamiltonian. Firstly, we assume that each nucleus is associated with the same number of vectors $\vec{r}_i$ following the Voronoi decomposition, which means that $\Ne^{(I)}=\Ne$. Secondly, we assume that the vectors $\vec{x}_{iI}$ are independent of any nuclear positions; in other words, $\vec{x}_{iI}$ can be written simply as $\vec{x}_i$.} By further transformation of the co-ordinate system such that the $I^\textrm{th}$ nucleus lies at the origin, $\bR_I=0$, the average Hamiltonian becomes
\begin{align}
\nonumber
   \Hat{H}_\textrm{av} = \Nn \times  &\Bigg \{ \sum_{i=1}^{\Ne} \Bigg[ -\frac{\nabla_i^2}{2} -\frac{Z}{|\vec{x}_i|} + \frac{1}{2} \sum_{\substack{j=1\\j\neq i}}^{\Ne} \frac{1}{|\bx_{i}-\bx_{j}|}\Bigg]\\
   \label{eq:Ham_3}
   &+\sum_{J=2}^{\Nn}\left[\frac{Z^2}{2|\vec{R}_J|} - \sum_{i=1}^{\Ne} \left( \frac{Z}{\clmb{\bR_J}{\bx_i}} - \frac{1}{2} \sum_{j=1}^{\Ne} 
     \frac{1}{|\bR_J+\bx_{j}-\bx_i|}\right)\right]\Bigg\}\,,
\end{align}

\end{widetext}

The inhomogeneous term, $\hat{H}_\textrm{in}$ is the difference between the full Hamiltonian $\hat{H}$ of Eq.~(\ref{eq:Ham_2}) and $\hat{H}_\textrm{av}$ of Eq.~(\ref{eq:Ham_3}). In our derivation, we choose to neglect it. This is a reasonable assumption if the nuclear distribution is relatively uniform (for the fictitious system considered in the previous paragraph in which the nuclear distribution is exactly uniform, $\mel{\Psi}{\hat{H}_\textrm{in}}{\Psi}=0$). The contribution of $\hat{H}_\textrm{in}$ for non-uniform systems can in principle be taken into account via perturbation theory. \newchange{As a conceptual remark, we note that the average Hamiltonian \eqref{eq:Ham_3} is akin to the Hamiltonian in a typical KS-DFT simulation for a periodic system, in which some electrons in a unit cell (here containing a single nucleus) interact between themselves and their periodically repeating images.}


We shall now split the re-formulated Hamiltonian \eqref{eq:Ham_3} into two parts as follows,
\begin{align}
\label{eq:Ham_4}
    \hat{H}_\textrm{av}&=\Nn \times \Bigg\{ \hat{H}_\textrm{el}^\textrm{at} +  \frac{Z^2}{2}\sum_{J=2}^{\Nn} \frac{\hat{W}_J}{|\vec{R}_J|}\Bigg\},\ \textrm{with}\\
    \hat{H}_\textrm{el}^\textrm{at} &= \sum_{i=1}^{\Ne} \Bigg[ -\frac{\nabla_i^2}{2}
    \label{eq:H_el_at}
   -\frac{Z}{|\vec{x}_i|} 
   + \frac{1}{2}\sum_{\substack{j=1\\j\neq i}}^{\Ne} \frac{1}{|\bx_i-\bx_j|}\Bigg],\\
    \label{eq:W_J}
    \hat{W}_J &=  1 - \frac{1}{Z^2}\sum_{i=1}^{\Ne}\Bigg[ \frac{2Z}{|\hat{\bR}_J-\vec{y}_{iJ}|}\nonumber\\
    &\hspace{8em}- \sum_{j=1}^{\Ne}\frac{1}{|\hat{\bR}_J+\vec{y}_{jJ}-\vec{y}_{iJ}|}\Bigg],
\end{align}
where $\vec{y}_{iJ}=\bx_i/|\bR_J|$ and $\hat{\bR}_J=\bR_J/|\bR_J|$. 

The first component $\hat{H}_\textrm{el}^\textrm{at}$ \eqref{eq:H_el_at} is equivalent to the electronic Hamiltonian of a single atom and we shall return to it later in the paper. 
\change{We first treat} the terms $\hat{W}_J$ \eqref{eq:W_J} which make up the second component of the Hamiltonian~\eqref{eq:Ham_4}. We shall expand perturbatively in powers of $|\yiJ|$, since by the Voronoi decomposition of space, the inequality $|\yiJ|\leq \frac{1}{2}$ holds strictly, 
\change{because} the distance between an electron and its nearest nucleus cannot exceed half the distance between two nuclei. 
Moreover, for electrons located in cells far from the central cell, and for electrons tightly bound to the central nucleus, $|\yiJ|$ is much lower than 1, which further justifies a power expansion.

We re-write $\hat{W}_J$ in the form
\begin{multline}
    \hat{W}_J=1 - \frac{1}{Z^2}\sum_{i=1}^{\Ne}\Bigg[ \frac{2Z}{\sqrt{1-2\hat{\bR}_J\cdot\yiJ+|\yiJ|^2}} \\
    -\sum_{j=1}^{\Ne} \frac{1}{\sqrt{1-2\hat{\bR}_J\cdot(\yiJ-\yiJ[j])+(\yiJ-\yiJ[j])^2}}\Bigg]
\end{multline}
\change{and} expand both terms in the above expression using the binomial expansion for $1/\sqrt{1+\epsilon}$. 
\change{A full derivation can be found in Appendix~\ref{App:WJ}.} The zeroth-order term in this expansion is equal to
\begin{align} \label{eq:W0}
    \hat{W}_J^{(0)}&=\left(\frac{Z-\Ne}{Z}\right)^2 \overset{Z=\Ne}{=}0,
\end{align}
where the final equality holds for systems with neutral charge, $\Ne=Z$. 
We now consider the first and second order terms in the expansion of $\hat{W}_J$, which turn out to be generally non-vanishing and equal to
\begin{align}
    \hat{W}_J^{(1)}&=-\frac{2}{Z}\hat{\vec{R}}_J \cdot \vec{Y}_J; \\
    \nonumber
    \hat{W}_J^{(2)}&=\frac{Z-\Ne}{Z^2}\isum\left[|\yiJ|^2-3(\hat{\vec{R}}_J \cdot\yiJ)^2 \right] \\
    &\hspace{3em}+\frac{1}{Z^2} \left[|\vec{Y}_J|^2 - 3(\hat{\vec{R}}_J\cdot\vec{Y}_J)^2\right] \\
    \hat{W}_J^{(2)}& \overset{Z=\Ne}{=}\frac{1}{Z^2}\left[|\vec{Y}_J|^2 - 3(\hat{\vec{R}}_J\cdot\vec{Y}_J)^2\right]\,,
\end{align}
where we introduced the notation
\begin{equation}
    \vec{Y}_J=\isum \yiJ\,.
\end{equation}
In principle, one could continue this perturbative expansion to include even higher order terms. However, in our AA model, we neglect all the coupling terms \change{$\hat{W}_J^{(k)}$, with $k \geqslant 1$,} leaving only an atomic Hamiltonian. Considering the interaction terms we neglect in our model already provides insight regarding the limits under which we might expect it to be accurate. Of course, we would expect high accuracy in the limit of a dilute gas ($|\yiJ|\ll1$). Moreover, for an approximately uniform nuclear distribution, which implies a highly symmetric electronic distribution, we would expect the expectation value \change{$\left< \vec{Y}_J \right>$} 
to be close to zero. This suggests the model is likely to even be accurate in the high-density limit, in which the Voronoi cells and their enclosed electronic distribution will be spherically symmetric to a large extent. We can also expect accurate results at high temperatures, when there is significant ionization and the electron kinetic energy dominates over interaction effects. In addition, the detailed derivation performed here shows how the AA model can be made more accurate in the future, by including higher-order terms of $\hat{W}_J$ \change{and perturbatively treating the inhomogeneous term, $\hat{H}_\textrm{in}$}.

We have therefore reduced the full many-body Hamiltonian of electrons and nuclei, defined by equations (\ref{eq:H_tot}--\ref{eq:H_el_nuc}), to $\Nn\times\Hatom$, where $\Hatom$ is the Hamiltonian for a system of electrons interacting with a single fixed nucleus. We note that, although we ignore inter-cell interaction terms in the Hamiltonian, these interactions will be partially accounted for by the choice of boundary conditions; this is what distinguishes this model from a single isolated atom. In the literature, such a reduction is commonly referred to as an AA model: in neglecting completely the $\hat{W}_J$ terms, our model falls within a class of AA models known as ion-sphere models \cite{R72,JGB06,SGSKR08,JNC12}; other so-called ion-correlation models \cite{R91,MWHD13} attempt to model to some extent these coupling contributions via the introduction of a uniform background field \cite{Li_1979,DP82,STJZS14,SSADC_14}. Under the umbrella of ion-sphere models there exists an abundance of models, differing for example in how the atomic Schr\"odinger equation is solved for the bound and unbound electrons, choice of boundary conditions and more besides; we shall draw comparisons with appropriate examples of such models later in this paper.


\subsection{Finite-temperature quantum systems}\label{subsec:theory_pt_2}

We are interested in applying the above model to systems at finite temperatures; such systems are described by a statistical ensemble of states. In this sub-section, we review some basic theory regarding quantum statistical ensembles.

We restrict our analysis to systems at some fixed temperature in thermal equilibrium with a reservoir, which (like the nuclei) is treated classically. In such an ensemble, known as the grand canonical ensemble, the grand canonical Hamiltonian $\hat{\Omega}$ and associated grand canonical potential (or grand free energy) $\free$ play the role of the Hamiltonian and energy in a zero-temperature calculation,
\begin{align}
\label{eq:gr_can_1}
    \hat{\Omega}&=\hat{H} - \T\hat{S} - \mu\hat{N};\\
    \label{eq:gr_can}   
    \free &= \Tr [\hat{\Omega}\hat\Gamma] = E - \T S -\mu N,
\end{align}
where $\hat{S}$ and $\hat{N}$ are the entropy and number particle operators; $\hat{\Gamma}$ is a statistical operator which can be used to determine average observable values of operators; and $\mu$ is the chemical potential, defined as the change in energy when a particle is added or removed from the system {\footnote{If different species were admitted in the Hamiltonian, or a magnetic field present, then the term $\mu\hat{N}$ would be denoted $\sum_s\mu_s\hat{N}_s$, where $s$ denotes the species (or electron spin). Since we only consider the electronic Hamiltonian in the absence of a magnetic field in our model, we use the simplified form of Eq.~\eqref{eq:gr_can_1}}}. $\hat{S}$ and $\hat{\Gamma}$ are defined by
\begin{align}
    \hat{S} &= -\kb \ln(\hat\Gamma),\ \textrm{with} \\
    \hat\Gamma & = \sum_{k,N} w_{k,N} \ket{\Psi_{k,N}}\bra{\Psi_{k,N}},
\end{align}
where $\ket{\Psi_{k,N}}$ are orthonormal eigenstates of the operator $\hat{\Gamma}$ 
, where $k$ is the principal quantum number and $N$ is the number of electrons in the system. $w_{k,N}$ are statistical weights, which satisfy $\sum_{k,N}w_{k,N}=1$. 

One example for the use of the operator $\hat{\Gamma}$ is for finding the ensemble density:
\begin{align} \label{eq:n_ens}
    n(\rr) = \Tr \left[ \hat\Gamma \hat{n}(\rr) \right] = \sum_{k,N} w_{k,N} n_{k,N}(\rr),
\end{align}
where
$\hat{n}(\rr) = \sum_{i=1}^N \delta(\rr-\rr_i)$ is the density operator and $n_{k,N} (\rr)= \bra{\Psi_{k,N}} \hat{n}(\rr) \ket{\Psi_{k,N}}$ is the density of the $(k,N)^\textrm{th}$ state.

When the system of electrons and nuclei is in thermal equilibrium with the reservoir, the grand canonical potential is minimized with respect to $\hat{\Gamma}$. In this case the equilibrium weights $w_{k,N}$ are given by
\begin{align}
    w_{k,N} = \frac{1}{Z}e^{-\beta (E_{k,N}-\mu N)},
\end{align}
with \change{$E_{k,N} = \bra{\Psi_{k,N}} \hat{H} \ket{\Psi_{k,N}}$} the equilibrium eigenvalues of the states $\ket{\Psi_{k,N}}$, $\beta=1/\kb \T$, and $Z$ is the partition function,
\begin{equation}
    Z = \sum_{k,N} e^{-\beta (E_{k,N}-\mu N)}.
\end{equation}
We also note that the grand canonical potential is often expressed in terms of the partition function,
\begin{equation}
    \free = -\kb\ln(Z).
\end{equation}

\subsection{Finite-temperature KS-DFT}\label{subsec:theory_pt_3}

Whilst there are many first-principles techniques to determine ground-state electronic properties, the majority of these are not computationally feasible at finite temperatures, even for atomic systems. We shall use KS-DFT as it has the best balance between accuracy and speed, particularly in the low-temperature part of the WDM regime. In its original formulation \cite{HK64}, DFT establishes a one-to-one mapping between the ground-state electronic density and external potential $v_\textrm{ext}(\br)$; this means the ground-state density is (in principle) sufficient to compute all observables. Mermin \cite{M65} extended the DFT formalism to ensembles at finite temperatures; in this case, there is a mapping between the equilibrium ensemble density $n_0(\br)$ and the external potential minus the chemical potential, $v_\textrm{ext}(\br)-\mu$, and hence $n_0(\br)$ can be used to compute observable averages. The grand canonical potential $\free=\free[n]$ is thus a functional of the density \cite{PPFS11},
\begin{align}
    \free[n] &= F^{\T}[n] + \int \dd{\vec{r}} n(\br) [v\ext (\br) - \mu],\ \textrm{with} \\
    F^{\T}[n] &= T^{\T}[n] + V^{\T}_\textrm{ee}[n] - \T S^{\T}[n],
\end{align}
where $T^{\T}[n]$ denotes the electronic kinetic energy at temperature $\T$, $ V^{\T}_\textrm{ee}[n]$ the electron-electron interaction energy, and $v\ext(\br)$ the external potential (which in our case is just the electron-nuclear attractive field). $F^{\T}[n]$ is denoted the universal functional because it has no dependence on the external potential; minimizing $\free[n]$ with respect to the density yields the equilibrium free energy.

Like in the ground-state case, it is convenient to introduce an auxiliary system of non-interacting electrons which has the same density and temperature as the fully-interacting system. This is known as the KS system \cite{KS65,KV83}. The definition of the grand free energy within KS theory is given by
\begin{align}
\label{eq:KS_free}
\nonumber
    \free [n] = T_s^ \T [n] &- \T S_s^\T [n] + U [n] + F_\textrm{xc}^\T [n] \\
    &+\int \dd{\vec{r}} n(\br) [v\ext (\br) - \mu],
\end{align}
where $v_\textrm{ext}(\br)$ is the external potential experienced by the electrons (in our model, the electron-nuclear potential); and $ F_\textrm{xc}^\T$ is the exchange-correlation (XC) free energy functional, which is defined as
\begin{equation}
    F_\textrm{xc}^\T [n] = F^\T [n] - \left(T_s^\T [n] - \T S_s^\T [n]\right) - U[n],
\end{equation}
and $U[n]$ is the usual Hartree energy, given by
\begin{equation} \label{eq:Hartree}
    U[\n]=\frac{1}{2}\int\int\dd{\vec{r}} \dd{\vec{r'}} \frac{\n(\vec{r})\n(\vec{r'})}{|\vec{r}-\vec{r'}|}.
\end{equation}

The exact XC free energy functional contains all the information about electron-electron interactions. The KS orbitals, and hence the density, are obtained by solving the finite-temperature KS equations,
\begin{gather}
\label{eq:thermal_KS}
    \left[-\frac{\grad^2}{2} +\vs (\br) \right] \wf(\br) = \epsilon_{(i)\sigma}^{\T} \wf(\br),\ \textrm{with} \\
    \vs(\vec{r}) = v\ext (\br) + v_\textrm{H}(\br) + \fdv{F_\textrm{xc}^\T [\nup,\ndown]}{n^{\sigma}(\br)}, \label{eq:vs}
\end{gather}
where $v_\textrm{H}(\vec{r})$ is the Hartree potential, given by the functional derivative of the Hartree energy \eqref{eq:Hartree} with respect to the density. In the above, we have moved from a pure density-functional formalism to a spin-density-functional formalism \cite{BF72}, which allows for a formal treatment of magnetic systems, and often yields more accurate results even in the absence of an external magnetic field \cite{Parr_Yang}. The variable $\sigma=\uparrow,\downarrow$ represents the spin-channel. Strictly speaking, we note that the KS potential, and thus by extension the KS eigenvalues, should depend on the chemical potential due to the aforementioned mapping between the density and the external potential minus the chemical potential. However, since this makes no practical difference in the static case aside from a shifting of the orbital energies, we adopt the more common convention seen in the equations above in order to better connect with existing AA and KS-DFT literature. However, if one wanted to extend the model to compute linear response quantities (for example), the formally correct definition for the KS potential and orbital energies should be used.

The notation $\wf(\br)$ denotes respectively bound orbitals $\wf[i](\br)$, with discrete energy levels $\epsilon^\T_{i\sigma}$, and unbound orbitals $\wf[\epsilon](\br)$, with continuum energies $\epsilon^\T_\sigma$. The spin-densities are given by
\begin{equation}
\label{eq:KS_dens}
    n^\sigma(\br) = \sum_i f_{i\sigma} |\wf[i](\br)|^2 + \int \dd{\epsilon} g_\sigma(\epsilon) f_{\sigma}(\epsilon)|\wf[\epsilon](\br)|^2,
\end{equation}
where $f_{i\sigma}$ denotes the occupation number of the $i^{\textrm{th}}$ energy level $\epsilon_{i\sigma}^\T$, $f_{\sigma}(\epsilon)$ the continuum distribution function, and $g_\sigma(\epsilon)$ the density of states.
The total density is simply equal to the sum over the spin-densities, $\n(\br)=\nup(\br) +\ndown(\br)$. Since the KS system is a non-interacting system of electrons, the occupation numbers are determined by the Fermi--Dirac (FD) distribution,
\begin{equation}\label{eq:fd_occ_1}
    f_{i\sigma} = \frac{1}{1+e^{\beta(\epsilon_{i\sigma}^{\T}-\mu^\sigma)}},
\end{equation}
with $f_{\sigma}(\epsilon)$ defined in a similar way.

As discussed, finite-temperature KS-DFT is formally an exact method; however, the XC functional $F^\T_\textrm{xc}[\n]$ has to be approximated in practice. Formally, $F^\T_\textrm{xc}[\n]$ should depend explicitly on the temperature (a dependence which is absent in many approximations), and satisfy various exact conditions \cite{PPFS11}. Furthermore, even XC approximations which satisfy these requirements can suffer from various errors.
\change{We outline two such errors, particularly relevant in the context of this work, below:}


(i) \emph{Self-interaction error}: The classical Hartree energy (\ref{eq:Hartree}) in ground-state KS-DFT introduces a spurious self-interaction (SI) --- repulsion of the electron from its own charge density~\cite{PZ81} --- which must be compensated by the XC term. However, for many approximate XC functionals the self-interaction is not compensated fully, and one is left with a self-interaction error (SIE). A prototypical example is the H atom, for which approximate-XC calculations can be compared against an analytic result, quantifying the SIE. 
Self-interaction causes an array of problems, among which is also the under-prediction of ionization potentials \cite{KK08,GL12,SKKK14,KK20}.  Development of methods to mitigate 
the SIE is a very active area of research \cite{KP03,MSCY06,PRP14,SKKK14,SKMKK14,YPP17,JPJWTSJ19,RYDBPJSP19,CPPLHG20,SFKKTL20}.
The SIE is also present in finite-temperature KS-DFT and also there it must be compensated by the XC functional; finite temperature makes this task even more difficult.  

(ii) \emph{Ghost-interaction error}: In finite-temperature KS-DFT, the density is constructed as a weighted ensemble of densities of KS determinants (see Eq.~(\ref{eq:n_ens})). Formally, the KS states which form this ensemble should not interact with each other; however, the Hartree energy is a functional of the total density and thus a ghost-interaction error is present due to repulsion between electrons in different KS states \cite{GPG02}. This error is less 
studied than the SIE, but some correction schemes have been proposed when the ensemble weightings are defined \emph{a priori} \cite{GD13,SP17}.

\subsection{Reduction to average-atom model in finite-temperature KS-DFT} \label{subsec:theory_pt_4}

We shall now apply the finite-temperature KS scheme detailed above to the reduced Hamiltonian in Eq.~\eqref{eq:H_el_at}. Using this Hamiltonian, the computational cost of solving the KS equations is significantly reduced, because the finite-temperature KS equations \eqref{eq:thermal_KS} need only to be solved for a single atom. In the single-atom picture, the electron-nuclear potential is spherically symmetric. The same is not necessarily true for the KS potential defined in Eq.~\eqref{eq:vs}; however, we make the spherical approximation (which is usually made in atomic KS calculations), in which the KS potential is assumed equal to its spherically-averaged value. This means the bound KS orbitals $\wf[i](\br)$ can be decomposed into a product of radial and angular components,
\begin{equation}\label{eq:psi_decomp}
    \wf[i](\br) = \rnl[\sigma] \Ylm,
\end{equation}
where the angular component $\Ylm$ is a spherical Harmonic function. The eigenstates are now characterised by the quantum numbers $n,l,m$, and $\sigma$, where $n$ denotes the energy level, $l$ and $m$ characterise the orbital angular momentum, and $\sigma$ refers to the spin channel.

The radial component $\rnl[\sigma]$ is determined by solving the one-dimensional differential equation given by
\begin{multline}\label{eq:ks_radial}
    \left[\dv[2]{\rnl[\sigma]}{r} + \frac{2}{r}\dv{\rnl[\sigma]}{r} - \frac{l(l+1)}{r^2} \rnl[\sigma]\right]\\ + 2 \left[\epsilon^{\T,\sigma}_{nl} - \vs[\n](r) \right] \rnl[\sigma] = 0,
\end{multline}
where the spherically symmetric KS potential $\vs[n](r)$ is equal to
\begin{equation}
    \vs[n](r) = -\frac{Z}{r} + 4\pi \int_0^{\RVS} \dd{x} \frac{\n(x)x^2}{r^>(x)} + \fdv{F_\textrm{xc}^\T [\nup,\ndown]}{n^{\sigma}(r)},
  \end{equation}
where $r^>(x)=\max(r,x)$. All integrals are performed within the Voronoi sphere, whose radius $\RVS$ \footnote{This radius is often denoted as the radius of the Wigner--Seitz cell, $R_\textrm{WS}$ in the literature. We use the notation $\Rws$ to clearly distinguish this quantity from the Wigner--Seitz radius $r_s$, which depends on the free electron density only.} is determined from the average density of the nuclei $n_i$,
\begin{equation}\label{eq:RVS_defn}
    \RVS=\left(\frac{3}{4\pi n_i}\right)^{1/3}\,.
\end{equation}

In the WDM regime, i.e.\ at high pressures and temperatures, the unbound electrons play an important role. Due to the nature of continuum states \cite{PMAG01}, it is difficult to compute the unbound density exactly and thus it is determined (in our model) in an approximate manner; we discuss this further in the following section. For now, we note that, without further approximation, the \change{total (spin) electron} density is split into bound and unbound components,
\begin{align}
    \n^\sigma(r)=\nb(r) + \nub(r),
\end{align}
with the bound component given by
\begin{equation}\label{eq:nb_el}
\nb(r)=\sum_{n,l} (2l+1) f_{nl}^\sigma |\rnl[\sigma]|^2,
\end{equation}
\change{where $f_{nl}^\sigma$ are the Fermi Dirac occupations, as defined by Eq.~\eqref{eq:fd_occ_1},} 
and with the KS orbitals normalized within the Voronoi sphere,
\begin{equation}\label{eq:KS_norm}
    4\pi\int_0^{\RVS}\dd{r} r^2 |\rnl[\sigma][r]|^2=1\,.
\end{equation}
Since we adopt the formalism of spin-DFT, the number of electrons in each spin-channel is fixed to some integer value $\Ne^\sigma$, which also fixes the total number of electrons in the Voronoi sphere \change{equal to} 
its average value $\Ne=\sum_\sigma \Ne^\sigma$. The (spin-dependent) chemical potential $\mu^\sigma$ in the FD distribution is thus determined from the condition that the number of electrons in each spin channel is fixed. In theory, the values of $\Ne^\sigma$ should be determined by whichever configuration minimizes the grand free energy \change{\footnote{In practise, it is often known \emph{a priori} which configuration will \newchange{likely} be most energetically favourable from experience and physical intuition. In the examples we consider later, we take $\Ne^\uparrow=1,\ \Ne^\downarrow=0$ for Hydrogen, and \textcolor{red}{$\Ne^\uparrow=\Ne^\downarrow=2$} for Beryllium}}.


We make a clarification here regarding the fact that there are two chemical potentials, $\mu^\uparrow$ and $\mu^\downarrow$. This is a necessary consequence of using spin-DFT; however, one could say that in the absence of a magnetic field, there should be only a single chemical potential for the electrons. In fact, this is not a problem, because only one of the chemical potentials is physically relevant in this scenario: when adding an electron, the lower chemical potential is relevant, and when removing an electron, the higher chemical potential is relevant. If non-spin-dependent DFT were used, as is quite common in AA models, then only the total electron number is conserved and a single chemical potential returned. 
However, it is well known in DFT literature that using spin-dependent DFT is advantageous, because it produces more accurate results for systems whose spin is not zero~\cite{Parr_Yang,Kotochigova_1997,callow2021density}, so we use the spin-dependent formulation in our model. 

In this KS-AA model, we impose the condition that the electron density is smooth 
at the boundary between neighbouring spheres. Physically, this is motivated by the fact that the real electron density (which is formally equal to the KS density) should be \change{smooth everywhere}. 
Of course, the true system cannot be split into identical spheres, so this condition is an approximation designed to mimic the real electron density. 
Mathematically, this means the following boundary condition should be imposed on the (spin) density,
\begin{gather}\label{eq:bc_ks}
    \dv{n^\sigma(r)}{r}\Bigg|_{r=\Rws}=0\,.
\end{gather}

\change{The above condition on the density is physically intuitive, and leads to boundary conditions on the radial KS orbitals at the Voronoi sphere's edge that are widely used in AA models. However, it is not a necessary condition, and alternative choices can be made to model the concept of an atom immersed in a plasma. For example, in Ref.~\onlinecite{STJZS14}, there is no boundary condition applied to the density or KS orbitals at the sphere's edge; instead, the potential is fixed to a constant value outside the sphere ($\vs(r>\RVS)=\vs(\RVS)$) and the KS orbitals are solved out to some radius $r_\textrm{max}\gg\RVS$. A similar approach (with the KS potential instead modified inside the Voronoi sphere) is used in the MUZE code \cite{MWHD13}, and we shall compare results obtained with that approach with the above condition on the density in \S~\ref{subsec:results_3}.}

\subsection{Boundary conditions and treatment of unbound electrons}\label{subsec:theory_pt_5}

As mentioned in the previous sub-section, we do not explicitly solve the KS equations for the continuum orbitals in our model. Instead, we treat the unbound electrons as being completely free (the ideal approximation), i.e., having uniform density, when we compute their contribution to the total density. For this distinction between bound and continuum states, we assume the continuous part of the energy spectrum starts at $\vs(R_\textrm{VS})$. This is equivalent to shifting the KS potential everywhere by the constant $\vs(R_\textrm{VS})$, in other words
\begin{equation}\label{eq:vs_shift}
    \bar{v}_\textrm{s}^{\tau, \sigma}(r) = {v}_\textrm{s}^{\tau, \sigma}(r) - \vs(R_\textrm{VS})\,,
\end{equation}
where $\bar{v}_\textrm{s}^{\tau, \sigma}(r)$ signifies the shifted potential, and then assuming the continuum starts at energies above zero (since $\bar{v}_\textrm{s}^{\tau, \sigma}(\RVS)=0$). We emphasize that shifting the potential by a constant has no effect on the KS orbitals, only their energy eigenvalues which shift by the same constant $\vs(R_\textrm{VS})$. Since it makes the notation easier, we assume the potential has been shifted in this way for the rest of the paper, unless specified otherwise.

 We note that, in modern AA codes, the unbound electrons are usually treated in a more sophisticated manner, either semi-classically (TF) or (more typically) in a fully quantum manner \cite{Li_1979,WSSI06,SHWI07}, for example by expanding the continuum states in a discrete set of normalizable states \change{\cite{PMAG03, P06}}. The TF approximation for the unbound density is known to have certain limitations, for example systematically over-estimating the chemical potential \cite{MWHD13}; these limitations are likely to be exacerbated using the even simpler ideal approximation. Nevertheless, our model will yield important insights into the comparison of different XC functionals and boundary conditions. We also compare this ideal treatment with TF and quantum unbound electrons in \S~\ref{sec:results}. Furthermore, we clarify that the density used to construct the KS free energy functional \eqref{eq:KS_free} and potential \eqref{eq:vs} is the \emph{full} density (bound and unbound), with the kinetic energy of the unbound electrons being given by the ideal expression.

The value of the unbound electron (spin)-density $\nub$ is determined from the number of unbound electrons, $\Nub=\int\dd{\br}\nub$, which implies $\nub=\Nub/V$, where $V = \frac{4}{3} \pi \RVS^3$ is the volume of the sphere. The total numbers of bound and unbound electrons are determined according to the FD distribution,

\begin{align}
\label{eq:Nb_el}
\Nb &= \sum_{n,l}^{\epsilon^{\T,\sigma}_{nl}\leq 0}(2l+1) f_{nl}^\sigma(\epsilon^{\T,\sigma}_{n\sigma},\mu^\sigma,\T); \\
\label{eq:Nub_el}
\Nub&=\frac{V}{2^{1/2}\pi^2} \int_{0}^\infty\dd\vec{\epsilon} \frac{\epsilon^{1/2}}{1+e^{\beta(\epsilon - \mu^\sigma)}}.
\end{align}
\change{We recall that the chemical potentials $\mu^\sigma$ are chosen such that the sum $\Nb+\Nub$ equals a preset value, $N_\textrm{e}^\sigma$.}

\change{We now proceed to discuss the question of boundary conditions. The boundary condition for the density has been specified by us in Eq.~\eqref{eq:bc_ks}. However, to solve the KS equations \eqref{eq:ks_radial}, one needs to specify the boundary conditions for the radial orbitals $X_{nl}^\sigma(r)$. Notably, Eq.~\eqref{eq:bc_ks} does not determine the orbital boundary conditions uniquely. For the unbound density, Eq.~\eqref{eq:bc_ks} is satisfied automatically. For the bound states, it}
implies the following equality for the radial orbitals $\rnl[\sigma]$ and their derivatives: 
\begin{equation}\label{eq:bc_ks_orbs}
    \sum_{nl} (2l+1)f_{nl}^\sigma X_{nl}^\sigma(R_\textrm{VS})\dv{\rnl[\sigma]}{r}\Bigg|_{r=\RVS}=0.
\end{equation}
There is no unique way to satisfy the above relation. The two most simple choices are either to require the radial wavefunctions to be zero at the boundary, or their derivatives to be zero, i.e.\
\begin{align}
    \label{eq:bc1}
    0 &= \rnl[\sigma][\RVS]\,,\ \textrm{or}\\
    \label{eq:bc2}
    0 &= \dv{\rnl[\sigma]}{r}\Bigg|_{r=\RVS}.
\end{align}
Mathematically, many other choices are also possible. Both of the above boundary conditions have been used in AA models (see for example Ref.~\onlinecite{SGSKR08} for an example of the former or Ref.~\onlinecite{JNC12} for an example of the latter), and it has been observed \cite{JNC12} that using the former choice \eqref{eq:bc1} yields markedly different results than the latter \eqref{eq:bc2} for the average ionization in Aluminum. We also observe that the choice of boundary conditions has a major impact on results in \S~\ref{sec:results} of this paper. \change{We additionally note that we will sometimes refer to the boundary conditions in Eqs.~\eqref{eq:bc1} and \eqref{eq:bc2} as b.c. (i) and b.c. (ii) respectively, in order to simplify notation.}

In spite of this, there has been limited analysis of the impact of choosing one of the above conditions over the other (perhaps because many models, instead or additionally, enforce boundary conditions on the potential as earlier discussed). However, Rozsnyai has conceptually identified \cite{R72,R91} these boundary conditions as corresponding to upper \eqref{eq:bc1} and lower \eqref{eq:bc2} limits of a band-structure, due to their association with bonding \eqref{eq:bc2} and anti-bonding \eqref{eq:bc1} molecular orbitals (MOs); Massacrier and co-workers have explored this band-structure interpretation further, interpolating between the band-structure limits via a Hubbard functional form for the density of states \cite{M94,PMC05,MBVSM21}.

\begin{figure}
    \centering
    \includegraphics{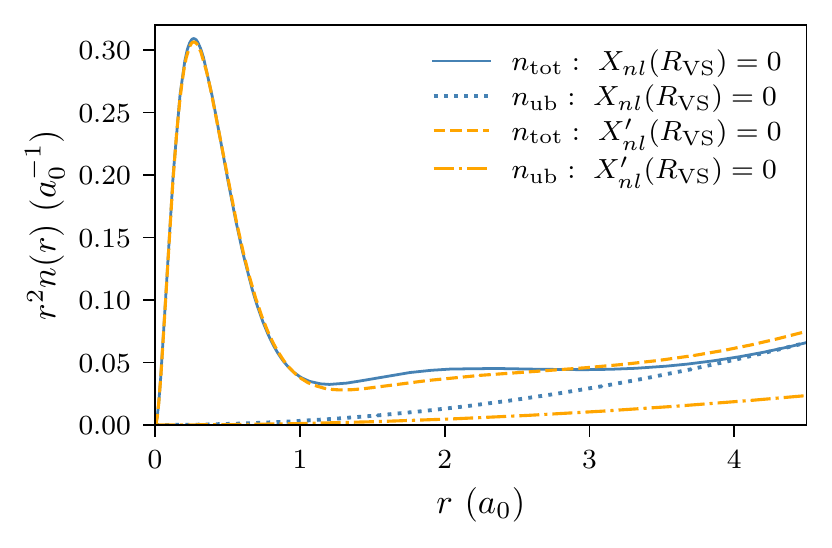}
    \caption{Electronic density distribution, \newchange{with unbound component ($n_\textrm{ub}$) and total ($n_\textrm{tot}$) plotted}, for Beryllium with mass density $\rho_m=0.266\ \textrm{g cm}^{-3}$, at temperature $\T=4\ \textrm{eV}$. We see the effect of different boundary conditions (\ref{eq:bc1},\ref{eq:bc2}) on the density profile.}
    \label{fig:Be_dens_comp}
\end{figure}

\change{In Fig.~\ref{fig:Be_dens_comp}, we compare the density distribution resulting from the two different boundary conditions for Beryllium with density $0.266\ \textrm{g\ cm}^{-3}$ and temperature $4\ \textrm{eV}$.}
\change{We find that near the origin ($r < 1 \, a_0$) the densities are the same, whereas close to the boundary, they differ significantly. Notably, using condition~\eqref{eq:bc1} leads to a higher unbound density: the requirement of the wavefunction vanishing at $\RVS$ ``pushes'' the electrons to the unbound states}. 


This completes our first-principles reduction from the intractable minimization of the grand canonical potential \eqref{eq:gr_can} for a fully-interacting system of electrons and nuclei \eqref{eq:H_tot} into a finite-temperature KS model for a single atom. In the remainder of this paper, we briefly discuss some numerical aspects and present some results to explore the behaviour of this AA model under different approximations.

\section{Numerical Aspects}\label{sec:numerics}

The majority of calculations were performed with the ORCHID code, an atomic DFT code which has been extended to include temperature and the boundary conditions described in \S~\ref{subsec:theory_pt_5}; ORCHID has been used previously in Refs.~\cite{KraislerMakovArgamanKelson09,KraislerMakovKelson10,Argaman13,KraislerSchild20,KraislerHodgson21}. 
An open-source successor to ORCHID, named atoMEC, is under development and can be used to reproduce many of the results in this paper \cite{callow_timothy_2021_5205719}. For the bound KS orbitals, the second-order differential equation \eqref{eq:ks_radial} is solved on a grid using the Numerov method in combination with a two-side shooting method \cite{B67,C72,Koonin}. The grid uses a logarithmic scale for the radial co-ordinate $r$, $x=\ln(Zr)$, where $Z$ is the atomic number, to ensure a higher density of grid points near the nucleus. To solve for the radial wave-functions $\rnl[\sigma][r]$, a transformation $\pnl[\sigma][x]=\rnl[\sigma][x]e^{x/2}$ is made, so that the differential equation \eqref{eq:ks_radial} becomes
\begin{gather}
\dv[2]{\pnl[\sigma][x]}{x} - 2e^{2x}\left[W(x)- \frac{\epstilde}{Z^2}\right]\pnl[\sigma][x]=0,\\
W(x)=\frac{\bar{v}_\textrm{s}^{\tau,\sigma}[n](x)}{Z^2}+\frac{1}{2}\left(l+\frac{1}{2}\right)^2e^{-2x},
\end{gather}
where the notation $\epstilde$ denotes the KS eigenvalues corresponding to the shifted KS potential $\bar{v}_\textrm{s}^{\tau,\sigma}$ \eqref{eq:vs_shift}.

Additionally, some calculations (involving different treatments of unbound electrons, to be later discussed) were performed using the MUZE code \cite{GHCH05,HFPF05,MWHD13}. Below, we discuss details of the implementation of our model in ORCHID, and refer readers to the aforementioned references for further details of the MUZE code.

The KS orbitals and their eigenvalues from the differential equation \eqref{eq:ks_radial} are used to determine the total density via Eqs.~\eqref{eq:nb_el}, \eqref{eq:Nb_el} and \eqref{eq:Nub_el}. The whole process proceeds self-consistently until the total energy is converged to within $10^{-6}$ Hartree, the spatially-averaged bound spin-densities are converged enough to satisfy $\left|\int dr (n_\sigma(r)-n_\sigma^f(r)) \right| < 10^{-6}$ a.u.\ (the superscript $f$ distinguishes the density of the previous iteration from the density of the last iteration) and the KS potentials (for both spin channels) satisfy $\left|\int dr (v_{\s,\sigma}(r)-v_{\s,\sigma}^f(r)) \right| < 10^{-6}$ a.u. Different guesses were trialled for initializing the KS potential in the self-consistent cycle; it was found the bare Coulomb potential is always a suitable initial choice for the KS potential.

For each atom, XC functional, density and temperature range \footnote{Starting from the lowest temperature $\T=0.001\ \textrm{Har}$, convergence was checked in approximate multiples of 3, i.e.  
$\T=0.001,0.003,0.01\dots\ \textrm{Har}$.}, convergence was checked with respect to the number of grid points $N_\textrm{grid}$, the leftmost grid-point $x_0=\EK{\ln}(Z r_0)$, 
and which bound states (characterised by the quantum numbers $n,l$) to account for. The required grid size depends strongly on the specific calculation, ranging between 1,000 to 40,000 points.
The number of bound states is also highly dependent on the particular calculation. However, a lower grid bound of $r_0=e^{-13}/Z\ a_0$ was found to be sufficient in almost all cases. 


\section{Results}\label{sec:results}

\begin{figure*}
    \centering
    \includegraphics{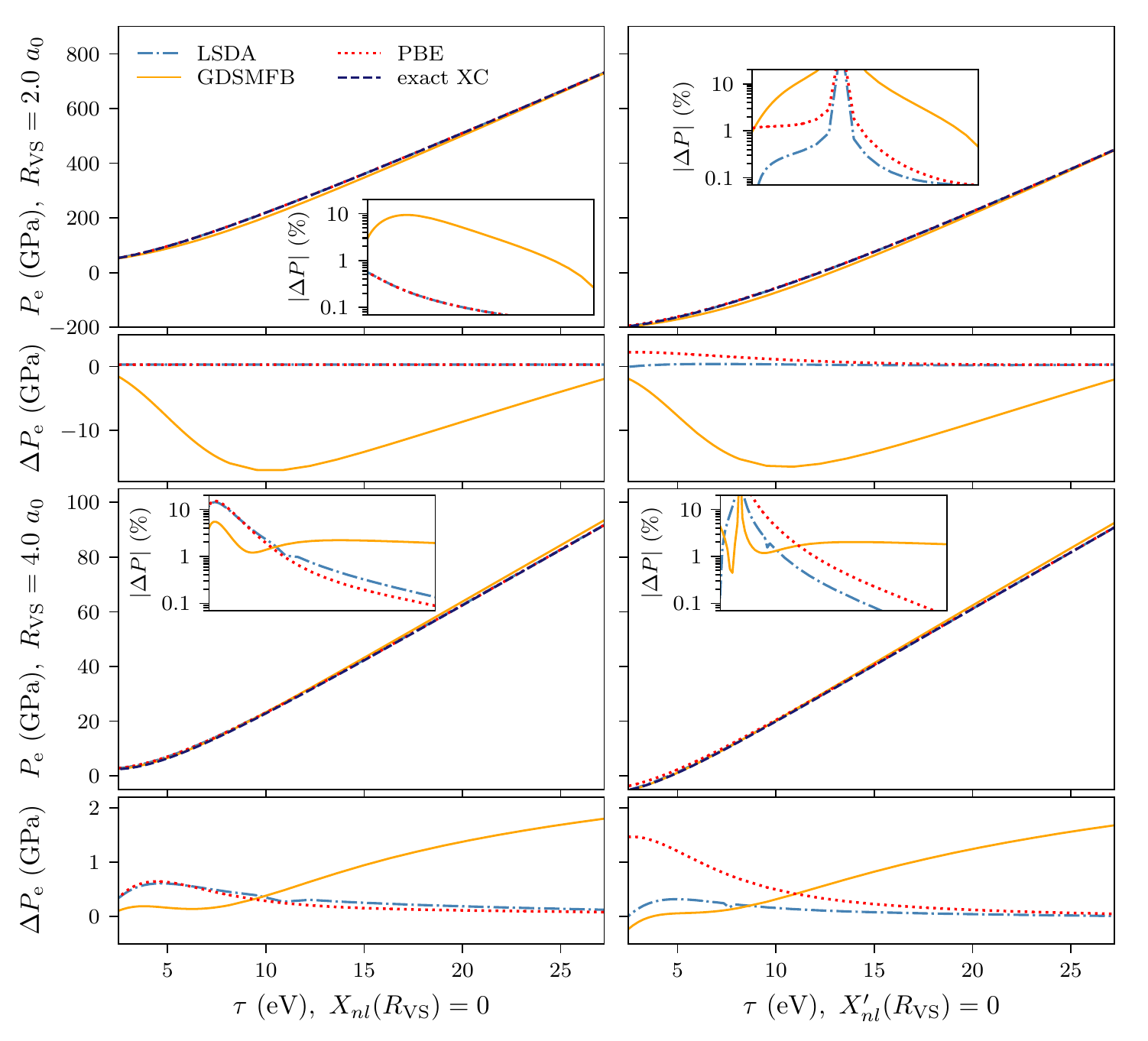}
    \caption{EOS ($P_\textrm{e}$ vs $\T$) for Hydrogen for (i) $\RVS=2.0a_0$ (top row) and (ii) $\RVS=4.0a_0$ (third from top row); results on the left are for the b.c. $\rnl[][\RVS]=0$ \eqref{eq:bc1} and on the right are for the b.c. $\rnld[\RVS]=0$. Insets: percentage difference in pressure (using a log scale for the $y$--axis) between the approximate functionals and the exact functional. The $x$--axes of the inset plots are identical to the (temperature) $x$--axes of the main plots. Smaller plots below main ones: actual difference in pressure between approximate and exact XC functionals, $\Delta P_\textrm{e}=P_\textrm{e}^\textrm{app}-P_\textrm{e}^\textrm{exact}$. 
    }
    \label{fig:H_EOS}
\end{figure*}

In the following, we use our KS-DFT AA model as a surrogate 
of AA models to probe their sensitivity to various choices of approximation. We focus mostly on the impact of XC functional and boundary conditions, but also briefly explore the influence of more advanced treatments of unbound electrons in \S~\ref{subsec:results_3}.

\change{We probe a range of temperatures from $0$ to $\sim 25$ eV, and densities from roughly $10^{-3}$ to $20\ \textrm{g\ cm}^{-3}$, corresponding to to Voronoi sphere radii in the range $2\ a_0\leq \RVS \leq 10\ a_0$; this roughly covers the lower-density and temperature region of the WDM phase space, as denoted in Fig.~\ref{fig:wdm_params}. For the temperature-dependent plots which are shown at fixed values of $\RVS$,} we connect these Voronoi sphere radii with their mass densities $\rho_m$, and the corresponding dimensionless parameters $r_s,\ \Gamma_i$, and $\Theta_e$ (the latter two computed at the mid-point of the temperature range we consider, $\T=13.1\ \textrm{eV}$) \change{in Table~\ref{tab:rs_theta_etc}}.

We note that the definitions of $r_s$, $\Gamma_i$, and $\Theta_\textrm{e}$ in Eqs.~\eqref{eq:Theta_e}--\eqref{eq:Gamma_i} use as input the free electron density, and furthermore that the definition for the Fermi energy is strictly valid for an ideal Fermi gas only. However, we of course deal with interacting electrons, some fraction of which are bound by the nuclei. For the numbers quoted in Table~\ref{tab:rs_theta_etc}, we assume for simplicity that the free electron density is a fixed quantity for a given ionic density (independent of temperature) and equal to the valence electron density; we also use the definition in Eq.~(\ref{eq:E_fermi}) for the Fermi energy regardless. These numbers should therefore be seen as rough indicators only rather than well-defined parameters. Based on these values, we note that the lowest density Hydrogen example ($\RVS=10.0\ a_0$) is at the limit of what is typically considered WDM conditions; however, it is a useful test case for functional comparison as the AA approximation (meaning the neglect of inter-cell interactions) should be very accurate in this case.

From our KS-DFT AA model, we directly obtain the free energy, KS orbitals and their energies, occupation numbers, and the number of unbound electrons. 
To access the electronic pressure $P_\textrm{e}$, we use the following relationship between the electronic free energy $F$ and the volume $V$ for a fixed temperature $\T$,
\begin{equation}
     P_\textrm{e}(V,\T)=-\pdv{F}{V}\Bigg|_\T,
\end{equation}
which we compute numerically via finite differences. The free energy $F$ is defined from the grand free potential $\Omega$ of Eq.~\eqref{eq:KS_free} via the relationship
\begin{equation}
    \Omega[n] = F[n] - \sum_{\sigma}\mu^\sigma \Ne^\sigma.
\end{equation}
Please see Appendix B for details of the construction of the free energy $F[n]$ in our AA model. We focus on the electronic pressure because KS-DFT does not give access to the ionic pressure. We have explored adding an approximate ionic pressure using the ideal gas law, $pV=nRT$, and observed this results in a noticeable increase in pressure. However, since we want to explore the impact of approximations which have no effect on the value of the ionic pressure at a given temperature and density, we only present results for the electronic pressure.

Furthermore, the number of unbound electrons, or equivalently (in our model) the mean ionization state (MIS), is an important property in dense plasmas \cite{MWHD13}, as are the KS orbital energies, which are used (for example) in the computation of thermal and electrical conductivities (\S~\ref{sec:intro}, Ref.~\onlinecite{HFR11}). We therefore focus on the aforementioned quantities.

We compare results for three XC functionals: firstly, the zero temperature local \change{spin}-density approximation (LSDA) \cite{KS65,PW92}, which is widely used in finite-temperature KS-DFT and AA models; \newchange{secondly}, the Perdew--Burke--Ernzerhof (PBE) generalized gradient approximation \cite{PBE96}, which is used extensively in ground-state DFT calculations \cite{PJGB15}. \newchange{Thirdly}, we consider the temperature-parameterized LDA by Groth \emph{et al.} (GDSMFB) \cite{GDSMFB17}; this functional retains the computational advantages of the ground-state LDA, and can easily be integrated into other AA or KS-DFT codes via the LIBXC package \cite{LSOM18}, as has been done here in ORCHID. We also investigated the temperature-parameterized LDA by Karasiev \emph{et al.} \cite{KSDT14}, and found that results were always in very close agreement with the GDSMFB functional (echoing similar observations in Refs.~\onlinecite{KTD19} and \onlinecite{RDV20}), and therefore we present results for LSDA, PBE and GDSMFB only.

In the following sections, we explore results for Hydrogen (\S~\ref{subsec:resultsA}) and Beryllium (\S~\ref{subsec:resultsB}); although the model we have derived is valid for systems with a macroscopic net charge, we consider only charge neutral examples ($\Ne=Z$). We drop the spin-dependent notation for all spin-dependent quantities such as the KS orbitals and their eigenvalues. Beryllium has an even number of electrons and thus the different spin channels give the same results. For Hydrogen, one of the spin channels is completely devoid of electrons (amounting to a constraint of $\Ne^\sigma=0$): since this spin channel does not contribute to many physically meaningful quantities, where we present results for spin-dependent quantities, these are for the occupied spin channel only.

\setlength{\tabcolsep}{8pt}

\begin{table}
    \centering
    \begin{tabular}{cccccc}
    \toprule
    Atom & $R_\textrm{VS}^*$ & $\rho_\textrm{m}^\dagger$ & $r_s^*$ & $\Gamma_i^\textrm{M}$ & $\Theta_e^\textrm{M}$ \\
    \midrule
    H & 2.0 & 0.337 & 2.0 & 0.50 & 1.09\\
    H & 4.0 & 0.042 & 4.0 & 0.25 & 4.34\\
    H & 10.0 & 0.0027 & 10.0 & 0.10 & 27.2\\
    Be & 2.0 & 3.04 & 1.59 & 8.0 & 0.68\\
    Be & 4.0 & 0.379 & 3.18 & 4.0 & 2.74\\      
    Be & 4.7 & 0.232 & 3.73 & 3.40 & 3.78\\
    \bottomrule
    \addlinespace[0.2em]
    \multicolumn{6}{l}{$^*$\footnotesize{Atomic units}\ \  $^\dagger$\footnotesize{g cm$^{-3}$}}\\
    \end{tabular}
    \caption{The values of mass density $\rho_\textrm{m}$, WS radius $r_s$, and coupling parameters $\Gamma_i^\textrm{M}$ and $\Theta_e^\textrm{M}$ (measured at the mid-point temperature of $\T=13.1$ eV) with the corresponding Voronoi sphere radii $R_\textrm{VS}$ on which we test our model.}
    \label{tab:rs_theta_etc}
\end{table}

\begin{figure*}
    \centering
    \includegraphics{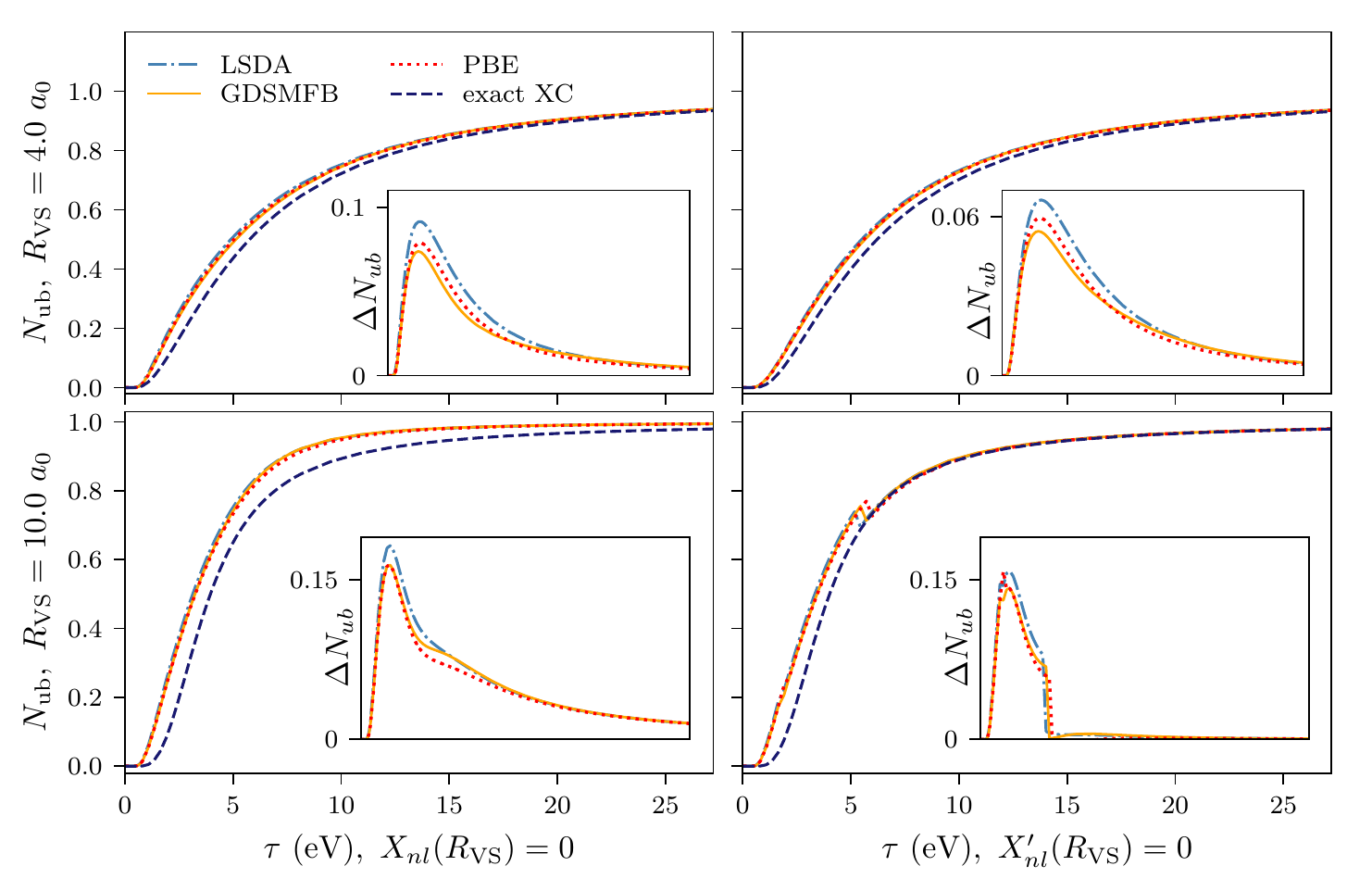}
    \caption{The number of unbound electrons $N_\textrm{ub}$ for Hydrogen as a function of temperature for (i) $\RVS=4.0\ a_0$ (top row) and (ii) $\RVS=10.0\ a_0$ (bottom row);  results on the left are for the b.c. $\rnl[][\RVS]=0$ \eqref{eq:bc1} and on the right are for the b.c. $\rnld[\RVS]=0$. Inset plots show difference in the number of unbound electrons as a function of temperature between the approximate functionals and the exact functional. Discontinuities in $ N_\textrm{ub}$ occur when energy levels transition from the continuum to the discrete part of the spectrum.}
    \label{fig:H_N_ub}
\end{figure*}

\subsection{Hydrogen}\label{subsec:resultsA}

We first apply our AA model to Hydrogen. Besides being an element of high interest in the WDM regime in its own right \cite{RDV20}, \change{for Hydrogen} we can solve 
\change{our AA model} exactly. Since we consider interactions only within the Voronoi sphere, which itself contains only one electron, there are no inter-electron interactions and thus the KS potential is given exactly by the electron-nuclear attraction,
\begin{equation}
   \vs(r)=v_\textrm{en}(r)=-\frac{1}{r}\,.
\end{equation}
The XC functional also cancels the Hartree energy,
\begin{equation}
    F_\textrm{xc}[n]=-U[n]\,,
\end{equation}
which is the exact XC functional in this case. We stress here that this is the exact result within the limits of the model we define, in which various approximations have already been made, such as the neglect of explicit intra-cell interactions; it therefore does not represent the truly exact limit for the Hydrogen plasma in general. \newchange{Another assumption that was mentioned but not discussed at length is that we take $\Ne^\uparrow=1,\Ne^\downarrow=0$. We fix this for convenience, but in principle one could search over all fractional $\Ne^{\uparrow,\downarrow}$, with $\Ne^\uparrow+\Ne^\downarrow=1$, and choose the configuration which minimizes the grand free energy.} However, \newchange{the exact XC functional} is a reference from which we can isolate the errors that result from approximations for the XC functional (as opposed to other approximations in the model). Henceforth, we use the 
term  
``exact'' when referring to this choice given the fact that we mean only to the exactness of the XC functional and potential.


In Fig.~\ref{fig:H_EOS}, we plot the $P_\textrm{e}\mbox{-}\T$ curve for Hydrogen at different values of $\RVS$. The most notable observation in this series of the plots is that the choice of boundary condition has a significant impact, particularly at lower temperatures and higher densities. Indeed, for $\RVS=2.0\ a_0$, the EOS data is qualitatively different, with the boundary condition $\rnld[\RVS]=0$ showing large negative pressures for all the functionals (including the exact one) at lower temperatures. The approximate functionals show good agreement with the exact result, rarely differing by more than $10\%$ and tending towards $<1\%$ for higher temperatures. An additional observation is that the LDA and PBE functionals yield almost identical results throughout, with the temperature-dependent GDSMFB functional tending to deviate slightly more from the exact result.

Next, in Fig.~\ref{fig:H_N_ub}, we compare the number of unbound electrons $N_\textrm{ub}$ as a function of temperature for two choices of density. For these (relatively low) densities, the two boundary conditions seem to be in quite good agreement. What is more interesting is that the approximate functionals tend to systematically over-predict the mean ionization state relative to the exact functional; furthermore, the approximate functionals differ minimally relative to each other. This is indicative of some kind of common error pertaining to semi-local XC functionals in general.

\begin{figure*}
    \centering
    \includegraphics{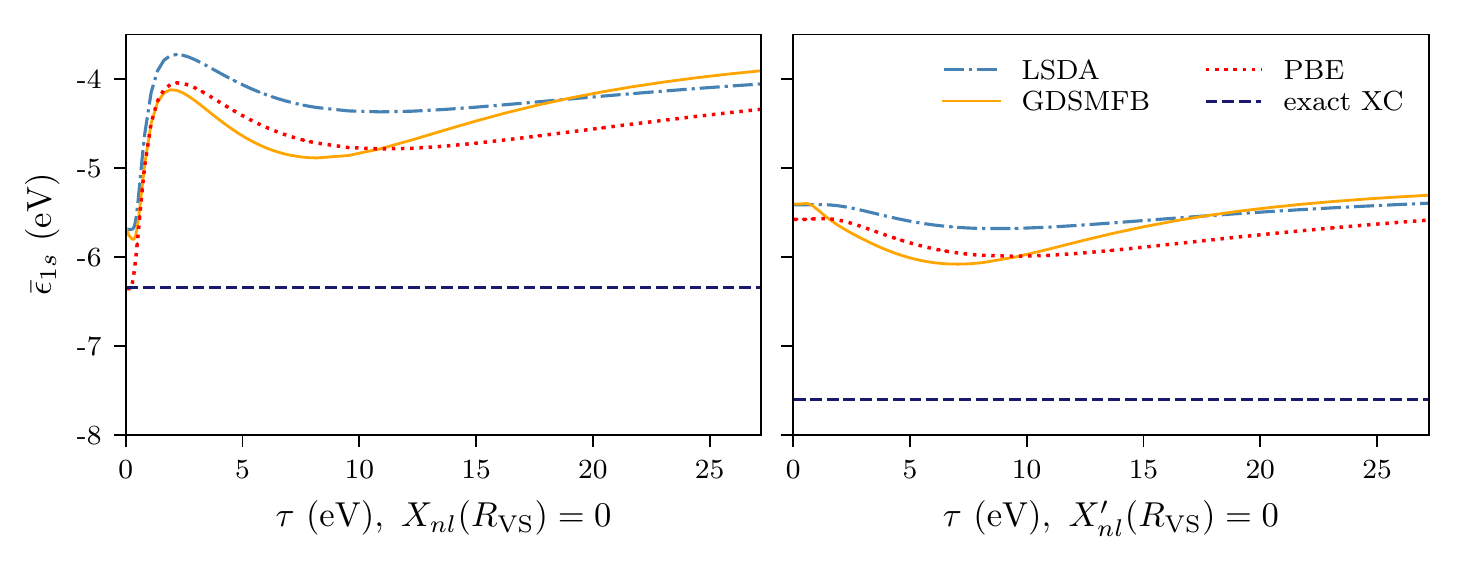}
    \caption{The energy level $\epstilde[1s][]$ for Hydrogen as a function of temperature  $\RVS=4.0\ a_0$; results on the left are for the b.c. $\rnl[][R_\textrm{VS}]=0$ \eqref{eq:bc1} and on the right are for the b.c. $\rnld[R_\textrm{VS}]=0$. Changing the b.c. for $\RVS=4.0\ a_0$ causes a shift in the energy level for both the approximate and exact functionals.}
    \label{fig:H_eps_1s}
\end{figure*}

In Fig.~\ref{fig:H_eps_1s}, we consider the $1s$ energy level $\epstilde[1s][]$ as a function of temperature for $\RVS=4.0\ a_0$. It is apparent that the approximate XC functionals systematically over-predict the $1s$ energy level (for both b.c.s), with once again minimal differences between the functionals themselves. The energy levels for the exact (bare Coulomb) XC functional are of course independent of temperature since $v_\textrm{en}(r)$ has no temperature dependence; interestingly, the energy levels from the approximate functionals also do not vary significantly across this temperature range (with an exception for very low temperatures for the b.c. $\rnl[][\RVS]=0$). An important observation from these curves is that the temperature-dependent GDSMFB functional does not seem to improve much the prediction of the $\epstilde[1s][]$ level: therefore, new XC functionals (most likely going beyond semi-local approximations) are required for finite-temperature KS-DFT.

Having explored the behaviour of various quantities as a function of temperature for fixed mass density, we now investigate the dependence of the same quantities on the mass density, for a fixed temperature 10 eV. We only compare results from the LSDA and exact XC functionals, since we already observed that all the approximate functionals gave very similar results; we also compare the two boundary conditions directly in the same plots. 

\begin{figure}
    \centering
    \includegraphics{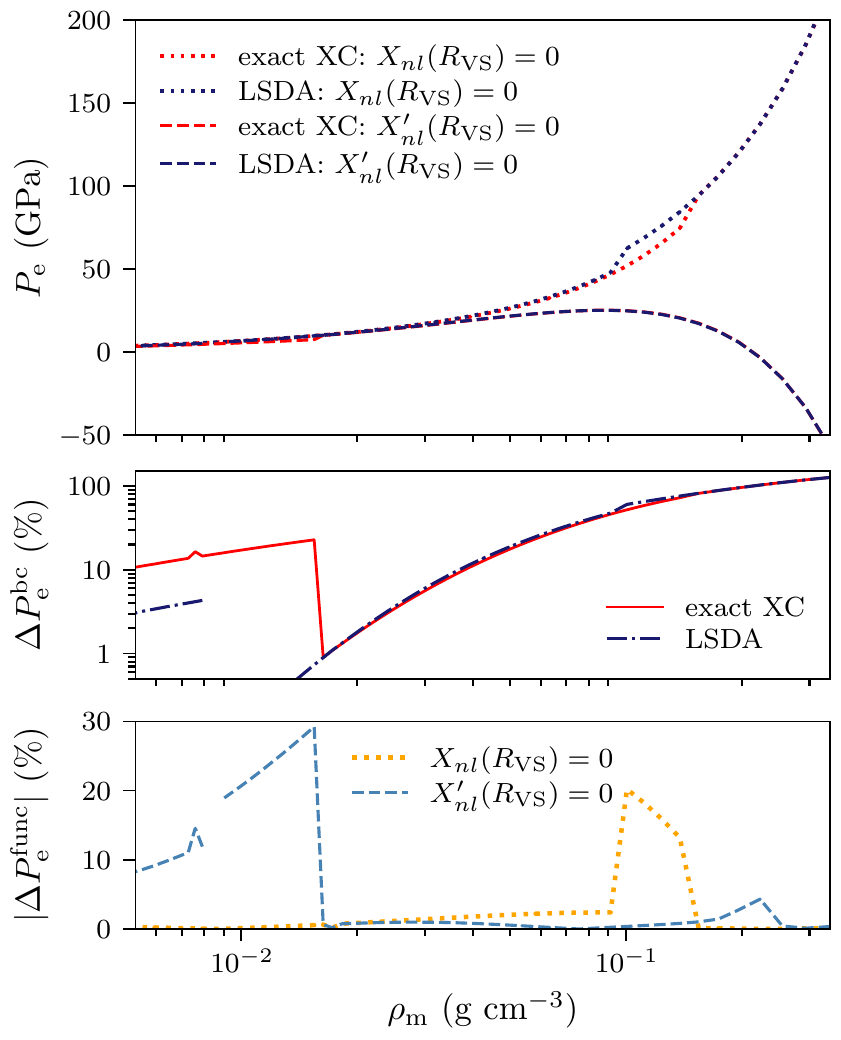}
    \caption{\change{Top: comparison of EOS ($P_\textrm{e}$ vs $\rho_\textrm{m}$) curves for Hydrogen as a function of density, at fixed temperature $\T=10\ \textrm{eV}$, for both boundary conditions and LSDA and exact XC functionals. Middle: percentage difference (logarithmic scale for $y$--axis) in pressure between the two boundary conditions, $\Delta P^\textrm{bc}_\textrm{e}=(P^\textrm{bc(i)}_\textrm{e}-P^\textrm{bc(ii)}_\textrm{e})/P^\textrm{bc(i)}_\textrm{e}$. Bottom: absolute percentage difference in pressure between LSDA and exact XC results, $|\Delta P^\textrm{func}_\textrm{e}|=|P^\textrm{LSDA}_\textrm{e}-P^\textrm{exact}_\textrm{e}|/|P^\textrm{exact}_\textrm{e}|$.}}
    \label{fig:H_rho_EOS_scan}
\end{figure}

In Fig.~\ref{fig:H_rho_EOS_scan}, we see the striking impact of the boundary condition on the electronic pressure as the density increases. As would be expected, for lower densities, when there are fewer interactions between neighbouring atoms, the two boundary conditions agree relatively well, with the difference between them (shown in the middle panel of Fig.~\ref{fig:H_rho_EOS_scan}) rarely exceeding 10$\%$ up to around $0.04\ \textrm{g cm}^{-3}$. However, at higher densities, the two boundary conditions diverge strongly; the boundary condition $\rnld[\RVS]=0$ actually has a turning point at which the pressure starts to decrease with increasing density and becomes strongly negative. The bottom panel of Fig.~\ref{fig:H_rho_EOS_scan} shows the difference between the functionals as a function of density (for both boundary conditions); we cannot draw any clear conclusions regarding systematic deficiencies of the LSDA functional from this plot. 

\begin{figure}
    \centering
    \includegraphics{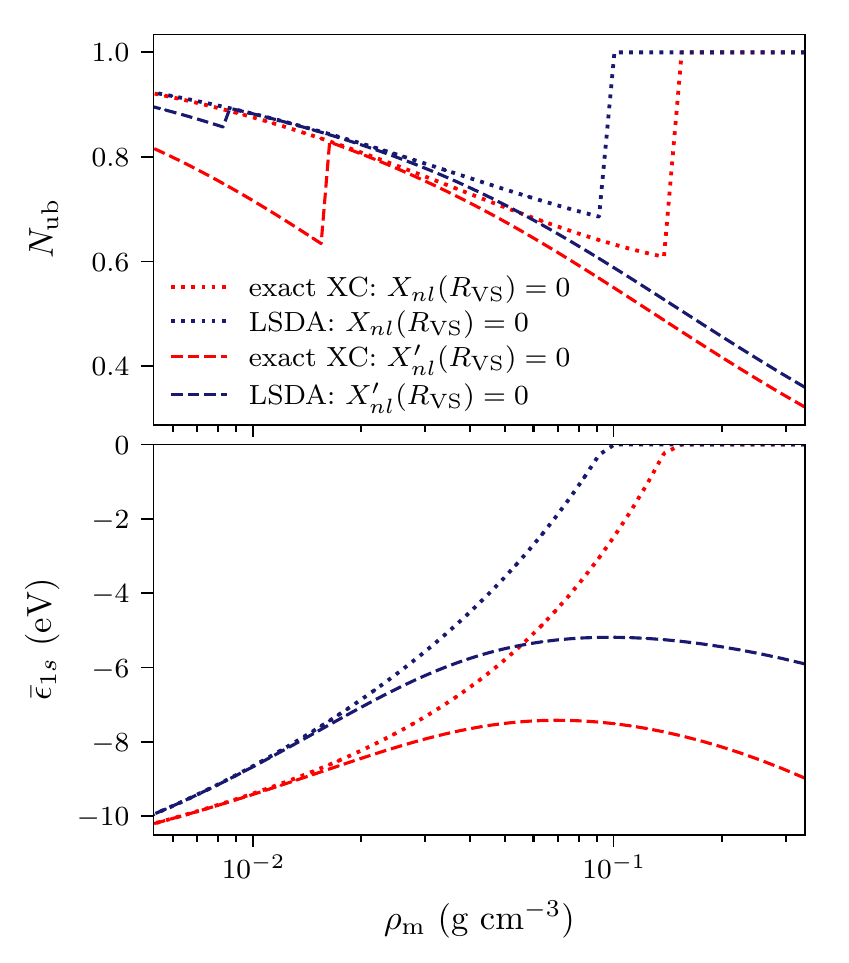}
    \caption{\change{Comparisons of (i) number of unbound electrons $N_\textrm{ub}$ (top) and (ii) energy level $\epstilde[1s][]$ (bottom) for Hydrogen as a function of density, at fixed temperature $\T=10\ \textrm{eV}$. Both boundary conditions and LSDA and exact XC results are shown for comparison.}}
    \label{fig:H_rho_Nub_scan}
\end{figure}

The main message from this figure is the huge impact of the boundary condition, and the limitations of choosing a single boundary condition as is frequently done in AA models. As discussed earlier, one possible solution is to consider a band-structure picture, either via some sensible approximation such as that employed by Massacrier and co-workers \cite{M94, PMC05, MBVSM21}, or even better using a first-principles approach which preserves the smoothness of the density at the sphere's edge \eqref{eq:bc_ks}. \newchange{Another quite striking feature of Fig.~\ref{fig:H_rho_EOS_scan} (also seen in Figs.~\ref{fig:H_N_ub} and~\ref{fig:H_rho_Nub_scan}) are sharp discontinuities, which appear due to the ionization of the $2s$ energy level at around $10^{-2}\ \textrm{g cm}^{-3}$ for the b.c. $\rnld[\RVS]=0$, and the ionization of the $1s$ energy level at around $10^{-1}\ \textrm{g cm}^{-3}$ for the b.c. $\rnl[\RVS]=0$. We shall later see similar discontinuities for Beryllium in \S~\ref{subsec:resultsB}, when we will discuss them in greater detail.}

The divergence of the pressure towards negative infinity for the $\rnld[\RVS]=0$ boundary condition is related to the ionization degree. In Fig.~\ref{fig:H_rho_Nub_scan}, we plot the number of unbound electrons (top) and the $1s$ energy level $\epstilde[1s][]$ as a function of the mass density, again for $\T=10\ \textrm{eV}$. As was observed for the electronic pressure, results from the two boundary conditions diverge with increasing density. In particular, the $\epstilde[1s][]$ level for the boundary condition $\rnld[\RVS]=0$, for both LSDA and the exact XC functionals, has a turning point and starts to decrease for densities above about $0.1\ \textrm{g cm}^{-3}$. This effect was also observed in Ref.~\onlinecite{MBVSM21} for Aluminium and Carbon. As a result, the number of unbound electrons with this boundary condition actually decreases (and seems to be approaching a value of zero) with increasing density. This results in decreasing pressure, because (for example) the kinetic energy decreases with a lower ionization degree, and thus the free energy decreases as the mass density increases. \newchange{It should therefore be noted that the extreme divergences in pressure between the two boundary conditions would likely be suppressed to some degree if a different definition of pressure is used, which is not so directly influenced by the ionization degree. Regardless, decreasing pressures and ionization degrees with increasing densities} is a counter-intuitive and seemingly unphysical result, since we expect greater ionization and pressures at higher densities. Furthermore, the behaviour of the $1s$ energy level raises important questions related to the concept of ionization potential depression (continuum lowering), a critical effect in materials under WDM conditions \cite{C14}. 

It is well documented in both experiments \cite{CVCB12,HJBH13,FKPM14} and theoretical models \cite{EK63,SP66,LA94} that ionization potentials --- defined as the energy required to excite a given bound electron into the continuum --- are lower for atoms immersed in a plasma relative to the isolated atom case, though there remains uncertainty regarding the precise nature of this effect \cite{C14,I14}. In KS-AA models, it is typical to associate the KS orbital energies with the actual electronic energy levels, and by extension (as the continuum levels are usually defined as those with positive energy, $\epsilon>0$) the orbital energies define the ionization potentials. As an aside, we note that there is in fact no formal relationship between the KS orbital energies, which belong to a fictitious system of non-interacting electrons, and the actual electronic energy levels, with the exception of the HOMO level whose (negative) value is equal to the ionization potential in ground-state KS-DFT only \cite{PPLB_82,LevyPerdewSahni84,Yang12,PerdewLevy97}. Nevertheless, it has been postulated that the KS orbitals are a reasonable surrogate for the molecular orbitals of the real interacting system \cite{SH99,HDCS02}, which justifies to some extent the association of the KS orbital energies with the real electronic energy levels. 

In light of the above, the density dependence of the $\epsilon_{1s}$ energy for the boundary condition $\rnld[\RVS]=0$ seems to be a strange result. This, and likewise the behaviour of the number of unbound electrons, points to the limitations of models such as our own which only take into account screening effects from the surrounding plasma in a coarse manner through boundary conditions on the density or potential, neglecting all explicit interactions between charge densities in the central sphere and its neighbours. In this sense, it is possible that the approximate XC functionals actually benefit from error cancellation within the model we define, as their errors relative to the exact XC functional may be partially compensated for by the opposing error induced by the neglect of inter-cell interactions.  However, this error cancellation is not omnipresent (it will not occur in the low-density limit for example); moreover, when more advanced models are constructed which include to a greater extent the effects of inter-cell interactions, the approximate XC functionals will no longer benefit from this error cancellation.

\begin{figure}
    \centering
    \includegraphics{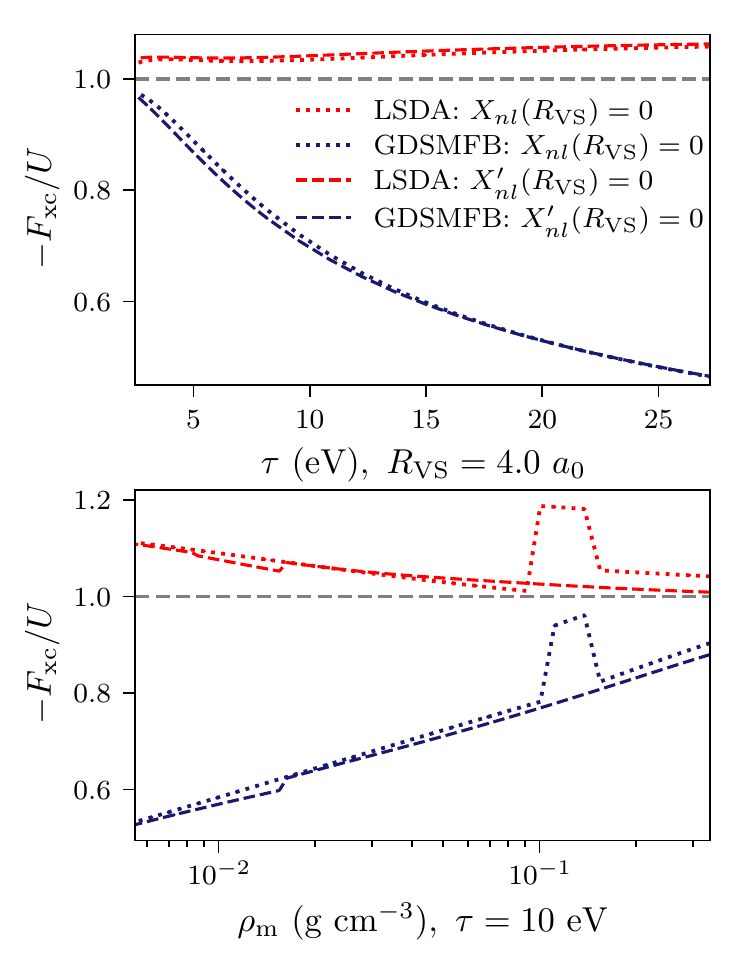}
    \caption{The ratio of XC energy to Hartree energy for Hydrogen, as a function of temperature (top) and as a function of density (bottom). The ratio is exactly equal to unity if the exact exchange functional is used, $-F_\textrm{xc}/U=1$ (shown in gray dashed line in the figure); when this value deviates from one it is a consequence of the SI error.}
    \label{fig:E_sic}
\end{figure}

\begin{figure}
    \centering
    \includegraphics{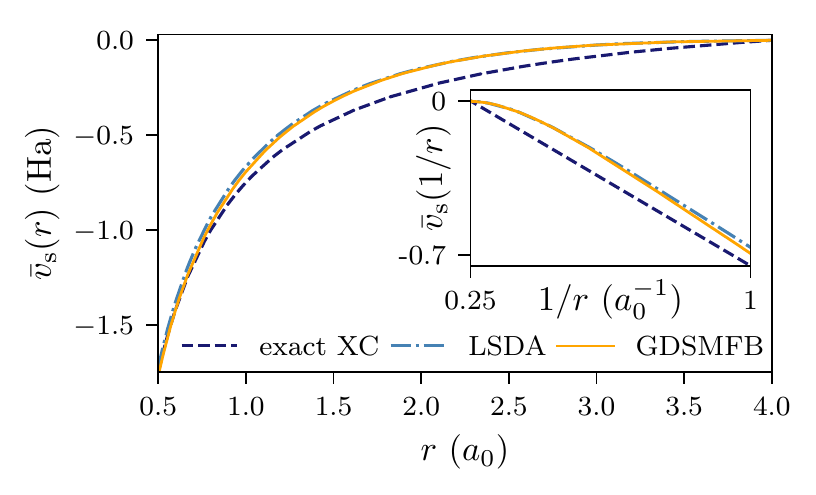}
    \caption{KS potentials for the exact XC ($\vs(r) = v_\textrm{en}(r)$), LSDA and GDSMFB functionals, for Hydrogen with density $\rho_\textrm{m}=0.042\ \textrm{g cm}^{-3}$ and temperature $\tau=8.16\ \textrm{eV}$, under the b.c. $\rnl[][\RVS]=0$. Inset: potentials plotted against $1/r$ to emphasize different asymptotic behaviours ($1/r\to1/\RVS$).}
    \label{fig:v_sic}
\end{figure}

Although it is clear that the choice of boundary condition typically far outweighs any error than the choice of XC functional, especially at higher densities, it is worth analysing in more detail the functional errors. The systematic under-binding of the electron density and over-prediction of the $1s$ energy level are both related to the SIE discussed in \S~\ref{subsec:theory_pt_3}.  We further analyse the SIE in Fig.~\ref{fig:E_sic}. Since there is only one electron 
in Hydrogen, the exact XC energy should exactly cancel the Hartree energy and therefore the ratio $-F_\textrm{xc}/U_\textrm{H}$ should equal 1. Interestingly, it appears that the GDSMFB functional is contaminated by a larger SI error (which increases with temperature and decreases with density) than the zero-temperature LSDA functional.
This may explain why the GDSMFB functional seems to yield slightly larger errors, relative to the exact XC result, for the electron pressure in Fig.~\ref{fig:H_EOS}. Of course, the SI error is particularly important for Hydrogen in the AA model, and may be overwhelmed by other factors in different examples. Nevertheless, this figure does not explain the failings of the LSDA functional relative to the exact result for smaller values of $\RVS$, since the ratio $-F_\textrm{xc}/E_\textrm{xc}$ does not significantly deviate from unity.

One of the ramifications of the SI error is the incorrect asymptotic behaviour of the KS potential, which (among other factors) contributes to the electron density being too delocalized; this error, known as the \emph{delocalization} error, is ubiquitous to (semi)-local XC functionals in DFT \cite{CMSY08,MSCY08}. In Fig.~\ref{fig:v_sic} we show an example of how both the LSDA and GDSMFB potentials differ from the exact result,with most notable differences in the asymptotic region in which they decay incorrectly. In \S~\ref{sec:discussion}, we consider some possibilities to mitigate against SI and delocalization errors.

\subsection{Beryllium}\label{subsec:resultsB}

We now apply our AA model to the Beryllium atom. Beryllium is used in ICF capsules \cite{LABGGHKLS04} and relevant to astrophysical processes \cite{AMO01}, and thus accurate simulations of Beryllium under WDM conditions are of high interest \cite{PSTDFDLGR12,LLZQQZY14}. Although there is no benchmark for the XC functional as in the case of Hydrogen, it is interesting nevertheless to compare choices of boundary conditions and XC functionals.


\begin{figure}
    \centering
    \includegraphics{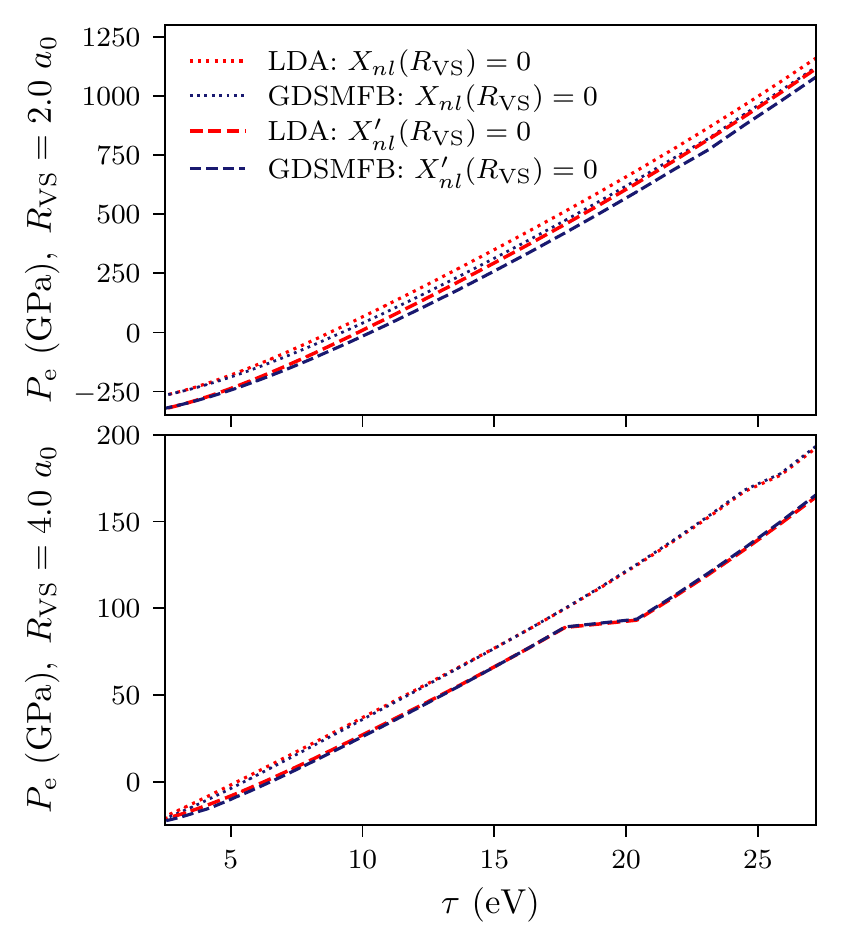}
    \caption{EOS data ($P_\textrm{e}$ vs $\tau$) for Beryllium for (i) $\RVS=2.0\ a_0$ (top) and (ii) $\RVS=4.0\ a_0$ (bottom). Both boundary conditions and the LDA and GDSMFB functionals are shown for comparison.}
    \label{fig:Be_EOS}
\end{figure}

\begin{figure}
    \centering
    \includegraphics{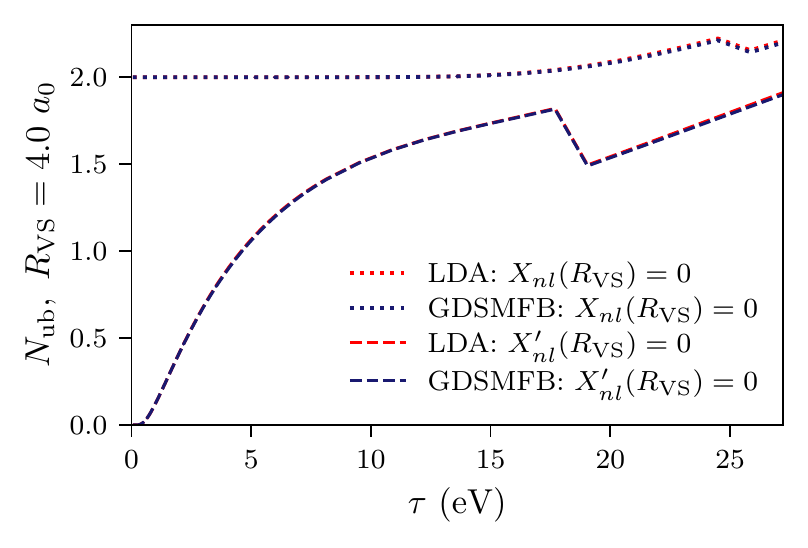}
    \caption{Number of unbound electrons for Beryllium for $\RVS=4.0\ a_0$, as a function of temperature. Both boundary conditions and the LDA and GDSMFB functionals are shown for comparison.}
    \label{fig:Be_N_ub}
\end{figure}

\begin{table}[h]
    \centering
    \begin{tabular}{ccccc}
    \toprule
    & \multicolumn{2}{c}{LDA} & \multicolumn{2}{c}{GDSMFB}\\
    \cmidrule(lr){2-3}\cmidrule(lr){4-5}
     $\T\ (\textrm{eV})$ & b.c. (i) & b.c. (ii) & b.c. (i) & b.c. (ii) \\
    \midrule
    & \multicolumn{4}{c}{$\epstilde[1s][]\ (\textrm{eV})$} \\
    \cmidrule(l){2-5}
    13.6	&	-104.6	&	-104.2	&	-106.0	&	-105.5	\\
    20.4	&	-108.3	&	-108.6	&	-109.8	&	-110.0	\\
    27.2	&	-117.3	&	-118.3	&	-118.8	&	-119.7	\\
    \midrule
    & \multicolumn{4}{c}{$\epstilde[2s][]\ (\textrm{eV})$} \\
    \cmidrule(l){2-5}   
    13.6	&	$>0$	&	-3.36	&	$>0$	&	-3.31	\\
    20.4	&	$>0$	&	-3.72	&	$>0$	&	-3.65	\\
    27.2	&	-0.74	&	-4.65	&	-0.57	&	-4.55	\\
    \midrule
    & \multicolumn{4}{c}{$\epstilde[2p][]\ (\textrm{eV})$} \\ 
    \cmidrule(l){2-5}   
    13.6	& \multicolumn{4}{c}{----------------- $>0$ -----------------}	\\
    20.4	&	$>0$	&	-0.14	&	$>0$	&	-0.18	\\
    27.2	&	$>0$	&	-1.00	&	$>0$	&	-1.00	\\
    \bottomrule
    \end{tabular}
    \caption{Comparison of KS orbital energies for Beryllium, with $\RVS=4.0\ a_0$.}
    \label{tab:Be_eigvals_funcs}
\end{table}

In Fig.~\ref{fig:Be_EOS}, we plot the pressure $P_\textrm{e}$ as a function of temperature $\tau$, 
for two values of the Voronoi sphere radius which correspond to mass densities of $3.04$ and $0.379\ \textrm{g cm}^{-3}$ respectively (for reference, the ambient solid density of Beryllium is $1.85\ \textrm{g cm}^{-3}$). Under these conditions, it seems the pressure for the two different boundary conditions is in relatively good agreement, though more significant for the lower density with $\RVS=4.0\ a_0$; furthermore, the LDA and GDSMFB functionals also agree very closely for both boundary conditions, with no observable differences between the functionals for the lower density. 

Next, in Fig.~\ref{fig:Be_N_ub}, we plot the mean ionization state $\Nubb$ for Beryllium with density $0.379\ \textrm{g cm}^{-3}$ as a function of temperature, again showing both boundary conditions and the LSDA and GDSMFB functionals in the same plot for comparison. Here, particularly at low temperatures, we see significant differences due to the boundary conditions, but the choice of functional has very little impact. The large deviation in the mean ionization state due to the choice of boundary conditions is explained by the eigenvalue spectrum. In Table~\ref{tab:Be_eigvals_funcs}, we see that the $2s$ energy level for the $\rnld[\RVS]=0$ boundary condition is consistently in the discrete part of the energy spectrum; by contrast, it is unbound up to around $\tau=25\ \textrm{eV}$ for the $\rnl[][\RVS]=0$ boundary condition. The eigenvalues in Table~\ref{tab:Be_eigvals_funcs} also explain the discontinuities in the $P_\textrm{e}$ vs $\T$ and $N_\textrm{ub}$ vs $\tau$ curves, since they arise when the $2s$ or $2p$ level (depending on the boundary condition) transitions from the continuum to the discrete part of the energy spectrum. 

The discontinuities observed in the pressure and number of unbound electrons relate to two limitations of the model. Firstly, the fact that we treat the unbound electron density as a constant means the physical problem being solved changes significantly when a new bound level emerges. In \S~\ref{subsec:results_3}, we shall explore the impact of treating the unbound electron density in a quantum manner, which should alleviate this problem. Secondly, our definition of ``unbound'' orbitals --- namely those orbitals with energies above the value of the KS potential at the sphere boundary --- is an oversimplification. In partially ionized plasmas like the ones we study, there may be some core states which are clearly bound to the nuclei, and likewise some clearly free electron density, but states with energies $\epsilon_{nl}^{\tau, \sigma}\sim \vs(\RVS)$ probably exhibit both bound and free characteristics and therefore cannot be neatly categorised as one or the other. More meaningful definitions for the mean ionization state than a simple energy threshold (which is commonly used in AA models) could make use of quantities such as the electron localization function \cite{SNWF97,FCS07}, the inverse participation ratio \cite{MWA11,GHPHWV20}, or electrical conductivity data \cite{BWSRDKGSR20}, but such an analysis is beyond the scope of this paper.

\begin{figure}
    \centering
    \includegraphics{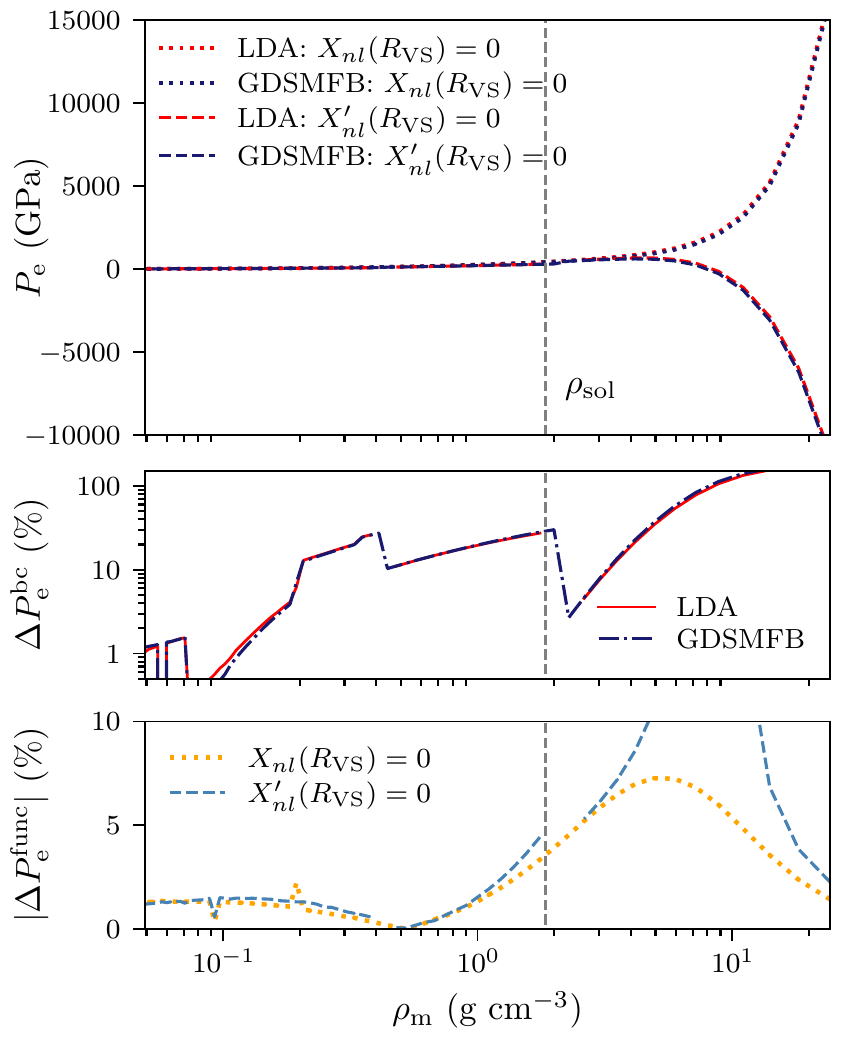}
    \caption{\change{Top: comparison of EOS ($P_\textrm{e}$ vs $\rho_\textrm{m}$) curves for Beryllium as a function of density, at fixed temperature $\T=20\ \textrm{eV}$, for both boundary conditions and LSDA and GDSMFB functionals (ambient density $\rho_\textrm{sol}$ indicated for reference). Middle: percentage difference (logarithmic scale for $y$--axis) in pressure between the two boundary conditions, $\Delta P^\textrm{bc}_\textrm{e}=(P^\textrm{bc(i)}_\textrm{e}-P^\textrm{bc(ii)}_\textrm{e})/P^\textrm{bc(i)}_\textrm{e}$. Bottom: absolute percentage difference in pressure between LSDA and GDSMFB results, $|\Delta P^\textrm{func}_\textrm{e}|=|P^\textrm{LSDA}_\textrm{e}-P^\textrm{GDSMFB}_\textrm{e}|/|P^\textrm{LSDA}_\textrm{e}|$.}}
    \label{fig:Be_dens_EOS}
\end{figure}

\begin{figure}
    \centering
    \includegraphics{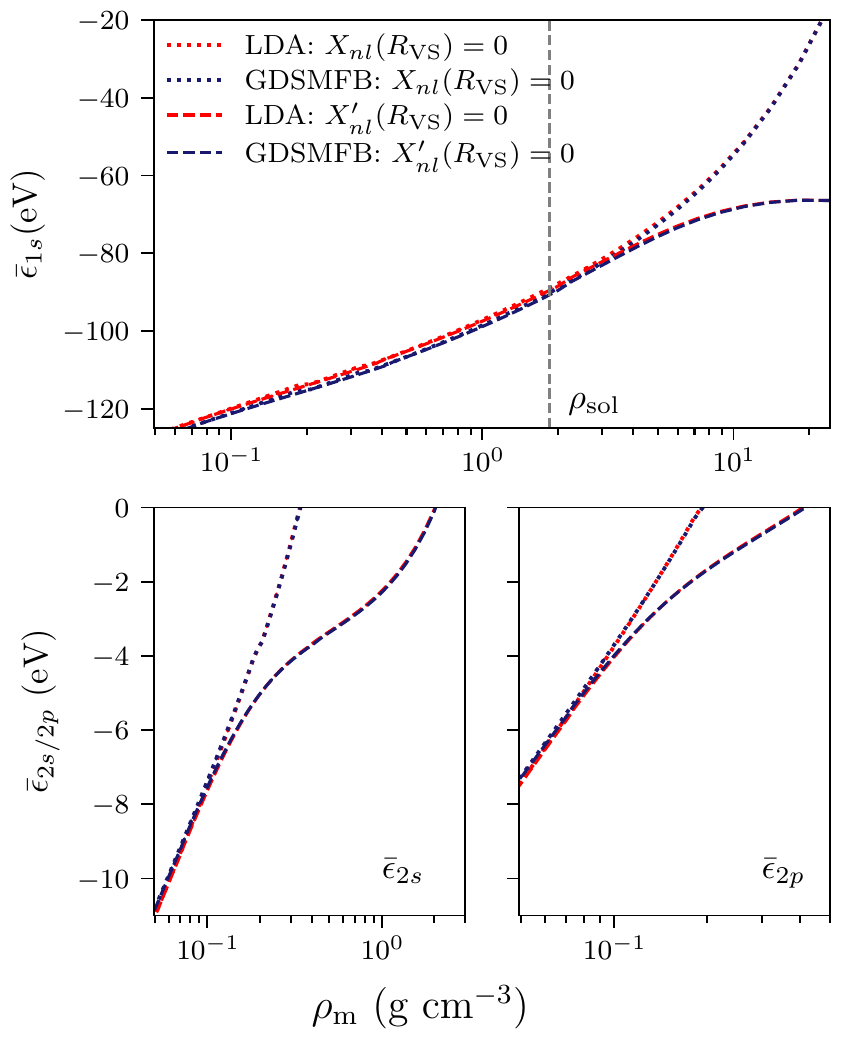}
    \caption{\change{Bound energy levels $\epstilde[1s][]$ (top), $\epstilde[2s][]$ (bottom left) and $\epstilde[2p][]$ (bottom right) for Beryllium as function of density, for fixed temperature $\T=20\ \textrm{eV}$ (ambient density $\rho_\textrm{s}$ shown for reference). Both boundary conditions and functionals are shown for comparison. Note the different scales on the $x$--axis, since the orbitals move into the continuum at different densities.}}
    \label{fig:Be_dens_eps}
\end{figure}

Next, we analyse the electronic pressure as a function of the mass density for fixed temperature $\tau=20\ \textrm{eV}$ in Fig.~\ref{fig:Be_dens_EOS}. The densities considered range from about $0.01$ to $10$ times the ambient solid density of Beryllium (indicated by $\rho_\textrm{sol}$ in the figure). The results in this plot are strongly reminiscent of what we observed for Hydrogen: namely, the two boundary conditions usually yield pressures within $<20\%$ of each other up to around twice the ambient density (at this temperature); after that, they diverge significantly, with very large negative pressures again observed for the $\rnld[\RVS]=0$ condition. The reasons for this unphysical behaviour were discussed already for Hydrogen. Comparison between the LSDA and GDSMFB functionals (shown in the lower panel of Fig.~\ref{fig:Be_dens_EOS}) indicate that the inclusion of temperature in the XC approximation is most important for Beryllium compressed to around 5 times its ambient density, with this being consistent for both boundary conditions. However, the functional effects are still dwarfed by the impact of the boundary condition.

Interestingly, almost the opposite effect is observed for the eigenvalues, shown in Fig.~\ref{fig:Be_dens_eps}. Here, we observe the functional has a small but non-trivial impact on the $\epstilde[1s][]$ level at densities lower than and including the ambient density. The boundary condition has very little impact on the $\epstilde[1s][]$ level up to that point, indicating these orbitals do not feel any effect from neighbouring spheres up to that density. The valence energy levels $\epstilde[2s][]$ and $\epstilde[2p][]$ separate at much lower densities, demonstrating that the choice of boundary condition is important for predicting ionization energies even for quite diffuse plasmas. The choice of functional has essentially no effect on these energy eigenvalues.

\subsection{Connection with other average-atom models}\label{subsec:results_3}

\change{In this section we make a connection to other existing AA models, by analysing how}
more sophisticated treatments of unbound electrons \change{and an alternative boundary condition} affect properties we \change{discussed so far.} 
We therefore hope to gain an understanding of how our analysis in previous sections might affect the development and usage of AA models more broadly. We use the MUZE code for this comparison.

\change{For the analysis of different treatments of unbound electrons}, we consider the transition
\begin{equation}
\nonumber
    \textrm{ideal (id)} \longrightarrow \textrm{Thomas--Fermi (TF)} \longrightarrow \textrm{quantum (qu)},
\end{equation}
where \textit{ideal} refers to the uniform approximation to unbound electrons we used so far, the TF unbound electron density \footnote{MUZE adopts a spin-restricted KS formalism, so both spin-up and spin-down orbitals share a common KS potential and are identical} is given by \cite{R72,FDM96}
\begin{equation}
    \n_\textrm{ub}(r) = \frac{\sqrt{2}}{\pi^2} \int^\infty_{-v_s^\T(r)}  \dd\vec{\epsilon} \frac{\epsilon^{1/2}}{1+e^{\beta(\epsilon - [\mu -v_s^\T(r)])}},
\end{equation}
and the \textit{quantum} unbound density is constructed in a similar way to the bound density \eqref{eq:KS_dens}, by solving explicitly the radial KS equations \eqref{eq:ks_radial} for continuum states $X_{\epsilon l}(r)$ discretized on the energy scale \cite{Li_1979,BI95}.
%

\change{The alternative boundary condition with which we compare does not impose any constraints on the bound KS orbitals at the edge of the Voronoi sphere. Instead, the KS potential is modified as follows,
\begin{equation} \label{eq:pot_cond}
    v_\textrm{s}^{\T,\textrm{M}}(r) = \Bigg\{\begin{array}{lr}
        \left(1-r/\RVS\right) v_\textrm{s}^\T(r), &  r\leq\RVS \\
        0, &  r>\RVS\\
        \end{array},
\end{equation}
where $v_\textrm{s}^{\T, \textrm{M}}(r)$ denotes that the KS potential is modified from its pure form $v_\textrm{s}^\T(r)$ as defined by Eq.~\eqref{eq:vs}. Using the above form for the KS potential, the radial KS orbitals are allowed to ``leak out'' of the Voronoi sphere and are thus computed up to an infinite radius, since they rapidly decay naturally to zero outside of the sphere. They are still normalized within the sphere according to \eqref{eq:KS_norm}. In MUZE, the unbound orbitals in the quantum treatment satisfy different boundary conditions which amount to continuity of the orbitals at the sphere's edge. Although the above modification to the potential is not strictly a boundary condition in the mathematical solution of the KS differential equations, we henceforth call it the ``potential condition'' because its role is essentially that of a boundary condition.}

For simplicity, we compare using only the LDA functional. Additionally, to make direct comparisons between the MUZE and ORCHID codes more straightforward, we switch from computing pressure with finite differences and instead use the following definition \cite{J00}:
\begin{equation} \label{eq:P_WJ}
    \tilde{P}_\textrm{e}(V,\tau)=\frac{2^{3/2}}{3\pi^2} \int_0^\infty \dd\vec{\epsilon} \frac{\epsilon^{3/2}}{1+e^{\beta(\epsilon - \mu)}}.
\end{equation}

\begin{figure}
    \centering
    \includegraphics{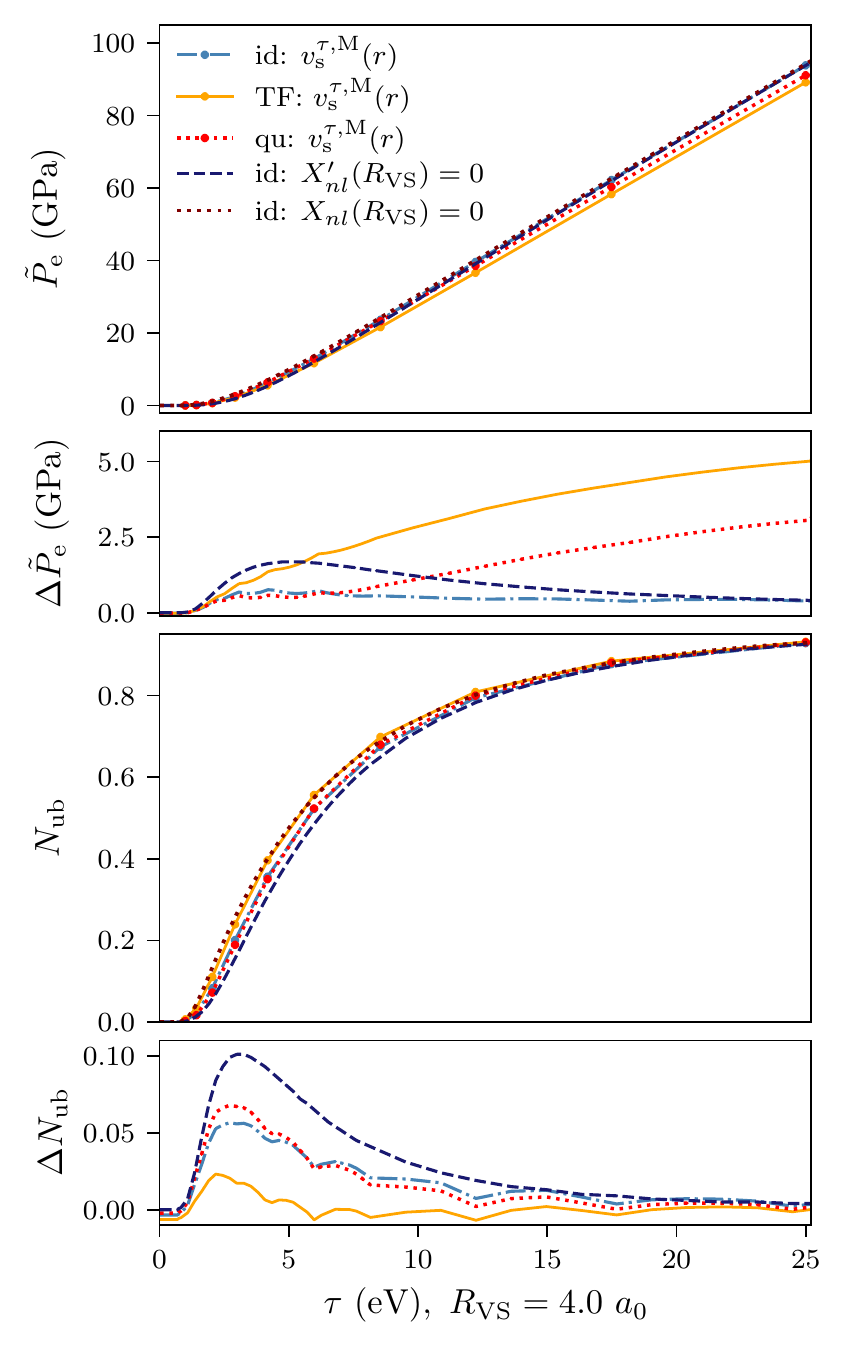}
    \caption{\change{Comparison of pressure $\tilde{P}_\textrm{e}$ (top) and number of unbound electrons $N_\textrm{ub}$ (second from bottom) for Hydrogen with $\RVS=4.0\ a_0$, for different boundary conditions and treatments of unbound electrons. The second from top panel shows the differences in $\tilde{P}_\textrm{e}$ relative to the result obtained from the b.c. \newchange{$\rnl[][\RVS]=0$}, and the bottom panel the equivalent difference for $N_\textrm{ub}$ (with differences between data points obtained via linear interpolation).}}
    \label{fig:H_MUZE_comp}
\end{figure}



\change{In Fig.~\ref{fig:H_MUZE_comp}, we compare the effect of both the boundary condition used and the treatment of unbound electrons for Hydrogen, on the pressure and MIS. Qualitatively, all results are in quite good agreement for this example, so we also plot the differences relative to the result obtained with ideal unbound electrons and the b.c. $\rnl[][\RVS]=0$, which we call the reference result. Specifically, $\Delta \tilde{P}_\textrm{e}=\tilde{P}^\textrm{ref}_\textrm{e} - \tilde{P}_\textrm{e}$, and $\Delta \Nubb = \Nubb^\textrm{ref} - \Nubb$, where the superscript ref denotes the reference result. We observe a few common trends from these plots. Firstly, the reference result is largely an upper bound for both the pressure and MIS. Additionally, the MIS results obtained via the potential condition seem to mostly lie in between the two orbital boundary conditions (regardless of the treatment of unbound electrons). Secondly, particularly with increasing temperature, the TF and quantum treatment of unbound electrons yields lower pressures than the ideal results, when compared with the same boundary condition on the potential or the orbital boundary conditions. Finally, especially at higher temperatures ($>10\ \textrm{eV}$), all approximations yield very similar predictions for the MIS.}

\begin{figure}
    \centering
    \includegraphics{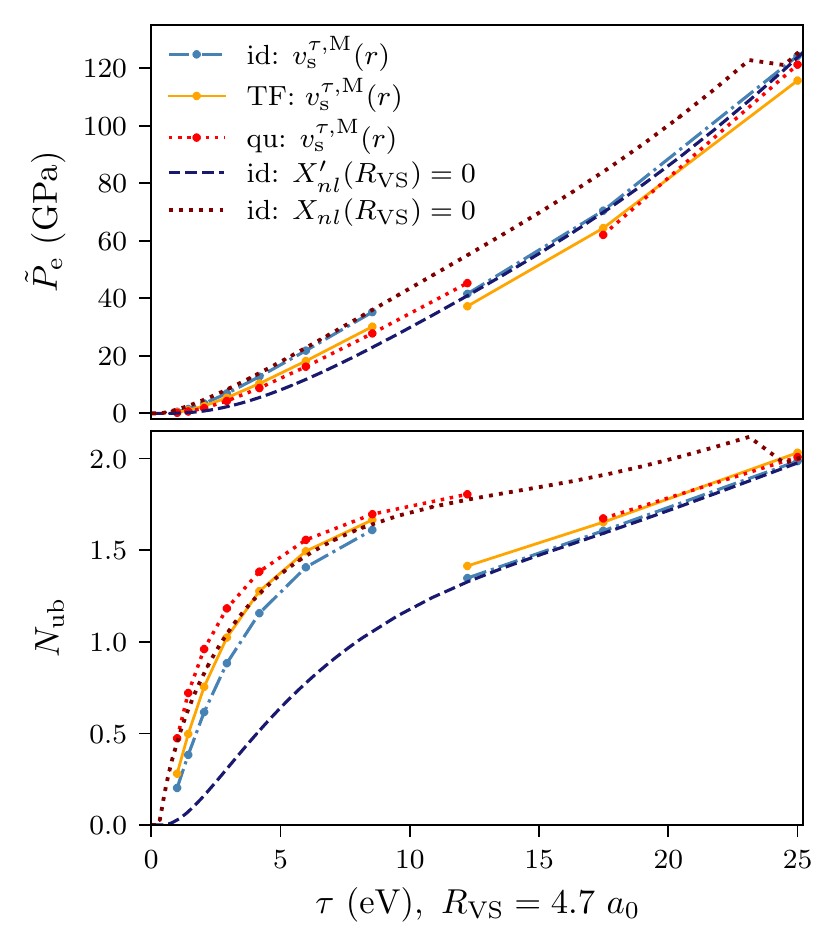}
    \caption{\change{Comparison of pressure $P_\textrm{e}$ (top) and number of unbound electrons $N_\textrm{ub}$ (second from bottom) for Beryllium with $\RVS=4.7\ a_0$, for different boundary conditions and treatments of unbound electrons. Gaps in plots are shown where there are discontinuities in the data.}}
    \label{fig:Be_MUZE_comp}
\end{figure}

\change{Next, in Fig.~\ref{fig:Be_MUZE_comp}, we plot an analogous set of results for Beryllium, this time with $\RVS=4.7\  a_0\ (\rho_\textrm{m}=0.232\ \textrm{g cm}^{-3}$). In this example, the differences between the various 
approximations are 
clearer. It is again the case that the boundary condition $\rnl[][\RVS]=0$ seems to consistently predict the highest pressures; also as was observed in Hydrogen, using the ideal approximation for the unbound electron density seems to yield higher pressures than the more advanced TF and quantum models. Of note here is that the choice of boundary condition, particularly for the MIS, seems to have more impact than the choice of treatment of the unbound electrons. Furthermore, it appears to be the case in this example that the MUZE solutions (using the potential boundary condition) jump from being roughly equal to solutions from our model using the b.c. $\rnl[][\RVS]=0$ to the b.c. $\rnld[\RVS]=0$ at a temperature around $10\sim15\ \textrm{eV}$.}


\change{The comparison of the $\epsilon_{2s}$ and $\epsilon_{2p}$ energy levels \footnote{\change{Here we denote the energy eigenvalues from both methods as $\epsilon_{nl}$. We stress that the ORCHID energy levels are still computed with the shifted potential $\bar{v}^{\T,\sigma}_\textrm{s}$, so in other words these are equal to the shifted levels $\epstilde$: we dropped the bar notation so as not to imply that MUZE shifts the potential by a constant. With either method, the KS potential used to determine the levels is equal to zero at the boundary so a direct comparison is appropriate.}} for Beryllium with $\RVS=4.7\ a_0$ in Table~\ref{tab:Be_MUZE_eigvals_comp} helps explain the MIS results seen in Fig.~\ref{fig:Be_MUZE_comp} and also (less directly) Fig.~\ref{fig:H_MUZE_comp}. In this table we observe that the eigenvalues given by the potential boundary condition always lie in between those given by the two orbital boundary conditions. This fits with the observation that the $\rnl[][\RVS]=0$ and $\rnld[\RVS]=0$ boundary conditions are respectively approximate upper and lower bounds for the MIS compared to the results obtained via the potential condition. Also in this table, it appears that changing the treatment of unbound electrons from ideal to quantum has only a small effect on the bound energy eigenvalues. 
}

\begin{table}[h]
    \centering
    \begin{tabular}{ccccc}
    \toprule
    & \multicolumn{3}{c}{ideal} & quantum\\
    \cmidrule(lr){2-4} \cmidrule(l){5-5}
     $\T\ (\textrm{eV})$ & b.c. (i) & b.c. (ii) & $v^{\T,\textrm{M}}_\textrm{s}$ &  $v^{\T,\textrm{M}}_\textrm{s}$\\
    \cmidrule(lr){1-4} \cmidrule(l){5-5}
    & \multicolumn{4}{c}{$\epsilon_{2s}\ (\textrm{eV})$} \\
    \cmidrule(l){2-5} 
    4.2	&	-1.27	&	-3.77	&	-2.6	&	-2.17	\\
    8.6	&	-1.70	&	-3.91	&	-2.76	&	-2.27	\\
    12.2	&	-1.86	&	-3.99	&	-3.16	&	-2.41	\\
    17.5	&	-2.31	&	-4.31	&	-3.3	&	-3.10	\\
    25.0	&	-4.01	&	-5.64	&	-4.73	&	-4.60	\\
    \midrule
    & \multicolumn{4}{c}{$\epsilon_{2p}\ (\textrm{eV})$} \\
    \cmidrule(l){2-5} 
    4.2	&	$>0$	&	-0.53	&	$>0$	&	$>0$	\\
    8.6	&	$>0$	&	-0.65	&	$>0$	&	$>0$	\\
    12.2	&	$>0$	&	-0.73	&	-0.00	&	$>0$	\\
    17.5	&	$>0$	&	-1.00	&	-0.11	&	-0.025	\\
    25.0	&	-0.162	&	-2.18	&	-1.18	&	-1.05	\\
    \bottomrule
    \end{tabular}
    \caption{\change{Comparison of KS orbital energies for Beryllium, with $\RVS=4.7\ a_0$, for different boundary conditions and treatments of unbound electrons.}}
    \label{tab:Be_MUZE_eigvals_comp}
\end{table}

\change{Based on the above results, the main conclusions that can be drawn are (i) that the ideal approximation has a tendency to overestimate pressure, particularly as the ionization degree increases, and (ii) that the potential boundary condition \eqref{eq:pot_cond} seems to yield results somewhere between the two orbital conditions \eqref{eq:bc1}, \eqref{eq:bc2}. In future works, it would be interesting to explore the nature of this relationship at higher densities, when the two orbital conditions yield significantly diverging results. It would also be insightful to compare the potential conditions with models, such as that developed in Ref.~\onlinecite{MBVSM21}, which use both orbital boundary conditions together. Eventually, evaluating the accuracy of different boundary conditions should be guided by comparisons with more advanced theoretical models, such as KS-DFT-MD models, neutral pseudo-atom AA models that constrain the potential within a correlation rather than Voronoi sphere \cite{SS_14,DKR_20}, and experimental results.}


\section{Discussion and Summary}\label{sec:discussion}

In this paper, we presented a fully first-principles derivation of a KS-AA model --- starting with the fully-interacting, many-body Hamiltonian of electrons and nuclei --- and ending with finite-temperature radial KS equations. We methodically considered the underlying assumptions and the interactions that are neglected in this model, yielding insight into the density and temperature limits under which the AA approximation is expected to be accurate. This analysis already yields some ideas regarding future directions for improving AA models: for example, one could go beyond the Born--Oppenheimer approximation and include non-adiabatic effects using the exact factorization method \cite{AMG10,GG14,CRG18}. Furthermore, through the inclusion (exact or approximate) of higher order terms in the perturbative expansion of the coupling terms $\hat{W}_J$ \eqref{eq:W_J}, there are possibilities to systematically improve AA models.

In our model, we impose the intuitive criterion that the KS density (which is formally equal to the real electronic density) must be smooth at the Voronoi sphere boundary, Eq.~\eqref{eq:bc_ks}. Imposing this criterion leads to the relation \eqref{eq:bc_ks_orbs} for the KS orbitals, which has no unique solution. We considered two options for satisfying this criterion --- Eqs.~\eqref{eq:bc1} and \eqref{eq:bc2} --- in our model, because they are the most simple options and are frequently imposed in AA models. 

We observed that these different boundary conditions had a significant impact on results, particularly for higher densities and lower temperatures (as expected), echoing the observation in Ref.~\onlinecite{JNC12}. This implies that AA models should carefully consider the choice of boundary condition under these limits; even better, further developing the analysis by Rozsynai \cite{R72,R91} (that these conditions represent band-structure limits) might remove the need to choose one particular condition, and instead one could envisage a scheme (such as the one applied in Refs.~\onlinecite{M94,PMC05,MBVSM21}) that interpolates between the two. \change{The ultimate goal is to deduce a more accurate boundary condition (or set of boundary conditions) from first-principles, perhaps by considering the effect of the terms neglected during the reduction of the Hamiltonian in \S~\ref{subsec:theory_pt_1}, such as the inhomogenous component $\hat{H}_\textrm{in}$, or the higher order terms in the perturbative expansion of the average component $\hat{H}_\textrm{av}$.}


We also compared results given by \change{these boundary conditions \eqref{eq:bc1} with an alternative condition on the potential \eqref{eq:pot_cond}, which is used in the MUZE code}; there were also significant differences between these results, further emphasizing the importance of boundary conditions in AA models and hence the benefit of further investigation on this subject.

\change{Furthermore, we also investigated the impact of using different approximations for the XC functional. In the case of Hydrogen, we compared with an `exact' benchmark ($F_\textrm{xc}[n] = -U[n]$). These comparisons indicated that some well-known errors of ground-state DFT, namely the self-interaction and delocalization errors, also affect the prediction of properties in our finite-temperature AA model, particularly the frontier energy levels. To mitigate the impact of these errors, one can borrow from the abundance of solutions suggested in ground-state KS-DFT, such as using hybrid functionals \cite{MSCY08,Perdew2801}. Whilst recognizing that the functional choice is generally much less significant than the choice of boundary condition in our AA model, existing temperature-dependent functionals (such as the GDSMFB functional \cite{GDSMFB17} which we have tested) do not offer serious improvement compared to the standard LSDA and PBE functionals. This motivates the development and use of more advanced temperature-dependent functionals in finite-temperature KS-DFT. Furthermore, the results in this paper offer some support for the observation in Ref.~\onlinecite{SJB16}, namely that going beyond semi-local approximations may be more important than including explicit temperature-dependence in XC functionals.}

To conclude, AA models are a crucial tool in the simulation of materials under WDM conditions. Their computational efficiency  not only facilitates calculations over large temperature and density ranges, but also offers an avenue to incorporate advanced features such as non-adiabatic or non-equilibrium effects; such effects are likely to be important in the WDM regime but are too complex to be included in full KS-DFT codes. The first-principles derivation and results presented in this paper yield insights regarding potential limitations of KS-DFT AA models: by understanding these limitations, and systematically improving the underlying approximations, there is scope to even further increase the usefulness of AA models.



\section*{Data availablity}

The data for all the figures in the paper can be downloaded from Ref.~\cite{paper_data}.

\section*{Acknowledgments}
We are grateful to Hardy Gross, \change{G\'erard Massacrier and Sang--Kil Son} for useful and detailed discussions. T.C. also thanks Zhandos Moldabekov \change{and Nikitas Gidopoulos} for helpful comments.
Part of A.C.'s initial work on this manuscript was supported by Sandia's Laboratory Directed Research and Development Project No. 200202. Sandia National Laboratories is a multimission laboratory managed and operated by National Technology and Engineering Solutions of Sandia, LLC., a wholly owned subsidiary of Honeywell International, Inc., for the U.S. Department of Energy’s National Nuclear Security Administration under contract DE--NA--0003525. This paper describes objective technical results and analysis. Any subjective views or opinions that might be expressed in the paper do not necessarily represent the views of the U.S. Department of Energy or the United States Government.
This work was partly funded by the Center for Advanced Systems Understanding (CASUS) which is financed by Germany's Federal Ministry of Education and Research (BMBF) and by the Saxon Ministry for Science, Culture and Tourism (SMWK) with tax funds on the basis of the budget approved by the Saxon State Parliament.

\begin{appendices}

\section{Derivation of inter--cell coupling terms $\hat{W}_J$ in many--body Hamiltonian}
\label{App:WJ}

We wish to expand the terms composing $\hat{W}_J$,
\begin{align}
\hat{W}_J&=1-\frac{1}{Z^2} \sum_{i=1}^{\Ne}\Big[{W}_{iJ}^a + \sum_{j=1}^{\Ne}{W}_{ijJ}^b\Big],\ \textrm{with} \\
    \hat{W}^a_{iJ}&=\frac{2Z}{\sqrt{1-2\hat{{\vec{R}}}_J\cdot\yiJ+|\yiJ|^2}} \\
    \hat{W}^b_{ijJ}&=- \frac{1}{\sqrt{1-2\hat{{\vec{R}}}_J\cdot(\yiJ-\yiJ[j])+(\yiJ-\yiJ[j])^2}}
\end{align}
in powers of $|\yiJ|$. We recall first the binomial expansion for $1/\sqrt{1+\epsilon}$,
\begin{equation}
    \frac{1}{\sqrt{1+\epsilon}} = 1 - \frac{1}{2}\epsilon + \frac{3}{8}\epsilon^2 - \frac{5}{{16}}\epsilon^3 + \mathcal{O}(\epsilon^4);
\end{equation}
we shall expand $\hat{W}_J$ up to second order only in $|\yiJ|$. We henceforth use the notation $\yi=\yiJ$ for simplicity. The expansions (ignoring higher order terms) for $\hat{W}_J$ are thus
\begin{align}
    \hat{W}_{iJ}^a =& Z\Big\{2 - \left[|\yi|^2 -2\hat{\vec{R}}_J\cdot \yi  \right] + \frac{3}{4}\left[|\yi|^2 -2\hat{\vec{R}}_J \cdot\yi  \right]^2\Big\} \\
    =& Z\Big\{2 + 2\hat{\vec{R}}_J \cdot\yi + \left[3(\hat{\vec{R}}_J \cdot\yi)^2 - |\yi|^2 \right]\Big\}\label{eq:Waij1} \\
    \nonumber
    =&\Bigg\{ (Z-\Ne)+\isum[j]\Bigg\}\Bigg\{2 + 2\hat{\vec{R}}_J \cdot\yi  \\
    \label{eq:Waij2}
    &\hspace{6em}+\left[ 3(\hat{\vec{R}}_J \cdot\yi)^2 -|\yi|^2\right]\Bigg\}\,,
\end{align}
where we have adopted a form that will be more convenient for expansions in going from Eq.~\eqref{eq:Waij1} to Eq.~\eqref{eq:Waij2}, using the fact that $\isum[j]=\Ne$. We can expand $\hat{W}_{ijJ}^b$ in a very similar manner,
\begin{align}
    \hat{W}_{ijJ}^b =& -\Bigg\{1 + \hat{\vec{R}}_J \cdot (\yi - \yi[j])  \\
    &+\frac{1}{2} \left[  3(\hat{\vec{R}}_J \cdot [\yi - \yi[j]])^2 - (\yi -\yi[j])^2  \right]\Bigg\}.
\end{align}

We now group terms of the same order in $|\yiJ|$ together. We start with the zeroth--order term $\hat{W}_J^{(0)}$,
\begin{align}
    \hat{W}_J^{(0)}&=-\frac{1}{Z^2}\isum\Bigg\{2(Z-\Ne)+2\isum[j]-\isum[j]\Bigg\}\\
    &=1 - \frac{\Ne(2Z-\Ne)}{Z^2}\\
    &=\left(\frac{Z-\Ne}{Z}\right)^2\,.
\end{align}
This term vanishes for charge neutral systems, $Z=\Ne$. Next, we consider the first--order term $\hat{W}_J^{(1)}$,
\begin{align}
\nonumber
    \hat{W}_J^{(1)} & =-\frac{1}{Z^2} \sum_{i=1}^{\Ne}\Bigg \{  2(Z-\Ne)\hat{\vec{R}}_J \cdot \yi \\
    &\hspace{3em} +2\isum[j]\hat{\vec{R}}_J \cdot \yi -\isum[j] \hat{\vec{R}}_J\cdot (\yi - \yi[j]) \Bigg\}\\
    &=-\frac{1}{Z^2} \hat{\vec{R}}_J \cdot\Bigg\{ \sum_{i=1}^{\Ne} 2(Z-\Ne)\ \yi
   + \isum[i,j](\yi + \yi[j])\Bigg\}\\
     &=-\frac{1}{Z^2} \hat{\vec{R}}_J \cdot \sum_{i=1}^{\Ne} \Bigg\{2(Z-\Ne)\ \yi
   + 2\isum[j]\yi\Bigg\}\\
   &=-\frac{2}{Z^2} \hat{\vec{R}}_J \cdot Z\sum_{i=1}^{\Ne}\yi\\
   &=-\frac{2}{Z}  \hat{\vec{R}}_J \cdot \hat{\vec{Y}}_J\,,
\end{align}
where we used the notation $\hat{\vec{Y}}_J=\sum_{i=1}^{\Ne}\vec{y}_{iJ}$.

\begin{widetext}
Finally, we consider the second--order term $\hat{W}_J^{(2)}$,
\begin{align}
    \hat{W}_J^{(2)}&=-\frac{1}{Z^2}\isum \Bigg\{ (Z-\Ne)\left[ 3(\hat{\vec{R}}_J \cdot\yi)^2 -|\yi|^2\right]+ \isum[j]\left(\left[ 3(\hat{\vec{R}}_J \cdot\yi)^2 -|\yi|^2\right]-\frac{1}{2}\left[  3(\hat{\vec{R}}_J \cdot [\yi - \yi[j]])^2 - (\yi -\yi[j])^2  \right]\right)\Bigg\}\\
    &=-\frac{1}{Z^2}\Bigg\{ (Z-\Ne)\isum \left[3(\hat{\vec{R}}_J \cdot\yi)^2 -|\yi|^2\right] \nonumber \\
   &\hspace{5em}+ \isum[i,j]\left( 3(\hat{\vec{R}}_J \cdot\yi)^2 + \frac{1}{2}\left[ |\yi|^2 + |\yi[j]|^2 - 2\yi\cdot\yi[j]\right] -\frac{3}{2}\left[ \RJ \cdot (\yiJ-\yiJ[j]) \right]^2\right)\Bigg\}\\
   &=-\frac{1}{Z^2}\Bigg\{ (Z-\Ne)\isum \left[3(\hat{\vec{R}}_J \cdot\yi)^2 -|\yi|^2\right] \nonumber \\
   &\hspace{5em}+\isum[i,j] \left[\frac{|\yi[j]|^2}{2} -\frac{|\yi|^2}{2}
    +\frac{3}{2}(\RJ\cdot\yi)^2 - \frac{3}{2}(\RJ\cdot\yi[j])^2-\yi\cdot\yi[j] + 3(\RJ\cdot\yi)(\RJ\cdot\yi[j])\right]\Bigg\}\\
    &=\frac{Z-\Ne}{Z^2}\isum\left[|\yi|^2-3(\hat{\vec{R}}_J \cdot\yi)^2\right] + \frac{1}{Z^2} \sum_{i,j=1}^{\Ne} \yiJ[j]\cdot \left[\yi - 3 \hat{\vec{R}}_J(\hat{\vec{R}}_J \cdot\yi)\right] \\
    &=\frac{Z-\Ne}{Z^2}\isum\left[|\yi|^2-3(\hat{\vec{R}}_J \cdot\yi)^2 \right] + \frac{1}{Z^2} \left[|\vec{Y}_J|^2 - 3(\hat{\vec{R}}_J\cdot\vec{Y}_J)^2\right]\,.
\end{align}
\end{widetext}
The first term in square brackets vanishes for $Z=\Ne$. This completes our derivation of the coupling terms up to second--order in $\hat{W}_J$.

\section{Construction of free energy in our AA model}

In finite-temperature KS-DFT, the free energy is equal to
\begin{equation}
    F[n] = E[n] - \tau S[n],
\end{equation}
where $S[n]$ is the (non-interacting) entropy and $E[n]$ is the internal energy functional,
\begin{equation}
    E[n] = T_\textrm{s}[n] + E_\textrm{en}[n] + U[n] + E_\textrm{xc}[n].
\end{equation}
In the above, $T_\textrm{s}[n]$ denotes the KS kinetic energy, $E_\textrm{en}[n]$ the electron-nuclear attraction energy, $U[n]$ the Hartree energy and $E_\textrm{xc}[n]$ the XC energy.

In our AA model, the unbound electron density is given by the ideal approximation and thus the usual orbital-based expressions for the KS kinetic energy and entropy cannot be applied. The kinetic energy and entropy are therefore split into bound and unbound components as follows,
\begin{align}
    T_\textrm{s}[n] &= T_\textrm{s}^\textrm{b}[\{\phi_i\}] + T_\textrm{s}^\textrm{ub} \\
   S[n] &= S^\textrm{b}[\{\phi_i\}] +S^\textrm{ub}\,,
\end{align}
where the superscripts b and ub denote bound and unbound terms respectively. In our AA model, these components are computed as
\pagebreak
\begin{align}
\nonumber
    T_\textrm{s}^\textrm{b}[\{\phi_i\}] &= - 2\pi  \sum_\sigma \sum_{l,n} (2l+1) f_{nl}^\sigma \\
    &\hspace{4em}\times\int_0^{\RVS} \dd{r} r^2 \rnl[\sigma] \dv[2]{\rnl[\sigma]}{r}\,,\\
    T_\textrm{s}^\textrm{ub} &= \sum_\sigma \frac{\Ne^\sigma V}{2^{1/2}\pi^2} \int_{0}^\infty\dd\vec{\epsilon} \frac{\epsilon^{3/2}}{1+e^{\beta(\epsilon - \mu^\sigma)}}\,,\\
    S^\textrm{b}[\{\phi_i\}] &= -\sum_{\sigma}\sum_{l,n} (2l+1) 
    \big[ f_{nl}^\sigma \log(f_{nl}^\sigma) \nonumber\\
    &\hspace{6em}+ (1-f_{nl}^\sigma) (\log(1-f_{nl}^\sigma) \big]\,, \\
    S^\textrm{ub} &= \sum_\sigma \frac{\Ne^\sigma V}{2^{1/2}\pi^2} 
    \int_{0}^\infty\dd\vec{\epsilon} \epsilon^{1/2} \big[f_{\epsilon}^\sigma\log{(f_{\epsilon}^\sigma)}\nonumber\\
    &\hspace{7em}+ (1-f_{\epsilon}^\sigma) \log(1-f_{\epsilon}^\sigma) \big]\,.
\end{align}

The remaining terms in the internal energy $E[n]$ take as input the full density (i.e. the sum of the bound and unbound components). In the AA model, these are given by
\begin{align}
     E_\textrm{en}[n] &= -4 \pi Z \int_0^{\RVS} \dd{r} r n(r)\,, \\
     U[n] &= \frac{1}{2}(4\pi)^2 \int_0^{\RVS} \dd{r} r^2 n(r)\int_0^{\RVS} \dd{x} \frac{\n(x)x^2}{r^>(x)}\,,\\
     E_\textrm{xc}[n] &= 4\pi  \int_0^{\RVS} \dd{r} r^2 e_\textrm{xc}[n^\uparrow,n^\downarrow](r)n(r)\,,
\end{align}
where $e_\textrm{xc}[n^\uparrow,n^\downarrow](r)$ is the XC energy density.

\end{appendices}


\begin{thebibliography}{212}%
\makeatletter
\providecommand \@ifxundefined [1]{%
 \@ifx{#1\undefined}
}%
\providecommand \@ifnum [1]{%
 \ifnum #1\expandafter \@firstoftwo
 \else \expandafter \@secondoftwo
 \fi
}%
\providecommand \@ifx [1]{%
 \ifx #1\expandafter \@firstoftwo
 \else \expandafter \@secondoftwo
 \fi
}%
\providecommand \natexlab [1]{#1}%
\providecommand \enquote  [1]{``#1''}%
\providecommand \bibnamefont  [1]{#1}%
\providecommand \bibfnamefont [1]{#1}%
\providecommand \citenamefont [1]{#1}%
\providecommand \href@noop [0]{\@secondoftwo}%
\providecommand \href [0]{\begingroup \@sanitize@url \@href}%
\providecommand \@href[1]{\@@startlink{#1}\@@href}%
\providecommand \@@href[1]{\endgroup#1\@@endlink}%
\providecommand \@sanitize@url [0]{\catcode `\\12\catcode `\$12\catcode
  `\&12\catcode `\#12\catcode `\^12\catcode `\_12\catcode `\%12\relax}%
\providecommand \@@startlink[1]{}%
\providecommand \@@endlink[0]{}%
\providecommand \url  [0]{\begingroup\@sanitize@url \@url }%
\providecommand \@url [1]{\endgroup\@href {#1}{\urlprefix }}%
\providecommand \urlprefix  [0]{URL }%
\providecommand \Eprint [0]{\href }%
\providecommand \doibase [0]{https://doi.org/}%
\providecommand \selectlanguage [0]{\@gobble}%
\providecommand \bibinfo  [0]{\@secondoftwo}%
\providecommand \bibfield  [0]{\@secondoftwo}%
\providecommand \translation [1]{[#1]}%
\providecommand \BibitemOpen [0]{}%
\providecommand \bibitemStop [0]{}%
\providecommand \bibitemNoStop [0]{.\EOS\space}%
\providecommand \EOS [0]{\spacefactor3000\relax}%
\providecommand \BibitemShut  [1]{\csname bibitem#1\endcsname}%
\let\auto@bib@innerbib\@empty
\bibitem [{\citenamefont {Graziani}\ \emph {et~al.}(2014)\citenamefont
  {Graziani}, \citenamefont {Desjarlais}, \citenamefont {Redmer},\ and\
  \citenamefont {Trickey}}]{GDRT14}%
  \BibitemOpen
  \bibinfo {editor} {\bibfnamefont {F.}~\bibnamefont {Graziani}}, \bibinfo
  {editor} {\bibfnamefont {M.~P.}\ \bibnamefont {Desjarlais}}, \bibinfo
  {editor} {\bibfnamefont {R.}~\bibnamefont {Redmer}},\ and\ \bibinfo {editor}
  {\bibfnamefont {S.~B.}\ \bibnamefont {Trickey}},\ eds.,\ \href@noop {} {\emph
  {\bibinfo {title} {Frontiers and Challenges in Warm Dense Matter}}},\
  \bibinfo {series} {Lecture Notes in Computational Science and Engineering},
  Vol.~\bibinfo {volume} {96}\ (\bibinfo  {publisher} {Springer International
  Publishing},\ \bibinfo {year} {2014})\BibitemShut {NoStop}%
\bibitem [{DOE(2009)}]{DOE09}%
  \BibitemOpen
  \href
  {https://science.osti.gov/-/media/fes/pdf/workshop-reports/Hedlp_brn_workshop_report_oct_2010.pdf}
  {\emph {\bibinfo {title} {Basic Research Needs for High Energy Density
  Laboratory Physics}}}\ (\bibinfo  {publisher} {U.S. DOE},\ \bibinfo {year}
  {2009})\BibitemShut {NoStop}%
\bibitem [{\citenamefont {Moses}\ \emph {et~al.}(2009)\citenamefont {Moses},
  \citenamefont {Boyd}, \citenamefont {Remington}, \citenamefont {Keane},\ and\
  \citenamefont {Al-Ayat}}]{MBRKA09}%
  \BibitemOpen
  \bibfield  {author} {\bibinfo {author} {\bibfnamefont {E.~I.}\ \bibnamefont
  {Moses}}, \bibinfo {author} {\bibfnamefont {R.~N.}\ \bibnamefont {Boyd}},
  \bibinfo {author} {\bibfnamefont {B.~A.}\ \bibnamefont {Remington}}, \bibinfo
  {author} {\bibfnamefont {C.~J.}\ \bibnamefont {Keane}},\ and\ \bibinfo
  {author} {\bibfnamefont {R.}~\bibnamefont {Al-Ayat}},\ }\bibfield  {title}
  {\bibinfo {title} {{The National Ignition Facility: Ushering in a new age for
  high energy density science}},\ }\href {https://doi.org/10.1063/1.3116505}
  {\bibfield  {journal} {\bibinfo  {journal} {Phys. Plasmas}\ }\textbf
  {\bibinfo {volume} {16}},\ \bibinfo {pages} {041006} (\bibinfo {year}
  {2009})}\BibitemShut {NoStop}%
\bibitem [{\citenamefont {Lindl}\ \emph {et~al.}(2004)\citenamefont {Lindl},
  \citenamefont {Amendt}, \citenamefont {Berger}, \citenamefont {Glendinning},
  \citenamefont {Glenzer}, \citenamefont {Haan}, \citenamefont {Kauffman},
  \citenamefont {Landen},\ and\ \citenamefont {Suter}}]{LABGGHKLS04}%
  \BibitemOpen
  \bibfield  {author} {\bibinfo {author} {\bibfnamefont {J.~D.}\ \bibnamefont
  {Lindl}}, \bibinfo {author} {\bibfnamefont {P.}~\bibnamefont {Amendt}},
  \bibinfo {author} {\bibfnamefont {R.~L.}\ \bibnamefont {Berger}}, \bibinfo
  {author} {\bibfnamefont {S.~G.}\ \bibnamefont {Glendinning}}, \bibinfo
  {author} {\bibfnamefont {S.~H.}\ \bibnamefont {Glenzer}}, \bibinfo {author}
  {\bibfnamefont {S.~W.}\ \bibnamefont {Haan}}, \bibinfo {author}
  {\bibfnamefont {R.~L.}\ \bibnamefont {Kauffman}}, \bibinfo {author}
  {\bibfnamefont {O.~L.}\ \bibnamefont {Landen}},\ and\ \bibinfo {author}
  {\bibfnamefont {L.~J.}\ \bibnamefont {Suter}},\ }\bibfield  {title} {\bibinfo
  {title} {The physics basis for ignition using indirect-drive targets on the
  {National Ignition Facility}},\ }\href {https://doi.org/10.1063/1.1578638}
  {\bibfield  {journal} {\bibinfo  {journal} {Phys. Plasmas}\ }\textbf
  {\bibinfo {volume} {11}},\ \bibinfo {pages} {339} (\bibinfo {year}
  {2004})}\BibitemShut {NoStop}%
\bibitem [{\citenamefont {Atzeni}\ and\ \citenamefont {Meyer-ter
  Vehn}(2004)}]{AM04}%
  \BibitemOpen
  \bibfield  {author} {\bibinfo {author} {\bibfnamefont {S.}~\bibnamefont
  {Atzeni}}\ and\ \bibinfo {author} {\bibfnamefont {J.}~\bibnamefont {Meyer-ter
  Vehn}},\ }\href@noop {} {\emph {\bibinfo {title} {The Physics of Inertial
  Fusion: Beam-Plasma Interaction, Hydrodynamics, Hot Dense Matter}}}\
  (\bibinfo  {publisher} {Clarendon Press},\ \bibinfo {year}
  {2004})\BibitemShut {NoStop}%
\bibitem [{\citenamefont {Matzen}\ \emph {et~al.}(2005)\citenamefont {Matzen},
  \citenamefont {Sweeney}, \citenamefont {Adams}, \citenamefont {Asay},
  \citenamefont {Bailey}, \citenamefont {Bennett}, \citenamefont {Bliss},
  \citenamefont {Bloomquist}, \citenamefont {Brunner}, \citenamefont
  {Campbell}, \citenamefont {Chandler}, \citenamefont {Coverdale},
  \citenamefont {Cuneo}, \citenamefont {Davis}, \citenamefont {Deeney},
  \citenamefont {Desjarlais}, \citenamefont {Donovan}, \citenamefont {Garasi},
  \citenamefont {Haill}, \citenamefont {Hall}, \citenamefont {Hanson},
  \citenamefont {Hurst}, \citenamefont {Jones}, \citenamefont {Knudson},
  \citenamefont {Leeper}, \citenamefont {Lemke}, \citenamefont {Mazarakis},
  \citenamefont {McDaniel}, \citenamefont {Mehlhorn}, \citenamefont {Nash},
  \citenamefont {Olson}, \citenamefont {Porter}, \citenamefont {Rambo},
  \citenamefont {Rosenthal}, \citenamefont {Rochau}, \citenamefont {Ruggles},
  \citenamefont {Ruiz}, \citenamefont {Sanford}, \citenamefont {Seamen},
  \citenamefont {Sinars}, \citenamefont {Slutz}, \citenamefont {Smith},
  \citenamefont {Struve}, \citenamefont {Stygar}, \citenamefont {Vesey},
  \citenamefont {Weinbrecht}, \citenamefont {Wenger},\ and\ \citenamefont
  {Yu}}]{MS05}%
  \BibitemOpen
  \bibfield  {author} {\bibinfo {author} {\bibfnamefont {M.~K.}\ \bibnamefont
  {Matzen}}, \bibinfo {author} {\bibfnamefont {M.~A.}\ \bibnamefont {Sweeney}},
  \bibinfo {author} {\bibfnamefont {R.~G.}\ \bibnamefont {Adams}}, \bibinfo
  {author} {\bibfnamefont {J.~R.}\ \bibnamefont {Asay}}, \bibinfo {author}
  {\bibfnamefont {J.~E.}\ \bibnamefont {Bailey}}, \bibinfo {author}
  {\bibfnamefont {G.~R.}\ \bibnamefont {Bennett}}, \bibinfo {author}
  {\bibfnamefont {D.~E.}\ \bibnamefont {Bliss}}, \bibinfo {author}
  {\bibfnamefont {D.~D.}\ \bibnamefont {Bloomquist}}, \bibinfo {author}
  {\bibfnamefont {T.~A.}\ \bibnamefont {Brunner}}, \bibinfo {author}
  {\bibfnamefont {R.~B.}\ \bibnamefont {Campbell}}, \bibinfo {author}
  {\bibfnamefont {G.~A.}\ \bibnamefont {Chandler}}, \bibinfo {author}
  {\bibfnamefont {C.~A.}\ \bibnamefont {Coverdale}}, \bibinfo {author}
  {\bibfnamefont {M.~E.}\ \bibnamefont {Cuneo}}, \bibinfo {author}
  {\bibfnamefont {J.-P.}\ \bibnamefont {Davis}}, \bibinfo {author}
  {\bibfnamefont {C.}~\bibnamefont {Deeney}}, \bibinfo {author} {\bibfnamefont
  {M.~P.}\ \bibnamefont {Desjarlais}}, \bibinfo {author} {\bibfnamefont
  {G.~L.}\ \bibnamefont {Donovan}}, \bibinfo {author} {\bibfnamefont {C.~J.}\
  \bibnamefont {Garasi}}, \bibinfo {author} {\bibfnamefont {T.~A.}\
  \bibnamefont {Haill}}, \bibinfo {author} {\bibfnamefont {C.~A.}\ \bibnamefont
  {Hall}}, \bibinfo {author} {\bibfnamefont {D.~L.}\ \bibnamefont {Hanson}},
  \bibinfo {author} {\bibfnamefont {M.~J.}\ \bibnamefont {Hurst}}, \bibinfo
  {author} {\bibfnamefont {B.}~\bibnamefont {Jones}}, \bibinfo {author}
  {\bibfnamefont {M.~D.}\ \bibnamefont {Knudson}}, \bibinfo {author}
  {\bibfnamefont {R.~J.}\ \bibnamefont {Leeper}}, \bibinfo {author}
  {\bibfnamefont {R.~W.}\ \bibnamefont {Lemke}}, \bibinfo {author}
  {\bibfnamefont {M.~G.}\ \bibnamefont {Mazarakis}}, \bibinfo {author}
  {\bibfnamefont {D.~H.}\ \bibnamefont {McDaniel}}, \bibinfo {author}
  {\bibfnamefont {T.~A.}\ \bibnamefont {Mehlhorn}}, \bibinfo {author}
  {\bibfnamefont {T.~J.}\ \bibnamefont {Nash}}, \bibinfo {author}
  {\bibfnamefont {C.~L.}\ \bibnamefont {Olson}}, \bibinfo {author}
  {\bibfnamefont {J.~L.}\ \bibnamefont {Porter}}, \bibinfo {author}
  {\bibfnamefont {P.~K.}\ \bibnamefont {Rambo}}, \bibinfo {author}
  {\bibfnamefont {S.~E.}\ \bibnamefont {Rosenthal}}, \bibinfo {author}
  {\bibfnamefont {G.~A.}\ \bibnamefont {Rochau}}, \bibinfo {author}
  {\bibfnamefont {L.~E.}\ \bibnamefont {Ruggles}}, \bibinfo {author}
  {\bibfnamefont {C.~L.}\ \bibnamefont {Ruiz}}, \bibinfo {author}
  {\bibfnamefont {T.~W.~L.}\ \bibnamefont {Sanford}}, \bibinfo {author}
  {\bibfnamefont {J.~F.}\ \bibnamefont {Seamen}}, \bibinfo {author}
  {\bibfnamefont {D.~B.}\ \bibnamefont {Sinars}}, \bibinfo {author}
  {\bibfnamefont {S.~A.}\ \bibnamefont {Slutz}}, \bibinfo {author}
  {\bibfnamefont {I.~C.}\ \bibnamefont {Smith}}, \bibinfo {author}
  {\bibfnamefont {K.~W.}\ \bibnamefont {Struve}}, \bibinfo {author}
  {\bibfnamefont {W.~A.}\ \bibnamefont {Stygar}}, \bibinfo {author}
  {\bibfnamefont {R.~A.}\ \bibnamefont {Vesey}}, \bibinfo {author}
  {\bibfnamefont {E.~A.}\ \bibnamefont {Weinbrecht}}, \bibinfo {author}
  {\bibfnamefont {D.~F.}\ \bibnamefont {Wenger}},\ and\ \bibinfo {author}
  {\bibfnamefont {E.~P.}\ \bibnamefont {Yu}},\ }\bibfield  {title} {\bibinfo
  {title} {Pulsed-power-driven high energy density physics and inertial
  confinement fusion research},\ }\href {https://doi.org/10.1063/1.1891746}
  {\bibfield  {journal} {\bibinfo  {journal} {Phys. Plasmas}\ }\textbf
  {\bibinfo {volume} {12}},\ \bibinfo {pages} {055503} (\bibinfo {year}
  {2005})}\BibitemShut {NoStop}%
\bibitem [{\citenamefont {Kritcher}\ \emph {et~al.}(2011)\citenamefont
  {Kritcher}, \citenamefont {D\"oppner}, \citenamefont {Fortmann},
  \citenamefont {Ma}, \citenamefont {Landen}, \citenamefont {Wallace},\ and\
  \citenamefont {Glenzer}}]{KDFMLWG11}%
  \BibitemOpen
  \bibfield  {author} {\bibinfo {author} {\bibfnamefont {A.~L.}\ \bibnamefont
  {Kritcher}}, \bibinfo {author} {\bibfnamefont {T.}~\bibnamefont {D\"oppner}},
  \bibinfo {author} {\bibfnamefont {C.}~\bibnamefont {Fortmann}}, \bibinfo
  {author} {\bibfnamefont {T.}~\bibnamefont {Ma}}, \bibinfo {author}
  {\bibfnamefont {O.~L.}\ \bibnamefont {Landen}}, \bibinfo {author}
  {\bibfnamefont {R.}~\bibnamefont {Wallace}},\ and\ \bibinfo {author}
  {\bibfnamefont {S.~H.}\ \bibnamefont {Glenzer}},\ }\bibfield  {title}
  {\bibinfo {title} {In-flight measurements of capsule shell adiabats in
  laser-driven implosions},\ }\href
  {https://doi.org/10.1103/PhysRevLett.107.015002} {\bibfield  {journal}
  {\bibinfo  {journal} {Phys. Rev. Lett.}\ }\textbf {\bibinfo {volume} {107}},\
  \bibinfo {pages} {015002} (\bibinfo {year} {2011})}\BibitemShut {NoStop}%
\bibitem [{\citenamefont {Hu}\ \emph {et~al.}(2011)\citenamefont {Hu},
  \citenamefont {Militzer}, \citenamefont {Goncharov},\ and\ \citenamefont
  {Skupsky}}]{HMGS11}%
  \BibitemOpen
  \bibfield  {author} {\bibinfo {author} {\bibfnamefont {S.~X.}\ \bibnamefont
  {Hu}}, \bibinfo {author} {\bibfnamefont {B.}~\bibnamefont {Militzer}},
  \bibinfo {author} {\bibfnamefont {V.~N.}\ \bibnamefont {Goncharov}},\ and\
  \bibinfo {author} {\bibfnamefont {S.}~\bibnamefont {Skupsky}},\ }\bibfield
  {title} {\bibinfo {title} {First-principles equation-of-state table of
  deuterium for inertial confinement fusion applications},\ }\href
  {https://doi.org/10.1103/PhysRevB.84.224109} {\bibfield  {journal} {\bibinfo
  {journal} {Phys. Rev. B}\ }\textbf {\bibinfo {volume} {84}},\ \bibinfo
  {pages} {224109} (\bibinfo {year} {2011})}\BibitemShut {NoStop}%
\bibitem [{\citenamefont {Betti}\ and\ \citenamefont {Hurricane}(2016)}]{BH16}%
  \BibitemOpen
  \bibfield  {author} {\bibinfo {author} {\bibfnamefont {R.}~\bibnamefont
  {Betti}}\ and\ \bibinfo {author} {\bibfnamefont {O.~A.}\ \bibnamefont
  {Hurricane}},\ }\bibfield  {title} {\bibinfo {title} {Inertial-confinement
  fusion with lasers},\ }\href {https://doi.org/10.1038/nphys3736} {\bibfield
  {journal} {\bibinfo  {journal} {Nat. Phys.}\ }\textbf {\bibinfo {volume}
  {12}},\ \bibinfo {pages} {435} (\bibinfo {year} {2016})}\BibitemShut
  {NoStop}%
\bibitem [{\citenamefont {Alf\`e}\ and\ \citenamefont {Gillan}(1998)}]{AG98}%
  \BibitemOpen
  \bibfield  {author} {\bibinfo {author} {\bibfnamefont {D.}~\bibnamefont
  {Alf\`e}}\ and\ \bibinfo {author} {\bibfnamefont {M.~J.}\ \bibnamefont
  {Gillan}},\ }\bibfield  {title} {\bibinfo {title} {First-principles
  calculation of transport coefficients},\ }\href
  {https://doi.org/10.1103/PhysRevLett.81.5161} {\bibfield  {journal} {\bibinfo
   {journal} {Phys. Rev. Lett.}\ }\textbf {\bibinfo {volume} {81}},\ \bibinfo
  {pages} {5161} (\bibinfo {year} {1998})}\BibitemShut {NoStop}%
\bibitem [{\citenamefont {Nguyen}\ and\ \citenamefont {Holmes}(2004)}]{NH04}%
  \BibitemOpen
  \bibfield  {author} {\bibinfo {author} {\bibfnamefont {J.~H.}\ \bibnamefont
  {Nguyen}}\ and\ \bibinfo {author} {\bibfnamefont {N.~C.}\ \bibnamefont
  {Holmes}},\ }\bibfield  {title} {\bibinfo {title} {Melting of iron at the
  physical conditions of the earth's core},\ }\href
  {https://doi.org/10.1038/nature02248} {\bibfield  {journal} {\bibinfo
  {journal} {Nature}\ }\textbf {\bibinfo {volume} {427}},\ \bibinfo {pages}
  {339} (\bibinfo {year} {2004})}\BibitemShut {NoStop}%
\bibitem [{\citenamefont {Remington}\ \emph {et~al.}(2006)\citenamefont
  {Remington}, \citenamefont {Drake},\ and\ \citenamefont {Ryutov}}]{RDR06}%
  \BibitemOpen
  \bibfield  {author} {\bibinfo {author} {\bibfnamefont {B.~A.}\ \bibnamefont
  {Remington}}, \bibinfo {author} {\bibfnamefont {R.~P.}\ \bibnamefont
  {Drake}},\ and\ \bibinfo {author} {\bibfnamefont {D.~D.}\ \bibnamefont
  {Ryutov}},\ }\bibfield  {title} {\bibinfo {title} {Experimental astrophysics
  with high power lasers and $z$ pinches},\ }\href
  {https://doi.org/10.1103/RevModPhys.78.755} {\bibfield  {journal} {\bibinfo
  {journal} {Rev. Mod. Phys.}\ }\textbf {\bibinfo {volume} {78}},\ \bibinfo
  {pages} {755} (\bibinfo {year} {2006})}\BibitemShut {NoStop}%
\bibitem [{\citenamefont {Fortov}(2009)}]{F09}%
  \BibitemOpen
  \bibfield  {author} {\bibinfo {author} {\bibfnamefont {V.~E.}\ \bibnamefont
  {Fortov}},\ }\bibfield  {title} {\bibinfo {title} {Extreme states of matter
  on earth and in space},\ }\href
  {https://doi.org/10.3367/ufne.0179.200906h.0653} {\bibfield  {journal}
  {\bibinfo  {journal} {Physics-Uspekhi}\ }\textbf {\bibinfo {volume} {52}},\
  \bibinfo {pages} {615} (\bibinfo {year} {2009})}\BibitemShut {NoStop}%
\bibitem [{\citenamefont {Nettelmann}\ \emph {et~al.}(2011)\citenamefont
  {Nettelmann}, \citenamefont {Fortney}, \citenamefont {Kramm},\ and\
  \citenamefont {Redmer}}]{NFKR11}%
  \BibitemOpen
  \bibfield  {author} {\bibinfo {author} {\bibfnamefont {N.}~\bibnamefont
  {Nettelmann}}, \bibinfo {author} {\bibfnamefont {J.~J.}\ \bibnamefont
  {Fortney}}, \bibinfo {author} {\bibfnamefont {U.}~\bibnamefont {Kramm}},\
  and\ \bibinfo {author} {\bibfnamefont {R.}~\bibnamefont {Redmer}},\
  }\bibfield  {title} {\bibinfo {title} {Thermal evolution and structure models
  of the transiting super-earth gj 1214b},\ }\href
  {http://stacks.iop.org/0004-637X/733/i=1/a=2} {\bibfield  {journal} {\bibinfo
   {journal} {Astrophys. J.}\ }\textbf {\bibinfo {volume} {733}},\ \bibinfo
  {pages} {2} (\bibinfo {year} {2011})}\BibitemShut {NoStop}%
\bibitem [{\citenamefont {Kramm}\ \emph {et~al.}(2012)\citenamefont {Kramm},
  \citenamefont {Nettelmann}, \citenamefont {Fortney}, \citenamefont
  {Neuh{\"a}user},\ and\ \citenamefont {Redmer}}]{KKNF12}%
  \BibitemOpen
  \bibfield  {author} {\bibinfo {author} {\bibfnamefont {U.}~\bibnamefont
  {Kramm}}, \bibinfo {author} {\bibfnamefont {N.}~\bibnamefont {Nettelmann}},
  \bibinfo {author} {\bibfnamefont {J.~J.}\ \bibnamefont {Fortney}}, \bibinfo
  {author} {\bibfnamefont {R.}~\bibnamefont {Neuh{\"a}user}},\ and\ \bibinfo
  {author} {\bibfnamefont {R.}~\bibnamefont {Redmer}},\ }\bibfield  {title}
  {\bibinfo {title} {{Constraining the interior of extrasolar giant planets
  with the tidal Love number $k_2$ using the example of HAT-P-13b"}},\ }\href
  {https://doi.org/10.1051/0004-6361/201118141} {\bibfield  {journal} {\bibinfo
   {journal} {A \& A}\ }\textbf {\bibinfo {volume} {538}},\ \bibinfo {pages}
  {8} (\bibinfo {year} {2012})}\BibitemShut {NoStop}%
\bibitem [{\citenamefont {Knudson}\ and\ \citenamefont
  {Desjarlais}(2009)}]{KD09}%
  \BibitemOpen
  \bibfield  {author} {\bibinfo {author} {\bibfnamefont {M.~D.}\ \bibnamefont
  {Knudson}}\ and\ \bibinfo {author} {\bibfnamefont {M.~P.}\ \bibnamefont
  {Desjarlais}},\ }\bibfield  {title} {\bibinfo {title} {Shock compression of
  quartz to 1.6 {T}{P}a: Redefining a pressure standard},\ }\href
  {https://link.aps.org/doi/10.1103/PhysRevLett.103.225501} {\bibfield
  {journal} {\bibinfo  {journal} {Phys. Rev. Lett.}\ }\textbf {\bibinfo
  {volume} {103}},\ \bibinfo {pages} {225501} (\bibinfo {year}
  {2009})}\BibitemShut {NoStop}%
\bibitem [{\citenamefont {Lorenzen}\ \emph {et~al.}(2009)\citenamefont
  {Lorenzen}, \citenamefont {Holst},\ and\ \citenamefont {Redmer}}]{LHR09}%
  \BibitemOpen
  \bibfield  {author} {\bibinfo {author} {\bibfnamefont {W.}~\bibnamefont
  {Lorenzen}}, \bibinfo {author} {\bibfnamefont {B.}~\bibnamefont {Holst}},\
  and\ \bibinfo {author} {\bibfnamefont {R.}~\bibnamefont {Redmer}},\
  }\bibfield  {title} {\bibinfo {title} {Demixing of hydrogen and helium at
  megabar pressures},\ }\href {https://doi.org/10.1103/PhysRevLett.102.115701}
  {\bibfield  {journal} {\bibinfo  {journal} {Phys. Rev. Lett.}\ }\textbf
  {\bibinfo {volume} {102}},\ \bibinfo {pages} {115701} (\bibinfo {year}
  {2009})}\BibitemShut {NoStop}%
\bibitem [{\citenamefont {Lorenzen}\ \emph {et~al.}(2011)\citenamefont
  {Lorenzen}, \citenamefont {Holst},\ and\ \citenamefont {Redmer}}]{LHR11}%
  \BibitemOpen
  \bibfield  {author} {\bibinfo {author} {\bibfnamefont {W.}~\bibnamefont
  {Lorenzen}}, \bibinfo {author} {\bibfnamefont {B.}~\bibnamefont {Holst}},\
  and\ \bibinfo {author} {\bibfnamefont {R.}~\bibnamefont {Redmer}},\
  }\bibfield  {title} {\bibinfo {title} {Metallization in hydrogen-helium
  mixtures},\ }\href {https://doi.org/10.1103/PhysRevB.84.235109} {\bibfield
  {journal} {\bibinfo  {journal} {Phys. Rev. B}\ }\textbf {\bibinfo {volume}
  {84}},\ \bibinfo {pages} {235109} (\bibinfo {year} {2011})}\BibitemShut
  {NoStop}%
\bibitem [{\citenamefont {Knudson}\ \emph {et~al.}(2015)\citenamefont
  {Knudson}, \citenamefont {Desjarlais}, \citenamefont {Becker}, \citenamefont
  {Lemke}, \citenamefont {Cochrane}, \citenamefont {Savage}, \citenamefont
  {Bliss}, \citenamefont {Mattsson},\ and\ \citenamefont
  {Redmer}}]{KDBLCSMR15}%
  \BibitemOpen
  \bibfield  {author} {\bibinfo {author} {\bibfnamefont {M.~D.}\ \bibnamefont
  {Knudson}}, \bibinfo {author} {\bibfnamefont {M.~P.}\ \bibnamefont
  {Desjarlais}}, \bibinfo {author} {\bibfnamefont {A.}~\bibnamefont {Becker}},
  \bibinfo {author} {\bibfnamefont {R.~W.}\ \bibnamefont {Lemke}}, \bibinfo
  {author} {\bibfnamefont {K.~R.}\ \bibnamefont {Cochrane}}, \bibinfo {author}
  {\bibfnamefont {M.~E.}\ \bibnamefont {Savage}}, \bibinfo {author}
  {\bibfnamefont {D.~E.}\ \bibnamefont {Bliss}}, \bibinfo {author}
  {\bibfnamefont {T.~R.}\ \bibnamefont {Mattsson}},\ and\ \bibinfo {author}
  {\bibfnamefont {R.}~\bibnamefont {Redmer}},\ }\bibfield  {title} {\bibinfo
  {title} {Direct observation of an abrupt insulator-to-metal transition in
  dense liquid deuterium},\ }\href {https://doi.org/10.1126/science.aaa7471}
  {\bibfield  {journal} {\bibinfo  {journal} {Science}\ }\textbf {\bibinfo
  {volume} {348}},\ \bibinfo {pages} {1455} (\bibinfo {year}
  {2015})}\BibitemShut {NoStop}%
\bibitem [{\citenamefont {Hubbard}\ \emph {et~al.}(1997)\citenamefont
  {Hubbard}, \citenamefont {Guillot}, \citenamefont {Lunine}, \citenamefont
  {Burrows}, \citenamefont {Saumon}, \citenamefont {Marley},\ and\
  \citenamefont {Freedman}}]{HGLBSMF97}%
  \BibitemOpen
  \bibfield  {author} {\bibinfo {author} {\bibfnamefont {W.~B.}\ \bibnamefont
  {Hubbard}}, \bibinfo {author} {\bibfnamefont {T.}~\bibnamefont {Guillot}},
  \bibinfo {author} {\bibfnamefont {J.~I.}\ \bibnamefont {Lunine}}, \bibinfo
  {author} {\bibfnamefont {A.}~\bibnamefont {Burrows}}, \bibinfo {author}
  {\bibfnamefont {D.}~\bibnamefont {Saumon}}, \bibinfo {author} {\bibfnamefont
  {M.~S.}\ \bibnamefont {Marley}},\ and\ \bibinfo {author} {\bibfnamefont
  {R.~S.}\ \bibnamefont {Freedman}},\ }\bibfield  {title} {\bibinfo {title}
  {Liquid metallic hydrogen and the structure of brown dwarfs and giant
  planets},\ }\href {https://doi.org/10.1063/1.872570} {\bibfield  {journal}
  {\bibinfo  {journal} {Phys. Plasmas}\ }\textbf {\bibinfo {volume} {4}},\
  \bibinfo {pages} {2011} (\bibinfo {year} {1997})}\BibitemShut {NoStop}%
\bibitem [{\citenamefont {Chabrier}\ \emph {et~al.}(2000)\citenamefont
  {Chabrier}, \citenamefont {Brassard}, \citenamefont {Fontaine},\ and\
  \citenamefont {Saumon}}]{CBFS00}%
  \BibitemOpen
  \bibfield  {author} {\bibinfo {author} {\bibfnamefont {G.}~\bibnamefont
  {Chabrier}}, \bibinfo {author} {\bibfnamefont {P.}~\bibnamefont {Brassard}},
  \bibinfo {author} {\bibfnamefont {G.}~\bibnamefont {Fontaine}},\ and\
  \bibinfo {author} {\bibfnamefont {D.}~\bibnamefont {Saumon}},\ }\bibfield
  {title} {\bibinfo {title} {Cooling sequences and color-magnitude diagrams for
  cool white dwarfs with hydrogen atmospheres},\ }\href
  {https://doi.org/10.1086/317092} {\bibfield  {journal} {\bibinfo  {journal}
  {Astrophys. J.}\ }\textbf {\bibinfo {volume} {543}},\ \bibinfo {pages} {216}
  (\bibinfo {year} {2000})}\BibitemShut {NoStop}%
\bibitem [{\citenamefont {Tamblyn}\ \emph {et~al.}(2008)\citenamefont
  {Tamblyn}, \citenamefont {Raty},\ and\ \citenamefont {Bonev}}]{TRB08}%
  \BibitemOpen
  \bibfield  {author} {\bibinfo {author} {\bibfnamefont {I.}~\bibnamefont
  {Tamblyn}}, \bibinfo {author} {\bibfnamefont {J.-Y.}\ \bibnamefont {Raty}},\
  and\ \bibinfo {author} {\bibfnamefont {S.~A.}\ \bibnamefont {Bonev}},\
  }\bibfield  {title} {\bibinfo {title} {Tetrahedral clustering in molten
  lithium under pressure},\ }\href
  {https://doi.org/10.1103/PhysRevLett.101.075703} {\bibfield  {journal}
  {\bibinfo  {journal} {Phys. Rev. Lett.}\ }\textbf {\bibinfo {volume} {101}},\
  \bibinfo {pages} {075703} (\bibinfo {year} {2008})}\BibitemShut {NoStop}%
\bibitem [{\citenamefont {Vorob'ev}\ and\ \citenamefont
  {Novikov}(2011)}]{VN11}%
  \BibitemOpen
  \bibfield  {author} {\bibinfo {author} {\bibfnamefont {V.~S.}\ \bibnamefont
  {Vorob'ev}}\ and\ \bibinfo {author} {\bibfnamefont {V.~G.}\ \bibnamefont
  {Novikov}},\ }\bibfield  {title} {\bibinfo {title} {Cell model of hydrogen
  liquid at megabar pressures},\ }\href
  {https://doi.org/http://dx.doi.org/10.1063/1.3563804} {\bibfield  {journal}
  {\bibinfo  {journal} {J. Chem. Phys.}\ }\textbf {\bibinfo {volume} {134}},\
  \bibinfo {eid} {114509} (\bibinfo {year} {2011})}\BibitemShut {NoStop}%
\bibitem [{\citenamefont {Ping}\ \emph {et~al.}(2006)\citenamefont {Ping},
  \citenamefont {Hanson}, \citenamefont {Koslow}, \citenamefont {Ogitsu},
  \citenamefont {Prendergast}, \citenamefont {Schwegler}, \citenamefont
  {Collins},\ and\ \citenamefont {Ng}}]{PHK06}%
  \BibitemOpen
  \bibfield  {author} {\bibinfo {author} {\bibfnamefont {Y.}~\bibnamefont
  {Ping}}, \bibinfo {author} {\bibfnamefont {D.}~\bibnamefont {Hanson}},
  \bibinfo {author} {\bibfnamefont {I.}~\bibnamefont {Koslow}}, \bibinfo
  {author} {\bibfnamefont {T.}~\bibnamefont {Ogitsu}}, \bibinfo {author}
  {\bibfnamefont {D.}~\bibnamefont {Prendergast}}, \bibinfo {author}
  {\bibfnamefont {E.}~\bibnamefont {Schwegler}}, \bibinfo {author}
  {\bibfnamefont {G.}~\bibnamefont {Collins}},\ and\ \bibinfo {author}
  {\bibfnamefont {A.}~\bibnamefont {Ng}},\ }\bibfield  {title} {\bibinfo
  {title} {Broadband dielectric function of nonequilibrium warm dense gold},\
  }\href {https://doi.org/10.1103/PhysRevLett.96.255003} {\bibfield  {journal}
  {\bibinfo  {journal} {Phys. Rev. Lett.}\ }\textbf {\bibinfo {volume} {96}},\
  \bibinfo {pages} {255003} (\bibinfo {year} {2006})}\BibitemShut {NoStop}%
\bibitem [{\citenamefont {Ernstorfer}\ \emph {et~al.}(2009)\citenamefont
  {Ernstorfer}, \citenamefont {Harb}, \citenamefont {Hebeisen}, \citenamefont
  {Sciaini}, \citenamefont {Dartigalongue},\ and\ \citenamefont
  {Miller}}]{EHH09}%
  \BibitemOpen
  \bibfield  {author} {\bibinfo {author} {\bibfnamefont {R.}~\bibnamefont
  {Ernstorfer}}, \bibinfo {author} {\bibfnamefont {M.}~\bibnamefont {Harb}},
  \bibinfo {author} {\bibfnamefont {C.~T.}\ \bibnamefont {Hebeisen}}, \bibinfo
  {author} {\bibfnamefont {G.}~\bibnamefont {Sciaini}}, \bibinfo {author}
  {\bibfnamefont {T.}~\bibnamefont {Dartigalongue}},\ and\ \bibinfo {author}
  {\bibfnamefont {R.~J.~D.}\ \bibnamefont {Miller}},\ }\bibfield  {title}
  {\bibinfo {title} {The formation of warm dense matter: Experimental evidence
  for electronic bond hardening in gold},\ }\href
  {https://doi.org/10.1126/science.1162697} {\bibfield  {journal} {\bibinfo
  {journal} {Science}\ }\textbf {\bibinfo {volume} {323}},\ \bibinfo {pages}
  {1033} (\bibinfo {year} {2009})}\BibitemShut {NoStop}%
\bibitem [{\citenamefont {Kandyla}\ \emph {et~al.}(2007)\citenamefont
  {Kandyla}, \citenamefont {Shih},\ and\ \citenamefont {Mazur}}]{KSM07}%
  \BibitemOpen
  \bibfield  {author} {\bibinfo {author} {\bibfnamefont {M.}~\bibnamefont
  {Kandyla}}, \bibinfo {author} {\bibfnamefont {T.}~\bibnamefont {Shih}},\ and\
  \bibinfo {author} {\bibfnamefont {E.}~\bibnamefont {Mazur}},\ }\bibfield
  {title} {\bibinfo {title} {Femtosecond dynamics of the laser-induced
  solid-to-liquid phase transition in aluminum},\ }\href
  {https://doi.org/10.1103/PhysRevB.75.214107} {\bibfield  {journal} {\bibinfo
  {journal} {Phys. Rev. B}\ }\textbf {\bibinfo {volume} {75}},\ \bibinfo
  {pages} {214107} (\bibinfo {year} {2007})}\BibitemShut {NoStop}%
\bibitem [{\citenamefont {Gericke}\ \emph {et~al.}(2010)\citenamefont
  {Gericke}, \citenamefont {W{\"u}nsch}, \citenamefont {Grinenko},\ and\
  \citenamefont {Vorberger}}]{GWGV10}%
  \BibitemOpen
  \bibfield  {author} {\bibinfo {author} {\bibfnamefont {D.~O.}\ \bibnamefont
  {Gericke}}, \bibinfo {author} {\bibfnamefont {K.}~\bibnamefont {W{\"u}nsch}},
  \bibinfo {author} {\bibfnamefont {A.}~\bibnamefont {Grinenko}},\ and\
  \bibinfo {author} {\bibfnamefont {J.}~\bibnamefont {Vorberger}},\ }\bibfield
  {title} {\bibinfo {title} {Structural properties of warm dense matter},\
  }\href {http://stacks.iop.org/1742-6596/220/i=1/a=012001} {\bibfield
  {journal} {\bibinfo  {journal} {J. Phys. Conf. Ser.}\ }\textbf {\bibinfo
  {volume} {220}},\ \bibinfo {pages} {012001} (\bibinfo {year}
  {2010})}\BibitemShut {NoStop}%
\bibitem [{\citenamefont {Glenzer}\ and\ \citenamefont {Redmer}(2009)}]{SR09}%
  \BibitemOpen
  \bibfield  {author} {\bibinfo {author} {\bibfnamefont {S.~H.}\ \bibnamefont
  {Glenzer}}\ and\ \bibinfo {author} {\bibfnamefont {R.}~\bibnamefont
  {Redmer}},\ }\bibfield  {title} {\bibinfo {title} {{X-ray Thomson scattering
  in high energy density plasmas}},\ }\href
  {https://doi.org/10.1103/RevModPhys.81.1625} {\bibfield  {journal} {\bibinfo
  {journal} {Rev. Mod. Phys.}\ }\textbf {\bibinfo {volume} {81}},\ \bibinfo
  {pages} {1625} (\bibinfo {year} {2009})}\BibitemShut {NoStop}%
\bibitem [{\citenamefont {Root}\ \emph {et~al.}(2010)\citenamefont {Root},
  \citenamefont {Magyar}, \citenamefont {Carpenter}, \citenamefont {Hanson},\
  and\ \citenamefont {Mattsson}}]{SMCHMT10}%
  \BibitemOpen
  \bibfield  {author} {\bibinfo {author} {\bibfnamefont {S.}~\bibnamefont
  {Root}}, \bibinfo {author} {\bibfnamefont {R.~J.}\ \bibnamefont {Magyar}},
  \bibinfo {author} {\bibfnamefont {J.~H.}\ \bibnamefont {Carpenter}}, \bibinfo
  {author} {\bibfnamefont {D.~L.}\ \bibnamefont {Hanson}},\ and\ \bibinfo
  {author} {\bibfnamefont {T.~R.}\ \bibnamefont {Mattsson}},\ }\bibfield
  {title} {\bibinfo {title} {Shock compression of a fifth period element:
  Liquid xenon to 840 gpa},\ }\href
  {https://doi.org/10.1103/PhysRevLett.105.085501} {\bibfield  {journal}
  {\bibinfo  {journal} {Phys. Rev. Lett.}\ }\textbf {\bibinfo {volume} {105}},\
  \bibinfo {pages} {085501} (\bibinfo {year} {2010})}\BibitemShut {NoStop}%
\bibitem [{\citenamefont {Smith}\ \emph {et~al.}(2014)\citenamefont {Smith},
  \citenamefont {Eggert}, \citenamefont {Jeanloz}, \citenamefont {Duffy},
  \citenamefont {Braun}, \citenamefont {Patterson}, \citenamefont {Rudd},
  \citenamefont {Biener}, \citenamefont {Lazicki}, \citenamefont {Hamza},
  \citenamefont {Wang}, \citenamefont {Braun}, \citenamefont {Benedict},
  \citenamefont {Celliers},\ and\ \citenamefont {Collins}}]{SEJ14}%
  \BibitemOpen
  \bibfield  {author} {\bibinfo {author} {\bibfnamefont {R.~F.}\ \bibnamefont
  {Smith}}, \bibinfo {author} {\bibfnamefont {J.~H.}\ \bibnamefont {Eggert}},
  \bibinfo {author} {\bibfnamefont {R.}~\bibnamefont {Jeanloz}}, \bibinfo
  {author} {\bibfnamefont {T.~S.}\ \bibnamefont {Duffy}}, \bibinfo {author}
  {\bibfnamefont {D.~G.}\ \bibnamefont {Braun}}, \bibinfo {author}
  {\bibfnamefont {J.~R.}\ \bibnamefont {Patterson}}, \bibinfo {author}
  {\bibfnamefont {R.~E.}\ \bibnamefont {Rudd}}, \bibinfo {author}
  {\bibfnamefont {J.}~\bibnamefont {Biener}}, \bibinfo {author} {\bibfnamefont
  {A.~E.}\ \bibnamefont {Lazicki}}, \bibinfo {author} {\bibfnamefont {A.~V.}\
  \bibnamefont {Hamza}}, \bibinfo {author} {\bibfnamefont {J.}~\bibnamefont
  {Wang}}, \bibinfo {author} {\bibfnamefont {T.}~\bibnamefont {Braun}},
  \bibinfo {author} {\bibfnamefont {L.~X.}\ \bibnamefont {Benedict}}, \bibinfo
  {author} {\bibfnamefont {P.~M.}\ \bibnamefont {Celliers}},\ and\ \bibinfo
  {author} {\bibfnamefont {G.~W.}\ \bibnamefont {Collins}},\ }\bibfield
  {title} {\bibinfo {title} {Ramp compression of diamond to five terapascals},\
  }\href {https://doi.org/10.1038/nature13526} {\bibfield  {journal} {\bibinfo
  {journal} {Nature}\ }\textbf {\bibinfo {volume} {511}},\ \bibinfo {pages}
  {330} (\bibinfo {year} {2014})}\BibitemShut {NoStop}%
\bibitem [{\citenamefont {Glenzer}\ \emph {et~al.}(2016)\citenamefont
  {Glenzer}, \citenamefont {Fletcher}, \citenamefont {Galtier}, \citenamefont
  {Nagler}, \citenamefont {Alonso-Mori}, \citenamefont {Barbrel}, \citenamefont
  {Brown}, \citenamefont {Chapman}, \citenamefont {Chen}, \citenamefont
  {Curry}, \citenamefont {Fiuza}, \citenamefont {Gamboa}, \citenamefont
  {Gauthier}, \citenamefont {Gericke}, \citenamefont {Gleason}, \citenamefont
  {Goede}, \citenamefont {Granados}, \citenamefont {Heimann}, \citenamefont
  {Kim}, \citenamefont {Kraus}, \citenamefont {MacDonald}, \citenamefont
  {Mackinnon}, \citenamefont {Mishra}, \citenamefont {Ravasio}, \citenamefont
  {Roedel}, \citenamefont {Sperling}, \citenamefont {Schumaker}, \citenamefont
  {Tsui}, \citenamefont {Vorberger}, \citenamefont {Zastrau}, \citenamefont
  {Fry}, \citenamefont {White}, \citenamefont {Hasting},\ and\ \citenamefont
  {Lee}}]{GFGN16}%
  \BibitemOpen
  \bibfield  {author} {\bibinfo {author} {\bibfnamefont {S.~H.}\ \bibnamefont
  {Glenzer}}, \bibinfo {author} {\bibfnamefont {L.~B.}\ \bibnamefont
  {Fletcher}}, \bibinfo {author} {\bibfnamefont {E.}~\bibnamefont {Galtier}},
  \bibinfo {author} {\bibfnamefont {B.}~\bibnamefont {Nagler}}, \bibinfo
  {author} {\bibfnamefont {R.}~\bibnamefont {Alonso-Mori}}, \bibinfo {author}
  {\bibfnamefont {B.}~\bibnamefont {Barbrel}}, \bibinfo {author} {\bibfnamefont
  {S.~B.}\ \bibnamefont {Brown}}, \bibinfo {author} {\bibfnamefont {D.~A.}\
  \bibnamefont {Chapman}}, \bibinfo {author} {\bibfnamefont {Z.}~\bibnamefont
  {Chen}}, \bibinfo {author} {\bibfnamefont {C.~B.}\ \bibnamefont {Curry}},
  \bibinfo {author} {\bibfnamefont {F.}~\bibnamefont {Fiuza}}, \bibinfo
  {author} {\bibfnamefont {E.}~\bibnamefont {Gamboa}}, \bibinfo {author}
  {\bibfnamefont {M.}~\bibnamefont {Gauthier}}, \bibinfo {author}
  {\bibfnamefont {D.~O.}\ \bibnamefont {Gericke}}, \bibinfo {author}
  {\bibfnamefont {A.}~\bibnamefont {Gleason}}, \bibinfo {author} {\bibfnamefont
  {S.}~\bibnamefont {Goede}}, \bibinfo {author} {\bibfnamefont
  {E.}~\bibnamefont {Granados}}, \bibinfo {author} {\bibfnamefont
  {P.}~\bibnamefont {Heimann}}, \bibinfo {author} {\bibfnamefont
  {J.}~\bibnamefont {Kim}}, \bibinfo {author} {\bibfnamefont {D.}~\bibnamefont
  {Kraus}}, \bibinfo {author} {\bibfnamefont {M.~J.}\ \bibnamefont
  {MacDonald}}, \bibinfo {author} {\bibfnamefont {A.~J.}\ \bibnamefont
  {Mackinnon}}, \bibinfo {author} {\bibfnamefont {R.}~\bibnamefont {Mishra}},
  \bibinfo {author} {\bibfnamefont {A.}~\bibnamefont {Ravasio}}, \bibinfo
  {author} {\bibfnamefont {C.}~\bibnamefont {Roedel}}, \bibinfo {author}
  {\bibfnamefont {P.}~\bibnamefont {Sperling}}, \bibinfo {author}
  {\bibfnamefont {W.}~\bibnamefont {Schumaker}}, \bibinfo {author}
  {\bibfnamefont {Y.~Y.}\ \bibnamefont {Tsui}}, \bibinfo {author}
  {\bibfnamefont {J.}~\bibnamefont {Vorberger}}, \bibinfo {author}
  {\bibfnamefont {U.}~\bibnamefont {Zastrau}}, \bibinfo {author} {\bibfnamefont
  {A.}~\bibnamefont {Fry}}, \bibinfo {author} {\bibfnamefont {W.~E.}\
  \bibnamefont {White}}, \bibinfo {author} {\bibfnamefont {J.~B.}\ \bibnamefont
  {Hasting}},\ and\ \bibinfo {author} {\bibfnamefont {H.~J.}\ \bibnamefont
  {Lee}},\ }\bibfield  {title} {\bibinfo {title} {{Matter under extreme
  conditions experiments at the Linac Coherent Light Source}},\ }\href
  {https://doi.org/10.1088/0953-4075/49/9/092001} {\bibfield  {journal}
  {\bibinfo  {journal} {J. Phys. B}\ }\textbf {\bibinfo {volume} {49}},\
  \bibinfo {pages} {092001} (\bibinfo {year} {2016})}\BibitemShut {NoStop}%
\bibitem [{\citenamefont {Tschentscher}\ \emph {et~al.}(2017)\citenamefont
  {Tschentscher}, \citenamefont {Bressler}, \citenamefont {Grünert},
  \citenamefont {Madsen}, \citenamefont {Mancuso}, \citenamefont {Meyer},
  \citenamefont {Scherz}, \citenamefont {Sinn},\ and\ \citenamefont
  {Zastrau}}]{TBG17}%
  \BibitemOpen
  \bibfield  {author} {\bibinfo {author} {\bibfnamefont {T.}~\bibnamefont
  {Tschentscher}}, \bibinfo {author} {\bibfnamefont {C.}~\bibnamefont
  {Bressler}}, \bibinfo {author} {\bibfnamefont {J.}~\bibnamefont {Grünert}},
  \bibinfo {author} {\bibfnamefont {A.}~\bibnamefont {Madsen}}, \bibinfo
  {author} {\bibfnamefont {A.}~\bibnamefont {Mancuso}}, \bibinfo {author}
  {\bibfnamefont {M.}~\bibnamefont {Meyer}}, \bibinfo {author} {\bibfnamefont
  {A.}~\bibnamefont {Scherz}}, \bibinfo {author} {\bibfnamefont
  {H.}~\bibnamefont {Sinn}},\ and\ \bibinfo {author} {\bibfnamefont
  {U.}~\bibnamefont {Zastrau}},\ }\bibfield  {title} {\bibinfo {title} {Photon
  beam transport and scientific instruments at the {European XFEL}},\ }\href
  {https://doi.org/10.3390/app7060592} {\bibfield  {journal} {\bibinfo
  {journal} {Appl. Sci.}\ }\textbf {\bibinfo {volume} {7}},\ \bibinfo {pages}
  {592} (\bibinfo {year} {2017})}\BibitemShut {NoStop}%
\bibitem [{Note1()}]{Note1}%
  \BibitemOpen
  \bibinfo {note} {Known also as the Wigner--Seitz radius or Brueckner
  parameter}\BibitemShut {NoStop}%
\bibitem [{\citenamefont {Bonitz}\ \emph {et~al.}(2020)\citenamefont {Bonitz},
  \citenamefont {Dornheim}, \citenamefont {Moldabekov}, \citenamefont {Zhang},
  \citenamefont {Hamann}, \citenamefont {Kählert}, \citenamefont {Filinov},
  \citenamefont {Ramakrishna},\ and\ \citenamefont {Vorberger}}]{BDMZHKFRV20}%
  \BibitemOpen
  \bibfield  {author} {\bibinfo {author} {\bibfnamefont {M.}~\bibnamefont
  {Bonitz}}, \bibinfo {author} {\bibfnamefont {T.}~\bibnamefont {Dornheim}},
  \bibinfo {author} {\bibfnamefont {Z.~A.}\ \bibnamefont {Moldabekov}},
  \bibinfo {author} {\bibfnamefont {S.}~\bibnamefont {Zhang}}, \bibinfo
  {author} {\bibfnamefont {P.}~\bibnamefont {Hamann}}, \bibinfo {author}
  {\bibfnamefont {H.}~\bibnamefont {Kählert}}, \bibinfo {author}
  {\bibfnamefont {A.}~\bibnamefont {Filinov}}, \bibinfo {author} {\bibfnamefont
  {K.}~\bibnamefont {Ramakrishna}},\ and\ \bibinfo {author} {\bibfnamefont
  {J.}~\bibnamefont {Vorberger}},\ }\bibfield  {title} {\bibinfo {title}
  {\textit{Ab initio} simulation of warm dense matter},\ }\href
  {https://doi.org/10.1063/1.5143225} {\bibfield  {journal} {\bibinfo
  {journal} {Phys. Plasmas}\ }\textbf {\bibinfo {volume} {27}},\ \bibinfo
  {pages} {042710} (\bibinfo {year} {2020})}\BibitemShut {NoStop}%
\bibitem [{Note2()}]{Note2}%
  \BibitemOpen
  \bibinfo {note} {We note there are sometimes different conventions for the
  definitions of $E_\protect \textrm {F}$ and $r_s$ (see for example Refs.~
  \cite {BDMZHKFRV20} and \cite {DGB18}), but where we refer to these
  parameters we use the above definitions and adopt the convention of
  Ref.~\cite {NFP99} in which the free electron density is equated with the
  valence electron density.}\BibitemShut {Stop}%
\bibitem [{\citenamefont {Dornheim}\ \emph {et~al.}(2018)\citenamefont
  {Dornheim}, \citenamefont {Groth},\ and\ \citenamefont {Bonitz}}]{DGB18}%
  \BibitemOpen
  \bibfield  {author} {\bibinfo {author} {\bibfnamefont {T.}~\bibnamefont
  {Dornheim}}, \bibinfo {author} {\bibfnamefont {S.}~\bibnamefont {Groth}},\
  and\ \bibinfo {author} {\bibfnamefont {M.}~\bibnamefont {Bonitz}},\
  }\bibfield  {title} {\bibinfo {title} {The uniform electron gas at warm dense
  matter conditions},\ }\href
  {https://doi.org/https://doi.org/10.1016/j.physrep.2018.04.001} {\bibfield
  {journal} {\bibinfo  {journal} {Phys. Rep.}\ }\textbf {\bibinfo {volume}
  {744}},\ \bibinfo {pages} {1 } (\bibinfo {year} {2018})}\BibitemShut
  {NoStop}%
\bibitem [{\citenamefont {Hohenberg}\ and\ \citenamefont {Kohn}(1964)}]{HK64}%
  \BibitemOpen
  \bibfield  {author} {\bibinfo {author} {\bibfnamefont {P.}~\bibnamefont
  {Hohenberg}}\ and\ \bibinfo {author} {\bibfnamefont {W.}~\bibnamefont
  {Kohn}},\ }\bibfield  {title} {\bibinfo {title} {Inhomogeneous electron
  gas},\ }\href {https://doi.org/10.1103/PhysRev.136.B864} {\bibfield
  {journal} {\bibinfo  {journal} {Phys. Rev.}\ }\textbf {\bibinfo {volume}
  {136}},\ \bibinfo {pages} {B864} (\bibinfo {year} {1964})}\BibitemShut
  {NoStop}%
\bibitem [{\citenamefont {Kohn}\ and\ \citenamefont {Sham}(1965)}]{KS65}%
  \BibitemOpen
  \bibfield  {author} {\bibinfo {author} {\bibfnamefont {W.}~\bibnamefont
  {Kohn}}\ and\ \bibinfo {author} {\bibfnamefont {L.~J.}\ \bibnamefont
  {Sham}},\ }\bibfield  {title} {\bibinfo {title} {Self-consistent equations
  including exchange and correlation effects},\ }\href
  {https://doi.org/10.1103/PhysRev.140.A1133} {\bibfield  {journal} {\bibinfo
  {journal} {Phys. Rev.}\ }\textbf {\bibinfo {volume} {140}},\ \bibinfo {pages}
  {A1133} (\bibinfo {year} {1965})}\BibitemShut {NoStop}%
\bibitem [{\citenamefont {Desjarlais}(2003)}]{D03}%
  \BibitemOpen
  \bibfield  {author} {\bibinfo {author} {\bibfnamefont {M.~P.}\ \bibnamefont
  {Desjarlais}},\ }\bibfield  {title} {\bibinfo {title} {Density-functional
  calculations of the liquid deuterium {Hugoniot}, reshock, and reverberation
  timing},\ }\href {https://doi.org/10.1103/PhysRevB.68.064204} {\bibfield
  {journal} {\bibinfo  {journal} {Phys. Rev. B}\ }\textbf {\bibinfo {volume}
  {68}},\ \bibinfo {pages} {064204} (\bibinfo {year} {2003})}\BibitemShut
  {NoStop}%
\bibitem [{\citenamefont {Holst}\ \emph {et~al.}(2008)\citenamefont {Holst},
  \citenamefont {Redmer},\ and\ \citenamefont {Desjarlais}}]{HRD08}%
  \BibitemOpen
  \bibfield  {author} {\bibinfo {author} {\bibfnamefont {B.}~\bibnamefont
  {Holst}}, \bibinfo {author} {\bibfnamefont {R.}~\bibnamefont {Redmer}},\ and\
  \bibinfo {author} {\bibfnamefont {M.~P.}\ \bibnamefont {Desjarlais}},\
  }\bibfield  {title} {\bibinfo {title} {Thermophysical properties of warm
  dense hydrogen using quantum molecular dynamics simulations},\ }\href
  {https://link.aps.org/doi/10.1103/PhysRevB.77.184201} {\bibfield  {journal}
  {\bibinfo  {journal} {Phys. Rev. B}\ }\textbf {\bibinfo {volume} {77}},\
  \bibinfo {pages} {184201} (\bibinfo {year} {2008})}\BibitemShut {NoStop}%
\bibitem [{\citenamefont {Dreizler}\ and\ \citenamefont {Gross}(1990)}]{DG90}%
  \BibitemOpen
  \bibfield  {author} {\bibinfo {author} {\bibfnamefont {R.}~\bibnamefont
  {Dreizler}}\ and\ \bibinfo {author} {\bibfnamefont {E.}~\bibnamefont
  {Gross}},\ }\href@noop {} {\emph {\bibinfo {title} {Density Functional
  Theory: An Approach to the Quantum Many-Body Problem}}}\ (\bibinfo
  {publisher} {Springer--Verlag},\ \bibinfo {year} {1990})\BibitemShut
  {NoStop}%
\bibitem [{\citenamefont {Parr}\ and\ \citenamefont {Yang}(1994)}]{PY94}%
  \BibitemOpen
  \bibfield  {author} {\bibinfo {author} {\bibfnamefont {R.}~\bibnamefont
  {Parr}}\ and\ \bibinfo {author} {\bibfnamefont {W.}~\bibnamefont {Yang}},\
  }\href {https://books.google.de/books?id=mxiOngEACAAJ} {\emph {\bibinfo
  {title} {Density-Functional Theory of Atoms and Molecules}}},\ International
  Series of Monographs on Chemistry\ (\bibinfo  {publisher} {Oxford University
  Press, USA},\ \bibinfo {year} {1994})\BibitemShut {NoStop}%
\bibitem [{\citenamefont {Fiolhais}\ \emph {et~al.}(2003)\citenamefont
  {Fiolhais}, \citenamefont {Nogueira},\ and\ \citenamefont {Marques}}]{FNM03}%
  \BibitemOpen
  \bibfield  {author} {\bibinfo {author} {\bibfnamefont {C.}~\bibnamefont
  {Fiolhais}}, \bibinfo {author} {\bibfnamefont {F.}~\bibnamefont {Nogueira}},\
  and\ \bibinfo {author} {\bibfnamefont {M.~A.}\ \bibnamefont {Marques}},\
  }\href@noop {} {\emph {\bibinfo {title} {A primer in density functional
  theory}}},\ Vol.\ \bibinfo {volume} {620}\ (\bibinfo  {publisher} {Springer
  Science \& Business Media},\ \bibinfo {year} {2003})\BibitemShut {NoStop}%
\bibitem [{\citenamefont {Hasnip}\ \emph {et~al.}(2014)\citenamefont {Hasnip},
  \citenamefont {Refson}, \citenamefont {Probert}, \citenamefont {Yates},
  \citenamefont {Clark},\ and\ \citenamefont {Pickard}}]{HRPYCP11}%
  \BibitemOpen
  \bibfield  {author} {\bibinfo {author} {\bibfnamefont {P.~J.}\ \bibnamefont
  {Hasnip}}, \bibinfo {author} {\bibfnamefont {K.}~\bibnamefont {Refson}},
  \bibinfo {author} {\bibfnamefont {M.~I.~J.}\ \bibnamefont {Probert}},
  \bibinfo {author} {\bibfnamefont {J.~R.}\ \bibnamefont {Yates}}, \bibinfo
  {author} {\bibfnamefont {S.~J.}\ \bibnamefont {Clark}},\ and\ \bibinfo
  {author} {\bibfnamefont {C.~J.}\ \bibnamefont {Pickard}},\ }\bibfield
  {title} {\bibinfo {title} {Density functional theory in the solid state},\
  }\href {https://doi.org/10.1098/rsta.2013.0270} {\bibfield  {journal}
  {\bibinfo  {journal} {Philos. Trans. R. Soc. A}\ }\textbf {\bibinfo {volume}
  {372}},\ \bibinfo {pages} {20130270} (\bibinfo {year} {2014})}\BibitemShut
  {NoStop}%
\bibitem [{\citenamefont {Burke}(2012)}]{B12}%
  \BibitemOpen
  \bibfield  {author} {\bibinfo {author} {\bibfnamefont {K.}~\bibnamefont
  {Burke}},\ }\bibfield  {title} {\bibinfo {title} {Perspective on density
  functional theory},\ }\href
  {https://doi.org/http://dx.doi.org/10.1063/1.4704546} {\bibfield  {journal}
  {\bibinfo  {journal} {J. Chem. Phys.}\ }\textbf {\bibinfo {volume} {136}},\
  \bibinfo {eid} {150901} (\bibinfo {year} {2012})}\BibitemShut {NoStop}%
\bibitem [{\citenamefont {Yu}\ \emph {et~al.}(2016)\citenamefont {Yu},
  \citenamefont {Li},\ and\ \citenamefont {Truhlar}}]{HST16}%
  \BibitemOpen
  \bibfield  {author} {\bibinfo {author} {\bibfnamefont {H.~S.}\ \bibnamefont
  {Yu}}, \bibinfo {author} {\bibfnamefont {S.~L.}\ \bibnamefont {Li}},\ and\
  \bibinfo {author} {\bibfnamefont {D.~G.}\ \bibnamefont {Truhlar}},\
  }\bibfield  {title} {\bibinfo {title} {{Perspective: Kohn--Sham density
  functional theory descending a staircase}},\ }\href
  {https://doi.org/10.1063/1.4963168} {\bibfield  {journal} {\bibinfo
  {journal} {J. Chem. Phys.}\ }\textbf {\bibinfo {volume} {145}},\ \bibinfo
  {pages} {130901} (\bibinfo {year} {2016})}\BibitemShut {NoStop}%
\bibitem [{\citenamefont {Mermin}(1965)}]{M65}%
  \BibitemOpen
  \bibfield  {author} {\bibinfo {author} {\bibfnamefont {N.~D.}\ \bibnamefont
  {Mermin}},\ }\bibfield  {title} {\bibinfo {title} {Thermal properties of the
  inhomogenous electron gas},\ }\href
  {https://link.aps.org/doi/10.1103/PhysRev.137.A1441} {\bibfield  {journal}
  {\bibinfo  {journal} {Phys. Rev.}\ }\textbf {\bibinfo {volume} {137}},\
  \bibinfo {pages} {A: 1441} (\bibinfo {year} {1965})}\BibitemShut {NoStop}%
\bibitem [{\citenamefont {Mattsson}\ \emph {et~al.}(2004)\citenamefont
  {Mattsson}, \citenamefont {Schultz}, \citenamefont {Desjarlais},
  \citenamefont {Mattsson},\ and\ \citenamefont {Leung}}]{MSDML04}%
  \BibitemOpen
  \bibfield  {author} {\bibinfo {author} {\bibfnamefont {A.~E.}\ \bibnamefont
  {Mattsson}}, \bibinfo {author} {\bibfnamefont {P.~A.}\ \bibnamefont
  {Schultz}}, \bibinfo {author} {\bibfnamefont {M.~P.}\ \bibnamefont
  {Desjarlais}}, \bibinfo {author} {\bibfnamefont {T.~R.}\ \bibnamefont
  {Mattsson}},\ and\ \bibinfo {author} {\bibfnamefont {K.}~\bibnamefont
  {Leung}},\ }\bibfield  {title} {\bibinfo {title} {Designing meaningful
  density functional theory calculations in materials science{\textemdash}a
  primer},\ }\href {https://doi.org/10.1088/0965-0393/13/1/r01} {\bibfield
  {journal} {\bibinfo  {journal} {Model. Simul. Mater. Sci. Eng}\ }\textbf
  {\bibinfo {volume} {13}},\ \bibinfo {pages} {R1} (\bibinfo {year}
  {2004})}\BibitemShut {NoStop}%
\bibitem [{\citenamefont {Holst}\ \emph {et~al.}(2011)\citenamefont {Holst},
  \citenamefont {French},\ and\ \citenamefont {Redmer}}]{HFR11}%
  \BibitemOpen
  \bibfield  {author} {\bibinfo {author} {\bibfnamefont {B.}~\bibnamefont
  {Holst}}, \bibinfo {author} {\bibfnamefont {M.}~\bibnamefont {French}},\ and\
  \bibinfo {author} {\bibfnamefont {R.}~\bibnamefont {Redmer}},\ }\bibfield
  {title} {\bibinfo {title} {Electronic transport coefficients from \textit{ab
  initio} simulations and application to dense liquid hydrogen},\ }\href
  {https://link.aps.org/doi/10.1103/PhysRevB.83.235120} {\bibfield  {journal}
  {\bibinfo  {journal} {Phys. Rev. B}\ }\textbf {\bibinfo {volume} {83}},\
  \bibinfo {pages} {235120} (\bibinfo {year} {2011})}\BibitemShut {NoStop}%
\bibitem [{\citenamefont {Pople}(1999)}]{P99}%
  \BibitemOpen
  \bibfield  {author} {\bibinfo {author} {\bibfnamefont {J.~A.}\ \bibnamefont
  {Pople}},\ }\bibfield  {title} {\bibinfo {title} {Nobel lecture: Quantum
  chemical models},\ }\href {https://doi.org/10.1103/RevModPhys.71.1267}
  {\bibfield  {journal} {\bibinfo  {journal} {Rev. Mod. Phys.}\ }\textbf
  {\bibinfo {volume} {71}},\ \bibinfo {pages} {1267} (\bibinfo {year}
  {1999})}\BibitemShut {NoStop}%
\bibitem [{\citenamefont {Perdew}\ and\ \citenamefont {Schmidt}(2001)}]{PS01}%
  \BibitemOpen
  \bibfield  {author} {\bibinfo {author} {\bibfnamefont {J.~P.}\ \bibnamefont
  {Perdew}}\ and\ \bibinfo {author} {\bibfnamefont {K.}~\bibnamefont
  {Schmidt}},\ }\bibfield  {title} {\bibinfo {title} {Jacob’s ladder of
  density functional approximations for the exchange-correlation energy},\
  }\href {https://doi.org/10.1063/1.1390175} {\bibfield  {journal} {\bibinfo
  {journal} {AIP Conference Proceedings}\ }\textbf {\bibinfo {volume} {577}},\
  \bibinfo {pages} {1} (\bibinfo {year} {2001})}\BibitemShut {NoStop}%
\bibitem [{\citenamefont {Cohen}\ \emph {et~al.}(2012)\citenamefont {Cohen},
  \citenamefont {Mori-Sánchez},\ and\ \citenamefont {Yang}}]{CMY12}%
  \BibitemOpen
  \bibfield  {author} {\bibinfo {author} {\bibfnamefont {A.~J.}\ \bibnamefont
  {Cohen}}, \bibinfo {author} {\bibfnamefont {P.}~\bibnamefont
  {Mori-Sánchez}},\ and\ \bibinfo {author} {\bibfnamefont {W.}~\bibnamefont
  {Yang}},\ }\bibfield  {title} {\bibinfo {title} {Challenges for density
  functional theory},\ }\href {https://doi.org/10.1021/cr200107z} {\bibfield
  {journal} {\bibinfo  {journal} {Chem. Rev.}\ }\textbf {\bibinfo {volume}
  {112}},\ \bibinfo {pages} {289} (\bibinfo {year} {2012})}\BibitemShut
  {NoStop}%
\bibitem [{\citenamefont {Medvedev}\ \emph {et~al.}(2017)\citenamefont
  {Medvedev}, \citenamefont {Bushmarinov}, \citenamefont {Sun}, \citenamefont
  {Perdew},\ and\ \citenamefont {Lyssenko}}]{MBSPL17}%
  \BibitemOpen
  \bibfield  {author} {\bibinfo {author} {\bibfnamefont {M.~G.}\ \bibnamefont
  {Medvedev}}, \bibinfo {author} {\bibfnamefont {I.~S.}\ \bibnamefont
  {Bushmarinov}}, \bibinfo {author} {\bibfnamefont {J.}~\bibnamefont {Sun}},
  \bibinfo {author} {\bibfnamefont {J.~P.}\ \bibnamefont {Perdew}},\ and\
  \bibinfo {author} {\bibfnamefont {K.~A.}\ \bibnamefont {Lyssenko}},\
  }\bibfield  {title} {\bibinfo {title} {Density functional theory is straying
  from the path toward the exact functional},\ }\href
  {https://doi.org/10.1126/science.aah5975} {\bibfield  {journal} {\bibinfo
  {journal} {Science}\ }\textbf {\bibinfo {volume} {355}},\ \bibinfo {pages}
  {49} (\bibinfo {year} {2017})}\BibitemShut {NoStop}%
\bibitem [{\citenamefont {Toulouse}(2021)}]{T21}%
  \BibitemOpen
  \bibfield  {author} {\bibinfo {author} {\bibfnamefont {J.}~\bibnamefont
  {Toulouse}},\ }\href@noop {} {\bibinfo {title} {Review of approximations for
  the exchange-correlation energy in density-functional theory}} (\bibinfo
  {year} {2021}),\ \Eprint {https://arxiv.org/abs/2103.02645} {arXiv:2103.02645
  [physics.chem-ph]} \BibitemShut {NoStop}%
\bibitem [{\citenamefont {Kohn}\ and\ \citenamefont {Vashista}(1983)}]{KV83}%
  \BibitemOpen
  \bibfield  {author} {\bibinfo {author} {\bibfnamefont {W.}~\bibnamefont
  {Kohn}}\ and\ \bibinfo {author} {\bibfnamefont {P.}~\bibnamefont
  {Vashista}},\ }\bibinfo {title} {General density functional theory},\ in\
  \href@noop {} {\emph {\bibinfo {booktitle} {Theory of the Inhomogeneous
  Electron Gas}}},\ \bibinfo {series} {Physics of Solids and Liquids},
  Vol.~\bibinfo {volume} {1},\ \bibinfo {editor} {edited by\ \bibinfo {editor}
  {\bibfnamefont {S.}~\bibnamefont {Lundqvist}}\ and\ \bibinfo {editor}
  {\bibfnamefont {N.~H.}\ \bibnamefont {March}}}\ (\bibinfo  {publisher}
  {Springer US},\ \bibinfo {year} {1983})\ pp.\ \bibinfo {pages}
  {79--147}\BibitemShut {NoStop}%
\bibitem [{\citenamefont {Karasiev}\ \emph
  {et~al.}(2014{\natexlab{a}})\citenamefont {Karasiev}, \citenamefont
  {Sjostrom}, \citenamefont {Dufty},\ and\ \citenamefont {Trickey}}]{KSDT14}%
  \BibitemOpen
  \bibfield  {author} {\bibinfo {author} {\bibfnamefont {V.~V.}\ \bibnamefont
  {Karasiev}}, \bibinfo {author} {\bibfnamefont {T.}~\bibnamefont {Sjostrom}},
  \bibinfo {author} {\bibfnamefont {J.}~\bibnamefont {Dufty}},\ and\ \bibinfo
  {author} {\bibfnamefont {S.~B.}\ \bibnamefont {Trickey}},\ }\bibfield
  {title} {\bibinfo {title} {Accurate homogeneous electron gas
  exchange-correlation free energy for local spin-density calculations},\
  }\href {https://doi.org/10.1103/PhysRevLett.112.076403} {\bibfield  {journal}
  {\bibinfo  {journal} {Phys. Rev. Lett.}\ }\textbf {\bibinfo {volume} {112}},\
  \bibinfo {pages} {076403} (\bibinfo {year} {2014}{\natexlab{a}})}\BibitemShut
  {NoStop}%
\bibitem [{\citenamefont {Smith}\ \emph {et~al.}(2016)\citenamefont {Smith},
  \citenamefont {Pribram-Jones},\ and\ \citenamefont {Burke}}]{SJB16}%
  \BibitemOpen
  \bibfield  {author} {\bibinfo {author} {\bibfnamefont {J.~C.}\ \bibnamefont
  {Smith}}, \bibinfo {author} {\bibfnamefont {A.}~\bibnamefont
  {Pribram-Jones}},\ and\ \bibinfo {author} {\bibfnamefont {K.}~\bibnamefont
  {Burke}},\ }\bibfield  {title} {\bibinfo {title} {Exact thermal density
  functional theory for a model system: Correlation components and accuracy of
  the zero-temperature exchange-correlation approximation},\ }\href
  {https://doi.org/10.1103/PhysRevB.93.245131} {\bibfield  {journal} {\bibinfo
  {journal} {Phys. Rev. B}\ }\textbf {\bibinfo {volume} {93}},\ \bibinfo
  {pages} {245131} (\bibinfo {year} {2016})}\BibitemShut {NoStop}%
\bibitem [{\citenamefont {Pittalis}\ \emph {et~al.}(2011)\citenamefont
  {Pittalis}, \citenamefont {Proetto}, \citenamefont {Floris}, \citenamefont
  {Sanna}, \citenamefont {Bersier}, \citenamefont {Burke},\ and\ \citenamefont
  {Gross}}]{PPFS11}%
  \BibitemOpen
  \bibfield  {author} {\bibinfo {author} {\bibfnamefont {S.}~\bibnamefont
  {Pittalis}}, \bibinfo {author} {\bibfnamefont {C.~R.}\ \bibnamefont
  {Proetto}}, \bibinfo {author} {\bibfnamefont {A.}~\bibnamefont {Floris}},
  \bibinfo {author} {\bibfnamefont {A.}~\bibnamefont {Sanna}}, \bibinfo
  {author} {\bibfnamefont {C.}~\bibnamefont {Bersier}}, \bibinfo {author}
  {\bibfnamefont {K.}~\bibnamefont {Burke}},\ and\ \bibinfo {author}
  {\bibfnamefont {E.~K.~U.}\ \bibnamefont {Gross}},\ }\bibfield  {title}
  {\bibinfo {title} {Exact conditions in finite-temperature density-functional
  theory},\ }\href {https://link.aps.org/doi/10.1103/PhysRevLett.107.163001}
  {\bibfield  {journal} {\bibinfo  {journal} {Phys. Rev. Lett.}\ }\textbf
  {\bibinfo {volume} {107}},\ \bibinfo {pages} {163001} (\bibinfo {year}
  {2011})}\BibitemShut {NoStop}%
\bibitem [{\citenamefont {Dufty}\ and\ \citenamefont {Trickey}(2011)}]{DT11}%
  \BibitemOpen
  \bibfield  {author} {\bibinfo {author} {\bibfnamefont {J.~W.}\ \bibnamefont
  {Dufty}}\ and\ \bibinfo {author} {\bibfnamefont {S.~B.}\ \bibnamefont
  {Trickey}},\ }\bibfield  {title} {\bibinfo {title} {Scaling, bounds, and
  inequalities for the noninteracting density functionals at finite
  temperature},\ }\href {https://link.aps.org/doi/10.1103/PhysRevB.84.125118}
  {\bibfield  {journal} {\bibinfo  {journal} {Phys. Rev. B}\ }\textbf {\bibinfo
  {volume} {84}},\ \bibinfo {pages} {125118} (\bibinfo {year}
  {2011})}\BibitemShut {NoStop}%
\bibitem [{\citenamefont {Pribram-Jones}\ \emph {et~al.}(2014)\citenamefont
  {Pribram-Jones}, \citenamefont {Pittalis}, \citenamefont {Gross},\ and\
  \citenamefont {Burke}}]{PPGB14}%
  \BibitemOpen
  \bibfield  {author} {\bibinfo {author} {\bibfnamefont {A.}~\bibnamefont
  {Pribram-Jones}}, \bibinfo {author} {\bibfnamefont {S.}~\bibnamefont
  {Pittalis}}, \bibinfo {author} {\bibfnamefont {E.}~\bibnamefont {Gross}},\
  and\ \bibinfo {author} {\bibfnamefont {K.}~\bibnamefont {Burke}},\ }\bibfield
   {title} {\bibinfo {title} {Thermal density functional theory in context},\
  }in\ \href {https://doi.org/10.1007/978-3-319-04912-0_2} {\emph {\bibinfo
  {booktitle} {Frontiers and Challenges in Warm Dense Matter}}},\ \bibinfo
  {series} {Lecture Notes in Computational Science and Engineering},
  Vol.~\bibinfo {volume} {96},\ \bibinfo {editor} {edited by\ \bibinfo {editor}
  {\bibfnamefont {F.}~\bibnamefont {Graziani}}, \bibinfo {editor}
  {\bibfnamefont {M.~P.}\ \bibnamefont {Desjarlais}}, \bibinfo {editor}
  {\bibfnamefont {R.}~\bibnamefont {Redmer}},\ and\ \bibinfo {editor}
  {\bibfnamefont {S.~B.}\ \bibnamefont {Trickey}}}\ (\bibinfo  {publisher}
  {Springer International Publishing},\ \bibinfo {year} {2014})\ pp.\ \bibinfo
  {pages} {25--60}\BibitemShut {NoStop}%
\bibitem [{\citenamefont {Dufty}\ and\ \citenamefont {Trickey}(2016)}]{DT16}%
  \BibitemOpen
  \bibfield  {author} {\bibinfo {author} {\bibfnamefont {J.~W.}\ \bibnamefont
  {Dufty}}\ and\ \bibinfo {author} {\bibfnamefont {S.}~\bibnamefont
  {Trickey}},\ }\bibfield  {title} {\bibinfo {title} {Finite temperature
  scaling in density functional theory},\ }\href
  {https://doi.org/10.1080/00268976.2015.1122844} {\bibfield  {journal}
  {\bibinfo  {journal} {Mol. Phys.}\ }\textbf {\bibinfo {volume} {114}},\
  \bibinfo {pages} {988} (\bibinfo {year} {2016})}\BibitemShut {NoStop}%
\bibitem [{\citenamefont {Pribram-Jones}\ and\ \citenamefont
  {Burke}(2016)}]{JB16}%
  \BibitemOpen
  \bibfield  {author} {\bibinfo {author} {\bibfnamefont {A.}~\bibnamefont
  {Pribram-Jones}}\ and\ \bibinfo {author} {\bibfnamefont {K.}~\bibnamefont
  {Burke}},\ }\bibfield  {title} {\bibinfo {title} {Connection formulas for
  thermal density functional theory},\ }\href
  {https://doi.org/10.1103/PhysRevB.93.205140} {\bibfield  {journal} {\bibinfo
  {journal} {Phys. Rev. B}\ }\textbf {\bibinfo {volume} {93}},\ \bibinfo
  {pages} {205140} (\bibinfo {year} {2016})}\BibitemShut {NoStop}%
\bibitem [{\citenamefont {Burke}\ \emph {et~al.}(2016)\citenamefont {Burke},
  \citenamefont {Smith}, \citenamefont {Grabowski},\ and\ \citenamefont
  {Pribram-Jones}}]{BSGJ16}%
  \BibitemOpen
  \bibfield  {author} {\bibinfo {author} {\bibfnamefont {K.}~\bibnamefont
  {Burke}}, \bibinfo {author} {\bibfnamefont {J.~C.}\ \bibnamefont {Smith}},
  \bibinfo {author} {\bibfnamefont {P.~E.}\ \bibnamefont {Grabowski}},\ and\
  \bibinfo {author} {\bibfnamefont {A.}~\bibnamefont {Pribram-Jones}},\
  }\bibfield  {title} {\bibinfo {title} {Exact conditions on the temperature
  dependence of density functionals},\ }\href
  {https://doi.org/10.1103/PhysRevB.93.195132} {\bibfield  {journal} {\bibinfo
  {journal} {Phys. Rev. B}\ }\textbf {\bibinfo {volume} {93}},\ \bibinfo
  {pages} {195132} (\bibinfo {year} {2016})}\BibitemShut {NoStop}%
\bibitem [{\citenamefont {Smith}\ \emph {et~al.}(2018)\citenamefont {Smith},
  \citenamefont {Sagredo},\ and\ \citenamefont {Burke}}]{SSB18}%
  \BibitemOpen
  \bibfield  {author} {\bibinfo {author} {\bibfnamefont {J.~C.}\ \bibnamefont
  {Smith}}, \bibinfo {author} {\bibfnamefont {F.}~\bibnamefont {Sagredo}},\
  and\ \bibinfo {author} {\bibfnamefont {K.}~\bibnamefont {Burke}},\ }\bibinfo
  {title} {Warming up density functional theory},\ in\ \href
  {https://doi.org/10.1007/978-981-10-5651-2_11} {\emph {\bibinfo {booktitle}
  {Frontiers of Quantum Chemistry}}},\ \bibinfo {editor} {edited by\ \bibinfo
  {editor} {\bibfnamefont {M.~J.}\ \bibnamefont {W{\'o}jcik}}, \bibinfo
  {editor} {\bibfnamefont {H.}~\bibnamefont {Nakatsuji}}, \bibinfo {editor}
  {\bibfnamefont {B.}~\bibnamefont {Kirtman}},\ and\ \bibinfo {editor}
  {\bibfnamefont {Y.}~\bibnamefont {Ozaki}}}\ (\bibinfo  {publisher} {Springer
  Singapore},\ \bibinfo {address} {Singapore},\ \bibinfo {year} {2018})\ pp.\
  \bibinfo {pages} {249--271}\BibitemShut {NoStop}%
\bibitem [{\citenamefont {Sagredo}\ and\ \citenamefont {Burke}(2020)}]{SB20}%
  \BibitemOpen
  \bibfield  {author} {\bibinfo {author} {\bibfnamefont {F.}~\bibnamefont
  {Sagredo}}\ and\ \bibinfo {author} {\bibfnamefont {K.}~\bibnamefont
  {Burke}},\ }\bibfield  {title} {\bibinfo {title} {Confirmation of the {PPLB}
  derivative discontinuity: Exact chemical potential at finite temperatures of
  a model system},\ }\href {https://doi.org/10.1021/acs.jctc.0c00711}
  {\bibfield  {journal} {\bibinfo  {journal} {J. Chem. Comput. Theory}\
  }\textbf {\bibinfo {volume} {16}},\ \bibinfo {pages} {7225} (\bibinfo {year}
  {2020})}\BibitemShut {NoStop}%
\bibitem [{\citenamefont {Perrot}\ and\ \citenamefont
  {Dharma-wardana}(2000)}]{PD00}%
  \BibitemOpen
  \bibfield  {author} {\bibinfo {author} {\bibfnamefont {F.}~\bibnamefont
  {Perrot}}\ and\ \bibinfo {author} {\bibfnamefont {M.~W.~C.}\ \bibnamefont
  {Dharma-wardana}},\ }\bibfield  {title} {\bibinfo {title} {Spin-polarized
  electron liquid at arbitrary
  temperatures:$\hskip0.3em${}$\hskip0.3em${}exchange-correlation energies,
  electron-distribution functions, and the static response functions},\ }\href
  {https://doi.org/10.1103/PhysRevB.62.16536} {\bibfield  {journal} {\bibinfo
  {journal} {Phys. Rev. B}\ }\textbf {\bibinfo {volume} {62}},\ \bibinfo
  {pages} {16536} (\bibinfo {year} {2000})}\BibitemShut {NoStop}%
\bibitem [{\citenamefont {Gupta}\ and\ \citenamefont {Rajagopal}(1980)}]{GR80}%
  \BibitemOpen
  \bibfield  {author} {\bibinfo {author} {\bibfnamefont {U.}~\bibnamefont
  {Gupta}}\ and\ \bibinfo {author} {\bibfnamefont {A.~K.}\ \bibnamefont
  {Rajagopal}},\ }\bibfield  {title} {\bibinfo {title} {Exchange-correlation
  potential for inhomogeneous electron systems at finite temperatures},\ }\href
  {https://doi.org/10.1103/PhysRevA.22.2792} {\bibfield  {journal} {\bibinfo
  {journal} {Phys. Rev. A}\ }\textbf {\bibinfo {volume} {22}},\ \bibinfo
  {pages} {2792} (\bibinfo {year} {1980})}\BibitemShut {NoStop}%
\bibitem [{\citenamefont {Dharma-wardana}\ and\ \citenamefont
  {Taylor}(1981)}]{DT81}%
  \BibitemOpen
  \bibfield  {author} {\bibinfo {author} {\bibfnamefont {M.~W.~C.}\
  \bibnamefont {Dharma-wardana}}\ and\ \bibinfo {author} {\bibfnamefont
  {R.}~\bibnamefont {Taylor}},\ }\bibfield  {title} {\bibinfo {title} {Exchange
  and correlation potentials for finite temperature quantum calculations at
  intermediate degeneracies},\ }\href
  {http://stacks.iop.org/0022-3719/14/i=5/a=011} {\bibfield  {journal}
  {\bibinfo  {journal} {J. Phys. C}\ }\textbf {\bibinfo {volume} {14}},\
  \bibinfo {pages} {629} (\bibinfo {year} {1981})}\BibitemShut {NoStop}%
\bibitem [{\citenamefont {Langreth}\ and\ \citenamefont {Mehl}(1983)}]{LM83}%
  \BibitemOpen
  \bibfield  {author} {\bibinfo {author} {\bibfnamefont {D.}~\bibnamefont
  {Langreth}}\ and\ \bibinfo {author} {\bibfnamefont {M.}~\bibnamefont
  {Mehl}},\ }\bibfield  {title} {\bibinfo {title} {Beyond the local-density
  approximation in calculations of ground-state electronic properties},\ }\href
  {https://link.aps.org/doi/10.1103/PhysRevB.28.1809} {\bibfield  {journal}
  {\bibinfo  {journal} {Phys. Rev. B}\ }\textbf {\bibinfo {volume} {28}},\
  \bibinfo {pages} {1809} (\bibinfo {year} {1983})}\BibitemShut {NoStop}%
\bibitem [{\citenamefont {Sjostrom}\ and\ \citenamefont {Dufty}(2013)}]{SD13}%
  \BibitemOpen
  \bibfield  {author} {\bibinfo {author} {\bibfnamefont {T.}~\bibnamefont
  {Sjostrom}}\ and\ \bibinfo {author} {\bibfnamefont {J.}~\bibnamefont
  {Dufty}},\ }\bibfield  {title} {\bibinfo {title} {Uniform electron gas at
  finite temperatures},\ }\href {https://doi.org/10.1103/PhysRevB.88.115123}
  {\bibfield  {journal} {\bibinfo  {journal} {Phys. Rev. B}\ }\textbf {\bibinfo
  {volume} {88}},\ \bibinfo {pages} {115123} (\bibinfo {year}
  {2013})}\BibitemShut {NoStop}%
\bibitem [{\citenamefont {Brown}\ \emph {et~al.}(2013)\citenamefont {Brown},
  \citenamefont {Clark}, \citenamefont {DuBois},\ and\ \citenamefont
  {Ceperley}}]{BCDC13}%
  \BibitemOpen
  \bibfield  {author} {\bibinfo {author} {\bibfnamefont {E.~W.}\ \bibnamefont
  {Brown}}, \bibinfo {author} {\bibfnamefont {B.~K.}\ \bibnamefont {Clark}},
  \bibinfo {author} {\bibfnamefont {J.~L.}\ \bibnamefont {DuBois}},\ and\
  \bibinfo {author} {\bibfnamefont {D.~M.}\ \bibnamefont {Ceperley}},\
  }\bibfield  {title} {\bibinfo {title} {Path-integral {Monte Carlo} simulation
  of the warm dense homogeneous electron gas},\ }\href
  {https://doi.org/10.1103/PhysRevLett.110.146405} {\bibfield  {journal}
  {\bibinfo  {journal} {Phys. Rev. Lett.}\ }\textbf {\bibinfo {volume} {110}},\
  \bibinfo {pages} {146405} (\bibinfo {year} {2013})}\BibitemShut {NoStop}%
\bibitem [{\citenamefont {Dornheim}\ \emph {et~al.}(2016)\citenamefont
  {Dornheim}, \citenamefont {Groth}, \citenamefont {Sjostrom}, \citenamefont
  {Malone}, \citenamefont {Foulkes},\ and\ \citenamefont {Bonitz}}]{DGSMFB16}%
  \BibitemOpen
  \bibfield  {author} {\bibinfo {author} {\bibfnamefont {T.}~\bibnamefont
  {Dornheim}}, \bibinfo {author} {\bibfnamefont {S.}~\bibnamefont {Groth}},
  \bibinfo {author} {\bibfnamefont {T.}~\bibnamefont {Sjostrom}}, \bibinfo
  {author} {\bibfnamefont {F.~D.}\ \bibnamefont {Malone}}, \bibinfo {author}
  {\bibfnamefont {W.~M.~C.}\ \bibnamefont {Foulkes}},\ and\ \bibinfo {author}
  {\bibfnamefont {M.}~\bibnamefont {Bonitz}},\ }\bibfield  {title} {\bibinfo
  {title} {\textit{Ab initio} quantum {Monte Carlo} simulation of the warm
  dense electron gas in the thermodynamic limit},\ }\href
  {https://doi.org/10.1103/PhysRevLett.117.156403} {\bibfield  {journal}
  {\bibinfo  {journal} {Phys. Rev. Lett.}\ }\textbf {\bibinfo {volume} {117}},\
  \bibinfo {pages} {156403} (\bibinfo {year} {2016})}\BibitemShut {NoStop}%
\bibitem [{\citenamefont {Groth}\ \emph {et~al.}(2017)\citenamefont {Groth},
  \citenamefont {Dornheim}, \citenamefont {Sjostrom}, \citenamefont {Malone},
  \citenamefont {Foulkes},\ and\ \citenamefont {Bonitz}}]{GDSMFB17}%
  \BibitemOpen
  \bibfield  {author} {\bibinfo {author} {\bibfnamefont {S.}~\bibnamefont
  {Groth}}, \bibinfo {author} {\bibfnamefont {T.}~\bibnamefont {Dornheim}},
  \bibinfo {author} {\bibfnamefont {T.}~\bibnamefont {Sjostrom}}, \bibinfo
  {author} {\bibfnamefont {F.~D.}\ \bibnamefont {Malone}}, \bibinfo {author}
  {\bibfnamefont {W.~M.~C.}\ \bibnamefont {Foulkes}},\ and\ \bibinfo {author}
  {\bibfnamefont {M.}~\bibnamefont {Bonitz}},\ }\bibfield  {title} {\bibinfo
  {title} {\textit{Ab initio} exchange-correlation free energy of the uniform
  electron gas at warm dense matter conditions},\ }\href
  {https://doi.org/10.1103/PhysRevLett.119.135001} {\bibfield  {journal}
  {\bibinfo  {journal} {Phys. Rev. Lett.}\ }\textbf {\bibinfo {volume} {119}},\
  \bibinfo {pages} {135001} (\bibinfo {year} {2017})}\BibitemShut {NoStop}%
\bibitem [{\citenamefont {Sjostrom}\ and\ \citenamefont
  {Daligault}(2014)}]{SD14}%
  \BibitemOpen
  \bibfield  {author} {\bibinfo {author} {\bibfnamefont {T.}~\bibnamefont
  {Sjostrom}}\ and\ \bibinfo {author} {\bibfnamefont {J.}~\bibnamefont
  {Daligault}},\ }\bibfield  {title} {\bibinfo {title} {Gradient corrections to
  the exchange-correlation free energy},\ }\href
  {https://doi.org/10.1103/PhysRevB.90.155109} {\bibfield  {journal} {\bibinfo
  {journal} {Phys. Rev. B}\ }\textbf {\bibinfo {volume} {90}},\ \bibinfo
  {pages} {155109} (\bibinfo {year} {2014})}\BibitemShut {NoStop}%
\bibitem [{\citenamefont {Karasiev}\ \emph {et~al.}(2018)\citenamefont
  {Karasiev}, \citenamefont {Dufty},\ and\ \citenamefont {Trickey}}]{KDT18}%
  \BibitemOpen
  \bibfield  {author} {\bibinfo {author} {\bibfnamefont {V.~V.}\ \bibnamefont
  {Karasiev}}, \bibinfo {author} {\bibfnamefont {J.~W.}\ \bibnamefont
  {Dufty}},\ and\ \bibinfo {author} {\bibfnamefont {S.~B.}\ \bibnamefont
  {Trickey}},\ }\bibfield  {title} {\bibinfo {title} {Nonempirical semilocal
  free-energy density functional for matter under extreme conditions},\ }\href
  {https://doi.org/10.1103/PhysRevLett.120.076401} {\bibfield  {journal}
  {\bibinfo  {journal} {Phys. Rev. Lett.}\ }\textbf {\bibinfo {volume} {120}},\
  \bibinfo {pages} {076401} (\bibinfo {year} {2018})}\BibitemShut {NoStop}%
\bibitem [{\citenamefont {Lippert}\ \emph {et~al.}(2006)\citenamefont
  {Lippert}, \citenamefont {Modine},\ and\ \citenamefont {Wright}}]{LMW06}%
  \BibitemOpen
  \bibfield  {author} {\bibinfo {author} {\bibfnamefont {R.~A.}\ \bibnamefont
  {Lippert}}, \bibinfo {author} {\bibfnamefont {N.~A.}\ \bibnamefont
  {Modine}},\ and\ \bibinfo {author} {\bibfnamefont {A.~F.}\ \bibnamefont
  {Wright}},\ }\bibfield  {title} {\bibinfo {title} {The optimized effective
  potential with finite temperature},\ }\href
  {https://doi.org/10.1088/0953-8984/18/17/016} {\bibfield  {journal} {\bibinfo
   {journal} {J. Phys. Condens. Matter}\ }\textbf {\bibinfo {volume} {18}},\
  \bibinfo {pages} {4295} (\bibinfo {year} {2006})}\BibitemShut {NoStop}%
\bibitem [{\citenamefont {Greiner}\ \emph {et~al.}(2010)\citenamefont
  {Greiner}, \citenamefont {Carrier},\ and\ \citenamefont {G\"orling}}]{GCG10}%
  \BibitemOpen
  \bibfield  {author} {\bibinfo {author} {\bibfnamefont {M.}~\bibnamefont
  {Greiner}}, \bibinfo {author} {\bibfnamefont {P.}~\bibnamefont {Carrier}},\
  and\ \bibinfo {author} {\bibfnamefont {A.}~\bibnamefont {G\"orling}},\
  }\bibfield  {title} {\bibinfo {title} {Extension of exact-exchange density
  functional theory of solids to finite temperatures},\ }\href
  {https://doi.org/10.1103/PhysRevB.81.155119} {\bibfield  {journal} {\bibinfo
  {journal} {Phys. Rev. B}\ }\textbf {\bibinfo {volume} {81}},\ \bibinfo
  {pages} {155119} (\bibinfo {year} {2010})}\BibitemShut {NoStop}%
\bibitem [{\citenamefont {Karasiev}\ \emph
  {et~al.}(2014{\natexlab{b}})\citenamefont {Karasiev}, \citenamefont
  {Sjostrom}, \citenamefont {Chakraborty}, \citenamefont {Dufty}, \citenamefont
  {Runge}, \citenamefont {Harris},\ and\ \citenamefont {Trickey}}]{KSCD14}%
  \BibitemOpen
  \bibfield  {author} {\bibinfo {author} {\bibfnamefont {V.~V.}\ \bibnamefont
  {Karasiev}}, \bibinfo {author} {\bibfnamefont {T.}~\bibnamefont {Sjostrom}},
  \bibinfo {author} {\bibfnamefont {D.}~\bibnamefont {Chakraborty}}, \bibinfo
  {author} {\bibfnamefont {J.~W.}\ \bibnamefont {Dufty}}, \bibinfo {author}
  {\bibfnamefont {K.}~\bibnamefont {Runge}}, \bibinfo {author} {\bibfnamefont
  {F.~E.}\ \bibnamefont {Harris}},\ and\ \bibinfo {author} {\bibfnamefont
  {S.}~\bibnamefont {Trickey}},\ }\bibinfo {title} {Innovations in
  finite-temperature density functionals},\ in\ \href@noop {} {\emph {\bibinfo
  {booktitle} {Frontiers and Challenges in Warm Dense Matter}}},\ \bibinfo
  {series} {Lecture Notes in Computational Science and Engineering},
  Vol.~\bibinfo {volume} {96},\ \bibinfo {editor} {edited by\ \bibinfo {editor}
  {\bibfnamefont {F.}~\bibnamefont {Graziani}}, \bibinfo {editor}
  {\bibfnamefont {M.~P.}\ \bibnamefont {Desjarlais}}, \bibinfo {editor}
  {\bibfnamefont {R.}~\bibnamefont {Redmer}},\ and\ \bibinfo {editor}
  {\bibfnamefont {S.~B.}\ \bibnamefont {Trickey}}}\ (\bibinfo  {publisher}
  {Springer International Publishing},\ \bibinfo {year} {2014})\ pp.\ \bibinfo
  {pages} {61--85}\BibitemShut {NoStop}%
\bibitem [{\citenamefont {Lign{\`e}res}\ and\ \citenamefont
  {Carter}(2005)}]{LC05}%
  \BibitemOpen
  \bibfield  {author} {\bibinfo {author} {\bibfnamefont {V.~L.}\ \bibnamefont
  {Lign{\`e}res}}\ and\ \bibinfo {author} {\bibfnamefont {E.~A.}\ \bibnamefont
  {Carter}},\ }\bibinfo {title} {An introduction to orbital-free density
  functional theory},\ in\ \href {https://doi.org/10.1007/978-1-4020-3286-8_9}
  {\emph {\bibinfo {booktitle} {Handbook of Materials Modeling: Methods}}}\
  (\bibinfo  {publisher} {Springer Netherlands},\ \bibinfo {address}
  {Dordrecht},\ \bibinfo {year} {2005})\ pp.\ \bibinfo {pages}
  {137--148}\BibitemShut {NoStop}%
\bibitem [{\citenamefont {Karasiev}\ \emph
  {et~al.}(2014{\natexlab{c}})\citenamefont {Karasiev}, \citenamefont
  {Sjostrom},\ and\ \citenamefont {Trickey}}]{kst14}%
  \BibitemOpen
  \bibfield  {author} {\bibinfo {author} {\bibfnamefont {V.~V.}\ \bibnamefont
  {Karasiev}}, \bibinfo {author} {\bibfnamefont {T.}~\bibnamefont {Sjostrom}},\
  and\ \bibinfo {author} {\bibfnamefont {S.}~\bibnamefont {Trickey}},\
  }\bibfield  {title} {\bibinfo {title} {{Finite-temperature orbital-free DFT
  molecular dynamics: Coupling Profess and Quantum Espresso}},\ }\href
  {https://doi.org/https://doi.org/10.1016/j.cpc.2014.08.023} {\bibfield
  {journal} {\bibinfo  {journal} {Comput, Phys, Commun.}\ }\textbf {\bibinfo
  {volume} {185}},\ \bibinfo {pages} {3240 } (\bibinfo {year}
  {2014}{\natexlab{c}})}\BibitemShut {NoStop}%
\bibitem [{\citenamefont {White}\ \emph {et~al.}(2013)\citenamefont {White},
  \citenamefont {Richardson}, \citenamefont {Crowley}, \citenamefont
  {Pattison}, \citenamefont {Harris},\ and\ \citenamefont
  {Gregori}}]{WRCPHG13}%
  \BibitemOpen
  \bibfield  {author} {\bibinfo {author} {\bibfnamefont {T.~G.}\ \bibnamefont
  {White}}, \bibinfo {author} {\bibfnamefont {S.}~\bibnamefont {Richardson}},
  \bibinfo {author} {\bibfnamefont {B.~J.~B.}\ \bibnamefont {Crowley}},
  \bibinfo {author} {\bibfnamefont {L.~K.}\ \bibnamefont {Pattison}}, \bibinfo
  {author} {\bibfnamefont {J.~W.~O.}\ \bibnamefont {Harris}},\ and\ \bibinfo
  {author} {\bibfnamefont {G.}~\bibnamefont {Gregori}},\ }\bibfield  {title}
  {\bibinfo {title} {Orbital-free density-functional theory simulations of the
  dynamic structure factor of warm dense aluminum},\ }\href
  {https://doi.org/10.1103/PhysRevLett.111.175002} {\bibfield  {journal}
  {\bibinfo  {journal} {Phys. Rev. Lett.}\ }\textbf {\bibinfo {volume} {111}},\
  \bibinfo {pages} {175002} (\bibinfo {year} {2013})}\BibitemShut {NoStop}%
\bibitem [{\citenamefont {Militzer}\ and\ \citenamefont
  {Ceperley}(2000)}]{MC00}%
  \BibitemOpen
  \bibfield  {author} {\bibinfo {author} {\bibfnamefont {B.}~\bibnamefont
  {Militzer}}\ and\ \bibinfo {author} {\bibfnamefont {D.~M.}\ \bibnamefont
  {Ceperley}},\ }\bibfield  {title} {\bibinfo {title} {Path integral {Monte
  Carlo} calculation of the deuterium hugoniot},\ }\href
  {https://doi.org/10.1103/PhysRevLett.85.1890} {\bibfield  {journal} {\bibinfo
   {journal} {Phys. Rev. Lett.}\ }\textbf {\bibinfo {volume} {85}},\ \bibinfo
  {pages} {1890} (\bibinfo {year} {2000})}\BibitemShut {NoStop}%
\bibitem [{\citenamefont {Filinov}\ \emph {et~al.}(2001)\citenamefont
  {Filinov}, \citenamefont {Bonitz}, \citenamefont {Ebeling},\ and\
  \citenamefont {Fortov}}]{FBEF01}%
  \BibitemOpen
  \bibfield  {author} {\bibinfo {author} {\bibfnamefont {V.~S.}\ \bibnamefont
  {Filinov}}, \bibinfo {author} {\bibfnamefont {M.}~\bibnamefont {Bonitz}},
  \bibinfo {author} {\bibfnamefont {W.}~\bibnamefont {Ebeling}},\ and\ \bibinfo
  {author} {\bibfnamefont {V.~E.}\ \bibnamefont {Fortov}},\ }\bibfield  {title}
  {\bibinfo {title} {{Thermodynamics of hot dense H-plasmas: path integral
  Monte Carlo simulations and analytical approximations}},\ }\href
  {https://doi.org/10.1088/0741-3335/43/6/301} {\bibfield  {journal} {\bibinfo
  {journal} {Plasma Phys. Control. Fusion}\ }\textbf {\bibinfo {volume} {43}},\
  \bibinfo {pages} {743} (\bibinfo {year} {2001})}\BibitemShut {NoStop}%
\bibitem [{\citenamefont {Militzer}(2009)}]{M09}%
  \BibitemOpen
  \bibfield  {author} {\bibinfo {author} {\bibfnamefont {B.}~\bibnamefont
  {Militzer}},\ }\bibfield  {title} {\bibinfo {title} {Path integral monte
  carlo and density functional molecular dynamics simulations of hot, dense
  helium},\ }\href {https://doi.org/10.1103/PhysRevB.79.155105} {\bibfield
  {journal} {\bibinfo  {journal} {Phys. Rev. B}\ }\textbf {\bibinfo {volume}
  {79}},\ \bibinfo {pages} {155105} (\bibinfo {year} {2009})}\BibitemShut
  {NoStop}%
\bibitem [{\citenamefont {Driver}\ and\ \citenamefont {Militzer}(2012)}]{DM12}%
  \BibitemOpen
  \bibfield  {author} {\bibinfo {author} {\bibfnamefont {K.~P.}\ \bibnamefont
  {Driver}}\ and\ \bibinfo {author} {\bibfnamefont {B.}~\bibnamefont
  {Militzer}},\ }\bibfield  {title} {\bibinfo {title} {All-electron path
  integral {Monte Carlo} simulations of warm dense matter: Application to water
  and carbon plasmas},\ }\href {https://doi.org/10.1103/PhysRevLett.108.115502}
  {\bibfield  {journal} {\bibinfo  {journal} {Phys. Rev. Lett.}\ }\textbf
  {\bibinfo {volume} {108}},\ \bibinfo {pages} {115502} (\bibinfo {year}
  {2012})}\BibitemShut {NoStop}%
\bibitem [{\citenamefont {Zhou}\ \emph {et~al.}(2005)\citenamefont {Zhou},
  \citenamefont {Ligneres},\ and\ \citenamefont {Carter}}]{BLC05}%
  \BibitemOpen
  \bibfield  {author} {\bibinfo {author} {\bibfnamefont {B.}~\bibnamefont
  {Zhou}}, \bibinfo {author} {\bibfnamefont {V.~L.}\ \bibnamefont {Ligneres}},\
  and\ \bibinfo {author} {\bibfnamefont {E.~A.}\ \bibnamefont {Carter}},\
  }\bibfield  {title} {\bibinfo {title} {Improving the orbital-free density
  functional theory description of covalent materials},\ }\href
  {https://doi.org/10.1063/1.1834563} {\bibfield  {journal} {\bibinfo
  {journal} {J. Chem. Phys.}\ }\textbf {\bibinfo {volume} {122}},\ \bibinfo
  {pages} {044103} (\bibinfo {year} {2005})}\BibitemShut {NoStop}%
\bibitem [{\citenamefont {Gao}\ \emph {et~al.}(2016)\citenamefont {Gao},
  \citenamefont {Zhang}, \citenamefont {Kang}, \citenamefont {Wang},
  \citenamefont {Zhang},\ and\ \citenamefont {He}}]{ZKWCZ16}%
  \BibitemOpen
  \bibfield  {author} {\bibinfo {author} {\bibfnamefont {C.}~\bibnamefont
  {Gao}}, \bibinfo {author} {\bibfnamefont {S.}~\bibnamefont {Zhang}}, \bibinfo
  {author} {\bibfnamefont {W.}~\bibnamefont {Kang}}, \bibinfo {author}
  {\bibfnamefont {C.}~\bibnamefont {Wang}}, \bibinfo {author} {\bibfnamefont
  {P.}~\bibnamefont {Zhang}},\ and\ \bibinfo {author} {\bibfnamefont {X.~T.}\
  \bibnamefont {He}},\ }\bibfield  {title} {\bibinfo {title} {Validity boundary
  of orbital-free molecular dynamics method corresponding to thermal ionization
  of shell structure},\ }\href {https://doi.org/10.1103/PhysRevB.94.205115}
  {\bibfield  {journal} {\bibinfo  {journal} {Phys. Rev. B}\ }\textbf {\bibinfo
  {volume} {94}},\ \bibinfo {pages} {205115} (\bibinfo {year}
  {2016})}\BibitemShut {NoStop}%
\bibitem [{\citenamefont {Ellis}\ \emph {et~al.}(2021)\citenamefont {Ellis},
  \citenamefont {Fiedler}, \citenamefont {Popoola}, \citenamefont {Modine},
  \citenamefont {Stephens}, \citenamefont {Thompson}, \citenamefont {Cangi},\
  and\ \citenamefont {Rajamanickam}}]{ECMSTR20}%
  \BibitemOpen
  \bibfield  {author} {\bibinfo {author} {\bibfnamefont {J.~A.}\ \bibnamefont
  {Ellis}}, \bibinfo {author} {\bibfnamefont {L.}~\bibnamefont {Fiedler}},
  \bibinfo {author} {\bibfnamefont {G.~A.}\ \bibnamefont {Popoola}}, \bibinfo
  {author} {\bibfnamefont {N.~A.}\ \bibnamefont {Modine}}, \bibinfo {author}
  {\bibfnamefont {J.~A.}\ \bibnamefont {Stephens}}, \bibinfo {author}
  {\bibfnamefont {A.~P.}\ \bibnamefont {Thompson}}, \bibinfo {author}
  {\bibfnamefont {A.}~\bibnamefont {Cangi}},\ and\ \bibinfo {author}
  {\bibfnamefont {S.}~\bibnamefont {Rajamanickam}},\ }\bibfield  {title}
  {\bibinfo {title} {Accelerating finite-temperature kohn-sham density
  functional theory with deep neural networks},\ }\href
  {https://doi.org/10.1103/PhysRevB.104.035120} {\bibfield  {journal} {\bibinfo
   {journal} {Phys. Rev. B}\ }\textbf {\bibinfo {volume} {104}},\ \bibinfo
  {pages} {035120} (\bibinfo {year} {2021})}\BibitemShut {NoStop}%
\bibitem [{\citenamefont {Cytter}\ \emph {et~al.}(2018)\citenamefont {Cytter},
  \citenamefont {Rabani}, \citenamefont {Neuhauser},\ and\ \citenamefont
  {Baer}}]{CRNB18}%
  \BibitemOpen
  \bibfield  {author} {\bibinfo {author} {\bibfnamefont {Y.}~\bibnamefont
  {Cytter}}, \bibinfo {author} {\bibfnamefont {E.}~\bibnamefont {Rabani}},
  \bibinfo {author} {\bibfnamefont {D.}~\bibnamefont {Neuhauser}},\ and\
  \bibinfo {author} {\bibfnamefont {R.}~\bibnamefont {Baer}},\ }\bibfield
  {title} {\bibinfo {title} {Stochastic density functional theory at finite
  temperatures},\ }\href {https://doi.org/10.1103/PhysRevB.97.115207}
  {\bibfield  {journal} {\bibinfo  {journal} {Phys. Rev. B}\ }\textbf {\bibinfo
  {volume} {97}},\ \bibinfo {pages} {115207} (\bibinfo {year}
  {2018})}\BibitemShut {NoStop}%
\bibitem [{\citenamefont {Cytter}\ \emph {et~al.}(2019)\citenamefont {Cytter},
  \citenamefont {Rabani}, \citenamefont {Neuhauser}, \citenamefont {Preising},
  \citenamefont {Redmer},\ and\ \citenamefont {Baer}}]{CRNPRB20}%
  \BibitemOpen
  \bibfield  {author} {\bibinfo {author} {\bibfnamefont {Y.}~\bibnamefont
  {Cytter}}, \bibinfo {author} {\bibfnamefont {E.}~\bibnamefont {Rabani}},
  \bibinfo {author} {\bibfnamefont {D.}~\bibnamefont {Neuhauser}}, \bibinfo
  {author} {\bibfnamefont {M.}~\bibnamefont {Preising}}, \bibinfo {author}
  {\bibfnamefont {R.}~\bibnamefont {Redmer}},\ and\ \bibinfo {author}
  {\bibfnamefont {R.}~\bibnamefont {Baer}},\ }\bibfield  {title} {\bibinfo
  {title} {Transition to metallization in warm dense helium-hydrogen mixtures
  using stochastic density functional theory within the {Kubo-Greenwood}
  formalism},\ }\href {https://doi.org/10.1103/PhysRevB.100.195101} {\bibfield
  {journal} {\bibinfo  {journal} {Phys. Rev. B}\ }\textbf {\bibinfo {volume}
  {100}},\ \bibinfo {pages} {195101} (\bibinfo {year} {2019})}\BibitemShut
  {NoStop}%
\bibitem [{\citenamefont {White}\ and\ \citenamefont {Collins}(2020)}]{WC20}%
  \BibitemOpen
  \bibfield  {author} {\bibinfo {author} {\bibfnamefont {A.~J.}\ \bibnamefont
  {White}}\ and\ \bibinfo {author} {\bibfnamefont {L.~A.}\ \bibnamefont
  {Collins}},\ }\bibfield  {title} {\bibinfo {title} {Fast and universal
  {Kohn-Sham} density functional theory algorithm for warm dense matter to hot
  dense plasma},\ }\href {https://doi.org/10.1103/PhysRevLett.125.055002}
  {\bibfield  {journal} {\bibinfo  {journal} {Phys. Rev. Lett.}\ }\textbf
  {\bibinfo {volume} {125}},\ \bibinfo {pages} {055002} (\bibinfo {year}
  {2020})}\BibitemShut {NoStop}%
\bibitem [{\citenamefont {Mazevet}\ and\ \citenamefont {Z\'erah}(2008)}]{MZ08}%
  \BibitemOpen
  \bibfield  {author} {\bibinfo {author} {\bibfnamefont {S.}~\bibnamefont
  {Mazevet}}\ and\ \bibinfo {author} {\bibfnamefont {G.}~\bibnamefont
  {Z\'erah}},\ }\bibfield  {title} {\bibinfo {title} {\textit{Ab initio}
  simulations of the {$K$}-edge shift along the aluminum {Hugoniot}},\ }\href
  {https://doi.org/10.1103/PhysRevLett.101.155001} {\bibfield  {journal}
  {\bibinfo  {journal} {Phys. Rev. Lett.}\ }\textbf {\bibinfo {volume} {101}},\
  \bibinfo {pages} {155001} (\bibinfo {year} {2008})}\BibitemShut {NoStop}%
\bibitem [{\citenamefont {Zhang}\ \emph {et~al.}(2016)\citenamefont {Zhang},
  \citenamefont {Zhao}, \citenamefont {Kang}, \citenamefont {Zhang},\ and\
  \citenamefont {He}}]{ZZKZH16}%
  \BibitemOpen
  \bibfield  {author} {\bibinfo {author} {\bibfnamefont {S.}~\bibnamefont
  {Zhang}}, \bibinfo {author} {\bibfnamefont {S.}~\bibnamefont {Zhao}},
  \bibinfo {author} {\bibfnamefont {W.}~\bibnamefont {Kang}}, \bibinfo {author}
  {\bibfnamefont {P.}~\bibnamefont {Zhang}},\ and\ \bibinfo {author}
  {\bibfnamefont {X.-T.}\ \bibnamefont {He}},\ }\bibfield  {title} {\bibinfo
  {title} {Link between {$K$} absorption edges and thermodynamic properties of
  warm dense plasmas established by an improved first-principles method},\
  }\href {https://doi.org/10.1103/PhysRevB.93.115114} {\bibfield  {journal}
  {\bibinfo  {journal} {Phys. Rev. B}\ }\textbf {\bibinfo {volume} {93}},\
  \bibinfo {pages} {115114} (\bibinfo {year} {2016})}\BibitemShut {NoStop}%
\bibitem [{\citenamefont {Wigner}\ and\ \citenamefont {Seitz}(1933)}]{WS33}%
  \BibitemOpen
  \bibfield  {author} {\bibinfo {author} {\bibfnamefont {E.}~\bibnamefont
  {Wigner}}\ and\ \bibinfo {author} {\bibfnamefont {F.}~\bibnamefont {Seitz}},\
  }\bibfield  {title} {\bibinfo {title} {On the constitution of metallic
  sodium},\ }\href {https://doi.org/10.1103/PhysRev.43.804} {\bibfield
  {journal} {\bibinfo  {journal} {Phys. Rev.}\ }\textbf {\bibinfo {volume}
  {43}},\ \bibinfo {pages} {804} (\bibinfo {year} {1933})}\BibitemShut
  {NoStop}%
\bibitem [{\citenamefont {Wigner}\ and\ \citenamefont {Seitz}(1934)}]{WS34}%
  \BibitemOpen
  \bibfield  {author} {\bibinfo {author} {\bibfnamefont {E.}~\bibnamefont
  {Wigner}}\ and\ \bibinfo {author} {\bibfnamefont {F.}~\bibnamefont {Seitz}},\
  }\bibfield  {title} {\bibinfo {title} {On the constitution of metallic
  sodium. ii},\ }\href {https://doi.org/10.1103/PhysRev.46.509} {\bibfield
  {journal} {\bibinfo  {journal} {Phys. Rev.}\ }\textbf {\bibinfo {volume}
  {46}},\ \bibinfo {pages} {509} (\bibinfo {year} {1934})}\BibitemShut
  {NoStop}%
\bibitem [{\citenamefont {Slater}\ and\ \citenamefont {Krutter}(1935)}]{SK35}%
  \BibitemOpen
  \bibfield  {author} {\bibinfo {author} {\bibfnamefont {J.~C.}\ \bibnamefont
  {Slater}}\ and\ \bibinfo {author} {\bibfnamefont {H.~M.}\ \bibnamefont
  {Krutter}},\ }\bibfield  {title} {\bibinfo {title} {The {Thomas-Fermi} method
  for metals},\ }\href {https://doi.org/10.1103/PhysRev.47.559} {\bibfield
  {journal} {\bibinfo  {journal} {Phys. Rev.}\ }\textbf {\bibinfo {volume}
  {47}},\ \bibinfo {pages} {559} (\bibinfo {year} {1935})}\BibitemShut
  {NoStop}%
\bibitem [{\citenamefont {Feynman}\ \emph {et~al.}(1949)\citenamefont
  {Feynman}, \citenamefont {Metropolis},\ and\ \citenamefont {Teller}}]{FMT49}%
  \BibitemOpen
  \bibfield  {author} {\bibinfo {author} {\bibfnamefont {R.~P.}\ \bibnamefont
  {Feynman}}, \bibinfo {author} {\bibfnamefont {N.}~\bibnamefont
  {Metropolis}},\ and\ \bibinfo {author} {\bibfnamefont {E.}~\bibnamefont
  {Teller}},\ }\bibfield  {title} {\bibinfo {title} {Equations of state of
  elements based on the generalized {Fermi-Thomas} theory},\ }\href
  {https://doi.org/10.1103/PhysRev.75.1561} {\bibfield  {journal} {\bibinfo
  {journal} {Phys. Rev.}\ }\textbf {\bibinfo {volume} {75}},\ \bibinfo {pages}
  {1561} (\bibinfo {year} {1949})}\BibitemShut {NoStop}%
\bibitem [{\citenamefont {Latter}(1955)}]{La55}%
  \BibitemOpen
  \bibfield  {author} {\bibinfo {author} {\bibfnamefont {R.}~\bibnamefont
  {Latter}},\ }\bibfield  {title} {\bibinfo {title} {Atomic energy levels for
  the {Thomas-Fermi and Thomas-Fermi-Dirac} potential},\ }\href
  {https://doi.org/10.1103/PhysRev.99.510} {\bibfield  {journal} {\bibinfo
  {journal} {Phys. Rev.}\ }\textbf {\bibinfo {volume} {99}},\ \bibinfo {pages}
  {510} (\bibinfo {year} {1955})}\BibitemShut {NoStop}%
\bibitem [{\citenamefont {Thomas}(1927)}]{T27}%
  \BibitemOpen
  \bibfield  {author} {\bibinfo {author} {\bibfnamefont {L.}~\bibnamefont
  {Thomas}},\ }\bibfield  {title} {\bibinfo {title} {The calculation of atomic
  fields},\ }\href {https://doi.org/10.1017/S0305004100011683} {\bibfield
  {journal} {\bibinfo  {journal} {Math. Proc. Camb. Phil. Soc.}\ }\textbf
  {\bibinfo {volume} {23}},\ \bibinfo {pages} {542} (\bibinfo {year}
  {1927})}\BibitemShut {NoStop}%
\bibitem [{\citenamefont {Fermi}(1927)}]{F27}%
  \BibitemOpen
  \bibfield  {author} {\bibinfo {author} {\bibfnamefont {E.}~\bibnamefont
  {Fermi}},\ }\href@noop {} {\bibfield  {journal} {\bibinfo  {journal} {Rend.
  Acc. Naz. Lincei}\ }\textbf {\bibinfo {volume} {6}} (\bibinfo {year}
  {1927})}\BibitemShut {NoStop}%
\bibitem [{\citenamefont {Rozsnyai}(1972)}]{R72}%
  \BibitemOpen
  \bibfield  {author} {\bibinfo {author} {\bibfnamefont {B.~F.}\ \bibnamefont
  {Rozsnyai}},\ }\bibfield  {title} {\bibinfo {title} {Relativistic
  {Hartree-Fock-Slater} calculations for arbitrary temperature and matter
  density},\ }\href {https://doi.org/10.1103/PhysRevA.5.1137} {\bibfield
  {journal} {\bibinfo  {journal} {Phys. Rev. A}\ }\textbf {\bibinfo {volume}
  {5}},\ \bibinfo {pages} {1137} (\bibinfo {year} {1972})}\BibitemShut
  {NoStop}%
\bibitem [{\citenamefont {Liberman}(1979)}]{Li_1979}%
  \BibitemOpen
  \bibfield  {author} {\bibinfo {author} {\bibfnamefont {D.~A.}\ \bibnamefont
  {Liberman}},\ }\bibfield  {title} {\bibinfo {title} {Self-consistent field
  model for condensed matter},\ }\href
  {https://doi.org/10.1103/PhysRevB.20.4981} {\bibfield  {journal} {\bibinfo
  {journal} {Phys. Rev. B}\ }\textbf {\bibinfo {volume} {20}},\ \bibinfo
  {pages} {4981} (\bibinfo {year} {1979})}\BibitemShut {NoStop}%
\bibitem [{\citenamefont {Dharma-wardana}\ and\ \citenamefont
  {Perrot}(1982)}]{DP82}%
  \BibitemOpen
  \bibfield  {author} {\bibinfo {author} {\bibfnamefont {M.~W.~C.}\
  \bibnamefont {Dharma-wardana}}\ and\ \bibinfo {author} {\bibfnamefont
  {F.~m.~c.}\ \bibnamefont {Perrot}},\ }\bibfield  {title} {\bibinfo {title}
  {Density-functional theory of hydrogen plasmas},\ }\href
  {https://doi.org/10.1103/PhysRevA.26.2096} {\bibfield  {journal} {\bibinfo
  {journal} {Phys. Rev. A}\ }\textbf {\bibinfo {volume} {26}},\ \bibinfo
  {pages} {2096} (\bibinfo {year} {1982})}\BibitemShut {NoStop}%
\bibitem [{\citenamefont {Chihara}(1985)}]{C85}%
  \BibitemOpen
  \bibfield  {author} {\bibinfo {author} {\bibfnamefont {J.}~\bibnamefont
  {Chihara}},\ }\bibfield  {title} {\bibinfo {title} {Liquid metals and plasmas
  as nucleus-electron mixtures},\ }\href
  {https://doi.org/10.1088/0022-3719/18/16/008} {\bibfield  {journal} {\bibinfo
   {journal} {Journal of Physics C: Solid State Physics}\ }\textbf {\bibinfo
  {volume} {18}},\ \bibinfo {pages} {3103} (\bibinfo {year}
  {1985})}\BibitemShut {NoStop}%
\bibitem [{\citenamefont {Perrot}\ \emph {et~al.}(1990)\citenamefont {Perrot},
  \citenamefont {Furutani},\ and\ \citenamefont {Dharma-wardana}}]{PFD90}%
  \BibitemOpen
  \bibfield  {author} {\bibinfo {author} {\bibfnamefont {F.}~\bibnamefont
  {Perrot}}, \bibinfo {author} {\bibfnamefont {Y.}~\bibnamefont {Furutani}},\
  and\ \bibinfo {author} {\bibfnamefont {M.~W.~C.}\ \bibnamefont
  {Dharma-wardana}},\ }\bibfield  {title} {\bibinfo {title} {{Electron-ion
  correlation potentials in the density-functional theory of H and He
  plasmas}},\ }\href {https://doi.org/10.1103/PhysRevA.41.1096} {\bibfield
  {journal} {\bibinfo  {journal} {Phys. Rev. A}\ }\textbf {\bibinfo {volume}
  {41}},\ \bibinfo {pages} {1096} (\bibinfo {year} {1990})}\BibitemShut
  {NoStop}%
\bibitem [{\citenamefont {Rozsnyai}(1991)}]{R91}%
  \BibitemOpen
  \bibfield  {author} {\bibinfo {author} {\bibfnamefont {B.~F.}\ \bibnamefont
  {Rozsnyai}},\ }\bibfield  {title} {\bibinfo {title} {Photoabsorption in hot
  plasmas based on the ion-sphere and ion-correlation models},\ }\href
  {https://doi.org/10.1103/PhysRevA.43.3035} {\bibfield  {journal} {\bibinfo
  {journal} {Phys. Rev. A}\ }\textbf {\bibinfo {volume} {43}},\ \bibinfo
  {pages} {3035} (\bibinfo {year} {1991})}\BibitemShut {NoStop}%
\bibitem [{\citenamefont {Yuan}(2002)}]{Y02}%
  \BibitemOpen
  \bibfield  {author} {\bibinfo {author} {\bibfnamefont {J.}~\bibnamefont
  {Yuan}},\ }\bibfield  {title} {\bibinfo {title} {Self-consistent average-atom
  scheme for electronic structure of hot and dense plasmas of mixture},\ }\href
  {https://doi.org/10.1103/PhysRevE.66.047401} {\bibfield  {journal} {\bibinfo
  {journal} {Phys. Rev. E}\ }\textbf {\bibinfo {volume} {66}},\ \bibinfo
  {pages} {047401} (\bibinfo {year} {2002})}\BibitemShut {NoStop}%
\bibitem [{\citenamefont {Blancard}\ and\ \citenamefont
  {Faussurier}(2004)}]{BF04}%
  \BibitemOpen
  \bibfield  {author} {\bibinfo {author} {\bibfnamefont {C.}~\bibnamefont
  {Blancard}}\ and\ \bibinfo {author} {\bibfnamefont {G.}~\bibnamefont
  {Faussurier}},\ }\bibfield  {title} {\bibinfo {title} {Equation of state and
  transport coefficients for dense plasmas},\ }\href
  {https://doi.org/10.1103/PhysRevE.69.016409} {\bibfield  {journal} {\bibinfo
  {journal} {Phys. Rev. E}\ }\textbf {\bibinfo {volume} {69}},\ \bibinfo
  {pages} {016409} (\bibinfo {year} {2004})}\BibitemShut {NoStop}%
\bibitem [{\citenamefont {Johnson}\ \emph {et~al.}(2006)\citenamefont
  {Johnson}, \citenamefont {Guet},\ and\ \citenamefont {Bertsch}}]{JGB06}%
  \BibitemOpen
  \bibfield  {author} {\bibinfo {author} {\bibfnamefont {W.}~\bibnamefont
  {Johnson}}, \bibinfo {author} {\bibfnamefont {C.}~\bibnamefont {Guet}},\ and\
  \bibinfo {author} {\bibfnamefont {G.}~\bibnamefont {Bertsch}},\ }\bibfield
  {title} {\bibinfo {title} {Optical properties of plasmas based on an
  average-atom model},\ }\href
  {https://doi.org/https://doi.org/10.1016/j.jqsrt.2005.05.026} {\bibfield
  {journal} {\bibinfo  {journal} {J. Quant. Spectrosc. Radiat. Transf.}\
  }\textbf {\bibinfo {volume} {99}},\ \bibinfo {pages} {327 } (\bibinfo {year}
  {2006})}\BibitemShut {NoStop}%
\bibitem [{\citenamefont {Sterne}\ \emph {et~al.}(2007)\citenamefont {Sterne},
  \citenamefont {Hansen}, \citenamefont {Wilson},\ and\ \citenamefont
  {Isaacs}}]{SHWI07}%
  \BibitemOpen
  \bibfield  {author} {\bibinfo {author} {\bibfnamefont {P.}~\bibnamefont
  {Sterne}}, \bibinfo {author} {\bibfnamefont {S.}~\bibnamefont {Hansen}},
  \bibinfo {author} {\bibfnamefont {B.}~\bibnamefont {Wilson}},\ and\ \bibinfo
  {author} {\bibfnamefont {W.}~\bibnamefont {Isaacs}},\ }\bibfield  {title}
  {\bibinfo {title} {{Equation of state, occupation probabilities and
  conductivities in the average atom Purgatorio code}},\ }\href
  {https://doi.org/https://doi.org/10.1016/j.hedp.2007.02.037} {\bibfield
  {journal} {\bibinfo  {journal} {High Energy Density Phys.}\ }\textbf
  {\bibinfo {volume} {3}},\ \bibinfo {pages} {278} (\bibinfo {year}
  {2007})}\BibitemShut {NoStop}%
\bibitem [{\citenamefont {Wilson}\ \emph {et~al.}(2006)\citenamefont {Wilson},
  \citenamefont {Sonnad}, \citenamefont {Sterne},\ and\ \citenamefont
  {Isaacs}}]{WSSI06}%
  \BibitemOpen
  \bibfield  {author} {\bibinfo {author} {\bibfnamefont {B.}~\bibnamefont
  {Wilson}}, \bibinfo {author} {\bibfnamefont {V.}~\bibnamefont {Sonnad}},
  \bibinfo {author} {\bibfnamefont {P.}~\bibnamefont {Sterne}},\ and\ \bibinfo
  {author} {\bibfnamefont {W.}~\bibnamefont {Isaacs}},\ }\bibfield  {title}
  {\bibinfo {title} {{Purgatorio — a new implementation of the Inferno
  algorithm}},\ }\href
  {https://doi.org/https://doi.org/10.1016/j.jqsrt.2005.05.053} {\bibfield
  {journal} {\bibinfo  {journal} {J. Quant. Spectrosc. Radiat. Transf.}\
  }\textbf {\bibinfo {volume} {99}},\ \bibinfo {pages} {658} (\bibinfo {year}
  {2006})}\BibitemShut {NoStop}%
\bibitem [{\citenamefont {Blenski}\ and\ \citenamefont
  {Cichocki}(2007)}]{BC07}%
  \BibitemOpen
  \bibfield  {author} {\bibinfo {author} {\bibfnamefont {T.}~\bibnamefont
  {Blenski}}\ and\ \bibinfo {author} {\bibfnamefont {B.}~\bibnamefont
  {Cichocki}},\ }\bibfield  {title} {\bibinfo {title} {Variational theory of
  average-atom and superconfigurations in quantum plasmas},\ }\href
  {https://doi.org/10.1103/PhysRevE.75.056402} {\bibfield  {journal} {\bibinfo
  {journal} {Phys. Rev. E}\ }\textbf {\bibinfo {volume} {75}},\ \bibinfo
  {pages} {056402} (\bibinfo {year} {2007})}\BibitemShut {NoStop}%
\bibitem [{\citenamefont {Faussurier}\ \emph {et~al.}(2010)\citenamefont
  {Faussurier}, \citenamefont {Blancard}, \citenamefont {Cossé},\ and\
  \citenamefont {Renaudin}}]{FBCR10}%
  \BibitemOpen
  \bibfield  {author} {\bibinfo {author} {\bibfnamefont {G.}~\bibnamefont
  {Faussurier}}, \bibinfo {author} {\bibfnamefont {C.}~\bibnamefont
  {Blancard}}, \bibinfo {author} {\bibfnamefont {P.}~\bibnamefont {Cossé}},\
  and\ \bibinfo {author} {\bibfnamefont {P.}~\bibnamefont {Renaudin}},\
  }\bibfield  {title} {\bibinfo {title} {Equation of state, transport
  coefficients, and stopping power of dense plasmas from the average-atom model
  self-consistent approach for astrophysical and laboratory plasmas},\ }\href
  {https://doi.org/10.1063/1.3420276} {\bibfield  {journal} {\bibinfo
  {journal} {Phys. Plasmas}\ }\textbf {\bibinfo {volume} {17}},\ \bibinfo
  {pages} {052707} (\bibinfo {year} {2010})}\BibitemShut {NoStop}%
\bibitem [{\citenamefont {Sahoo}\ \emph {et~al.}(2008)\citenamefont {Sahoo},
  \citenamefont {Gribakin}, \citenamefont {Shabbir~Naz}, \citenamefont
  {Kohanoff},\ and\ \citenamefont {Riley}}]{SGSKR08}%
  \BibitemOpen
  \bibfield  {author} {\bibinfo {author} {\bibfnamefont {S.}~\bibnamefont
  {Sahoo}}, \bibinfo {author} {\bibfnamefont {G.~F.}\ \bibnamefont {Gribakin}},
  \bibinfo {author} {\bibfnamefont {G.}~\bibnamefont {Shabbir~Naz}}, \bibinfo
  {author} {\bibfnamefont {J.}~\bibnamefont {Kohanoff}},\ and\ \bibinfo
  {author} {\bibfnamefont {D.}~\bibnamefont {Riley}},\ }\bibfield  {title}
  {\bibinfo {title} {Compton scatter profiles for warm dense matter},\ }\href
  {https://doi.org/10.1103/PhysRevE.77.046402} {\bibfield  {journal} {\bibinfo
  {journal} {Phys. Rev. E}\ }\textbf {\bibinfo {volume} {77}},\ \bibinfo
  {pages} {046402} (\bibinfo {year} {2008})}\BibitemShut {NoStop}%
\bibitem [{\citenamefont {Piron}\ and\ \citenamefont {Blenski}(2011)}]{PB11}%
  \BibitemOpen
  \bibfield  {author} {\bibinfo {author} {\bibfnamefont {R.}~\bibnamefont
  {Piron}}\ and\ \bibinfo {author} {\bibfnamefont {T.}~\bibnamefont
  {Blenski}},\ }\bibfield  {title} {\bibinfo {title}
  {{Variational-average-atom-in-quantum-plasmas (VAAQP) code and virial
  theorem: Equation-of-state and shock-Hugoniot calculations for warm dense Al,
  Fe, Cu, and Pb}},\ }\href {https://doi.org/10.1103/PhysRevE.83.026403}
  {\bibfield  {journal} {\bibinfo  {journal} {Phys. Rev. E}\ }\textbf {\bibinfo
  {volume} {83}},\ \bibinfo {pages} {026403} (\bibinfo {year}
  {2011})}\BibitemShut {NoStop}%
\bibitem [{\citenamefont {Johnson}\ \emph {et~al.}(2012)\citenamefont
  {Johnson}, \citenamefont {Nilsen},\ and\ \citenamefont {Cheng}}]{JNC12}%
  \BibitemOpen
  \bibfield  {author} {\bibinfo {author} {\bibfnamefont {W.~R.}\ \bibnamefont
  {Johnson}}, \bibinfo {author} {\bibfnamefont {J.}~\bibnamefont {Nilsen}},\
  and\ \bibinfo {author} {\bibfnamefont {K.~T.}\ \bibnamefont {Cheng}},\
  }\bibfield  {title} {\bibinfo {title} {Thomson scattering in the average-atom
  approximation},\ }\href {https://doi.org/10.1103/PhysRevE.86.036410}
  {\bibfield  {journal} {\bibinfo  {journal} {Phys. Rev. E}\ }\textbf {\bibinfo
  {volume} {86}},\ \bibinfo {pages} {036410} (\bibinfo {year}
  {2012})}\BibitemShut {NoStop}%
\bibitem [{\citenamefont {Starrett}\ and\ \citenamefont {Saumon}(2012)}]{SS12}%
  \BibitemOpen
  \bibfield  {author} {\bibinfo {author} {\bibfnamefont {C.~E.}\ \bibnamefont
  {Starrett}}\ and\ \bibinfo {author} {\bibfnamefont {D.}~\bibnamefont
  {Saumon}},\ }\bibfield  {title} {\bibinfo {title} {Fully variational average
  atom model with ion-ion correlations},\ }\href
  {https://doi.org/10.1103/PhysRevE.85.026403} {\bibfield  {journal} {\bibinfo
  {journal} {Phys. Rev. E}\ }\textbf {\bibinfo {volume} {85}},\ \bibinfo
  {pages} {026403} (\bibinfo {year} {2012})}\BibitemShut {NoStop}%
\bibitem [{\citenamefont {Starrett}\ and\ \citenamefont {Saumon}(2013)}]{SS13}%
  \BibitemOpen
  \bibfield  {author} {\bibinfo {author} {\bibfnamefont {C.~E.}\ \bibnamefont
  {Starrett}}\ and\ \bibinfo {author} {\bibfnamefont {D.}~\bibnamefont
  {Saumon}},\ }\bibfield  {title} {\bibinfo {title} {Electronic and ionic
  structures of warm and hot dense matter},\ }\href
  {https://doi.org/10.1103/PhysRevE.87.013104} {\bibfield  {journal} {\bibinfo
  {journal} {Phys. Rev. E}\ }\textbf {\bibinfo {volume} {87}},\ \bibinfo
  {pages} {013104} (\bibinfo {year} {2013})}\BibitemShut {NoStop}%
\bibitem [{\citenamefont {Murillo}\ \emph {et~al.}(2013)\citenamefont
  {Murillo}, \citenamefont {Weisheit}, \citenamefont {Hansen},\ and\
  \citenamefont {Dharma-wardana}}]{MWHD13}%
  \BibitemOpen
  \bibfield  {author} {\bibinfo {author} {\bibfnamefont {M.~S.}\ \bibnamefont
  {Murillo}}, \bibinfo {author} {\bibfnamefont {J.}~\bibnamefont {Weisheit}},
  \bibinfo {author} {\bibfnamefont {S.~B.}\ \bibnamefont {Hansen}},\ and\
  \bibinfo {author} {\bibfnamefont {M.~W.~C.}\ \bibnamefont {Dharma-wardana}},\
  }\bibfield  {title} {\bibinfo {title} {Partial ionization in dense plasmas:
  Comparisons among average-atom density functional models},\ }\href
  {https://doi.org/10.1103/PhysRevE.87.063113} {\bibfield  {journal} {\bibinfo
  {journal} {Phys. Rev. E}\ }\textbf {\bibinfo {volume} {87}},\ \bibinfo
  {pages} {063113} (\bibinfo {year} {2013})}\BibitemShut {NoStop}%
\bibitem [{\citenamefont {Son}\ \emph {et~al.}(2014)\citenamefont {Son},
  \citenamefont {Thiele}, \citenamefont {Jurek}, \citenamefont {Ziaja},\ and\
  \citenamefont {Santra}}]{STJZS14}%
  \BibitemOpen
  \bibfield  {author} {\bibinfo {author} {\bibfnamefont {S.-K.}\ \bibnamefont
  {Son}}, \bibinfo {author} {\bibfnamefont {R.}~\bibnamefont {Thiele}},
  \bibinfo {author} {\bibfnamefont {Z.}~\bibnamefont {Jurek}}, \bibinfo
  {author} {\bibfnamefont {B.}~\bibnamefont {Ziaja}},\ and\ \bibinfo {author}
  {\bibfnamefont {R.}~\bibnamefont {Santra}},\ }\bibfield  {title} {\bibinfo
  {title} {Quantum-mechanical calculation of ionization-potential lowering in
  dense plasmas},\ }\href {https://doi.org/10.1103/PhysRevX.4.031004}
  {\bibfield  {journal} {\bibinfo  {journal} {Phys. Rev. X}\ }\textbf {\bibinfo
  {volume} {4}},\ \bibinfo {pages} {031004} (\bibinfo {year}
  {2014})}\BibitemShut {NoStop}%
\bibitem [{\citenamefont {Saumon}\ \emph {et~al.}(2014)\citenamefont {Saumon},
  \citenamefont {Starrett}, \citenamefont {Anta}, \citenamefont {Daughton},\
  and\ \citenamefont {Chabrier}}]{SSADC_14}%
  \BibitemOpen
  \bibfield  {author} {\bibinfo {author} {\bibfnamefont {D.}~\bibnamefont
  {Saumon}}, \bibinfo {author} {\bibfnamefont {C.}~\bibnamefont {Starrett}},
  \bibinfo {author} {\bibfnamefont {J.}~\bibnamefont {Anta}}, \bibinfo {author}
  {\bibfnamefont {W.}~\bibnamefont {Daughton}},\ and\ \bibinfo {author}
  {\bibfnamefont {G.}~\bibnamefont {Chabrier}},\ }\bibinfo {title} {The
  structure of warm dense matter modeled with an average atom model with
  ion-ion correlations},\ in\ \href@noop {} {\emph {\bibinfo {booktitle}
  {Frontiers and Challenges in Warm Dense Matter}}},\ \bibinfo {series}
  {Lecture Notes in Computational Science and Engineering}, Vol.~\bibinfo
  {volume} {96},\ \bibinfo {editor} {edited by\ \bibinfo {editor}
  {\bibfnamefont {F.}~\bibnamefont {Graziani}}, \bibinfo {editor}
  {\bibfnamefont {M.~P.}\ \bibnamefont {Desjarlais}}, \bibinfo {editor}
  {\bibfnamefont {R.}~\bibnamefont {Redmer}},\ and\ \bibinfo {editor}
  {\bibfnamefont {S.~B.}\ \bibnamefont {Trickey}}}\ (\bibinfo  {publisher}
  {Springer International Publishing},\ \bibinfo {year} {2014})\ pp.\ \bibinfo
  {pages} {151--176}\BibitemShut {NoStop}%
\bibitem [{\citenamefont {Starrett}\ and\ \citenamefont
  {Saumon}(2014)}]{SS_14}%
  \BibitemOpen
  \bibfield  {author} {\bibinfo {author} {\bibfnamefont {C.}~\bibnamefont
  {Starrett}}\ and\ \bibinfo {author} {\bibfnamefont {D.}~\bibnamefont
  {Saumon}},\ }\bibfield  {title} {\bibinfo {title} {A simple method for
  determining the ionic structure of warm dense matter},\ }\href
  {https://doi.org/10.1016/j.hedp.2013.12.001} {\bibfield  {journal} {\bibinfo
  {journal} {High Energy Density Phys.}\ }\textbf {\bibinfo {volume} {10}},\
  \bibinfo {pages} {35} (\bibinfo {year} {2014})}\BibitemShut {NoStop}%
\bibitem [{\citenamefont {Starrett}\ \emph {et~al.}(2019)\citenamefont
  {Starrett}, \citenamefont {Gill}, \citenamefont {Sjostrom},\ and\
  \citenamefont {Greeff}}]{SGSG19}%
  \BibitemOpen
  \bibfield  {author} {\bibinfo {author} {\bibfnamefont {C.}~\bibnamefont
  {Starrett}}, \bibinfo {author} {\bibfnamefont {N.}~\bibnamefont {Gill}},
  \bibinfo {author} {\bibfnamefont {T.}~\bibnamefont {Sjostrom}},\ and\
  \bibinfo {author} {\bibfnamefont {C.}~\bibnamefont {Greeff}},\ }\bibfield
  {title} {\bibinfo {title} {{Wide ranging equation of state with Tartarus: A
  hybrid Green’s function/orbital based average atom code}},\ }\href
  {https://doi.org/https://doi.org/10.1016/j.cpc.2018.10.002} {\bibfield
  {journal} {\bibinfo  {journal} {Comput. Phys. Commun.}\ }\textbf {\bibinfo
  {volume} {235}},\ \bibinfo {pages} {50} (\bibinfo {year} {2019})}\BibitemShut
  {NoStop}%
\bibitem [{\citenamefont {Dharma-wardana}\ \emph {et~al.}(2020)\citenamefont
  {Dharma-wardana}, \citenamefont {Klug},\ and\ \citenamefont
  {Remsing}}]{DKR_20}%
  \BibitemOpen
  \bibfield  {author} {\bibinfo {author} {\bibfnamefont {M.~W.~C.}\
  \bibnamefont {Dharma-wardana}}, \bibinfo {author} {\bibfnamefont {D.~D.}\
  \bibnamefont {Klug}},\ and\ \bibinfo {author} {\bibfnamefont {R.~C.}\
  \bibnamefont {Remsing}},\ }\bibfield  {title} {\bibinfo {title}
  {Liquid-liquid phase transitions in silicon},\ }\href
  {https://doi.org/10.1103/PhysRevLett.125.075702} {\bibfield  {journal}
  {\bibinfo  {journal} {Phys. Rev. Lett.}\ }\textbf {\bibinfo {volume} {125}},\
  \bibinfo {pages} {075702} (\bibinfo {year} {2020})}\BibitemShut {NoStop}%
\bibitem [{\citenamefont {Massacrier}\ \emph {et~al.}(2021)\citenamefont
  {Massacrier}, \citenamefont {B\"ohme}, \citenamefont {Vorberger},
  \citenamefont {Soubiran},\ and\ \citenamefont {Militzer}}]{MBVSM21}%
  \BibitemOpen
  \bibfield  {author} {\bibinfo {author} {\bibfnamefont {G.}~\bibnamefont
  {Massacrier}}, \bibinfo {author} {\bibfnamefont {M.}~\bibnamefont {B\"ohme}},
  \bibinfo {author} {\bibfnamefont {J.}~\bibnamefont {Vorberger}}, \bibinfo
  {author} {\bibfnamefont {F.}~\bibnamefont {Soubiran}},\ and\ \bibinfo
  {author} {\bibfnamefont {B.}~\bibnamefont {Militzer}},\ }\bibfield  {title}
  {\bibinfo {title} {Reconciling ionization energies and band gaps of warm
  dense matter derived with \textit{ab initio} simulations and average atom
  models},\ }\href {https://doi.org/10.1103/PhysRevResearch.3.023026}
  {\bibfield  {journal} {\bibinfo  {journal} {Phys. Rev. Research}\ }\textbf
  {\bibinfo {volume} {3}},\ \bibinfo {pages} {023026} (\bibinfo {year}
  {2021})}\BibitemShut {NoStop}%
\bibitem [{\citenamefont {Bekx}\ \emph {et~al.}(2020)\citenamefont {Bekx},
  \citenamefont {Son}, \citenamefont {Ziaja},\ and\ \citenamefont
  {Santra}}]{BSZS20}%
  \BibitemOpen
  \bibfield  {author} {\bibinfo {author} {\bibfnamefont {J.~J.}\ \bibnamefont
  {Bekx}}, \bibinfo {author} {\bibfnamefont {S.-K.}\ \bibnamefont {Son}},
  \bibinfo {author} {\bibfnamefont {B.}~\bibnamefont {Ziaja}},\ and\ \bibinfo
  {author} {\bibfnamefont {R.}~\bibnamefont {Santra}},\ }\bibfield  {title}
  {\bibinfo {title} {Electronic-structure calculations for nonisothermal warm
  dense matter},\ }\href {https://doi.org/10.1103/PhysRevResearch.2.033061}
  {\bibfield  {journal} {\bibinfo  {journal} {Phys. Rev. Research}\ }\textbf
  {\bibinfo {volume} {2}},\ \bibinfo {pages} {033061} (\bibinfo {year}
  {2020})}\BibitemShut {NoStop}%
\bibitem [{\citenamefont {Starrett}\ and\ \citenamefont
  {Shaffer}(2020)}]{SS20}%
  \BibitemOpen
  \bibfield  {author} {\bibinfo {author} {\bibfnamefont {C.~E.}\ \bibnamefont
  {Starrett}}\ and\ \bibinfo {author} {\bibfnamefont {N.}~\bibnamefont
  {Shaffer}},\ }\bibfield  {title} {\bibinfo {title} {Multiple scattering
  theory for dense plasmas},\ }\href
  {https://doi.org/10.1103/PhysRevE.102.043211} {\bibfield  {journal} {\bibinfo
   {journal} {Phys. Rev. E}\ }\textbf {\bibinfo {volume} {102}},\ \bibinfo
  {pages} {043211} (\bibinfo {year} {2020})}\BibitemShut {NoStop}%
\bibitem [{\citenamefont {Lejaeghere}\ \emph {et~al.}(2016)\citenamefont
  {Lejaeghere}, \citenamefont {Bihlmayer}, \citenamefont {Bj{\"o}rkman},
  \citenamefont {Blaha}, \citenamefont {Bl{\"u}gel}, \citenamefont {Blum},
  \citenamefont {Caliste}, \citenamefont {Castelli}, \citenamefont {Clark},
  \citenamefont {Dal~Corso}, \citenamefont {de~Gironcoli}, \citenamefont
  {Deutsch}, \citenamefont {Dewhurst}, \citenamefont {Di~Marco}, \citenamefont
  {Draxl}, \citenamefont {Dulak}, \citenamefont {Eriksson}, \citenamefont
  {Flores-Livas}, \citenamefont {Garrity}, \citenamefont {Genovese},
  \citenamefont {Giannozzi}, \citenamefont {Giantomassi}, \citenamefont
  {Goedecker}, \citenamefont {Gonze}, \citenamefont {Gr{\r a}n{\"a}s},
  \citenamefont {Gross}, \citenamefont {Gulans}, \citenamefont {Gygi},
  \citenamefont {Hamann}, \citenamefont {Hasnip}, \citenamefont {Holzwarth},
  \citenamefont {Iu{\c s}an}, \citenamefont {Jochym}, \citenamefont {Jollet},
  \citenamefont {Jones}, \citenamefont {Kresse}, \citenamefont {Koepernik},
  \citenamefont {K{\"u}{\c c}{\"u}kbenli}, \citenamefont {Kvashnin},
  \citenamefont {Locht}, \citenamefont {Lubeck}, \citenamefont {Marsman},
  \citenamefont {Marzari}, \citenamefont {Nitzsche}, \citenamefont
  {Nordstr{\"o}m}, \citenamefont {Ozaki}, \citenamefont {Paulatto},
  \citenamefont {Pickard}, \citenamefont {Poelmans}, \citenamefont {Probert},
  \citenamefont {Refson}, \citenamefont {Richter}, \citenamefont {Rignanese},
  \citenamefont {Saha}, \citenamefont {Scheffler}, \citenamefont {Schlipf},
  \citenamefont {Schwarz}, \citenamefont {Sharma}, \citenamefont {Tavazza},
  \citenamefont {Thunstr{\"o}m}, \citenamefont {Tkatchenko}, \citenamefont
  {Torrent}, \citenamefont {Vanderbilt}, \citenamefont {van Setten},
  \citenamefont {Van~Speybroeck}, \citenamefont {Wills}, \citenamefont {Yates},
  \citenamefont {Zhang},\ and\ \citenamefont {Cottenier}}]{Lejaeghereaad3000}%
  \BibitemOpen
  \bibfield  {author} {\bibinfo {author} {\bibfnamefont {K.}~\bibnamefont
  {Lejaeghere}}, \bibinfo {author} {\bibfnamefont {G.}~\bibnamefont
  {Bihlmayer}}, \bibinfo {author} {\bibfnamefont {T.}~\bibnamefont
  {Bj{\"o}rkman}}, \bibinfo {author} {\bibfnamefont {P.}~\bibnamefont {Blaha}},
  \bibinfo {author} {\bibfnamefont {S.}~\bibnamefont {Bl{\"u}gel}}, \bibinfo
  {author} {\bibfnamefont {V.}~\bibnamefont {Blum}}, \bibinfo {author}
  {\bibfnamefont {D.}~\bibnamefont {Caliste}}, \bibinfo {author} {\bibfnamefont
  {I.~E.}\ \bibnamefont {Castelli}}, \bibinfo {author} {\bibfnamefont {S.~J.}\
  \bibnamefont {Clark}}, \bibinfo {author} {\bibfnamefont {A.}~\bibnamefont
  {Dal~Corso}}, \bibinfo {author} {\bibfnamefont {S.}~\bibnamefont
  {de~Gironcoli}}, \bibinfo {author} {\bibfnamefont {T.}~\bibnamefont
  {Deutsch}}, \bibinfo {author} {\bibfnamefont {J.~K.}\ \bibnamefont
  {Dewhurst}}, \bibinfo {author} {\bibfnamefont {I.}~\bibnamefont {Di~Marco}},
  \bibinfo {author} {\bibfnamefont {C.}~\bibnamefont {Draxl}}, \bibinfo
  {author} {\bibfnamefont {M.}~\bibnamefont {Dulak}}, \bibinfo {author}
  {\bibfnamefont {O.}~\bibnamefont {Eriksson}}, \bibinfo {author}
  {\bibfnamefont {J.~A.}\ \bibnamefont {Flores-Livas}}, \bibinfo {author}
  {\bibfnamefont {K.~F.}\ \bibnamefont {Garrity}}, \bibinfo {author}
  {\bibfnamefont {L.}~\bibnamefont {Genovese}}, \bibinfo {author}
  {\bibfnamefont {P.}~\bibnamefont {Giannozzi}}, \bibinfo {author}
  {\bibfnamefont {M.}~\bibnamefont {Giantomassi}}, \bibinfo {author}
  {\bibfnamefont {S.}~\bibnamefont {Goedecker}}, \bibinfo {author}
  {\bibfnamefont {X.}~\bibnamefont {Gonze}}, \bibinfo {author} {\bibfnamefont
  {O.}~\bibnamefont {Gr{\r a}n{\"a}s}}, \bibinfo {author} {\bibfnamefont
  {E.~K.~U.}\ \bibnamefont {Gross}}, \bibinfo {author} {\bibfnamefont
  {A.}~\bibnamefont {Gulans}}, \bibinfo {author} {\bibfnamefont
  {F.}~\bibnamefont {Gygi}}, \bibinfo {author} {\bibfnamefont {D.~R.}\
  \bibnamefont {Hamann}}, \bibinfo {author} {\bibfnamefont {P.~J.}\
  \bibnamefont {Hasnip}}, \bibinfo {author} {\bibfnamefont {N.~A.~W.}\
  \bibnamefont {Holzwarth}}, \bibinfo {author} {\bibfnamefont {D.}~\bibnamefont
  {Iu{\c s}an}}, \bibinfo {author} {\bibfnamefont {D.~B.}\ \bibnamefont
  {Jochym}}, \bibinfo {author} {\bibfnamefont {F.}~\bibnamefont {Jollet}},
  \bibinfo {author} {\bibfnamefont {D.}~\bibnamefont {Jones}}, \bibinfo
  {author} {\bibfnamefont {G.}~\bibnamefont {Kresse}}, \bibinfo {author}
  {\bibfnamefont {K.}~\bibnamefont {Koepernik}}, \bibinfo {author}
  {\bibfnamefont {E.}~\bibnamefont {K{\"u}{\c c}{\"u}kbenli}}, \bibinfo
  {author} {\bibfnamefont {Y.~O.}\ \bibnamefont {Kvashnin}}, \bibinfo {author}
  {\bibfnamefont {I.~L.~M.}\ \bibnamefont {Locht}}, \bibinfo {author}
  {\bibfnamefont {S.}~\bibnamefont {Lubeck}}, \bibinfo {author} {\bibfnamefont
  {M.}~\bibnamefont {Marsman}}, \bibinfo {author} {\bibfnamefont
  {N.}~\bibnamefont {Marzari}}, \bibinfo {author} {\bibfnamefont
  {U.}~\bibnamefont {Nitzsche}}, \bibinfo {author} {\bibfnamefont
  {L.}~\bibnamefont {Nordstr{\"o}m}}, \bibinfo {author} {\bibfnamefont
  {T.}~\bibnamefont {Ozaki}}, \bibinfo {author} {\bibfnamefont
  {L.}~\bibnamefont {Paulatto}}, \bibinfo {author} {\bibfnamefont {C.~J.}\
  \bibnamefont {Pickard}}, \bibinfo {author} {\bibfnamefont {W.}~\bibnamefont
  {Poelmans}}, \bibinfo {author} {\bibfnamefont {M.~I.~J.}\ \bibnamefont
  {Probert}}, \bibinfo {author} {\bibfnamefont {K.}~\bibnamefont {Refson}},
  \bibinfo {author} {\bibfnamefont {M.}~\bibnamefont {Richter}}, \bibinfo
  {author} {\bibfnamefont {G.-M.}\ \bibnamefont {Rignanese}}, \bibinfo {author}
  {\bibfnamefont {S.}~\bibnamefont {Saha}}, \bibinfo {author} {\bibfnamefont
  {M.}~\bibnamefont {Scheffler}}, \bibinfo {author} {\bibfnamefont
  {M.}~\bibnamefont {Schlipf}}, \bibinfo {author} {\bibfnamefont
  {K.}~\bibnamefont {Schwarz}}, \bibinfo {author} {\bibfnamefont
  {S.}~\bibnamefont {Sharma}}, \bibinfo {author} {\bibfnamefont
  {F.}~\bibnamefont {Tavazza}}, \bibinfo {author} {\bibfnamefont
  {P.}~\bibnamefont {Thunstr{\"o}m}}, \bibinfo {author} {\bibfnamefont
  {A.}~\bibnamefont {Tkatchenko}}, \bibinfo {author} {\bibfnamefont
  {M.}~\bibnamefont {Torrent}}, \bibinfo {author} {\bibfnamefont
  {D.}~\bibnamefont {Vanderbilt}}, \bibinfo {author} {\bibfnamefont {M.~J.}\
  \bibnamefont {van Setten}}, \bibinfo {author} {\bibfnamefont
  {V.}~\bibnamefont {Van~Speybroeck}}, \bibinfo {author} {\bibfnamefont
  {J.~M.}\ \bibnamefont {Wills}}, \bibinfo {author} {\bibfnamefont {J.~R.}\
  \bibnamefont {Yates}}, \bibinfo {author} {\bibfnamefont {G.-X.}\ \bibnamefont
  {Zhang}},\ and\ \bibinfo {author} {\bibfnamefont {S.}~\bibnamefont
  {Cottenier}},\ }\bibfield  {title} {\bibinfo {title} {Reproducibility in
  density functional theory calculations of solids},\ }\href
  {https://science.sciencemag.org/content/351/6280/aad3000} {\bibfield
  {journal} {\bibinfo  {journal} {Science}\ }\textbf {\bibinfo {volume} {351}}
  (\bibinfo {year} {2016})}\BibitemShut {NoStop}%
\bibitem [{\citenamefont {Born}\ and\ \citenamefont
  {Oppenheimer}(1927)}]{BO27}%
  \BibitemOpen
  \bibfield  {author} {\bibinfo {author} {\bibfnamefont {M.}~\bibnamefont
  {Born}}\ and\ \bibinfo {author} {\bibfnamefont {R.}~\bibnamefont
  {Oppenheimer}},\ }\bibfield  {title} {\bibinfo {title} {Zur quantentheorie
  der molekeln},\ }\href {https://doi.org/10.1002/andp.19273892002} {\bibfield
  {journal} {\bibinfo  {journal} {Annalen der Physik}\ }\textbf {\bibinfo
  {volume} {389}},\ \bibinfo {pages} {457} (\bibinfo {year}
  {1927})}\BibitemShut {NoStop}%
\bibitem [{\citenamefont {Abedi}\ \emph {et~al.}(2010)\citenamefont {Abedi},
  \citenamefont {Maitra},\ and\ \citenamefont {Gross}}]{AMG10}%
  \BibitemOpen
  \bibfield  {author} {\bibinfo {author} {\bibfnamefont {A.}~\bibnamefont
  {Abedi}}, \bibinfo {author} {\bibfnamefont {N.~T.}\ \bibnamefont {Maitra}},\
  and\ \bibinfo {author} {\bibfnamefont {E.~K.~U.}\ \bibnamefont {Gross}},\
  }\bibfield  {title} {\bibinfo {title} {Exact factorization of the
  time-dependent electron-nuclear wave function},\ }\href
  {https://doi.org/10.1103/PhysRevLett.105.123002} {\bibfield  {journal}
  {\bibinfo  {journal} {Phys. Rev. Lett.}\ }\textbf {\bibinfo {volume} {105}},\
  \bibinfo {pages} {123002} (\bibinfo {year} {2010})}\BibitemShut {NoStop}%
\bibitem [{\citenamefont {Larder}\ \emph {et~al.}(2019)\citenamefont {Larder},
  \citenamefont {Gericke}, \citenamefont {Richardson}, \citenamefont {Mabey},
  \citenamefont {White},\ and\ \citenamefont {Gregori}}]{LGRMWG19}%
  \BibitemOpen
  \bibfield  {author} {\bibinfo {author} {\bibfnamefont {B.}~\bibnamefont
  {Larder}}, \bibinfo {author} {\bibfnamefont {D.~O.}\ \bibnamefont {Gericke}},
  \bibinfo {author} {\bibfnamefont {S.}~\bibnamefont {Richardson}}, \bibinfo
  {author} {\bibfnamefont {P.}~\bibnamefont {Mabey}}, \bibinfo {author}
  {\bibfnamefont {T.~G.}\ \bibnamefont {White}},\ and\ \bibinfo {author}
  {\bibfnamefont {G.}~\bibnamefont {Gregori}},\ }\bibfield  {title} {\bibinfo
  {title} {Fast nonadiabatic dynamics of many-body quantum systems},\ }\href
  {https://doi.org/10.1126/sciadv.aaw1634} {\bibfield  {journal} {\bibinfo
  {journal} {Science Advances}\ }\textbf {\bibinfo {volume} {5}},\ \bibinfo
  {pages} {1634} (\bibinfo {year} {2019})}\BibitemShut {NoStop}%
\bibitem [{Note3()}]{Note3}%
  \BibitemOpen
  \bibinfo {note} {If different species were admitted in the Hamiltonian, or a
  magnetic field present, then the term $\mu \protect \hat {N}$ would be
  denoted $\DOTSB \sum@ \slimits@ _s\mu _s\protect \hat {N}_s$, where $s$
  denotes the species (or electron spin). Since we only consider the electronic
  Hamiltonian in the absence of a magnetic field in our model, we use the
  simplified form of Eq.~\protect \textup {\hbox {\mathsurround \z@ \protect
  \normalfont (\ignorespaces \ref {eq:gr_can_1}\unskip \@@italiccorr
  )}}}\BibitemShut {NoStop}%
\bibitem [{\citenamefont {von Barth}\ and\ \citenamefont {Hedin}(1972)}]{BF72}%
  \BibitemOpen
  \bibfield  {author} {\bibinfo {author} {\bibfnamefont {U.}~\bibnamefont {von
  Barth}}\ and\ \bibinfo {author} {\bibfnamefont {L.}~\bibnamefont {Hedin}},\
  }\bibfield  {title} {\bibinfo {title} {A local exchange-correlation potential
  for the spin polarized case. i},\ }\href
  {https://doi.org/10.1088/0022-3719/5/13/012} {\bibfield  {journal} {\bibinfo
  {journal} {J. Phys. C}\ }\textbf {\bibinfo {volume} {5}},\ \bibinfo {pages}
  {1629} (\bibinfo {year} {1972})}\BibitemShut {NoStop}%
\bibitem [{\citenamefont {Parr}\ and\ \citenamefont {Yang}(1989)}]{Parr_Yang}%
  \BibitemOpen
  \bibfield  {author} {\bibinfo {author} {\bibfnamefont {R.~G.}\ \bibnamefont
  {Parr}}\ and\ \bibinfo {author} {\bibfnamefont {W.}~\bibnamefont {Yang}},\
  }\bibinfo {title} {Density functional theory of atoms and molecules}\
  (\bibinfo  {publisher} {Oxford University Press},\ \bibinfo {year} {1989})\
  Chap.~\bibinfo {chapter} {8}, pp.\ \bibinfo {pages} {173--174}\BibitemShut
  {NoStop}%
\bibitem [{\citenamefont {Perdew}\ and\ \citenamefont {Zunger}(1981)}]{PZ81}%
  \BibitemOpen
  \bibfield  {author} {\bibinfo {author} {\bibfnamefont {J.~P.}\ \bibnamefont
  {Perdew}}\ and\ \bibinfo {author} {\bibfnamefont {A.}~\bibnamefont
  {Zunger}},\ }\bibfield  {title} {\bibinfo {title} {Self-interaction
  correction to density-functional approximations for many-electron systems},\
  }\href {https://link.aps.org/doi/10.1103/PhysRevB.23.5048} {\bibfield
  {journal} {\bibinfo  {journal} {Phys. Rev. B}\ }\textbf {\bibinfo {volume}
  {23}},\ \bibinfo {pages} {5048} (\bibinfo {year} {1981})}\BibitemShut
  {NoStop}%
\bibitem [{\citenamefont {K\"ummel}\ and\ \citenamefont {Kronik}(2008)}]{KK08}%
  \BibitemOpen
  \bibfield  {author} {\bibinfo {author} {\bibfnamefont {S.}~\bibnamefont
  {K\"ummel}}\ and\ \bibinfo {author} {\bibfnamefont {L.}~\bibnamefont
  {Kronik}},\ }\bibfield  {title} {\bibinfo {title} {Orbital-dependent density
  functionals: Theory and applications},\ }\href
  {https://doi.org/10.1103/RevModPhys.80.3} {\bibfield  {journal} {\bibinfo
  {journal} {Rev. Mod. Phys.}\ }\textbf {\bibinfo {volume} {80}},\ \bibinfo
  {pages} {3} (\bibinfo {year} {2008})}\BibitemShut {NoStop}%
\bibitem [{\citenamefont {Gidopoulos}\ and\ \citenamefont
  {Lathiotakis}(2012)}]{GL12}%
  \BibitemOpen
  \bibfield  {author} {\bibinfo {author} {\bibfnamefont {N.}~\bibnamefont
  {Gidopoulos}}\ and\ \bibinfo {author} {\bibfnamefont {N.}~\bibnamefont
  {Lathiotakis}},\ }\bibfield  {title} {\bibinfo {title} {Constraining density
  functional approximations to yield self-interaction free potentials},\ }\href
  {https://doi.org/10.1063/1.4728156} {\bibfield  {journal} {\bibinfo
  {journal} {J. Chem. Phys.}\ }\textbf {\bibinfo {volume} {136}},\ \bibinfo
  {pages} {224109} (\bibinfo {year} {2012})}\BibitemShut {NoStop}%
\bibitem [{\citenamefont {Schmidt}\ \emph
  {et~al.}(2014{\natexlab{a}})\citenamefont {Schmidt}, \citenamefont
  {Kraisler}, \citenamefont {Kronik},\ and\ \citenamefont
  {K{\"u}mmel}}]{SKKK14}%
  \BibitemOpen
  \bibfield  {author} {\bibinfo {author} {\bibfnamefont {T.}~\bibnamefont
  {Schmidt}}, \bibinfo {author} {\bibfnamefont {E.}~\bibnamefont {Kraisler}},
  \bibinfo {author} {\bibfnamefont {L.}~\bibnamefont {Kronik}},\ and\ \bibinfo
  {author} {\bibfnamefont {S.}~\bibnamefont {K{\"u}mmel}},\ }\bibfield  {title}
  {\bibinfo {title} {{One-electron self-interaction and the asymptotics of the
  Kohn--Sham potential: an impaired relation}},\ }\href
  {https://doi.org/10.1039/C3CP55433C} {\bibfield  {journal} {\bibinfo
  {journal} {Phys. Chem. Chem. Phys.}\ }\textbf {\bibinfo {volume} {16}},\
  \bibinfo {pages} {14357} (\bibinfo {year} {2014}{\natexlab{a}})}\BibitemShut
  {NoStop}%
\bibitem [{\citenamefont {Kronik}\ and\ \citenamefont {Kümmel}(2020)}]{KK20}%
  \BibitemOpen
  \bibfield  {author} {\bibinfo {author} {\bibfnamefont {L.}~\bibnamefont
  {Kronik}}\ and\ \bibinfo {author} {\bibfnamefont {S.}~\bibnamefont
  {Kümmel}},\ }\bibfield  {title} {\bibinfo {title} {Piecewise linearity{,}
  freedom from self-interaction{,} and a coulomb asymptotic potential: three
  related yet inequivalent properties of the exact density functional},\ }\href
  {https://doi.org/10.1039/D0CP02564J} {\bibfield  {journal} {\bibinfo
  {journal} {Phys. Chem. Chem. Phys.}\ }\textbf {\bibinfo {volume} {22}},\
  \bibinfo {pages} {16467} (\bibinfo {year} {2020})}\BibitemShut {NoStop}%
\bibitem [{\citenamefont {K\"{u}mmel}\ and\ \citenamefont
  {Perdew}(2003)}]{KP03}%
  \BibitemOpen
  \bibfield  {author} {\bibinfo {author} {\bibfnamefont {S.}~\bibnamefont
  {K\"{u}mmel}}\ and\ \bibinfo {author} {\bibfnamefont {J.~P.}\ \bibnamefont
  {Perdew}},\ }\bibfield  {title} {\bibinfo {title} {{Two avenues to
  self-interaction correction within Kohn--Sham theory: unitary invariance is
  the shortcut}},\ }\href {https://doi.org/10.1080/0026897031000094506}
  {\bibfield  {journal} {\bibinfo  {journal} {Mol. Phys.}\ }\textbf {\bibinfo
  {volume} {101}},\ \bibinfo {pages} {1363} (\bibinfo {year}
  {2003})}\BibitemShut {NoStop}%
\bibitem [{\citenamefont {Mori-Sánchez}\ \emph {et~al.}(2006)\citenamefont
  {Mori-Sánchez}, \citenamefont {Cohen},\ and\ \citenamefont {Yang}}]{MSCY06}%
  \BibitemOpen
  \bibfield  {author} {\bibinfo {author} {\bibfnamefont {P.}~\bibnamefont
  {Mori-Sánchez}}, \bibinfo {author} {\bibfnamefont {A.~J.}\ \bibnamefont
  {Cohen}},\ and\ \bibinfo {author} {\bibfnamefont {W.}~\bibnamefont {Yang}},\
  }\bibfield  {title} {\bibinfo {title} {Many-electron self-interaction error
  in approximate density functionals},\ }\href
  {https://doi.org/10.1063/1.2403848} {\bibfield  {journal} {\bibinfo
  {journal} {J. Chem. Phys.}\ }\textbf {\bibinfo {volume} {125}},\ \bibinfo
  {pages} {201102} (\bibinfo {year} {2006})}\BibitemShut {NoStop}%
\bibitem [{\citenamefont {Pederson}\ \emph {et~al.}(2014)\citenamefont
  {Pederson}, \citenamefont {Ruzsinszky},\ and\ \citenamefont
  {Perdew}}]{PRP14}%
  \BibitemOpen
  \bibfield  {author} {\bibinfo {author} {\bibfnamefont {M.~R.}\ \bibnamefont
  {Pederson}}, \bibinfo {author} {\bibfnamefont {A.}~\bibnamefont
  {Ruzsinszky}},\ and\ \bibinfo {author} {\bibfnamefont {J.~P.}\ \bibnamefont
  {Perdew}},\ }\bibfield  {title} {\bibinfo {title} {Communication:
  Self-interaction correction with unitary invariance in density functional
  theory},\ }\href {https://doi.org/10.1063/1.4869581} {\bibfield  {journal}
  {\bibinfo  {journal} {J. Chem. Phys.}\ }\textbf {\bibinfo {volume} {140}},\
  \bibinfo {pages} {121103} (\bibinfo {year} {2014})}\BibitemShut {NoStop}%
\bibitem [{\citenamefont {Schmidt}\ \emph
  {et~al.}(2014{\natexlab{b}})\citenamefont {Schmidt}, \citenamefont
  {Kraisler}, \citenamefont {Makmal}, \citenamefont {Kronik},\ and\
  \citenamefont {Kümmel}}]{SKMKK14}%
  \BibitemOpen
  \bibfield  {author} {\bibinfo {author} {\bibfnamefont {T.}~\bibnamefont
  {Schmidt}}, \bibinfo {author} {\bibfnamefont {E.}~\bibnamefont {Kraisler}},
  \bibinfo {author} {\bibfnamefont {A.}~\bibnamefont {Makmal}}, \bibinfo
  {author} {\bibfnamefont {L.}~\bibnamefont {Kronik}},\ and\ \bibinfo {author}
  {\bibfnamefont {S.}~\bibnamefont {Kümmel}},\ }\bibfield  {title} {\bibinfo
  {title} {{A self-interaction-free local hybrid functional: Accurate binding
  energies vis-à-vis accurate ionization potentials from Kohn-Sham
  eigenvalues}},\ }\href {https://doi.org/10.1063/1.4865942} {\bibfield
  {journal} {\bibinfo  {journal} {J. Chem. Phys.}\ }\textbf {\bibinfo {volume}
  {140}},\ \bibinfo {pages} {18A510} (\bibinfo {year}
  {2014}{\natexlab{b}})}\BibitemShut {NoStop}%
\bibitem [{\citenamefont {Yang}\ \emph {et~al.}(2017)\citenamefont {Yang},
  \citenamefont {Pederson},\ and\ \citenamefont {Perdew}}]{YPP17}%
  \BibitemOpen
  \bibfield  {author} {\bibinfo {author} {\bibfnamefont {Z.-h.}\ \bibnamefont
  {Yang}}, \bibinfo {author} {\bibfnamefont {M.~R.}\ \bibnamefont {Pederson}},\
  and\ \bibinfo {author} {\bibfnamefont {J.~P.}\ \bibnamefont {Perdew}},\
  }\bibfield  {title} {\bibinfo {title} {{Full self-consistency in the
  Fermi-orbital self-interaction correction}},\ }\href
  {https://doi.org/10.1103/PhysRevA.95.052505} {\bibfield  {journal} {\bibinfo
  {journal} {Phys. Rev. A}\ }\textbf {\bibinfo {volume} {95}},\ \bibinfo
  {pages} {052505} (\bibinfo {year} {2017})}\BibitemShut {NoStop}%
\bibitem [{\citenamefont {Jackson}\ \emph {et~al.}(2019)\citenamefont
  {Jackson}, \citenamefont {Peralta}, \citenamefont {Joshi}, \citenamefont
  {Withanage}, \citenamefont {Trepte}, \citenamefont {Sharkas},\ and\
  \citenamefont {Johnson}}]{JPJWTSJ19}%
  \BibitemOpen
  \bibfield  {author} {\bibinfo {author} {\bibfnamefont {K.~A.}\ \bibnamefont
  {Jackson}}, \bibinfo {author} {\bibfnamefont {J.~E.}\ \bibnamefont
  {Peralta}}, \bibinfo {author} {\bibfnamefont {R.~P.}\ \bibnamefont {Joshi}},
  \bibinfo {author} {\bibfnamefont {K.~P.}\ \bibnamefont {Withanage}}, \bibinfo
  {author} {\bibfnamefont {K.}~\bibnamefont {Trepte}}, \bibinfo {author}
  {\bibfnamefont {K.}~\bibnamefont {Sharkas}},\ and\ \bibinfo {author}
  {\bibfnamefont {A.~I.}\ \bibnamefont {Johnson}},\ }\bibfield  {title}
  {\bibinfo {title} {{Towards efficient density functional theory calculations
  without self-interaction: The Fermi-Löwdin orbital self-interaction
  correction}},\ }\href {https://doi.org/10.1088/1742-6596/1290/1/012002}
  {\bibfield  {journal} {\bibinfo  {journal} {J. Phys. Conf. Ser.}\ }\textbf
  {\bibinfo {volume} {1290}},\ \bibinfo {pages} {012002} (\bibinfo {year}
  {2019})}\BibitemShut {NoStop}%
\bibitem [{\citenamefont {Zope}\ \emph {et~al.}(2019)\citenamefont {Zope},
  \citenamefont {Yamamoto}, \citenamefont {Diaz}, \citenamefont {Baruah},
  \citenamefont {Peralta}, \citenamefont {Jackson}, \citenamefont {Santra},\
  and\ \citenamefont {Perdew}}]{RYDBPJSP19}%
  \BibitemOpen
  \bibfield  {author} {\bibinfo {author} {\bibfnamefont {R.~R.}\ \bibnamefont
  {Zope}}, \bibinfo {author} {\bibfnamefont {Y.}~\bibnamefont {Yamamoto}},
  \bibinfo {author} {\bibfnamefont {C.~M.}\ \bibnamefont {Diaz}}, \bibinfo
  {author} {\bibfnamefont {T.}~\bibnamefont {Baruah}}, \bibinfo {author}
  {\bibfnamefont {J.~E.}\ \bibnamefont {Peralta}}, \bibinfo {author}
  {\bibfnamefont {K.~A.}\ \bibnamefont {Jackson}}, \bibinfo {author}
  {\bibfnamefont {B.}~\bibnamefont {Santra}},\ and\ \bibinfo {author}
  {\bibfnamefont {J.~P.}\ \bibnamefont {Perdew}},\ }\bibfield  {title}
  {\bibinfo {title} {A step in the direction of resolving the paradox of
  perdew-zunger self-interaction correction},\ }\href
  {https://doi.org/10.1063/1.5129533} {\bibfield  {journal} {\bibinfo
  {journal} {J. Chem. Phys.}\ }\textbf {\bibinfo {volume} {151}},\ \bibinfo
  {pages} {214108} (\bibinfo {year} {2019})}\BibitemShut {NoStop}%
\bibitem [{\citenamefont {Callow}\ \emph {et~al.}(2020)\citenamefont {Callow},
  \citenamefont {Pearce}, \citenamefont {Pitts}, \citenamefont {Lathiotakis},
  \citenamefont {Hodgson},\ and\ \citenamefont {Gidopoulos}}]{CPPLHG20}%
  \BibitemOpen
  \bibfield  {author} {\bibinfo {author} {\bibfnamefont {T.~J.}\ \bibnamefont
  {Callow}}, \bibinfo {author} {\bibfnamefont {B.~J.}\ \bibnamefont {Pearce}},
  \bibinfo {author} {\bibfnamefont {T.}~\bibnamefont {Pitts}}, \bibinfo
  {author} {\bibfnamefont {N.~N.}\ \bibnamefont {Lathiotakis}}, \bibinfo
  {author} {\bibfnamefont {M.~J.~P.}\ \bibnamefont {Hodgson}},\ and\ \bibinfo
  {author} {\bibfnamefont {N.~I.}\ \bibnamefont {Gidopoulos}},\ }\bibfield
  {title} {\bibinfo {title} {Improving the exchange and correlation potential
  in density-functional approximations through constraints},\ }\href
  {https://doi.org/10.1039/D0FD00069H} {\bibfield  {journal} {\bibinfo
  {journal} {Faraday Discuss.}\ }\textbf {\bibinfo {volume} {224}},\ \bibinfo
  {pages} {126} (\bibinfo {year} {2020})}\BibitemShut {NoStop}%
\bibitem [{\citenamefont {Schwalbe}\ \emph {et~al.}(2020)\citenamefont
  {Schwalbe}, \citenamefont {Fiedler}, \citenamefont {Kraus}, \citenamefont
  {Kortus}, \citenamefont {Trepte},\ and\ \citenamefont {Lehtola}}]{SFKKTL20}%
  \BibitemOpen
  \bibfield  {author} {\bibinfo {author} {\bibfnamefont {S.}~\bibnamefont
  {Schwalbe}}, \bibinfo {author} {\bibfnamefont {L.}~\bibnamefont {Fiedler}},
  \bibinfo {author} {\bibfnamefont {J.}~\bibnamefont {Kraus}}, \bibinfo
  {author} {\bibfnamefont {J.}~\bibnamefont {Kortus}}, \bibinfo {author}
  {\bibfnamefont {K.}~\bibnamefont {Trepte}},\ and\ \bibinfo {author}
  {\bibfnamefont {S.}~\bibnamefont {Lehtola}},\ }\bibfield  {title} {\bibinfo
  {title} {{PyFLOSIC: Python-based Fermi–Löwdin orbital self-interaction
  correction}},\ }\href {https://doi.org/10.1063/5.0012519} {\bibfield
  {journal} {\bibinfo  {journal} {J. Chem. Phys.}\ }\textbf {\bibinfo {volume}
  {153}},\ \bibinfo {pages} {084104} (\bibinfo {year} {2020})}\BibitemShut
  {NoStop}%
\bibitem [{\citenamefont {Gidopoulos}\ \emph {et~al.}(2002)\citenamefont
  {Gidopoulos}, \citenamefont {Papaconstantinou},\ and\ \citenamefont
  {Gross}}]{GPG02}%
  \BibitemOpen
  \bibfield  {author} {\bibinfo {author} {\bibfnamefont {N.~I.}\ \bibnamefont
  {Gidopoulos}}, \bibinfo {author} {\bibfnamefont {P.~G.}\ \bibnamefont
  {Papaconstantinou}},\ and\ \bibinfo {author} {\bibfnamefont {E.~K.~U.}\
  \bibnamefont {Gross}},\ }\bibfield  {title} {\bibinfo {title} {Spurious
  interactions, and their correction, in the ensemble-kohn-sham scheme for
  excited states},\ }\href {https://doi.org/10.1103/PhysRevLett.88.033003}
  {\bibfield  {journal} {\bibinfo  {journal} {Phys. Rev. Lett.}\ }\textbf
  {\bibinfo {volume} {88}},\ \bibinfo {pages} {033003} (\bibinfo {year}
  {2002})}\BibitemShut {NoStop}%
\bibitem [{\citenamefont {Gould}\ and\ \citenamefont {Dobson}(2013)}]{GD13}%
  \BibitemOpen
  \bibfield  {author} {\bibinfo {author} {\bibfnamefont {T.}~\bibnamefont
  {Gould}}\ and\ \bibinfo {author} {\bibfnamefont {J.~F.}\ \bibnamefont
  {Dobson}},\ }\bibfield  {title} {\bibinfo {title} {The flexible nature of
  exchange, correlation, and hartree physics: Resolving “delocalization”
  errors in a “correlation free” density functional},\ }\href
  {https://doi.org/10.1063/1.4773284} {\bibfield  {journal} {\bibinfo
  {journal} {J. Chem. Phys.}\ }\textbf {\bibinfo {volume} {138}},\ \bibinfo
  {pages} {014103} (\bibinfo {year} {2013})}\BibitemShut {NoStop}%
\bibitem [{\citenamefont {Gould}\ and\ \citenamefont {Pittalis}(2017)}]{SP17}%
  \BibitemOpen
  \bibfield  {author} {\bibinfo {author} {\bibfnamefont {T.}~\bibnamefont
  {Gould}}\ and\ \bibinfo {author} {\bibfnamefont {S.}~\bibnamefont
  {Pittalis}},\ }\bibfield  {title} {\bibinfo {title} {Hartree and exchange in
  ensemble density functional theory: Avoiding the nonuniqueness disaster},\
  }\href {https://doi.org/10.1103/PhysRevLett.119.243001} {\bibfield  {journal}
  {\bibinfo  {journal} {Phys. Rev. Lett.}\ }\textbf {\bibinfo {volume} {119}},\
  \bibinfo {pages} {243001} (\bibinfo {year} {2017})}\BibitemShut {NoStop}%
\bibitem [{Note4()}]{Note4}%
  \BibitemOpen
  \bibinfo {note} {This radius is often denoted as the radius of the
  Wigner--Seitz cell, $R_\protect \textrm {WS}$ in the literature. We use the
  notation $R_\protect \textrm {VS}$ to clearly distinguish this quantity from
  the Wigner--Seitz radius $r_s$, which depends on the free electron density
  only.}\BibitemShut {Stop}%
\bibitem [{\citenamefont {P\'erez-Bernal}\ \emph {et~al.}(2001)\citenamefont
  {P\'erez-Bernal}, \citenamefont {Martel}, \citenamefont {Arias},\ and\
  \citenamefont {G\'omez-Camacho}}]{PMAG01}%
  \BibitemOpen
  \bibfield  {author} {\bibinfo {author} {\bibfnamefont {F.}~\bibnamefont
  {P\'erez-Bernal}}, \bibinfo {author} {\bibfnamefont {I.}~\bibnamefont
  {Martel}}, \bibinfo {author} {\bibfnamefont {J.~M.}\ \bibnamefont {Arias}},\
  and\ \bibinfo {author} {\bibfnamefont {J.}~\bibnamefont {G\'omez-Camacho}},\
  }\bibfield  {title} {\bibinfo {title} {Continuum discretization in a basis of
  transformed harmonic-oscillator states},\ }\href
  {https://doi.org/10.1103/PhysRevA.63.052111} {\bibfield  {journal} {\bibinfo
  {journal} {Phys. Rev. A}\ }\textbf {\bibinfo {volume} {63}},\ \bibinfo
  {pages} {052111} (\bibinfo {year} {2001})}\BibitemShut {NoStop}%
\bibitem [{Note5()}]{Note5}%
  \BibitemOpen
  \bibinfo {note} {In practise, it is often known \protect \emph {a priori}
  which configuration will \protect \leavevmode {\protect \color {black}likely}
  be most energetically favourable from experience and physical intuition. In
  the examples we consider later, we take $N_\protect \textrm {e}^\uparrow =1,\
  N_\protect \textrm {e}^\downarrow =0$ for Hydrogen, and \protect \leavevmode
  {\protect \color {red}$N_\protect \textrm {e}^\uparrow =N_\protect \textrm
  {e}^\downarrow =2$} for Beryllium}\BibitemShut {NoStop}%
\bibitem [{\citenamefont {Kotochigova}\ \emph {et~al.}(1997)\citenamefont
  {Kotochigova}, \citenamefont {Levine}, \citenamefont {Shirley}, \citenamefont
  {Stiles},\ and\ \citenamefont {Clark}}]{Kotochigova_1997}%
  \BibitemOpen
  \bibfield  {author} {\bibinfo {author} {\bibfnamefont {S.}~\bibnamefont
  {Kotochigova}}, \bibinfo {author} {\bibfnamefont {Z.~H.}\ \bibnamefont
  {Levine}}, \bibinfo {author} {\bibfnamefont {E.~L.}\ \bibnamefont {Shirley}},
  \bibinfo {author} {\bibfnamefont {M.~D.}\ \bibnamefont {Stiles}},\ and\
  \bibinfo {author} {\bibfnamefont {C.~W.}\ \bibnamefont {Clark}},\ }\bibfield
  {title} {\bibinfo {title} {Local-density-functional calculations of the
  energy of atoms},\ }\href {https://doi.org/10.1103/PhysRevA.55.191}
  {\bibfield  {journal} {\bibinfo  {journal} {Phys. Rev. A}\ }\textbf {\bibinfo
  {volume} {55}},\ \bibinfo {pages} {191} (\bibinfo {year} {1997})}\BibitemShut
  {NoStop}%
\bibitem [{\citenamefont {Callow}\ \emph
  {et~al.}(2021{\natexlab{a}})\citenamefont {Callow}, \citenamefont {Pearce},\
  and\ \citenamefont {Gidopoulos}}]{callow2021density}%
  \BibitemOpen
  \bibfield  {author} {\bibinfo {author} {\bibfnamefont {T.}~\bibnamefont
  {Callow}}, \bibinfo {author} {\bibfnamefont {B.}~\bibnamefont {Pearce}},\
  and\ \bibinfo {author} {\bibfnamefont {N.}~\bibnamefont {Gidopoulos}},\
  }\href@noop {} {\bibinfo {title} {Density functionals with spin-density
  accuracy for open shells}} (\bibinfo {year} {2021}{\natexlab{a}}),\ \Eprint
  {https://arxiv.org/abs/2110.00969} {arXiv:2110.00969 [physics.chem-ph]}
  \BibitemShut {NoStop}%
\bibitem [{\citenamefont {P\'erez-Bernal}\ \emph {et~al.}(2003)\citenamefont
  {P\'erez-Bernal}, \citenamefont {Martel}, \citenamefont {Arias},\ and\
  \citenamefont {G\'omez-Camacho}}]{PMAG03}%
  \BibitemOpen
  \bibfield  {author} {\bibinfo {author} {\bibfnamefont {F.}~\bibnamefont
  {P\'erez-Bernal}}, \bibinfo {author} {\bibfnamefont {I.}~\bibnamefont
  {Martel}}, \bibinfo {author} {\bibfnamefont {J.~M.}\ \bibnamefont {Arias}},\
  and\ \bibinfo {author} {\bibfnamefont {J.}~\bibnamefont {G\'omez-Camacho}},\
  }\bibfield  {title} {\bibinfo {title} {Continuum discretization using
  orthogonal polynomials},\ }\href {https://doi.org/10.1103/PhysRevA.67.052108}
  {\bibfield  {journal} {\bibinfo  {journal} {Phys. Rev. A}\ }\textbf {\bibinfo
  {volume} {67}},\ \bibinfo {pages} {052108} (\bibinfo {year}
  {2003})}\BibitemShut {NoStop}%
\bibitem [{\citenamefont {Peyrusse}(2006)}]{P06}%
  \BibitemOpen
  \bibfield  {author} {\bibinfo {author} {\bibfnamefont {O.}~\bibnamefont
  {Peyrusse}},\ }\bibfield  {title} {\bibinfo {title} {{The use of B-splines
  for calculating the electronic properties of atoms in plasmas}},\ }\href
  {https://doi.org/https://doi.org/10.1016/j.jqsrt.2005.05.037} {\bibfield
  {journal} {\bibinfo  {journal} {J. Quant. Spectrosc. Radiat. Transf.}\
  }\textbf {\bibinfo {volume} {99}},\ \bibinfo {pages} {469} (\bibinfo {year}
  {2006})},\ \bibinfo {note} {radiative Properties of Hot Dense
  Matter}\BibitemShut {NoStop}%
\bibitem [{\citenamefont {Massacrier}(1994)}]{M94}%
  \BibitemOpen
  \bibfield  {author} {\bibinfo {author} {\bibfnamefont {G.}~\bibnamefont
  {Massacrier}},\ }\bibfield  {title} {\bibinfo {title} {Self-consistent
  schemes for the calculation of ionic structures and populations in dense
  plasmas},\ }\href
  {https://doi.org/https://doi.org/10.1016/0022-4073(94)90083-3} {\bibfield
  {journal} {\bibinfo  {journal} {J. Quant. Spectrosc. Radiat. Transf.}\
  }\textbf {\bibinfo {volume} {51}},\ \bibinfo {pages} {221} (\bibinfo {year}
  {1994})}\BibitemShut {NoStop}%
\bibitem [{\citenamefont {Potekhin}\ \emph {et~al.}(2005)\citenamefont
  {Potekhin}, \citenamefont {Massacrier},\ and\ \citenamefont
  {Chabrier}}]{PMC05}%
  \BibitemOpen
  \bibfield  {author} {\bibinfo {author} {\bibfnamefont {A.~Y.}\ \bibnamefont
  {Potekhin}}, \bibinfo {author} {\bibfnamefont {G.}~\bibnamefont
  {Massacrier}},\ and\ \bibinfo {author} {\bibfnamefont {G.}~\bibnamefont
  {Chabrier}},\ }\bibfield  {title} {\bibinfo {title} {Equation of state for
  partially ionized carbon at high temperatures},\ }\href
  {https://doi.org/10.1103/PhysRevE.72.046402} {\bibfield  {journal} {\bibinfo
  {journal} {Phys. Rev. E}\ }\textbf {\bibinfo {volume} {72}},\ \bibinfo
  {pages} {046402} (\bibinfo {year} {2005})}\BibitemShut {NoStop}%
\bibitem [{\citenamefont {Kraisler}\ \emph {et~al.}(2009)\citenamefont
  {Kraisler}, \citenamefont {Makov}, \citenamefont {Argaman},\ and\
  \citenamefont {Kelson}}]{KraislerMakovArgamanKelson09}%
  \BibitemOpen
  \bibfield  {author} {\bibinfo {author} {\bibfnamefont {E.}~\bibnamefont
  {Kraisler}}, \bibinfo {author} {\bibfnamefont {G.}~\bibnamefont {Makov}},
  \bibinfo {author} {\bibfnamefont {N.}~\bibnamefont {Argaman}},\ and\ \bibinfo
  {author} {\bibfnamefont {I.}~\bibnamefont {Kelson}},\ }\bibfield  {title}
  {\bibinfo {title} {{Fractional occupation in Kohn--Sham density-functional
  theory and the treatment of non-pure-state $v$-representable densities}},\
  }\href {https://doi.org/10.1103/PhysRevA.80.032115} {\bibfield  {journal}
  {\bibinfo  {journal} {Phys. Rev. A}\ }\textbf {\bibinfo {volume} {80}},\
  \bibinfo {pages} {032115} (\bibinfo {year} {2009})}\BibitemShut {NoStop}%
\bibitem [{\citenamefont {Kraisler}\ \emph {et~al.}(2010)\citenamefont
  {Kraisler}, \citenamefont {Makov},\ and\ \citenamefont
  {Kelson}}]{KraislerMakovKelson10}%
  \BibitemOpen
  \bibfield  {author} {\bibinfo {author} {\bibfnamefont {E.}~\bibnamefont
  {Kraisler}}, \bibinfo {author} {\bibfnamefont {G.}~\bibnamefont {Makov}},\
  and\ \bibinfo {author} {\bibfnamefont {I.}~\bibnamefont {Kelson}},\
  }\bibfield  {title} {\bibinfo {title} {Ensemble $v$-representable \textit{ab
  initio} density-functional calculation of energy and spin in atoms: A test of
  exchange-correlation approximations},\ }\href
  {https://link.aps.org/doi/10.1103/PhysRevA.82.042516} {\bibfield  {journal}
  {\bibinfo  {journal} {Phys. Rev. A}\ }\textbf {\bibinfo {volume} {82}},\
  \bibinfo {pages} {042516} (\bibinfo {year} {2010})}\BibitemShut {NoStop}%
\bibitem [{\citenamefont {Argaman}\ \emph {et~al.}(2013)\citenamefont
  {Argaman}, \citenamefont {Makov},\ and\ \citenamefont
  {Kraisler}}]{Argaman13}%
  \BibitemOpen
  \bibfield  {author} {\bibinfo {author} {\bibfnamefont {U.}~\bibnamefont
  {Argaman}}, \bibinfo {author} {\bibfnamefont {G.}~\bibnamefont {Makov}},\
  and\ \bibinfo {author} {\bibfnamefont {E.}~\bibnamefont {Kraisler}},\
  }\bibfield  {title} {\bibinfo {title} {Higher ionization energies of atoms in
  density-functional theory},\ }\href
  {https://doi.org/10.1103/PhysRevA.88.042504} {\bibfield  {journal} {\bibinfo
  {journal} {Phys. Rev. A}\ }\textbf {\bibinfo {volume} {88}},\ \bibinfo
  {pages} {042504} (\bibinfo {year} {2013})}\BibitemShut {NoStop}%
\bibitem [{\citenamefont {Kraisler}\ and\ \citenamefont
  {Schild}(2020)}]{KraislerSchild20}%
  \BibitemOpen
  \bibfield  {author} {\bibinfo {author} {\bibfnamefont {E.}~\bibnamefont
  {Kraisler}}\ and\ \bibinfo {author} {\bibfnamefont {A.}~\bibnamefont
  {Schild}},\ }\bibfield  {title} {\bibinfo {title} {Discontinuous behavior of
  the pauli potential in density functional theory as a function of the
  electron number},\ }\href {https://doi.org/10.1103/PhysRevResearch.2.013159}
  {\bibfield  {journal} {\bibinfo  {journal} {Phys. Rev. Research}\ }\textbf
  {\bibinfo {volume} {2}},\ \bibinfo {pages} {013159} (\bibinfo {year}
  {2020})}\BibitemShut {NoStop}%
\bibitem [{\citenamefont {Kraisler}\ \emph {et~al.}(2021)\citenamefont
  {Kraisler}, \citenamefont {Hodgson},\ and\ \citenamefont
  {Gross}}]{KraislerHodgson21}%
  \BibitemOpen
  \bibfield  {author} {\bibinfo {author} {\bibfnamefont {E.}~\bibnamefont
  {Kraisler}}, \bibinfo {author} {\bibfnamefont {M.~J.~P.}\ \bibnamefont
  {Hodgson}},\ and\ \bibinfo {author} {\bibfnamefont {E.~K.~U.}\ \bibnamefont
  {Gross}},\ }\bibfield  {title} {\bibinfo {title} {{From Kohn–Sham to
  Many-Electron Energies via Step Structures in the Exchange-Correlation
  Potential}},\ }\href {https://doi.org/10.1021/acs.jctc.0c01093} {\bibfield
  {journal} {\bibinfo  {journal} {J. Chem. Theory Comput.}\ }\textbf {\bibinfo
  {volume} {17}},\ \bibinfo {pages} {1390} (\bibinfo {year}
  {2021})}\BibitemShut {NoStop}%
\bibitem [{\citenamefont {Callow}\ \emph
  {et~al.}(2021{\natexlab{b}})\citenamefont {Callow}, \citenamefont {Kotik},
  \citenamefont {Tsvetoslavova~Stankulova}, \citenamefont {Kraisler},\ and\
  \citenamefont {Cangi}}]{callow_timothy_2021_5205719}%
  \BibitemOpen
  \bibfield  {author} {\bibinfo {author} {\bibfnamefont {T.}~\bibnamefont
  {Callow}}, \bibinfo {author} {\bibfnamefont {D.}~\bibnamefont {Kotik}},
  \bibinfo {author} {\bibfnamefont {E.}~\bibnamefont
  {Tsvetoslavova~Stankulova}}, \bibinfo {author} {\bibfnamefont
  {E.}~\bibnamefont {Kraisler}},\ and\ \bibinfo {author} {\bibfnamefont
  {A.}~\bibnamefont {Cangi}},\ }\href {https://doi.org/10.5281/zenodo.5205718}
  {\bibinfo {title} {{atoMEC}}} (\bibinfo {year} {2021}{\natexlab{b}}),\
  \bibinfo {note} {\url{https://doi.org/10.5281/zenodo.5205718}}\BibitemShut
  {NoStop}%
\bibitem [{\citenamefont {Blatt}(1967)}]{B67}%
  \BibitemOpen
  \bibfield  {author} {\bibinfo {author} {\bibfnamefont {J.~M.}\ \bibnamefont
  {Blatt}},\ }\bibfield  {title} {\bibinfo {title} {{Practical points
  concerning the solution of the Schr{\"o}dinger equation}},\ }\href
  {https://www.sciencedirect.com/science/article/pii/0021999167900460}
  {\bibfield  {journal} {\bibinfo  {journal} {J. Comput. Phys.}\ }\textbf
  {\bibinfo {volume} {1}},\ \bibinfo {pages} {382} (\bibinfo {year}
  {1967})}\BibitemShut {NoStop}%
\bibitem [{\citenamefont {Chow}(1972)}]{C72}%
  \BibitemOpen
  \bibfield  {author} {\bibinfo {author} {\bibfnamefont {P.~C.}\ \bibnamefont
  {Chow}},\ }\bibfield  {title} {\bibinfo {title} {{Computer Solutions to the
  Schrödinger Equation}},\ }\href {https://doi.org/10.1119/1.1986627}
  {\bibfield  {journal} {\bibinfo  {journal} {Am. J. Phys.}\ }\textbf {\bibinfo
  {volume} {40}},\ \bibinfo {pages} {730} (\bibinfo {year} {1972})}\BibitemShut
  {NoStop}%
\bibitem [{\citenamefont {Koonin}(1986)}]{Koonin}%
  \BibitemOpen
  \bibfield  {author} {\bibinfo {author} {\bibfnamefont {S.~E.}\ \bibnamefont
  {Koonin}},\ }\href@noop {} {\emph {\bibinfo {title} {Computational
  Physics}}}\ (\bibinfo  {publisher} {Addison-Wesley},\ \bibinfo {year}
  {1986})\BibitemShut {NoStop}%
\bibitem [{\citenamefont {Gregori}\ \emph {et~al.}(2005)\citenamefont
  {Gregori}, \citenamefont {Hansen}, \citenamefont {Clarke}, \citenamefont
  {Heathcote}, \citenamefont {Key}, \citenamefont {King}, \citenamefont
  {Klein}, \citenamefont {Izumi}, \citenamefont {Mackinnon}, \citenamefont
  {Moon}, \citenamefont {Park}, \citenamefont {Pasley}, \citenamefont {Patel},
  \citenamefont {Patel}, \citenamefont {Remington}, \citenamefont {Ryutov},
  \citenamefont {Shepherd}, \citenamefont {Snavely}, \citenamefont {Wilks},
  \citenamefont {Zhang},\ and\ \citenamefont {Glenzer}}]{GHCH05}%
  \BibitemOpen
  \bibfield  {author} {\bibinfo {author} {\bibfnamefont {G.}~\bibnamefont
  {Gregori}}, \bibinfo {author} {\bibfnamefont {S.~B.}\ \bibnamefont {Hansen}},
  \bibinfo {author} {\bibfnamefont {R.}~\bibnamefont {Clarke}}, \bibinfo
  {author} {\bibfnamefont {R.}~\bibnamefont {Heathcote}}, \bibinfo {author}
  {\bibfnamefont {M.~H.}\ \bibnamefont {Key}}, \bibinfo {author} {\bibfnamefont
  {J.}~\bibnamefont {King}}, \bibinfo {author} {\bibfnamefont {R.~I.}\
  \bibnamefont {Klein}}, \bibinfo {author} {\bibfnamefont {N.}~\bibnamefont
  {Izumi}}, \bibinfo {author} {\bibfnamefont {A.~J.}\ \bibnamefont
  {Mackinnon}}, \bibinfo {author} {\bibfnamefont {S.~J.}\ \bibnamefont {Moon}},
  \bibinfo {author} {\bibfnamefont {H.-S.}\ \bibnamefont {Park}}, \bibinfo
  {author} {\bibfnamefont {J.}~\bibnamefont {Pasley}}, \bibinfo {author}
  {\bibfnamefont {N.}~\bibnamefont {Patel}}, \bibinfo {author} {\bibfnamefont
  {P.~K.}\ \bibnamefont {Patel}}, \bibinfo {author} {\bibfnamefont {B.~A.}\
  \bibnamefont {Remington}}, \bibinfo {author} {\bibfnamefont {D.~D.}\
  \bibnamefont {Ryutov}}, \bibinfo {author} {\bibfnamefont {R.}~\bibnamefont
  {Shepherd}}, \bibinfo {author} {\bibfnamefont {R.~A.}\ \bibnamefont
  {Snavely}}, \bibinfo {author} {\bibfnamefont {S.~C.}\ \bibnamefont {Wilks}},
  \bibinfo {author} {\bibfnamefont {B.~B.}\ \bibnamefont {Zhang}},\ and\
  \bibinfo {author} {\bibfnamefont {S.~H.}\ \bibnamefont {Glenzer}},\
  }\bibfield  {title} {\bibinfo {title} {Experimental characterization of a
  strongly coupled solid density plasma generated in a short-pulse laser target
  interaction},\ }\href
  {https://doi.org/https://doi.org/10.1002/ctpp.200510032} {\bibfield
  {journal} {\bibinfo  {journal} {Contrib. Plasma Phys.}\ }\textbf {\bibinfo
  {volume} {45}},\ \bibinfo {pages} {284} (\bibinfo {year} {2005})}\BibitemShut
  {NoStop}%
\bibitem [{\citenamefont {Hansen}\ \emph {et~al.}(2005)\citenamefont {Hansen},
  \citenamefont {Faenov}, \citenamefont {Pikuz}, \citenamefont {Fournier},
  \citenamefont {Shepherd}, \citenamefont {Chen}, \citenamefont {Widmann},
  \citenamefont {Wilks}, \citenamefont {Ping}, \citenamefont {Chung},
  \citenamefont {Niles}, \citenamefont {Hunter}, \citenamefont {Dyer},\ and\
  \citenamefont {Ditmire}}]{HFPF05}%
  \BibitemOpen
  \bibfield  {author} {\bibinfo {author} {\bibfnamefont {S.~B.}\ \bibnamefont
  {Hansen}}, \bibinfo {author} {\bibfnamefont {A.~Y.}\ \bibnamefont {Faenov}},
  \bibinfo {author} {\bibfnamefont {T.~A.}\ \bibnamefont {Pikuz}}, \bibinfo
  {author} {\bibfnamefont {K.~B.}\ \bibnamefont {Fournier}}, \bibinfo {author}
  {\bibfnamefont {R.}~\bibnamefont {Shepherd}}, \bibinfo {author}
  {\bibfnamefont {H.}~\bibnamefont {Chen}}, \bibinfo {author} {\bibfnamefont
  {K.}~\bibnamefont {Widmann}}, \bibinfo {author} {\bibfnamefont {S.~C.}\
  \bibnamefont {Wilks}}, \bibinfo {author} {\bibfnamefont {Y.}~\bibnamefont
  {Ping}}, \bibinfo {author} {\bibfnamefont {H.~K.}\ \bibnamefont {Chung}},
  \bibinfo {author} {\bibfnamefont {A.}~\bibnamefont {Niles}}, \bibinfo
  {author} {\bibfnamefont {J.~R.}\ \bibnamefont {Hunter}}, \bibinfo {author}
  {\bibfnamefont {G.}~\bibnamefont {Dyer}},\ and\ \bibinfo {author}
  {\bibfnamefont {T.}~\bibnamefont {Ditmire}},\ }\bibfield  {title} {\bibinfo
  {title} {{Temperature determination using $K\ensuremath{\alpha}$ spectra from
  $M$-shell Ti ions}},\ }\href {https://doi.org/10.1103/PhysRevE.72.036408}
  {\bibfield  {journal} {\bibinfo  {journal} {Phys. Rev. E}\ }\textbf {\bibinfo
  {volume} {72}},\ \bibinfo {pages} {036408} (\bibinfo {year}
  {2005})}\BibitemShut {NoStop}%
\bibitem [{Note6()}]{Note6}%
  \BibitemOpen
  \bibinfo {note} {Starting from the lowest temperature $\tau =0.001\ \protect
  \textrm {Har}$, convergence was checked in approximate multiples of 3, i.e.
  $\tau =0.001,0.003,0.01\protect \dots \ \protect \textrm {Har}$.}\BibitemShut
  {Stop}%
\bibitem [{\citenamefont {Perdew}\ and\ \citenamefont {Wang}(1992)}]{PW92}%
  \BibitemOpen
  \bibfield  {author} {\bibinfo {author} {\bibfnamefont {J.~P.}\ \bibnamefont
  {Perdew}}\ and\ \bibinfo {author} {\bibfnamefont {Y.}~\bibnamefont {Wang}},\
  }\bibfield  {title} {\bibinfo {title} {Accurate and simple analytic
  representation of the electron-gas correlation energy},\ }\href
  {https://doi.org/10.1103/PhysRevB.45.13244} {\bibfield  {journal} {\bibinfo
  {journal} {Phys. Rev. B}\ }\textbf {\bibinfo {volume} {45}},\ \bibinfo
  {pages} {13244} (\bibinfo {year} {1992})}\BibitemShut {NoStop}%
\bibitem [{\citenamefont {Perdew}\ \emph {et~al.}(1996)\citenamefont {Perdew},
  \citenamefont {Burke},\ and\ \citenamefont {Ernzerhof}}]{PBE96}%
  \BibitemOpen
  \bibfield  {author} {\bibinfo {author} {\bibfnamefont {J.~P.}\ \bibnamefont
  {Perdew}}, \bibinfo {author} {\bibfnamefont {K.}~\bibnamefont {Burke}},\ and\
  \bibinfo {author} {\bibfnamefont {M.}~\bibnamefont {Ernzerhof}},\ }\bibfield
  {title} {\bibinfo {title} {Generalized gradient approximation made simple},\
  }\href {https://doi.org/10.1103/PhysRevLett.77.3865} {\bibfield  {journal}
  {\bibinfo  {journal} {Phys. Rev. Lett.}\ }\textbf {\bibinfo {volume} {77}},\
  \bibinfo {pages} {3865} (\bibinfo {year} {1996})}\BibitemShut {NoStop}%
\bibitem [{\citenamefont {Pribram-Jones}\ \emph {et~al.}(2015)\citenamefont
  {Pribram-Jones}, \citenamefont {Gross},\ and\ \citenamefont
  {Burke}}]{PJGB15}%
  \BibitemOpen
  \bibfield  {author} {\bibinfo {author} {\bibfnamefont {A.}~\bibnamefont
  {Pribram-Jones}}, \bibinfo {author} {\bibfnamefont {D.~A.}\ \bibnamefont
  {Gross}},\ and\ \bibinfo {author} {\bibfnamefont {K.}~\bibnamefont {Burke}},\
  }\bibfield  {title} {\bibinfo {title} {{DFT}: A theory full of holes?},\
  }\href {https://doi.org/10.1146/annurev-physchem-040214-121420} {\bibfield
  {journal} {\bibinfo  {journal} {Ann. Rev. Phys. Chem.}\ }\textbf {\bibinfo
  {volume} {66}},\ \bibinfo {pages} {283} (\bibinfo {year} {2015})}\BibitemShut
  {NoStop}%
\bibitem [{\citenamefont {Lehtola}\ \emph {et~al.}(2018)\citenamefont
  {Lehtola}, \citenamefont {Steigemann}, \citenamefont {Oliveira},\ and\
  \citenamefont {Marques}}]{LSOM18}%
  \BibitemOpen
  \bibfield  {author} {\bibinfo {author} {\bibfnamefont {S.}~\bibnamefont
  {Lehtola}}, \bibinfo {author} {\bibfnamefont {C.}~\bibnamefont {Steigemann}},
  \bibinfo {author} {\bibfnamefont {M.~J.}\ \bibnamefont {Oliveira}},\ and\
  \bibinfo {author} {\bibfnamefont {M.~A.}\ \bibnamefont {Marques}},\
  }\bibfield  {title} {\bibinfo {title} {Recent developments in libxc — a
  comprehensive library of functionals for density functional theory},\ }\href
  {https://doi.org/https://doi.org/10.1016/j.softx.2017.11.002} {\bibfield
  {journal} {\bibinfo  {journal} {SoftwareX}\ }\textbf {\bibinfo {volume}
  {7}},\ \bibinfo {pages} {1 } (\bibinfo {year} {2018})}\BibitemShut {NoStop}%
\bibitem [{\citenamefont {Karasiev}\ \emph {et~al.}(2019)\citenamefont
  {Karasiev}, \citenamefont {Trickey},\ and\ \citenamefont {Dufty}}]{KTD19}%
  \BibitemOpen
  \bibfield  {author} {\bibinfo {author} {\bibfnamefont {V.~V.}\ \bibnamefont
  {Karasiev}}, \bibinfo {author} {\bibfnamefont {S.~B.}\ \bibnamefont
  {Trickey}},\ and\ \bibinfo {author} {\bibfnamefont {J.~W.}\ \bibnamefont
  {Dufty}},\ }\bibfield  {title} {\bibinfo {title} {Status of free-energy
  representations for the homogeneous electron gas},\ }\href
  {https://doi.org/10.1103/PhysRevB.99.195134} {\bibfield  {journal} {\bibinfo
  {journal} {Phys. Rev. B}\ }\textbf {\bibinfo {volume} {99}},\ \bibinfo
  {pages} {195134} (\bibinfo {year} {2019})}\BibitemShut {NoStop}%
\bibitem [{\citenamefont {Ramakrishna}\ \emph {et~al.}(2020)\citenamefont
  {Ramakrishna}, \citenamefont {Dornheim},\ and\ \citenamefont
  {Vorberger}}]{RDV20}%
  \BibitemOpen
  \bibfield  {author} {\bibinfo {author} {\bibfnamefont {K.}~\bibnamefont
  {Ramakrishna}}, \bibinfo {author} {\bibfnamefont {T.}~\bibnamefont
  {Dornheim}},\ and\ \bibinfo {author} {\bibfnamefont {J.}~\bibnamefont
  {Vorberger}},\ }\bibfield  {title} {\bibinfo {title} {Influence of finite
  temperature exchange-correlation effects in hydrogen},\ }\href
  {https://doi.org/10.1103/PhysRevB.101.195129} {\bibfield  {journal} {\bibinfo
   {journal} {Phys. Rev. B}\ }\textbf {\bibinfo {volume} {101}},\ \bibinfo
  {pages} {195129} (\bibinfo {year} {2020})}\BibitemShut {NoStop}%
\bibitem [{\citenamefont {Crowley}(2014)}]{C14}%
  \BibitemOpen
  \bibfield  {author} {\bibinfo {author} {\bibfnamefont {B.}~\bibnamefont
  {Crowley}},\ }\bibfield  {title} {\bibinfo {title} {Continuum lowering – a
  new perspective},\ }\href
  {https://doi.org/https://doi.org/10.1016/j.hedp.2014.04.003} {\bibfield
  {journal} {\bibinfo  {journal} {High Energy Density Phys.}\ }\textbf
  {\bibinfo {volume} {13}},\ \bibinfo {pages} {84} (\bibinfo {year}
  {2014})}\BibitemShut {NoStop}%
\bibitem [{\citenamefont {Ciricosta}\ \emph {et~al.}(2012)\citenamefont
  {Ciricosta}, \citenamefont {Vinko}, \citenamefont {Chung}, \citenamefont
  {Cho}, \citenamefont {Brown}, \citenamefont {Burian}, \citenamefont
  {Chalupsk\'y}, \citenamefont {Engelhorn}, \citenamefont {Falcone},
  \citenamefont {Graves}, \citenamefont {H\'ajkov\'a}, \citenamefont
  {Higginbotham}, \citenamefont {Juha}, \citenamefont {Krzywinski},
  \citenamefont {Lee}, \citenamefont {Messerschmidt}, \citenamefont {Murphy},
  \citenamefont {Ping}, \citenamefont {Rackstraw}, \citenamefont {Scherz},
  \citenamefont {Schlotter}, \citenamefont {Toleikis}, \citenamefont {Turner},
  \citenamefont {Vysin}, \citenamefont {Wang}, \citenamefont {Wu},
  \citenamefont {Zastrau}, \citenamefont {Zhu}, \citenamefont {Lee},
  \citenamefont {Heimann}, \citenamefont {Nagler},\ and\ \citenamefont
  {Wark}}]{CVCB12}%
  \BibitemOpen
  \bibfield  {author} {\bibinfo {author} {\bibfnamefont {O.}~\bibnamefont
  {Ciricosta}}, \bibinfo {author} {\bibfnamefont {S.~M.}\ \bibnamefont
  {Vinko}}, \bibinfo {author} {\bibfnamefont {H.-K.}\ \bibnamefont {Chung}},
  \bibinfo {author} {\bibfnamefont {B.-I.}\ \bibnamefont {Cho}}, \bibinfo
  {author} {\bibfnamefont {C.~R.~D.}\ \bibnamefont {Brown}}, \bibinfo {author}
  {\bibfnamefont {T.}~\bibnamefont {Burian}}, \bibinfo {author} {\bibfnamefont
  {J.}~\bibnamefont {Chalupsk\'y}}, \bibinfo {author} {\bibfnamefont
  {K.}~\bibnamefont {Engelhorn}}, \bibinfo {author} {\bibfnamefont {R.~W.}\
  \bibnamefont {Falcone}}, \bibinfo {author} {\bibfnamefont {C.}~\bibnamefont
  {Graves}}, \bibinfo {author} {\bibfnamefont {V.}~\bibnamefont {H\'ajkov\'a}},
  \bibinfo {author} {\bibfnamefont {A.}~\bibnamefont {Higginbotham}}, \bibinfo
  {author} {\bibfnamefont {L.}~\bibnamefont {Juha}}, \bibinfo {author}
  {\bibfnamefont {J.}~\bibnamefont {Krzywinski}}, \bibinfo {author}
  {\bibfnamefont {H.~J.}\ \bibnamefont {Lee}}, \bibinfo {author} {\bibfnamefont
  {M.}~\bibnamefont {Messerschmidt}}, \bibinfo {author} {\bibfnamefont {C.~D.}\
  \bibnamefont {Murphy}}, \bibinfo {author} {\bibfnamefont {Y.}~\bibnamefont
  {Ping}}, \bibinfo {author} {\bibfnamefont {D.~S.}\ \bibnamefont {Rackstraw}},
  \bibinfo {author} {\bibfnamefont {A.}~\bibnamefont {Scherz}}, \bibinfo
  {author} {\bibfnamefont {W.}~\bibnamefont {Schlotter}}, \bibinfo {author}
  {\bibfnamefont {S.}~\bibnamefont {Toleikis}}, \bibinfo {author}
  {\bibfnamefont {J.~J.}\ \bibnamefont {Turner}}, \bibinfo {author}
  {\bibfnamefont {L.}~\bibnamefont {Vysin}}, \bibinfo {author} {\bibfnamefont
  {T.}~\bibnamefont {Wang}}, \bibinfo {author} {\bibfnamefont {B.}~\bibnamefont
  {Wu}}, \bibinfo {author} {\bibfnamefont {U.}~\bibnamefont {Zastrau}},
  \bibinfo {author} {\bibfnamefont {D.}~\bibnamefont {Zhu}}, \bibinfo {author}
  {\bibfnamefont {R.~W.}\ \bibnamefont {Lee}}, \bibinfo {author} {\bibfnamefont
  {P.}~\bibnamefont {Heimann}}, \bibinfo {author} {\bibfnamefont
  {B.}~\bibnamefont {Nagler}},\ and\ \bibinfo {author} {\bibfnamefont {J.~S.}\
  \bibnamefont {Wark}},\ }\bibfield  {title} {\bibinfo {title} {Direct
  measurements of the ionization potential depression in a dense plasma},\
  }\href {https://doi.org/10.1103/PhysRevLett.109.065002} {\bibfield  {journal}
  {\bibinfo  {journal} {Phys. Rev. Lett.}\ }\textbf {\bibinfo {volume} {109}},\
  \bibinfo {pages} {065002} (\bibinfo {year} {2012})}\BibitemShut {NoStop}%
\bibitem [{\citenamefont {Hoarty}\ \emph {et~al.}(2013)\citenamefont {Hoarty},
  \citenamefont {Allan}, \citenamefont {James}, \citenamefont {Brown},
  \citenamefont {Hobbs}, \citenamefont {Hill}, \citenamefont {Harris},
  \citenamefont {Morton}, \citenamefont {Brookes}, \citenamefont {Shepherd},
  \citenamefont {Dunn}, \citenamefont {Chen}, \citenamefont {Von~Marley},
  \citenamefont {Beiersdorfer}, \citenamefont {Chung}, \citenamefont {Lee},
  \citenamefont {Brown},\ and\ \citenamefont {Emig}}]{HJBH13}%
  \BibitemOpen
  \bibfield  {author} {\bibinfo {author} {\bibfnamefont {D.~J.}\ \bibnamefont
  {Hoarty}}, \bibinfo {author} {\bibfnamefont {P.}~\bibnamefont {Allan}},
  \bibinfo {author} {\bibfnamefont {S.~F.}\ \bibnamefont {James}}, \bibinfo
  {author} {\bibfnamefont {C.~R.~D.}\ \bibnamefont {Brown}}, \bibinfo {author}
  {\bibfnamefont {L.~M.~R.}\ \bibnamefont {Hobbs}}, \bibinfo {author}
  {\bibfnamefont {M.~P.}\ \bibnamefont {Hill}}, \bibinfo {author}
  {\bibfnamefont {J.~W.~O.}\ \bibnamefont {Harris}}, \bibinfo {author}
  {\bibfnamefont {J.}~\bibnamefont {Morton}}, \bibinfo {author} {\bibfnamefont
  {M.~G.}\ \bibnamefont {Brookes}}, \bibinfo {author} {\bibfnamefont
  {R.}~\bibnamefont {Shepherd}}, \bibinfo {author} {\bibfnamefont
  {J.}~\bibnamefont {Dunn}}, \bibinfo {author} {\bibfnamefont {H.}~\bibnamefont
  {Chen}}, \bibinfo {author} {\bibfnamefont {E.}~\bibnamefont {Von~Marley}},
  \bibinfo {author} {\bibfnamefont {P.}~\bibnamefont {Beiersdorfer}}, \bibinfo
  {author} {\bibfnamefont {H.~K.}\ \bibnamefont {Chung}}, \bibinfo {author}
  {\bibfnamefont {R.~W.}\ \bibnamefont {Lee}}, \bibinfo {author} {\bibfnamefont
  {G.}~\bibnamefont {Brown}},\ and\ \bibinfo {author} {\bibfnamefont
  {J.}~\bibnamefont {Emig}},\ }\bibfield  {title} {\bibinfo {title}
  {Observations of the effect of ionization-potential depression in hot dense
  plasma},\ }\href {https://doi.org/10.1103/PhysRevLett.110.265003} {\bibfield
  {journal} {\bibinfo  {journal} {Phys. Rev. Lett.}\ }\textbf {\bibinfo
  {volume} {110}},\ \bibinfo {pages} {265003} (\bibinfo {year}
  {2013})}\BibitemShut {NoStop}%
\bibitem [{\citenamefont {Fletcher}\ \emph {et~al.}(2014)\citenamefont
  {Fletcher}, \citenamefont {Kritcher}, \citenamefont {Pak}, \citenamefont
  {Ma}, \citenamefont {D\"oppner}, \citenamefont {Fortmann}, \citenamefont
  {Divol}, \citenamefont {Jones}, \citenamefont {Landen}, \citenamefont
  {Scott}, \citenamefont {Vorberger}, \citenamefont {Chapman}, \citenamefont
  {Gericke}, \citenamefont {Mattern}, \citenamefont {Seidler}, \citenamefont
  {Gregori}, \citenamefont {Falcone},\ and\ \citenamefont {Glenzer}}]{FKPM14}%
  \BibitemOpen
  \bibfield  {author} {\bibinfo {author} {\bibfnamefont {L.~B.}\ \bibnamefont
  {Fletcher}}, \bibinfo {author} {\bibfnamefont {A.~L.}\ \bibnamefont
  {Kritcher}}, \bibinfo {author} {\bibfnamefont {A.}~\bibnamefont {Pak}},
  \bibinfo {author} {\bibfnamefont {T.}~\bibnamefont {Ma}}, \bibinfo {author}
  {\bibfnamefont {T.}~\bibnamefont {D\"oppner}}, \bibinfo {author}
  {\bibfnamefont {C.}~\bibnamefont {Fortmann}}, \bibinfo {author}
  {\bibfnamefont {L.}~\bibnamefont {Divol}}, \bibinfo {author} {\bibfnamefont
  {O.~S.}\ \bibnamefont {Jones}}, \bibinfo {author} {\bibfnamefont {O.~L.}\
  \bibnamefont {Landen}}, \bibinfo {author} {\bibfnamefont {H.~A.}\
  \bibnamefont {Scott}}, \bibinfo {author} {\bibfnamefont {J.}~\bibnamefont
  {Vorberger}}, \bibinfo {author} {\bibfnamefont {D.~A.}\ \bibnamefont
  {Chapman}}, \bibinfo {author} {\bibfnamefont {D.~O.}\ \bibnamefont
  {Gericke}}, \bibinfo {author} {\bibfnamefont {B.~A.}\ \bibnamefont
  {Mattern}}, \bibinfo {author} {\bibfnamefont {G.~T.}\ \bibnamefont
  {Seidler}}, \bibinfo {author} {\bibfnamefont {G.}~\bibnamefont {Gregori}},
  \bibinfo {author} {\bibfnamefont {R.~W.}\ \bibnamefont {Falcone}},\ and\
  \bibinfo {author} {\bibfnamefont {S.~H.}\ \bibnamefont {Glenzer}},\
  }\bibfield  {title} {\bibinfo {title} {Observations of continuum depression
  in warm dense matter with {X-Ray Thomson} scattering},\ }\href
  {https://doi.org/10.1103/PhysRevLett.112.145004} {\bibfield  {journal}
  {\bibinfo  {journal} {Phys. Rev. Lett.}\ }\textbf {\bibinfo {volume} {112}},\
  \bibinfo {pages} {145004} (\bibinfo {year} {2014})}\BibitemShut {NoStop}%
\bibitem [{\citenamefont {Ecker}\ and\ \citenamefont {Kröll}(1963)}]{EK63}%
  \BibitemOpen
  \bibfield  {author} {\bibinfo {author} {\bibfnamefont {G.}~\bibnamefont
  {Ecker}}\ and\ \bibinfo {author} {\bibfnamefont {W.}~\bibnamefont {Kröll}},\
  }\bibfield  {title} {\bibinfo {title} {Lowering of the ionization energy for
  a plasma in thermodynamic equilibrium},\ }\href
  {https://doi.org/10.1063/1.1724509} {\bibfield  {journal} {\bibinfo
  {journal} {Phys. Fluids}\ }\textbf {\bibinfo {volume} {6}},\ \bibinfo {pages}
  {62} (\bibinfo {year} {1963})}\BibitemShut {NoStop}%
\bibitem [{\citenamefont {Stewart}\ and\ \citenamefont
  {Pyatt~Jr}(1966)}]{SP66}%
  \BibitemOpen
  \bibfield  {author} {\bibinfo {author} {\bibfnamefont {J.~C.}\ \bibnamefont
  {Stewart}}\ and\ \bibinfo {author} {\bibfnamefont {K.~D.}\ \bibnamefont
  {Pyatt~Jr}},\ }\bibfield  {title} {\bibinfo {title} {Lowering of ionization
  potentials in plasmas},\ }\href
  {http://adsabs.harvard.edu/pdf/1966ApJ...144.1203S} {\bibfield  {journal}
  {\bibinfo  {journal} {The Astrophysical Journal}\ }\textbf {\bibinfo {volume}
  {144}},\ \bibinfo {pages} {1203} (\bibinfo {year} {1966})}\BibitemShut
  {NoStop}%
\bibitem [{\citenamefont {Liberman}\ and\ \citenamefont
  {Albritton}(1994)}]{LA94}%
  \BibitemOpen
  \bibfield  {author} {\bibinfo {author} {\bibfnamefont {D.}~\bibnamefont
  {Liberman}}\ and\ \bibinfo {author} {\bibfnamefont {J.}~\bibnamefont
  {Albritton}},\ }\bibfield  {title} {\bibinfo {title} {Dense plasma equation
  of state model},\ }\href
  {https://doi.org/https://doi.org/10.1016/0022-4073(94)90080-9} {\bibfield
  {journal} {\bibinfo  {journal} {J. Quant. Spectrosc. Radiat. Transf.}\
  }\textbf {\bibinfo {volume} {51}},\ \bibinfo {pages} {197} (\bibinfo {year}
  {1994})},\ \bibinfo {note} {special Issue Radiative Properties of Hot Dense
  Matter}\BibitemShut {NoStop}%
\bibitem [{\citenamefont {Iglesias}(2014)}]{I14}%
  \BibitemOpen
  \bibfield  {author} {\bibinfo {author} {\bibfnamefont {C.~A.}\ \bibnamefont
  {Iglesias}},\ }\bibfield  {title} {\bibinfo {title} {A plea for a
  reexamination of ionization potential depression measurements},\ }\href
  {https://doi.org/https://doi.org/10.1016/j.hedp.2014.04.002} {\bibfield
  {journal} {\bibinfo  {journal} {High Energy Density Phys.}\ }\textbf
  {\bibinfo {volume} {12}},\ \bibinfo {pages} {5} (\bibinfo {year}
  {2014})}\BibitemShut {NoStop}%
\bibitem [{\citenamefont {Perdew}\ \emph {et~al.}(1982)\citenamefont {Perdew},
  \citenamefont {Parr}, \citenamefont {Levy},\ and\ \citenamefont
  {Balduz}}]{PPLB_82}%
  \BibitemOpen
  \bibfield  {author} {\bibinfo {author} {\bibfnamefont {J.~P.}\ \bibnamefont
  {Perdew}}, \bibinfo {author} {\bibfnamefont {R.~G.}\ \bibnamefont {Parr}},
  \bibinfo {author} {\bibfnamefont {M.}~\bibnamefont {Levy}},\ and\ \bibinfo
  {author} {\bibfnamefont {J.~L.}\ \bibnamefont {Balduz}},\ }\bibfield  {title}
  {\bibinfo {title} {Density-functional theory for fractional particle number:
  Derivative discontinuities of the energy},\ }\href
  {https://doi.org/10.1103/PhysRevLett.49.1691} {\bibfield  {journal} {\bibinfo
   {journal} {Phys. Rev. Lett.}\ }\textbf {\bibinfo {volume} {49}},\ \bibinfo
  {pages} {1691} (\bibinfo {year} {1982})}\BibitemShut {NoStop}%
\bibitem [{\citenamefont {Levy}\ \emph {et~al.}(1984)\citenamefont {Levy},
  \citenamefont {Perdew},\ and\ \citenamefont {Sahni}}]{LevyPerdewSahni84}%
  \BibitemOpen
  \bibfield  {author} {\bibinfo {author} {\bibfnamefont {M.}~\bibnamefont
  {Levy}}, \bibinfo {author} {\bibfnamefont {J.~P.}\ \bibnamefont {Perdew}},\
  and\ \bibinfo {author} {\bibfnamefont {V.}~\bibnamefont {Sahni}},\ }\bibfield
   {title} {\bibinfo {title} {Exact differential equation for the density and
  ionization energy of a many-particle system},\ }\href
  {https://doi.org/10.1103/PhysRevA.30.2745} {\bibfield  {journal} {\bibinfo
  {journal} {Phys. Rev. A}\ }\textbf {\bibinfo {volume} {30}},\ \bibinfo
  {pages} {2745} (\bibinfo {year} {1984})}\BibitemShut {NoStop}%
\bibitem [{\citenamefont {Yang}\ \emph {et~al.}(2012)\citenamefont {Yang},
  \citenamefont {Cohen},\ and\ \citenamefont {Mori-S\'{a}nchez}}]{Yang12}%
  \BibitemOpen
  \bibfield  {author} {\bibinfo {author} {\bibfnamefont {W.}~\bibnamefont
  {Yang}}, \bibinfo {author} {\bibfnamefont {A.~J.}\ \bibnamefont {Cohen}},\
  and\ \bibinfo {author} {\bibfnamefont {P.}~\bibnamefont {Mori-S\'{a}nchez}},\
  }\bibfield  {title} {\bibinfo {title} {{Derivative discontinuity, bandgap and
  lowest unoccupied molecular orbital in density functional theory.}},\ }\href
  {https://doi.org/10.1063/1.3702391} {\bibfield  {journal} {\bibinfo
  {journal} {J. Chem. Phys.}\ }\textbf {\bibinfo {volume} {136}},\ \bibinfo
  {pages} {204111} (\bibinfo {year} {2012})}\BibitemShut {NoStop}%
\bibitem [{\citenamefont {Perdew}\ and\ \citenamefont
  {Levy}(1997)}]{PerdewLevy97}%
  \BibitemOpen
  \bibfield  {author} {\bibinfo {author} {\bibfnamefont {J.~P.}\ \bibnamefont
  {Perdew}}\ and\ \bibinfo {author} {\bibfnamefont {M.}~\bibnamefont {Levy}},\
  }\bibfield  {title} {\bibinfo {title} {{Comment on ``Significance of the
  highest occupied Kohn-Sham eigenvalue''}},\ }\href
  {https://doi.org/10.1103/PhysRevB.56.16021} {\bibfield  {journal} {\bibinfo
  {journal} {Phys. Rev. B}\ }\textbf {\bibinfo {volume} {56}},\ \bibinfo
  {pages} {16021} (\bibinfo {year} {1997})}\BibitemShut {NoStop}%
\bibitem [{\citenamefont {Stowasser}\ and\ \citenamefont
  {Hoffmann}(1999)}]{SH99}%
  \BibitemOpen
  \bibfield  {author} {\bibinfo {author} {\bibfnamefont {R.}~\bibnamefont
  {Stowasser}}\ and\ \bibinfo {author} {\bibfnamefont {R.}~\bibnamefont
  {Hoffmann}},\ }\bibfield  {title} {\bibinfo {title} {What do the {Kohn--Sham}
  orbitals and eigenvalues mean},\ }\href {https://doi.org/10.1021/ja9826892}
  {\bibfield  {journal} {\bibinfo  {journal} {J. Am. Chem. Soc.}\ }\textbf
  {\bibinfo {volume} {121}},\ \bibinfo {pages} {3414} (\bibinfo {year}
  {1999})}\BibitemShut {NoStop}%
\bibitem [{\citenamefont {Hamel}\ \emph {et~al.}(2002)\citenamefont {Hamel},
  \citenamefont {Duffy}, \citenamefont {Casida},\ and\ \citenamefont
  {Salahub}}]{HDCS02}%
  \BibitemOpen
  \bibfield  {author} {\bibinfo {author} {\bibfnamefont {S.}~\bibnamefont
  {Hamel}}, \bibinfo {author} {\bibfnamefont {P.}~\bibnamefont {Duffy}},
  \bibinfo {author} {\bibfnamefont {M.~E.}\ \bibnamefont {Casida}},\ and\
  \bibinfo {author} {\bibfnamefont {D.~R.}\ \bibnamefont {Salahub}},\
  }\bibfield  {title} {\bibinfo {title} {Kohn–sham orbitals and orbital
  energies: fictitious constructs but good approximations all the same},\
  }\href {https://doi.org/https://doi.org/10.1016/S0368-2048(02)00032-4}
  {\bibfield  {journal} {\bibinfo  {journal} {J. Electron Spectros. Relat.
  Phenomena.}\ }\textbf {\bibinfo {volume} {123}},\ \bibinfo {pages} {345}
  (\bibinfo {year} {2002})}\BibitemShut {NoStop}%
\bibitem [{\citenamefont {Cohen}\ \emph {et~al.}(2008)\citenamefont {Cohen},
  \citenamefont {Mori-S{\'a}nchez},\ and\ \citenamefont {Yang}}]{CMSY08}%
  \BibitemOpen
  \bibfield  {author} {\bibinfo {author} {\bibfnamefont {A.~J.}\ \bibnamefont
  {Cohen}}, \bibinfo {author} {\bibfnamefont {P.}~\bibnamefont
  {Mori-S{\'a}nchez}},\ and\ \bibinfo {author} {\bibfnamefont {W.}~\bibnamefont
  {Yang}},\ }\bibfield  {title} {\bibinfo {title} {Insights into current
  limitations of density functional theory},\ }\href
  {https://doi.org/10.1126/science.1158722} {\bibfield  {journal} {\bibinfo
  {journal} {Science}\ }\textbf {\bibinfo {volume} {321}},\ \bibinfo {pages}
  {792} (\bibinfo {year} {2008})}\BibitemShut {NoStop}%
\bibitem [{\citenamefont {Mori-S\'anchez}\ \emph {et~al.}(2008)\citenamefont
  {Mori-S\'anchez}, \citenamefont {Cohen},\ and\ \citenamefont
  {Yang}}]{MSCY08}%
  \BibitemOpen
  \bibfield  {author} {\bibinfo {author} {\bibfnamefont {P.}~\bibnamefont
  {Mori-S\'anchez}}, \bibinfo {author} {\bibfnamefont {A.~J.}\ \bibnamefont
  {Cohen}},\ and\ \bibinfo {author} {\bibfnamefont {W.}~\bibnamefont {Yang}},\
  }\bibfield  {title} {\bibinfo {title} {Localization and delocalization errors
  in density functional theory and implications for band-gap prediction},\
  }\href {https://doi.org/10.1103/PhysRevLett.100.146401} {\bibfield  {journal}
  {\bibinfo  {journal} {Phys. Rev. Lett.}\ }\textbf {\bibinfo {volume} {100}},\
  \bibinfo {pages} {146401} (\bibinfo {year} {2008})}\BibitemShut {NoStop}%
\bibitem [{\citenamefont {Altmann}\ \emph {et~al.}(2001)\citenamefont
  {Altmann}, \citenamefont {Mö{\ss}bauer},\ and\ \citenamefont
  {Oberauer}}]{AMO01}%
  \BibitemOpen
  \bibfield  {author} {\bibinfo {author} {\bibfnamefont {M.~F.}\ \bibnamefont
  {Altmann}}, \bibinfo {author} {\bibfnamefont {R.~L.}\ \bibnamefont
  {Mö{\ss}bauer}},\ and\ \bibinfo {author} {\bibfnamefont {L.~J.~N.}\
  \bibnamefont {Oberauer}},\ }\bibfield  {title} {\bibinfo {title} {Solar
  neutrinos},\ }\href {https://doi.org/10.1088/0034-4885/64/1/203} {\bibfield
  {journal} {\bibinfo  {journal} {Rep. Prog. Phys.}\ }\textbf {\bibinfo
  {volume} {64}},\ \bibinfo {pages} {97} (\bibinfo {year} {2001})}\BibitemShut
  {NoStop}%
\bibitem [{\citenamefont {Plagemann}\ \emph {et~al.}(2012)\citenamefont
  {Plagemann}, \citenamefont {Sperling}, \citenamefont {Thiele}, \citenamefont
  {Desjarlais}, \citenamefont {Fortmann}, \citenamefont {Döppner},
  \citenamefont {Lee}, \citenamefont {Glenzer},\ and\ \citenamefont
  {Redmer}}]{PSTDFDLGR12}%
  \BibitemOpen
  \bibfield  {author} {\bibinfo {author} {\bibfnamefont {K.-U.}\ \bibnamefont
  {Plagemann}}, \bibinfo {author} {\bibfnamefont {P.}~\bibnamefont {Sperling}},
  \bibinfo {author} {\bibfnamefont {R.}~\bibnamefont {Thiele}}, \bibinfo
  {author} {\bibfnamefont {M.~P.}\ \bibnamefont {Desjarlais}}, \bibinfo
  {author} {\bibfnamefont {C.}~\bibnamefont {Fortmann}}, \bibinfo {author}
  {\bibfnamefont {T.}~\bibnamefont {Döppner}}, \bibinfo {author}
  {\bibfnamefont {H.~J.}\ \bibnamefont {Lee}}, \bibinfo {author} {\bibfnamefont
  {S.~H.}\ \bibnamefont {Glenzer}},\ and\ \bibinfo {author} {\bibfnamefont
  {R.}~\bibnamefont {Redmer}},\ }\bibfield  {title} {\bibinfo {title} {Dynamic
  structure factor in warm dense beryllium},\ }\href
  {https://doi.org/10.1088/1367-2630/14/5/055020} {\bibfield  {journal}
  {\bibinfo  {journal} {New J. Phys.}\ }\textbf {\bibinfo {volume} {14}},\
  \bibinfo {pages} {055020} (\bibinfo {year} {2012})}\BibitemShut {NoStop}%
\bibitem [{\citenamefont {Li}\ \emph {et~al.}(2014)\citenamefont {Li},
  \citenamefont {Liu}, \citenamefont {Zeng}, \citenamefont {Wang},
  \citenamefont {Wu}, \citenamefont {Zhang},\ and\ \citenamefont
  {Yan}}]{LLZQQZY14}%
  \BibitemOpen
  \bibfield  {author} {\bibinfo {author} {\bibfnamefont {D.}~\bibnamefont
  {Li}}, \bibinfo {author} {\bibfnamefont {H.}~\bibnamefont {Liu}}, \bibinfo
  {author} {\bibfnamefont {S.}~\bibnamefont {Zeng}}, \bibinfo {author}
  {\bibfnamefont {C.}~\bibnamefont {Wang}}, \bibinfo {author} {\bibfnamefont
  {Z.}~\bibnamefont {Wu}}, \bibinfo {author} {\bibfnamefont {P.}~\bibnamefont
  {Zhang}},\ and\ \bibinfo {author} {\bibfnamefont {J.}~\bibnamefont {Yan}},\
  }\bibfield  {title} {\bibinfo {title} {Quantum molecular dynamics study of
  expanded beryllium: Evolution from warm dense matter to atomic fluid},\
  }\href {https://www.nature.com/articles/srep05898} {\bibfield  {journal}
  {\bibinfo  {journal} {Sci. Rep.}\ }\textbf {\bibinfo {volume} {4}},\ \bibinfo
  {pages} {1} (\bibinfo {year} {2014})}\BibitemShut {NoStop}%
\bibitem [{\citenamefont {Savin}\ \emph {et~al.}(1997)\citenamefont {Savin},
  \citenamefont {Nesper}, \citenamefont {Wengert},\ and\ \citenamefont
  {Fässler}}]{SNWF97}%
  \BibitemOpen
  \bibfield  {author} {\bibinfo {author} {\bibfnamefont {A.}~\bibnamefont
  {Savin}}, \bibinfo {author} {\bibfnamefont {R.}~\bibnamefont {Nesper}},
  \bibinfo {author} {\bibfnamefont {S.}~\bibnamefont {Wengert}},\ and\ \bibinfo
  {author} {\bibfnamefont {T.~F.}\ \bibnamefont {Fässler}},\ }\bibfield
  {title} {\bibinfo {title} {{ELF}: The electron localization function},\
  }\href {https://doi.org/https://doi.org/10.1002/anie.199718081} {\bibfield
  {journal} {\bibinfo  {journal} {Angewandte Chemie International Edition in
  English}\ }\textbf {\bibinfo {volume} {36}},\ \bibinfo {pages} {1808}
  (\bibinfo {year} {1997})}\BibitemShut {NoStop}%
\bibitem [{\citenamefont {Fuentealba}\ \emph {et~al.}(2007)\citenamefont
  {Fuentealba}, \citenamefont {Chamorro},\ and\ \citenamefont
  {Santos}}]{FCS07}%
  \BibitemOpen
  \bibfield  {author} {\bibinfo {author} {\bibfnamefont {P.}~\bibnamefont
  {Fuentealba}}, \bibinfo {author} {\bibfnamefont {E.}~\bibnamefont
  {Chamorro}},\ and\ \bibinfo {author} {\bibfnamefont {J.~C.}\ \bibnamefont
  {Santos}},\ }\bibfield  {title} {\bibinfo {title} {Chapter 5: Understanding
  and using the electron localization function},\ }in\ \href
  {https://doi.org/https://doi.org/10.1016/S1380-7323(07)80006-9} {\emph
  {\bibinfo {booktitle} {Theoretical Aspects of Chemical Reactivity}}},\
  \bibinfo {series} {Theoretical and Computational Chemistry}, Vol.~\bibinfo
  {volume} {19},\ \bibinfo {editor} {edited by\ \bibinfo {editor}
  {\bibfnamefont {A.}~\bibnamefont {Toro-Labbé}}}\ (\bibinfo  {publisher}
  {Elsevier},\ \bibinfo {year} {2007})\ pp.\ \bibinfo {pages}
  {57--85}\BibitemShut {NoStop}%
\bibitem [{\citenamefont {Murphy}\ \emph {et~al.}(2011)\citenamefont {Murphy},
  \citenamefont {Wortis},\ and\ \citenamefont {Atkinson}}]{MWA11}%
  \BibitemOpen
  \bibfield  {author} {\bibinfo {author} {\bibfnamefont {N.~C.}\ \bibnamefont
  {Murphy}}, \bibinfo {author} {\bibfnamefont {R.}~\bibnamefont {Wortis}},\
  and\ \bibinfo {author} {\bibfnamefont {W.~A.}\ \bibnamefont {Atkinson}},\
  }\bibfield  {title} {\bibinfo {title} {Generalized inverse participation
  ratio as a possible measure of localization for interacting systems},\ }\href
  {https://doi.org/10.1103/PhysRevB.83.184206} {\bibfield  {journal} {\bibinfo
  {journal} {Phys. Rev. B}\ }\textbf {\bibinfo {volume} {83}},\ \bibinfo
  {pages} {184206} (\bibinfo {year} {2011})}\BibitemShut {NoStop}%
\bibitem [{\citenamefont {Gawne}\ \emph {et~al.}(2020)\citenamefont {Gawne},
  \citenamefont {Hollebon}, \citenamefont {Perez-Callejo}, \citenamefont
  {Humphries}, \citenamefont {Wark},\ and\ \citenamefont {Vinko}}]{GHPHWV20}%
  \BibitemOpen
  \bibfield  {author} {\bibinfo {author} {\bibfnamefont {T.}~\bibnamefont
  {Gawne}}, \bibinfo {author} {\bibfnamefont {P.}~\bibnamefont {Hollebon}},
  \bibinfo {author} {\bibfnamefont {G.}~\bibnamefont {Perez-Callejo}}, \bibinfo
  {author} {\bibfnamefont {O.}~\bibnamefont {Humphries}}, \bibinfo {author}
  {\bibfnamefont {J.}~\bibnamefont {Wark}},\ and\ \bibinfo {author}
  {\bibfnamefont {S.}~\bibnamefont {Vinko}},\ }\bibfield  {title} {\bibinfo
  {title} {Investigating mechanisms of state (de) localisation in highly
  ionized, dense plasmas},\ }\href
  {https://meetings.aps.org/Meeting/DPP20/Session/VO05.3} {\bibfield  {journal}
  {\bibinfo  {journal} {Bulletin of the American Physical Society}\ }\textbf
  {\bibinfo {volume} {65}} (\bibinfo {year} {2020})}\BibitemShut {NoStop}%
\bibitem [{\citenamefont {Bethkenhagen}\ \emph {et~al.}(2020)\citenamefont
  {Bethkenhagen}, \citenamefont {Witte}, \citenamefont {Sch\"orner},
  \citenamefont {R\"opke}, \citenamefont {D\"oppner}, \citenamefont {Kraus},
  \citenamefont {Glenzer}, \citenamefont {Sterne},\ and\ \citenamefont
  {Redmer}}]{BWSRDKGSR20}%
  \BibitemOpen
  \bibfield  {author} {\bibinfo {author} {\bibfnamefont {M.}~\bibnamefont
  {Bethkenhagen}}, \bibinfo {author} {\bibfnamefont {B.~B.~L.}\ \bibnamefont
  {Witte}}, \bibinfo {author} {\bibfnamefont {M.}~\bibnamefont {Sch\"orner}},
  \bibinfo {author} {\bibfnamefont {G.}~\bibnamefont {R\"opke}}, \bibinfo
  {author} {\bibfnamefont {T.}~\bibnamefont {D\"oppner}}, \bibinfo {author}
  {\bibfnamefont {D.}~\bibnamefont {Kraus}}, \bibinfo {author} {\bibfnamefont
  {S.~H.}\ \bibnamefont {Glenzer}}, \bibinfo {author} {\bibfnamefont {P.~A.}\
  \bibnamefont {Sterne}},\ and\ \bibinfo {author} {\bibfnamefont
  {R.}~\bibnamefont {Redmer}},\ }\bibfield  {title} {\bibinfo {title} {Carbon
  ionization at gigabar pressures: An ab initio perspective on astrophysical
  high-density plasmas},\ }\href
  {https://doi.org/10.1103/PhysRevResearch.2.023260} {\bibfield  {journal}
  {\bibinfo  {journal} {Phys. Rev. Research}\ }\textbf {\bibinfo {volume}
  {2}},\ \bibinfo {pages} {023260} (\bibinfo {year} {2020})}\BibitemShut
  {NoStop}%
\bibitem [{Note7()}]{Note7}%
  \BibitemOpen
  \bibinfo {note} {MUZE adopts a spin-restricted KS formalism, so both spin-up
  and spin-down orbitals share a common KS potential and are
  identical}\BibitemShut {NoStop}%
\bibitem [{\citenamefont {Fromy}\ \emph {et~al.}(1996)\citenamefont {Fromy},
  \citenamefont {Deutsch},\ and\ \citenamefont {Maynard}}]{FDM96}%
  \BibitemOpen
  \bibfield  {author} {\bibinfo {author} {\bibfnamefont {P.}~\bibnamefont
  {Fromy}}, \bibinfo {author} {\bibfnamefont {C.}~\bibnamefont {Deutsch}},\
  and\ \bibinfo {author} {\bibfnamefont {G.}~\bibnamefont {Maynard}},\
  }\bibfield  {title} {\bibinfo {title} {{Thomas–Fermi}‐like and average
  atom models for dense and hot matter},\ }\href
  {https://doi.org/10.1063/1.871806} {\bibfield  {journal} {\bibinfo  {journal}
  {Phys. Plasmas}\ }\textbf {\bibinfo {volume} {3}},\ \bibinfo {pages} {714}
  (\bibinfo {year} {1996})}\BibitemShut {NoStop}%
\bibitem [{\citenamefont {Blenski}\ and\ \citenamefont
  {Ishikawa}(1995)}]{BI95}%
  \BibitemOpen
  \bibfield  {author} {\bibinfo {author} {\bibfnamefont {T.}~\bibnamefont
  {Blenski}}\ and\ \bibinfo {author} {\bibfnamefont {K.}~\bibnamefont
  {Ishikawa}},\ }\bibfield  {title} {\bibinfo {title} {Pressure ionization in
  the spherical ion-cell model of dense plasmas and a pressure formula in the
  relativistic pauli approximation},\ }\href
  {https://doi.org/10.1103/PhysRevE.51.4869} {\bibfield  {journal} {\bibinfo
  {journal} {Phys. Rev. E}\ }\textbf {\bibinfo {volume} {51}},\ \bibinfo
  {pages} {4869} (\bibinfo {year} {1995})}\BibitemShut {NoStop}%
\bibitem [{\citenamefont {Johnson}(2000)}]{J00}%
  \BibitemOpen
  \bibfield  {author} {\bibinfo {author} {\bibfnamefont {W.}~\bibnamefont
  {Johnson}},\ }\href {https://www3.nd.edu/~johnson/Publications/hottf.pdf}
  {\bibinfo {title} {Atoms at finite temperatures}} (\bibinfo {year} {2000}),\
  \bibinfo {note} {[Online; last accessed 13.03.21]}\BibitemShut {NoStop}%
\bibitem [{Note8()}]{Note8}%
  \BibitemOpen
  \bibinfo {note} {\protect \leavevmode {\protect \color {black}Here we denote
  the energy eigenvalues from both methods as $\epsilon _{nl}$. We stress that
  the ORCHID energy levels are still computed with the shifted potential
  $\protect \bar {v}^{\tau ,\sigma }_\protect \textrm {s}$, so in other words
  these are equal to the shifted levels $\epstilde $: we dropped the bar
  notation so as not to imply that MUZE shifts the potential by a constant.
  With either method, the KS potential used to determine the levels is equal to
  zero at the boundary so a direct comparison is appropriate.}}\BibitemShut
  {Stop}%
\bibitem [{\citenamefont {Gidopoulos}\ and\ \citenamefont
  {Gross}(2014)}]{GG14}%
  \BibitemOpen
  \bibfield  {author} {\bibinfo {author} {\bibfnamefont {N.~I.}\ \bibnamefont
  {Gidopoulos}}\ and\ \bibinfo {author} {\bibfnamefont {E.~K.~U.}\ \bibnamefont
  {Gross}},\ }\bibfield  {title} {\bibinfo {title} {Electronic non-adiabatic
  states: towards a density functional theory beyond the born--oppenheimer
  approximation},\ }\href {https://doi.org/10.1098/rsta.2013.0059} {\bibfield
  {journal} {\bibinfo  {journal} {Philos. Trans. R. Soc.}\ }\textbf {\bibinfo
  {volume} {372}},\ \bibinfo {pages} {20130059} (\bibinfo {year}
  {2014})}\BibitemShut {NoStop}%
\bibitem [{\citenamefont {Li}\ \emph {et~al.}(2018)\citenamefont {Li},
  \citenamefont {Requist},\ and\ \citenamefont {Gross}}]{CRG18}%
  \BibitemOpen
  \bibfield  {author} {\bibinfo {author} {\bibfnamefont {C.}~\bibnamefont
  {Li}}, \bibinfo {author} {\bibfnamefont {R.}~\bibnamefont {Requist}},\ and\
  \bibinfo {author} {\bibfnamefont {E.~K.~U.}\ \bibnamefont {Gross}},\
  }\bibfield  {title} {\bibinfo {title} {Density functional theory of electron
  transfer beyond the born-oppenheimer approximation: Case study of lif},\
  }\href {https://doi.org/10.1063/1.5011663} {\bibfield  {journal} {\bibinfo
  {journal} {J. Chem. Phys.}\ }\textbf {\bibinfo {volume} {148}},\ \bibinfo
  {pages} {084110} (\bibinfo {year} {2018})}\BibitemShut {NoStop}%
\bibitem [{\citenamefont {Perdew}\ \emph {et~al.}(2017)\citenamefont {Perdew},
  \citenamefont {Yang}, \citenamefont {Burke}, \citenamefont {Yang},
  \citenamefont {Gross}, \citenamefont {Scheffler}, \citenamefont {Scuseria},
  \citenamefont {Henderson}, \citenamefont {Zhang}, \citenamefont {Ruzsinszky},
  \citenamefont {Peng}, \citenamefont {Sun}, \citenamefont {Trushin},\ and\
  \citenamefont {G{\"o}rling}}]{Perdew2801}%
  \BibitemOpen
  \bibfield  {author} {\bibinfo {author} {\bibfnamefont {J.~P.}\ \bibnamefont
  {Perdew}}, \bibinfo {author} {\bibfnamefont {W.}~\bibnamefont {Yang}},
  \bibinfo {author} {\bibfnamefont {K.}~\bibnamefont {Burke}}, \bibinfo
  {author} {\bibfnamefont {Z.}~\bibnamefont {Yang}}, \bibinfo {author}
  {\bibfnamefont {E.~K.~U.}\ \bibnamefont {Gross}}, \bibinfo {author}
  {\bibfnamefont {M.}~\bibnamefont {Scheffler}}, \bibinfo {author}
  {\bibfnamefont {G.~E.}\ \bibnamefont {Scuseria}}, \bibinfo {author}
  {\bibfnamefont {T.~M.}\ \bibnamefont {Henderson}}, \bibinfo {author}
  {\bibfnamefont {I.~Y.}\ \bibnamefont {Zhang}}, \bibinfo {author}
  {\bibfnamefont {A.}~\bibnamefont {Ruzsinszky}}, \bibinfo {author}
  {\bibfnamefont {H.}~\bibnamefont {Peng}}, \bibinfo {author} {\bibfnamefont
  {J.}~\bibnamefont {Sun}}, \bibinfo {author} {\bibfnamefont {E.}~\bibnamefont
  {Trushin}},\ and\ \bibinfo {author} {\bibfnamefont {A.}~\bibnamefont
  {G{\"o}rling}},\ }\bibfield  {title} {\bibinfo {title} {Understanding band
  gaps of solids in generalized kohn{\textendash}sham theory},\ }\href
  {https://doi.org/10.1073/pnas.1621352114} {\bibfield  {journal} {\bibinfo
  {journal} {Proc. Natl. Acad. Sci.}\ }\textbf {\bibinfo {volume} {114}},\
  \bibinfo {pages} {2801} (\bibinfo {year} {2017})}\BibitemShut {NoStop}%
\bibitem [{\citenamefont {Callow}(2022)}]{paper_data}%
  \BibitemOpen
  \bibfield  {author} {\bibinfo {author} {\bibfnamefont {T.}~\bibnamefont
  {Callow}},\ }\bibfield  {title} {\bibinfo {title} {{Data publication for
  "First-principles derivation and properties of density-functional
  average-atom models"}},\ }\href {https://doi.org/10.5281/zenodo.6259746}
  {10.5281/zenodo.6259746} (\bibinfo {year} {2022})\BibitemShut {NoStop}%
\bibitem [{\citenamefont {Nogueira}\ \emph {et~al.}(1999)\citenamefont
  {Nogueira}, \citenamefont {Fiolhais},\ and\ \citenamefont {Perdew}}]{NFP99}%
  \BibitemOpen
  \bibfield  {author} {\bibinfo {author} {\bibfnamefont {F.}~\bibnamefont
  {Nogueira}}, \bibinfo {author} {\bibfnamefont {C.}~\bibnamefont {Fiolhais}},\
  and\ \bibinfo {author} {\bibfnamefont {J.~P.}\ \bibnamefont {Perdew}},\
  }\bibfield  {title} {\bibinfo {title} {Trends in the properties and
  structures of the simple metals from a universal local pseudopotential},\
  }\href {https://doi.org/10.1103/PhysRevB.59.2570} {\bibfield  {journal}
  {\bibinfo  {journal} {Phys. Rev. B}\ }\textbf {\bibinfo {volume} {59}},\
  \bibinfo {pages} {2570} (\bibinfo {year} {1999})}\BibitemShut {NoStop}%
\end{thebibliography}
%

\end{document}